\documentclass[twocolumn]{aastex631}
\usepackage{amsmath}
\usepackage{xfrac}
\usepackage{multirow}

\newcommand{\p}{$^\prime$}

\newcommand{\hi}{H\,{\small I}}

\newcommand{\Lya}{Ly$\alpha$}
\newcommand{\kms}{km~s$^{-1}$}
\newcommand{\ha}{H$\alpha$}
\newcommand{\um}{$\mu$m}
\newcommand{\sfr}{$\Sigma_{\rm{SFR}}$}
\newcommand{\sm}{$\Sigma_{\star}$}
\newcommand{\mhi}{$\Sigma_{\rm{H\,{\small I}}}$}

\newcommand{\sdssr}{{\em r}}
\newcommand{\sdssg}{{\em g}}
\newcommand{\mips}{24$\mu$m}
\newcommand{\wise}{22$\mu$m}
\newcommand{\mstar}{M$_\star$}
\newcommand{\higas}{M$_{HI}$}
\newcommand{\ci}{R$_{90}$/R$_{50}$}
\newcommand{\re}{R$_{eff}$}
\newcommand{\dsfr}{$\Delta_{sSFR}$}
\newcommand{\tdep}{$\tau_{dep}$}

\accepted{October 10, 2023}

\shorttitle{Stellar Disk Growth in Galaxies}
\shortauthors{Padave et al. 2023}
\graphicspath{{./}{figures/}}

\begin{document}

\title{DIISC-III: Signatures of Stellar Disk Growth in Nearby Galaxies}

\author[0000-0002-3472-0490]{Mansi Padave}
\affiliation{School of Earth \& Space Exploration, Arizona State University, Tempe, AZ 85287-1404, USA}

\author[0000-0002-2724-8298]{Sanchayeeta Borthakur}
\affiliation{School of Earth \& Space Exploration, Arizona State University, Tempe, AZ 85287-1404, USA}

\author[0000-0003-1268-5230]{Hansung B. Gim}
\affiliation{Department of Physics, Montana State University, P.O. Box 173840, Bozeman, MT 59717, USA}

\author[0000-0002-8528-7340]{David Thilker}
\affiliation{Department of Physics \& Astronomy, Johns Hopkins University, Baltimore, MD, 21218, USA}

\author[0000-0003-1268-5230]{Rolf A. Jansen}
\affiliation{School of Earth \& Space Exploration, Arizona State University, Tempe, AZ 85287-1404, USA}

\author[0000-0000-0000-0000]{Jacqueline Monckiewicz}
\affiliation{School of Earth \& Space Exploration, Arizona State University, Tempe, AZ 85287-1404, USA}

\author[0000-0001-5448-1821]{Robert C. Kennicutt}
\affiliation{Department of Astronomy and Steward Observatory, University of Arizona,  Tucson, AZ, 85721, USA}
\affiliation{Department of Physics and Astronomy,Texas A\&M University, College Station,TX, 77843,USA}

\author[0000-0000-0000-0000]{Guinevere Kauffmann}
\affiliation{Max-Planck-Institute f$\ddot{u}$r Astrophysik, Karl-Schwarzschild-Stra$\beta$e 1, D-85740 Garching, German}

\author[0000-0003-0724-4115]{Andrew J. Fox}
\affiliation{AURA for ESA, Space Telescope Science Institute, 3700 San Martin Drive, Baltimore, MD, 21218, USA}

\author[0000-0003-3168-5922]{Emmanuel Momjian}
\affiliation{National Radio Astronomy Observatory, 1003 Lopezville Rd, Socorro, NM 87801, USA}

\author[0000-0001-6670-6370]{Timothy Heckman}
\affil{Department of Physics \& Astronomy, Johns Hopkins University, Baltimore, MD, 21218, USA}



\begin{abstract}
We explore the growth of the stellar disks in 14 nearby spiral galaxies as part of the Deciphering the Interplay between the Interstellar medium, Stars, and the Circumgalactic medium (DIISC) survey. We study the radial distribution of specific star formation rates (sSFR) and investigate the ratio of the difference in the outer and inner sSFR (\dsfr$~={\rm sSFR}_{out}-{\rm sSFR}_{in}$) of the disk and the total sSFR, \dsfr/sSFR to quantify disk growth. We find \dsfr/sSFR and the \hi\ gas fraction to show a mild correlation of Spearman's $\rho=0.30$, indicating that star formation and disk growth are likely to proceed outward in galactic disks with high \hi\ gas fractions. 
The \hi\ gas fractions and \dsfr/sSFR of the galaxies also increase with the distance to the nearest L$_\star$ neighbor, suggesting that galaxies are likely to sustain their ISM cold gas and exhibit inside-out growth in isolated environments.  
However, the \hi\ content in their circumgalactic medium, probed by the \Lya\ equivalent width (W$_{Ly\alpha}$) excess, 
is observed to be suppressed in isolated environments, apparent from the strong anti-correlation between the W$_{Ly\alpha}$ excess and the distance to the 5$^{\rm th}$ nearest L$_\star$ neighbor (Spearman's $\rho=-0.62$). As expected, W$_{Ly\alpha}$ is also found to be suppressed in cluster galaxies. 
We find no relation between the W$_{Ly\alpha}$ excess of the detected CGM absorber and \dsfr/sSFR implying that the enhancement and suppression of the circumgalactic \hi\ gas does not affect the direction in which star formation proceeds in a galactic disk or vice-versa.   
\end{abstract}




\section{Introduction} \label{sec:intro}
The growth of galactic stellar disks is a well-studied quantitative tracer of galaxy evolution. 
Disk growth requires the accretion of gas \citep{birn03, kere05} onto the galactic disk, much of which is eventually turned into stars. According to the classical paradigm, cold low-metallicity gas from the intergalactic medium travels through the circumgalactic medium (CGM) of the galaxy and condenses onto the \hi\ disk \citep[][and references therein]{sancisi08}. There, it nurtures and sustains star formation, leading to stellar build-up and disk growth.  

If the gas accretion at later times occurs predominantly at large radii, the galactic disk is then said to grow {\it{inside-out}}, i.e., the outer stellar disk forms at a later stage than the inner disk, causing an increase in scale lengths \citep{lars76, fallefs80, whitefrenk91, mo98, naabo06, som08, dutton11}. Simulations indeed show that higher angular momentum gas is being accreted onto the outer disk at later times and increasing the star formation timescale with galactocentric radius \citep{somla01, sam03, brook06, brook12, rosk08, rosk10, pikling12, vlad18}.

On the observational front, the stellar mass--size relation shows that for the same stellar mass bin, galaxies at high redshift ($z\gtrsim2$) were smaller \citep{ferg04, bard05, buit08, franx08, vand13, rodrp17}. In the local Universe ($z\lesssim0.5$), observations of radial color and metallicity gradients show that young stars are spatially extended compared to the more centrally-concentrated old stellar populations \citep{mc04, muno07, muno11, wang11, pezz15, lian17}. Studies of the star formation history of the Milky Way \citep{frankel19}, and the Local Group \citep[NGC 300, M33, NGC 7793;][]{goga09, will09, sacchi19} also advocate for the inside-out mode of disk growth. Multiple studies of resolved stellar photometry that modeled the star formation history stipulate that younger stellar populations extend farther out into the galactic disk \citep{bell00, brown08, gonz14, dale16, dale20, godd17, smith22}. 

The presence of an extended ultraviolet disk (XUV) in nearby galaxies \citep{thilk07, lemo11}, indicative of relatively recent star formation beyond the optical disk ($\gtrsim$R$_{25}$), provides evidence of the inside-out mode of growth. In \cite{padave21}, we inspected the XUV disk galaxy, NGC 3344, and observed a radial increase in the specific star formation rates from $10^{-10}$~yr$^{-1}$ to $10^{-8}$~yr$^{-1}$, hinting towards an outward growing disk. We also detected a strong absorption component with a velocity offset of 113~km~s$^{-1}$ in the CGM which was interpreted as inflowing gas that may gradually accrete onto the galaxy's \hi\ disk and serve as the fuel for star formation. The approach of probing the CGM alongside the ISM and star formation is crucial to understanding disk growth.

In this study, we investigate the radial distribution of star formation, stellar mass, and \hi\ gas in 14 nearby galaxies from the Deciphering the Interplay between the Interstellar medium, Stars, and the Circumgalactic medium survey \citep[DIISC;][in prep]{borthinprep}. The objective is to connect the growth of the stellar disk to the interstellar medium (ISM), CGM, and local environment utilizing panchromatic data. We have ultraviolet (UV) absorption spectroscopy from the Cosmic Origins Spectrograph \citep[COS;][]{green12} aboard the Hubble Space
Telescope (HST) combined with archival far-UV (FUV) and near-UV (NUV) imaging from Galaxy Evolution Explorer (GALEX), obtained deep optical \sdssg, \sdssr, and \ha\ imaging with the Vatican Advanced Technology Telescope (VATT), and \hi-21 cm imaging using the 
NSF’s Karl G. Jansky Very Large Array \footnote{The National Radio Astronomy Observatory is a facility of the National Science Foundation operated under a \\cooperative agreement by Associated Universities, Inc.} (VLA). Together, these data provide information on the absorbers in the CGM, star formation rates, stellar masses, and cold \hi\ gas in the ISM for the 14 low-z galaxies, enabling us to analyze the nature of stellar disk growth in the context of the cosmic baryon cycle.  

The paper is organized as follows. A detailed description of the sample and the data used in this study is presented in Section \ref{sec:data}, and the adopted methods to derive the star formation rates, stellar and gas mass are discussed in Section \ref{sec:it}. In Section \ref{sec:prof}, we present the surface-brightness profiles and radial profiles of the surface density of stellar mass (\mstar), \hi\ mass (\mhi), and SFR (\sfr). We discuss the connection of stellar disk growth with atomic gas and galaxy environment in Section \ref{sec:discussion}. Lastly, we summarize our results in Section \ref{sec:conc}. 

\section{Sample \& Data} \label{sec:data}

Our sample consists of 14 spiral galaxies selected from the DIISC survey \citep[][in prep]{borthinprep} based on the following criteria: (1) The galaxies lie within 120 Mpc of the Milky Way, (2) have a background ultraviolet (UV) bright Quasi-Stellar Object (QSO) within 3.5 times the size of the \hi\ disk, and (3) observed by GALEX (see Section \ref{sec:uvdata}). The size of the \hi\ disk is defined as the radius deprojected at \mhi = $1$~M$_\odot$~pc$^{-2}$.
Table \ref{tbl:sample} lists the galaxies, their redshift, our adopted distances in Mpc, and scale in pc/\arcsec\ along with other properties (see Section \ref{sec:cos} and \ref{sec:it} for details). 
References for distances are listed in the notes to the table. For galaxies where direct measurements of distances were not available, we estimated the distances using the Cosmicflows-3 Distance–Velocity Calculator \citep{flowmodelcalc2020}.

\begin{deluxetable*}{lcccccccc}[!ht]
\tablecaption{Information on the Sample\label{tbl:sample}}
\tablehead{
 \colhead{Galaxy} & \colhead{z}   & \colhead{Distance}  & \colhead{Scale}  &  \colhead{$\log_{10}$ \mstar}  &  \colhead{$\log_{10}$ SFR} &   \colhead{$\log_{10}$ \higas} &\colhead{$\rho$} & \colhead{$\log_{10}(\sfrac{W}{\overline{W}})_{Ly\alpha}$}   \\
   \colhead{}  & \colhead{} & \colhead{(Mpc)}  & \colhead{(pc/\arcsec)}  &  \colhead{($M_\odot$)}  & \colhead{($M_\odot$~yr$^{-1}$)}  &   \colhead{($M_\odot$)} &   \colhead{(kpc)}  &   \colhead{} \\
\colhead{(1)}  & \colhead{(2)} & \colhead{(3)}  & \colhead{(4)}  & \colhead{(5)} &  \colhead{(6)}  & \colhead{(7)}  &   \colhead{(8)} &   \colhead{(9)}  } 
\startdata
 NGC 99  &  0.01772 & 79.4$^a$ & 378.17  &  10.62 & 0.39  & 10.35 & 163 & 0.107\\ 
 NGC 3344  & 0.00196  & 8.28$^b$  & 40.14  &   10.18 & -0.33 & 9.44 & 31 & -0.248 \\ 
 NGC 3351  & 0.00259 & 9.29$^b$  & 45.04  &  10.90 & -0.10  & 9.00 & 29 & 0.320\\ 
 NGC 3433   & 0.00908 & 44.59  & 216.17   &  10.85  & 0.08 & 10.16 & 128 & -0.207\\ 
NGC 3485  & 0.00477 & 29.43 & 142.68 & 10.45 & -0.08 & 9.75 & 83 & -0.134 \\
  NGC 3666 & 0.00353 & 17.1$^c$ & 82.90 & 10.31 & -0.49 & 9.44 & 56 & -0.086\\ 
NGC 3810 & 0.00331 & 15.3$^c$  & 74.18 &  10.44 & 0.10 & 9.43 & 41 & 0.053\\ 
 NGC 4303  &  0.00522 & 18.7$^d$ & 90.66 &  11.39 & 0.72 & 9.88 & 26 & 0.561\\ 
  NGC 4321   & 0.00525 & 13.93$^e$ & 67.53  &  10.58 & 0.40 & 9.38 & 39 & -0.787\\ 
  IC 3440   & 0.02597 & 120.46 &  584.01 &  10.22 & -0.31 & 9.73 & 83 & 0.140\\ 
 NGC 4921  & 0.01822 & 88.30$^f$ & 360.12  & 11.80 & -0.25 & 9.20 & 53 & -1.206\\ 
  UGC 9120  & 0.01918 & 77.26 & 374.56 & 11.11 & 0.13 & 9.65 & 90 & -0.140\\ 
NGC 5669  & 0.00456 & 14.5$^g$ & 70.30 &  9.52 & -0.54 & 9.23 & 26 & ...\\ 
 NGC 5951 & 0.00593 & 27.1$^g$  & 131.38  &  10.10 & -0.65 & 9.53 & 55 & -0.037
\enddata
\tablecomments
{Column (1): Galaxies in our sample. Column(2)  and Column (3): Redshift and adopted distances in Mpc. Column (4): Scale in pc/\arcsec. Column (5), (6), and (7): Star formation rate, stellar mass, and \hi\ mass for our galaxies described in Section \ref{sec:it}. \higas\ estimates adopted from ALFALFA \citep{alfalfa}. Column (8) and (9): Impact parameters of the QSO sightline and the \Lya\ equivalent width excess described in Section \ref{sec:cos}. $^a$ \cite{sanchez12}; $^b$ \cite{sabb18}; $^c$ \cite{tully16}; $^d$ \cite{m61dist}; $^e$ \cite{m100dist}; $^f$ \cite{ngc4921dist}; $^g$ \cite{sorce14}}
\end{deluxetable*}

\subsection{Observations} \label{sec:data2}
\subsubsection{Ultraviolet Imaging data}\label{sec:uvdata}
Archival FUV and NUV images from the Galaxy Evolution Explorer \citep[GALEX;][]{mart05, morr07} were obtained from the Mikulski Archive for Space Telescopes. GALEX carried out the All-Sky Imaging Survey (AIS) with typical exposures of $\sim100$~s, the Medium Imaging Survey (MIS) and Nearby Galaxy Survey (NGS), and the Guest Investigator (GII) program with typical exposures of $\sim1500$~s each \citep{bianchi03, morr07}. 
The selection of the galaxies was based on both the availability of deep UV imaging and their UV flux.  
The galaxies either have apparent FUV magnitudes $\leq16$ or deep MIS/NGS/GII observations to be able to investigate the flux distribution within their disk. 
IC 3440 and NGC 4921 have deep NGS and GII imaging but FUV magnitudes of 17.5 and 17.0, respectively. All galaxies, except NGC 3810, have MIS/NGS/GII observations in at least one of the bands.

\begin{figure*}
  \centering
   \includegraphics[trim = 13cm 12.5cm 0cm 0cm, clip,scale=0.04]{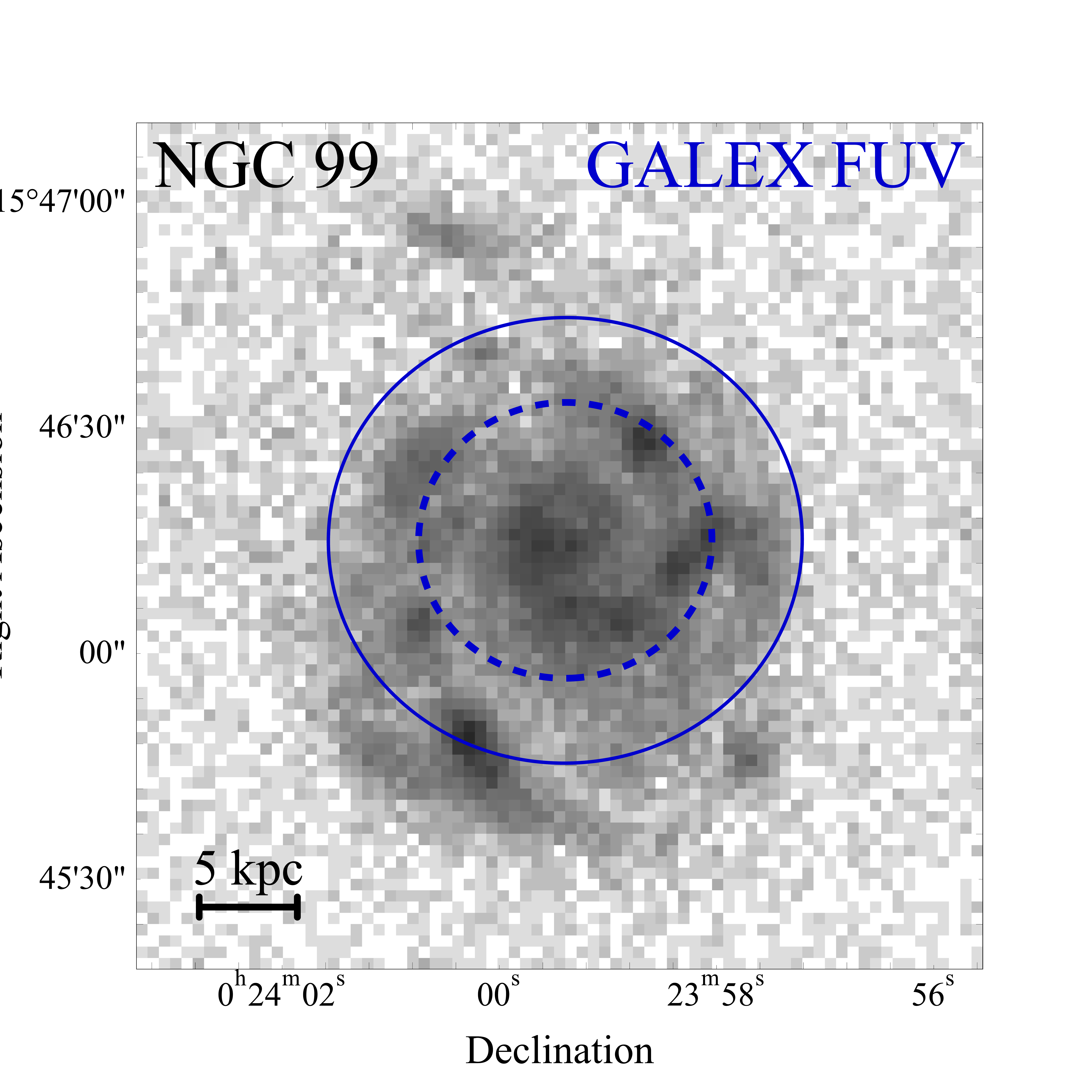}
\includegraphics[trim = 13cm 12.5cm 0cm 0cm, clip,scale=0.04]{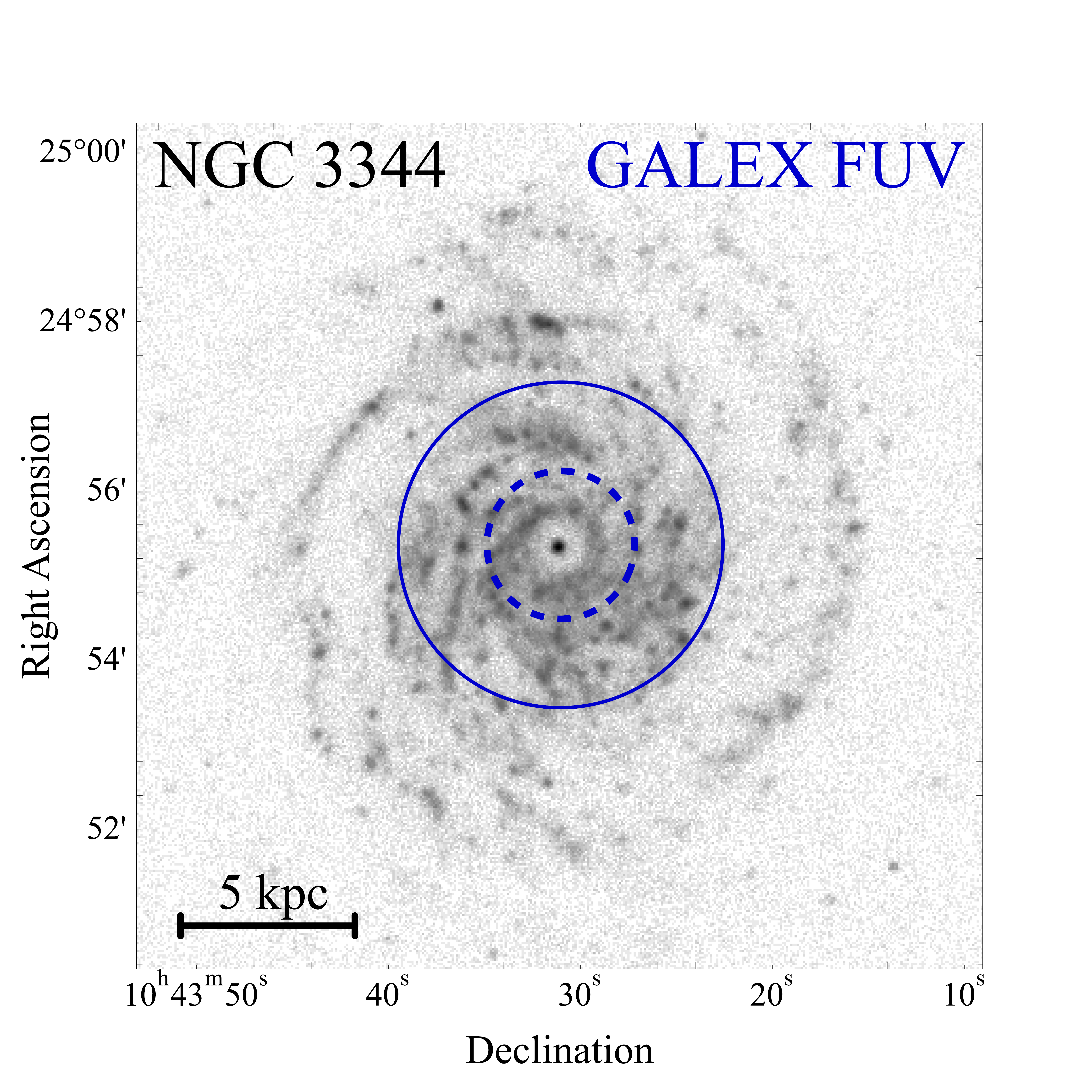}
\includegraphics[trim = 13cm 12.5cm 0cm 0cm, clip,scale=0.04]{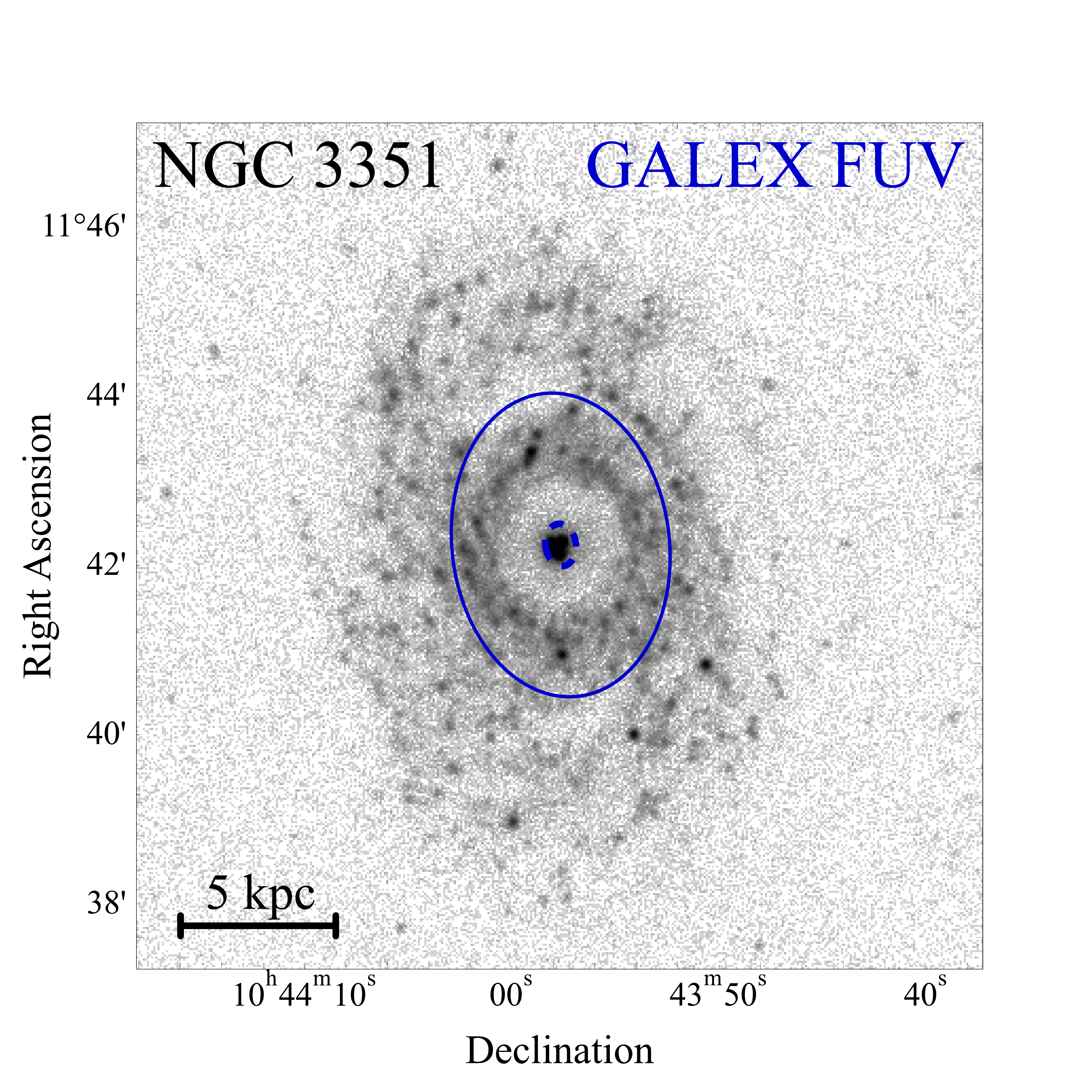}
\includegraphics[trim = 13cm 12.5cm 0cm 0cm, clip,scale=0.04]{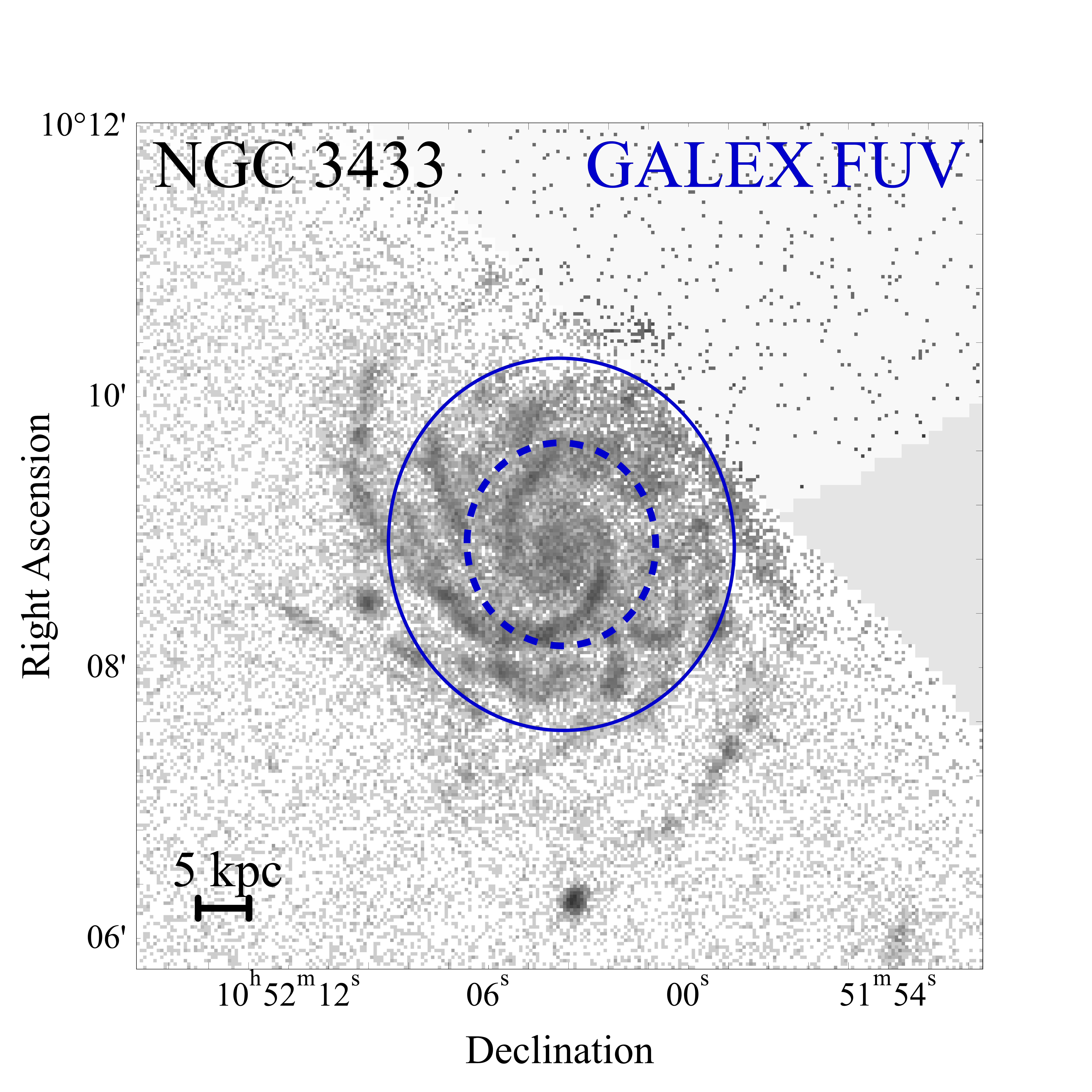}
\includegraphics[trim = 13cm 12.5cm 0cm 0cm, clip,scale=0.04]{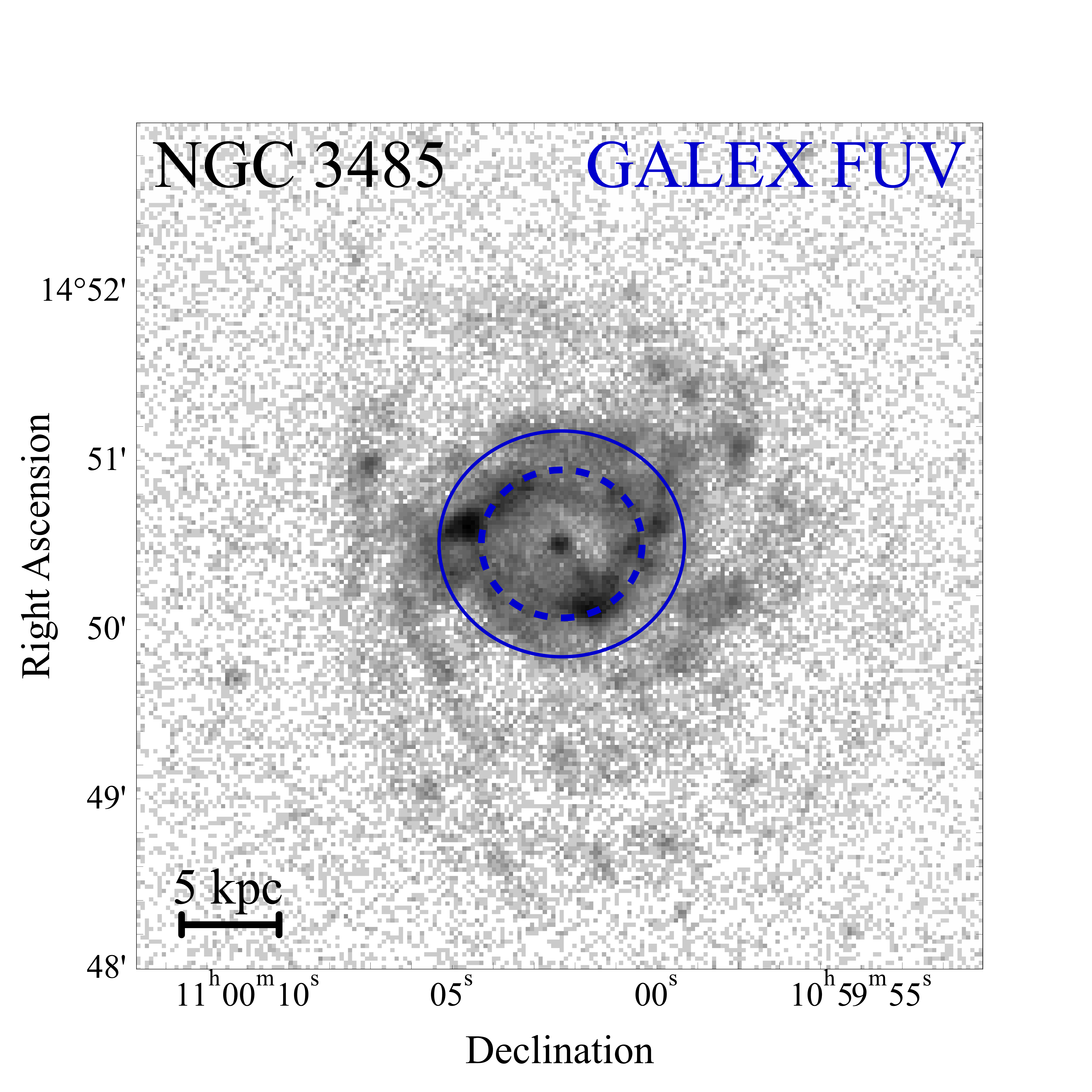}
\includegraphics[trim = 13cm 12.5cm 0cm 0cm, clip,scale=0.04]{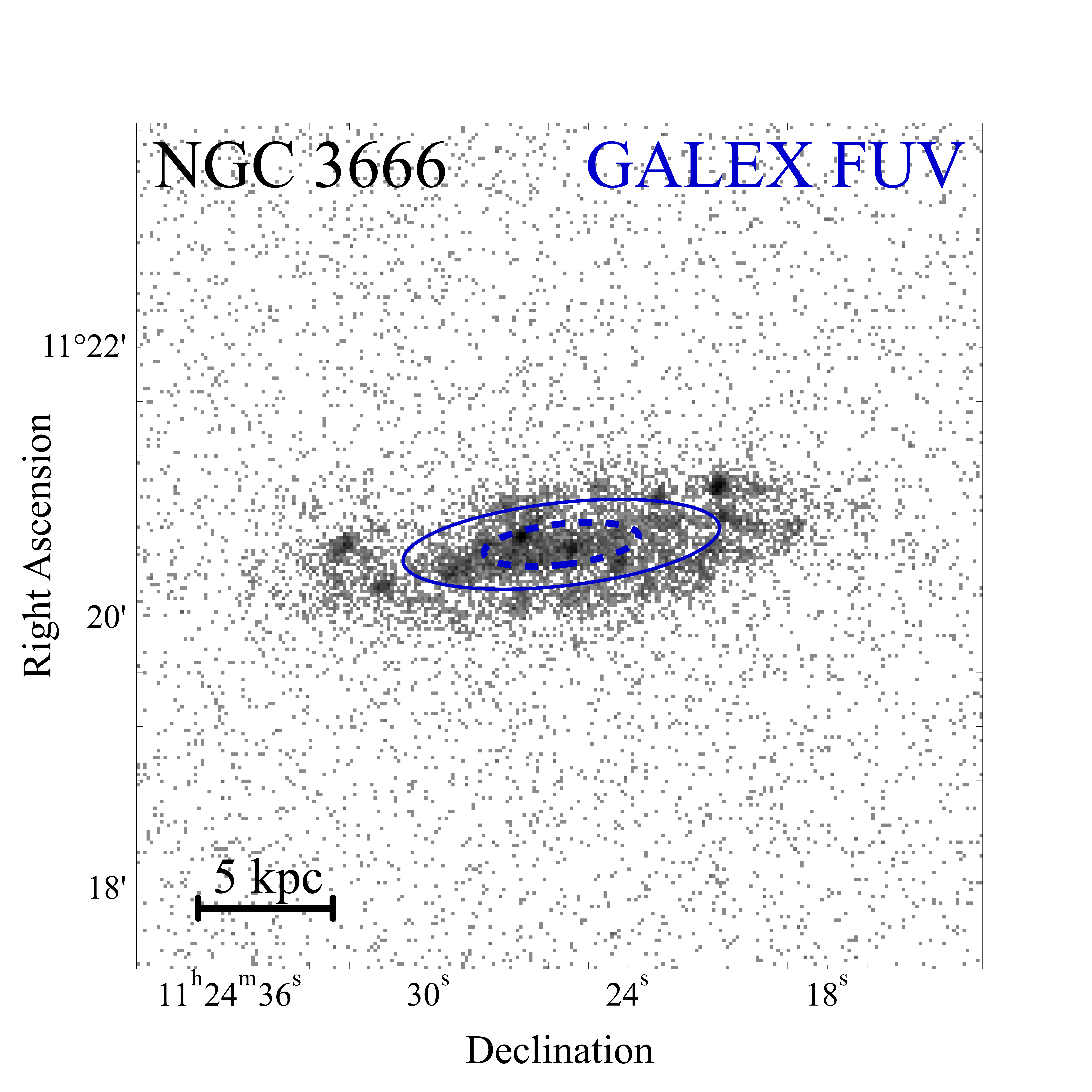}
\includegraphics[trim = 13cm 12.5cm 0cm 0cm, clip,scale=0.04]{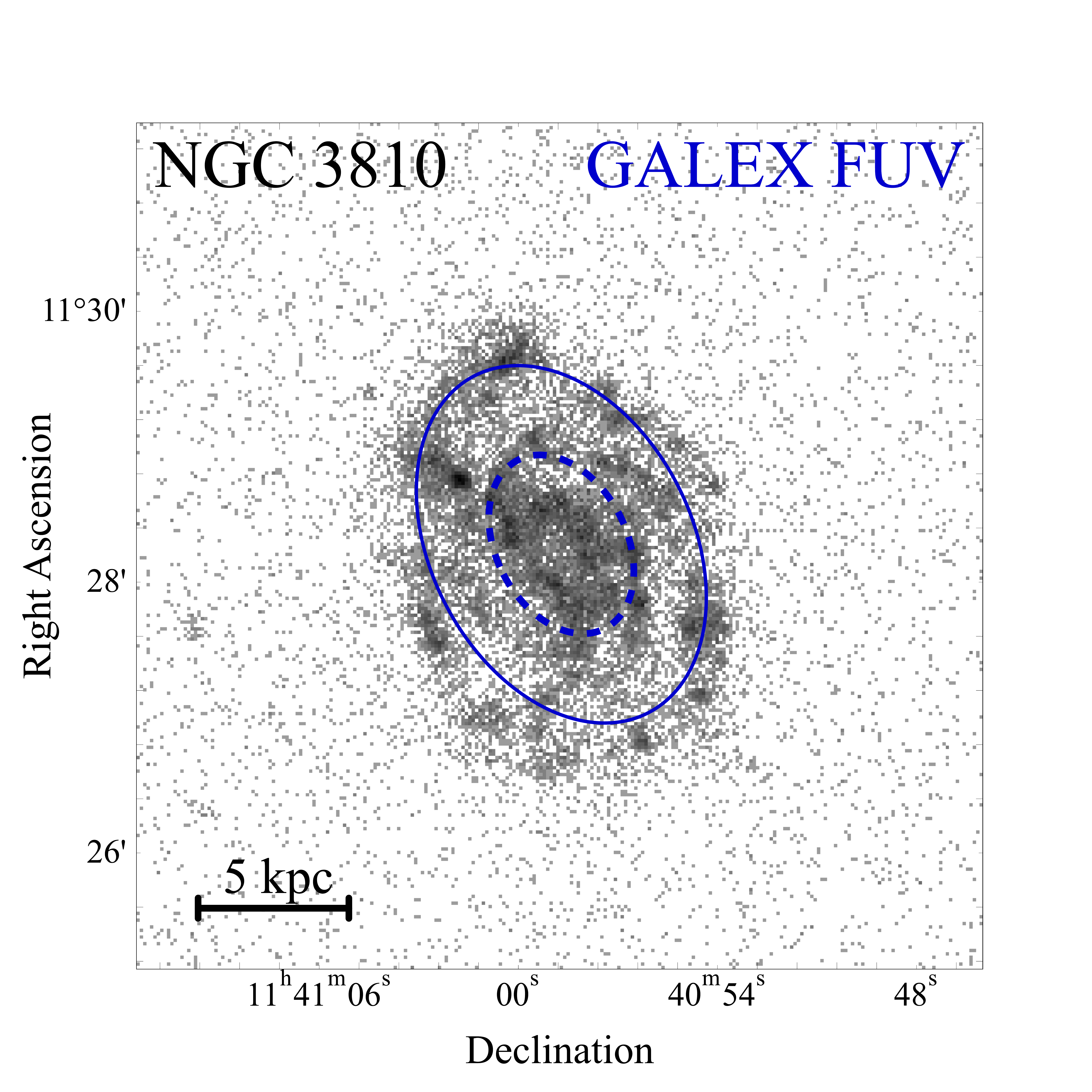}
\includegraphics[trim = 13cm 12.5cm 0cm 0cm, clip,scale=0.04]{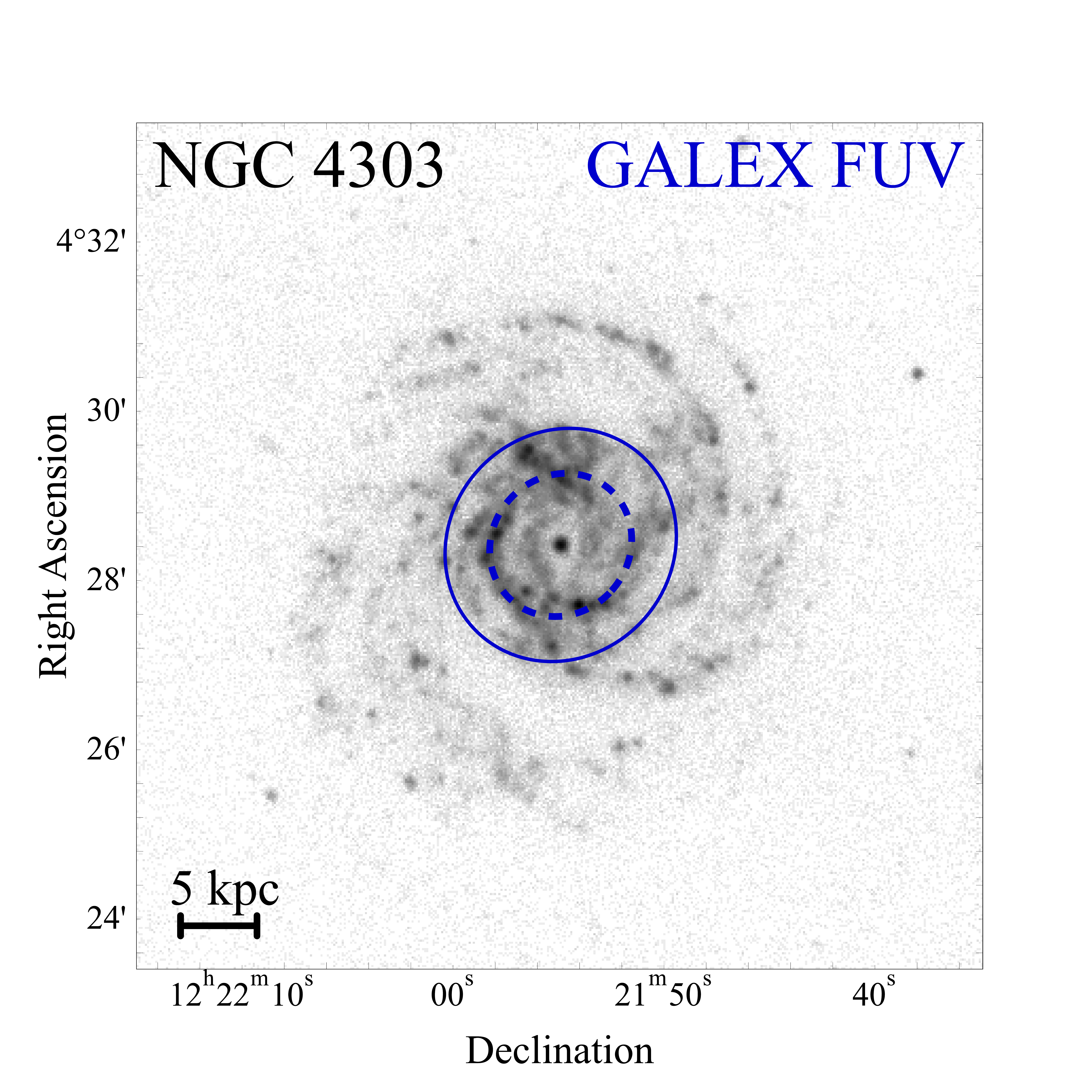}
\includegraphics[trim = 13cm 12.5cm 0cm 0cm, clip,scale=0.04]{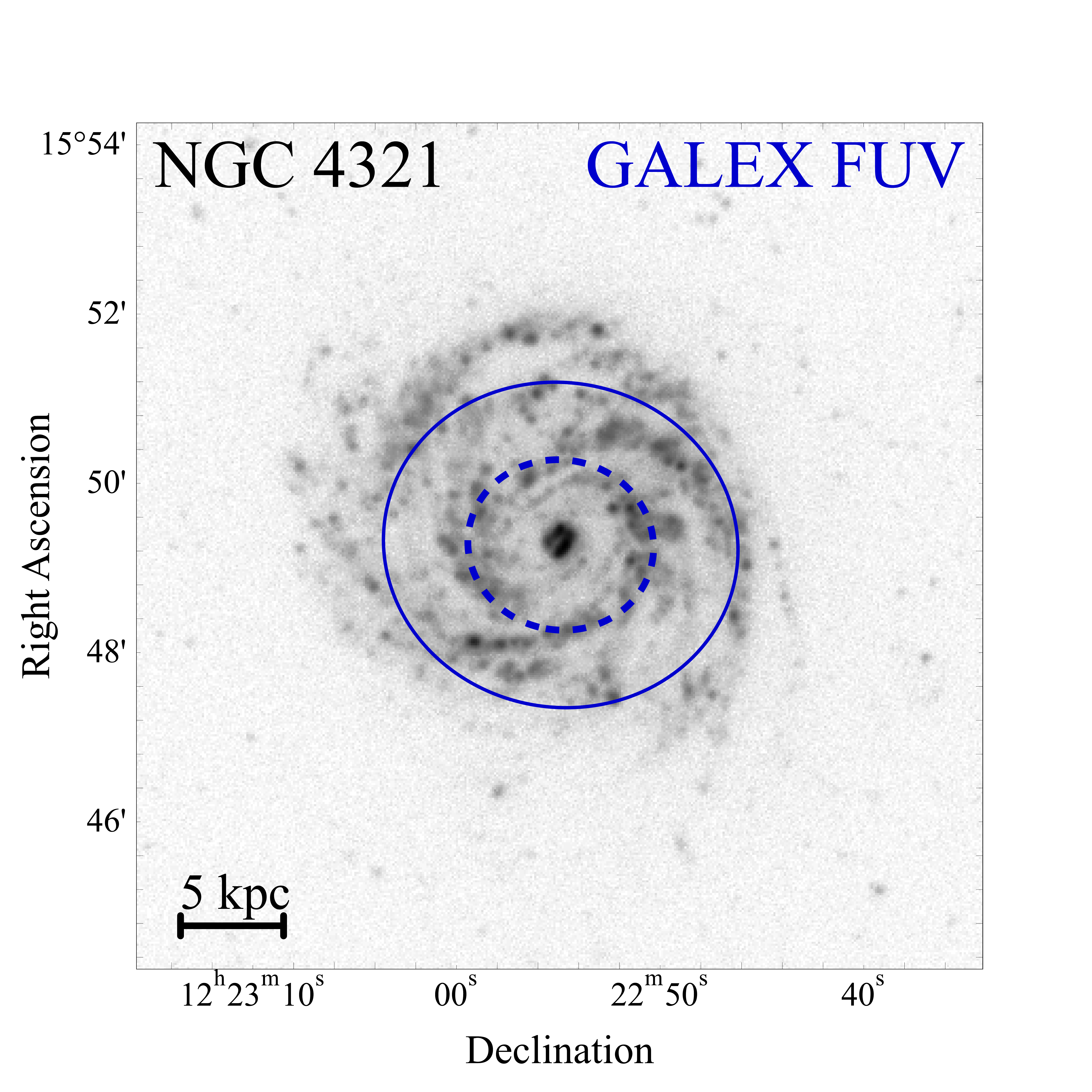}
\includegraphics[trim = 13cm 12.5cm 0cm 0cm, clip,scale=0.04]{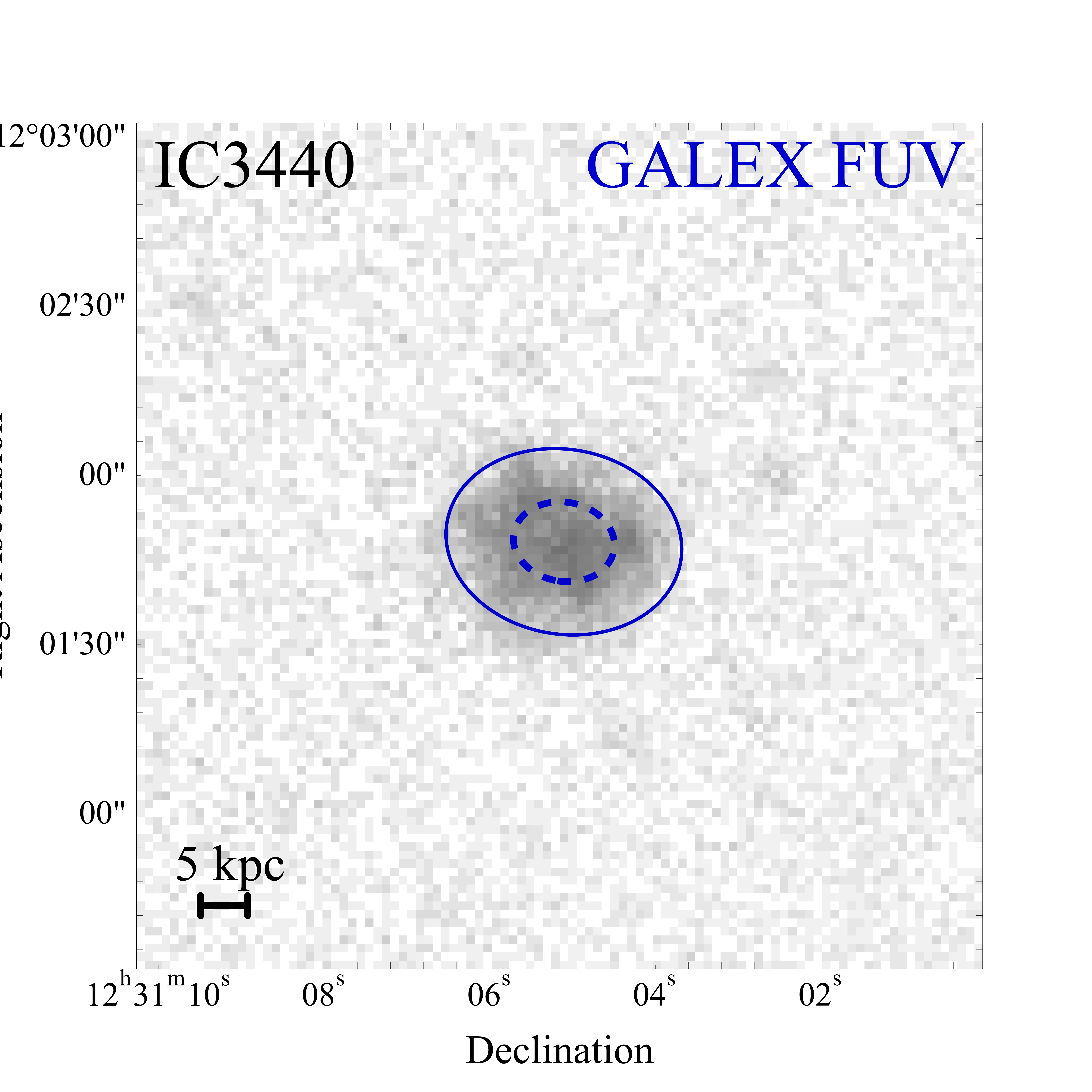}
\includegraphics[trim = 13cm 12.5cm 0cm 0cm, clip,scale=0.04]{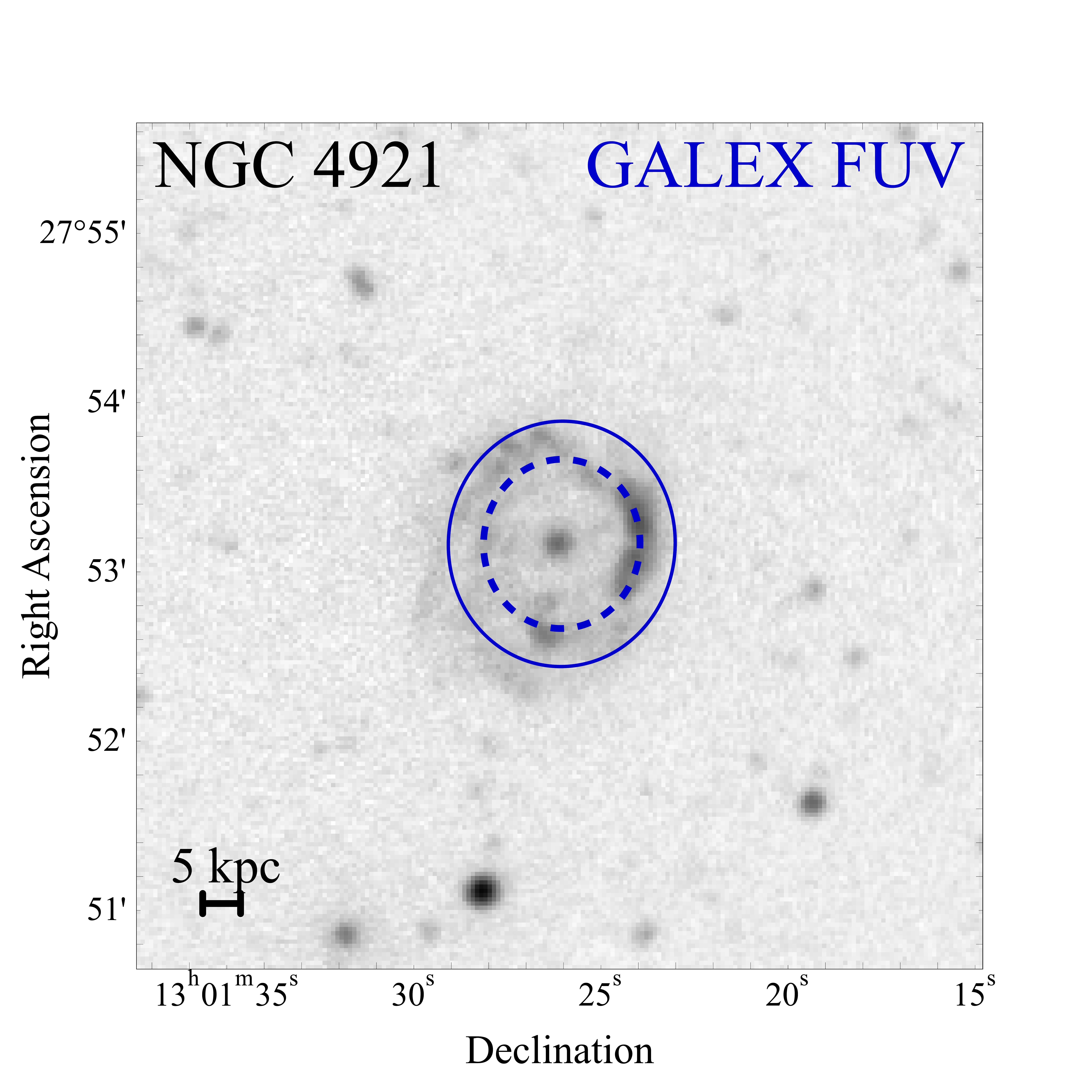}
\includegraphics[trim = 13cm 12.5cm 0cm 0cm, clip,scale=0.04]{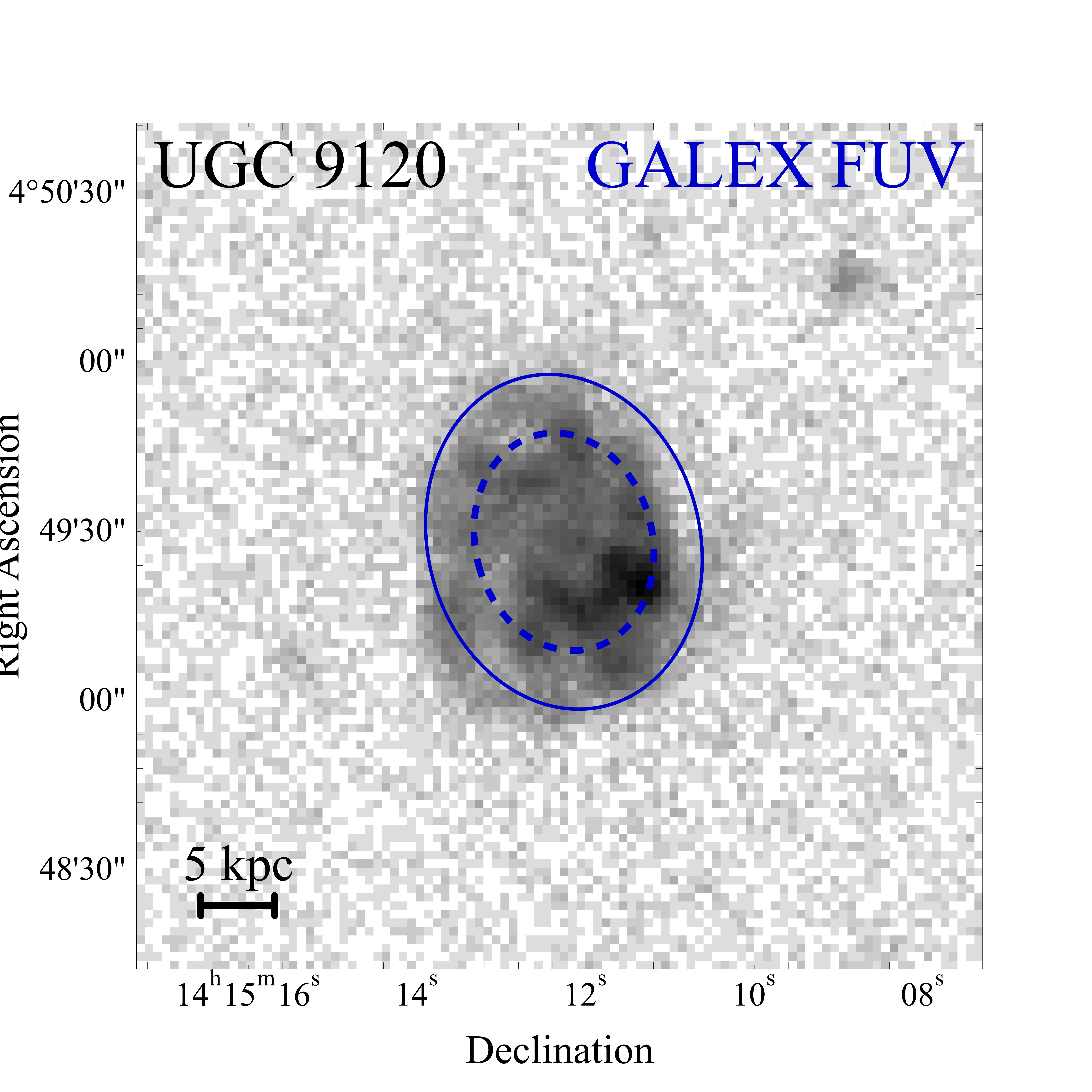}
\includegraphics[trim = 13cm 12.5cm 0cm 0cm, clip,scale=0.04]{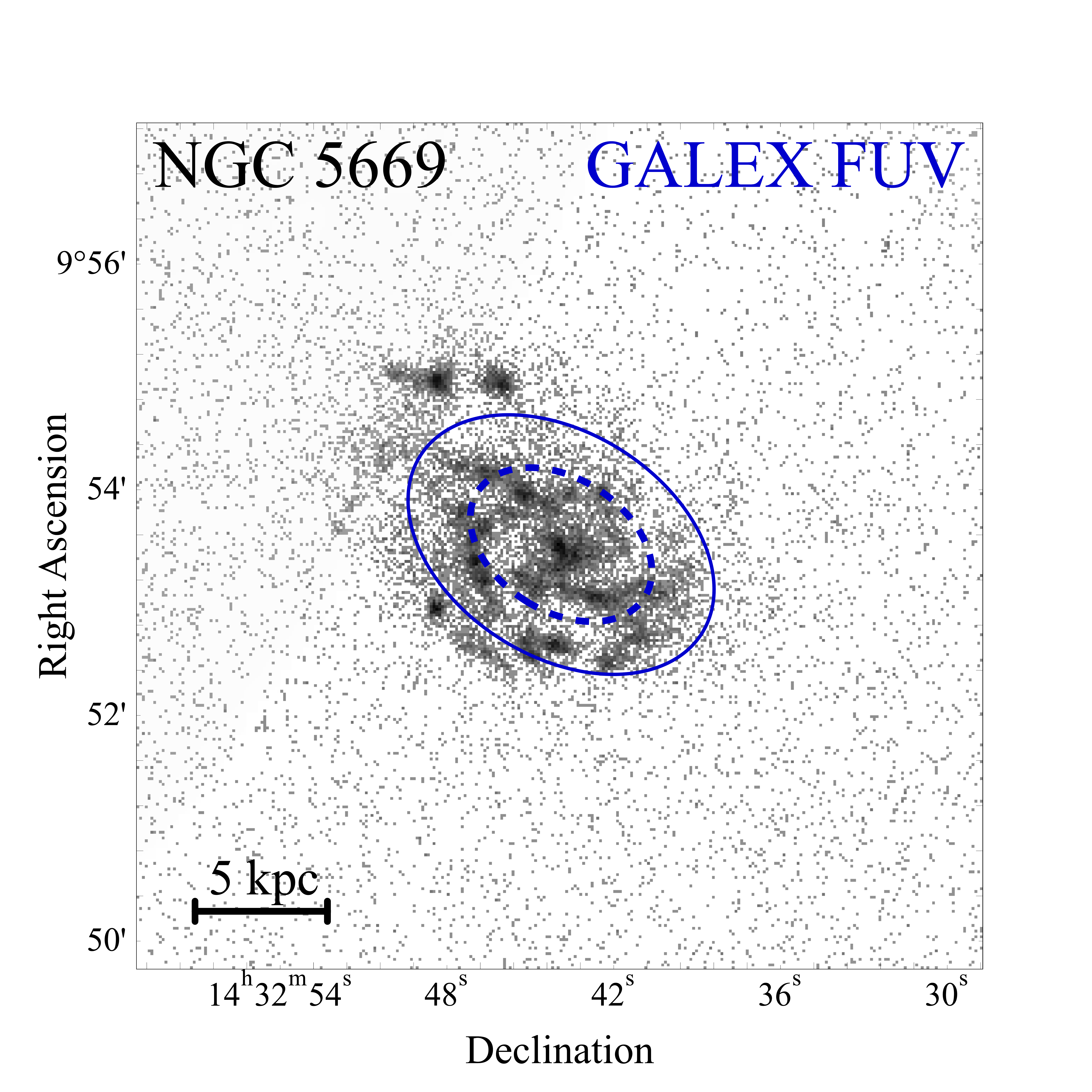}
\includegraphics[trim = 13cm 12.5cm 0cm 0cm, clip,scale=0.04]{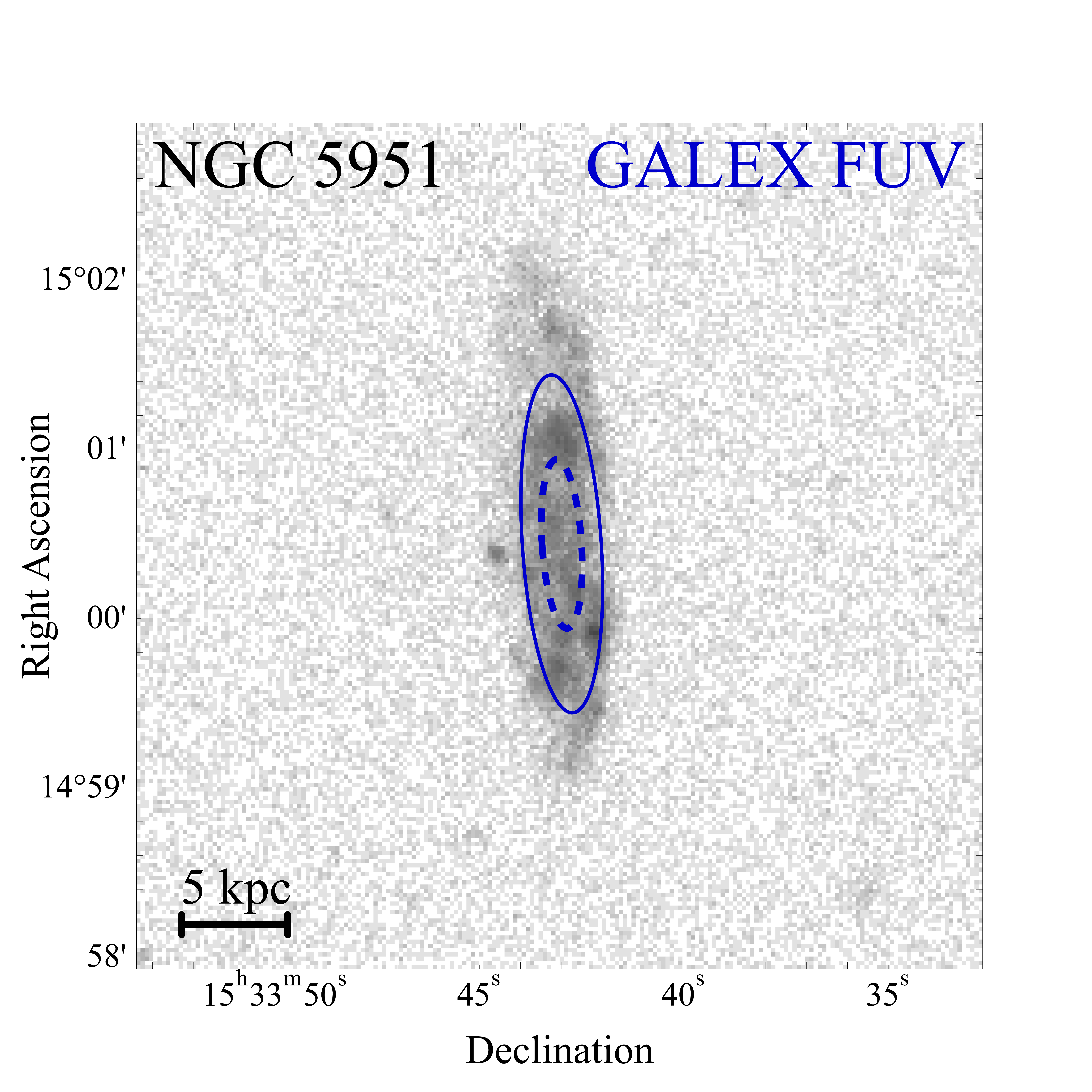}
\caption{GALEX FUV imagery for our galaxies. The blue ellipses mark the apertures containing 50\% (dashed) and 90\% (solid) of the total SFR estimated from the combination of FUV and infrared (\mips\ or \wise) maps (see Section \ref{sec:it} and \ref{sec:radii} for details). A physical scale of 5.0~kpc is shown at the bottom left corner of each map.}  \label{fig:galuv}
\end{figure*}

The FUV and NUV stacks for each galaxy were produced by coadding all available individual datasets mentioned above. Final count maps were created by multiplying the intensity (count/s) map with each galaxy's high-resolution relative response (s) map. The sky background was estimated from source-free regions and removed from each stack. The resulting FUV maps are shown in Figure \ref{fig:galuv}. FUV and NUV flux densities were also corrected for foreground reddening using the E(\bv) values from \cite{schl11} and A$_{FUV}=7.9\times$E(\bv) and A$_{F=NUV}=8.0\times$E(\bv) derived by \cite{card89}.
The 1$\sigma$ noise equivalent surface brightness for the FUV and NUV maps are 27.20~mag\,arcsec$^{-2}$ and 26.84~mag\,arcsec$^{-2}$, respectively. The FUV ($\lambda_{\rm eff}\sim1538.6$\AA) and
NUV ($\lambda_{\rm eff}\sim2315.7$\AA) maps have angular resolutions (FWHM) of 4\farcs2 and 5\farcs3, respectively, and a pixel scale of 1\farcs5.

\subsection{Optical Imaging data}
We imaged each of our sample galaxies in Sloan {\em g} and {\em r} filters (henceforth, \sdssg\ and \sdssr) with the VATT4k CCD imager on the 1.8 m {\em Vatican Advanced Technology Telescope} (VATT) at the Mt. 
Graham Observatory. These observations were carried out along with other observations for the DIISC survey under the H$\alpha$-DIISC program (Padave et al. in prep) and were conducted over a period of 19 nights from 2019 March to 2021 March. The seeing FWHM on these nights was between 0\farcs8 and 1\farcs8. 

\begin{figure*}
  \centering
   \includegraphics[trim = 13cm 12.5cm 0cm 0cm, clip,scale=0.04]{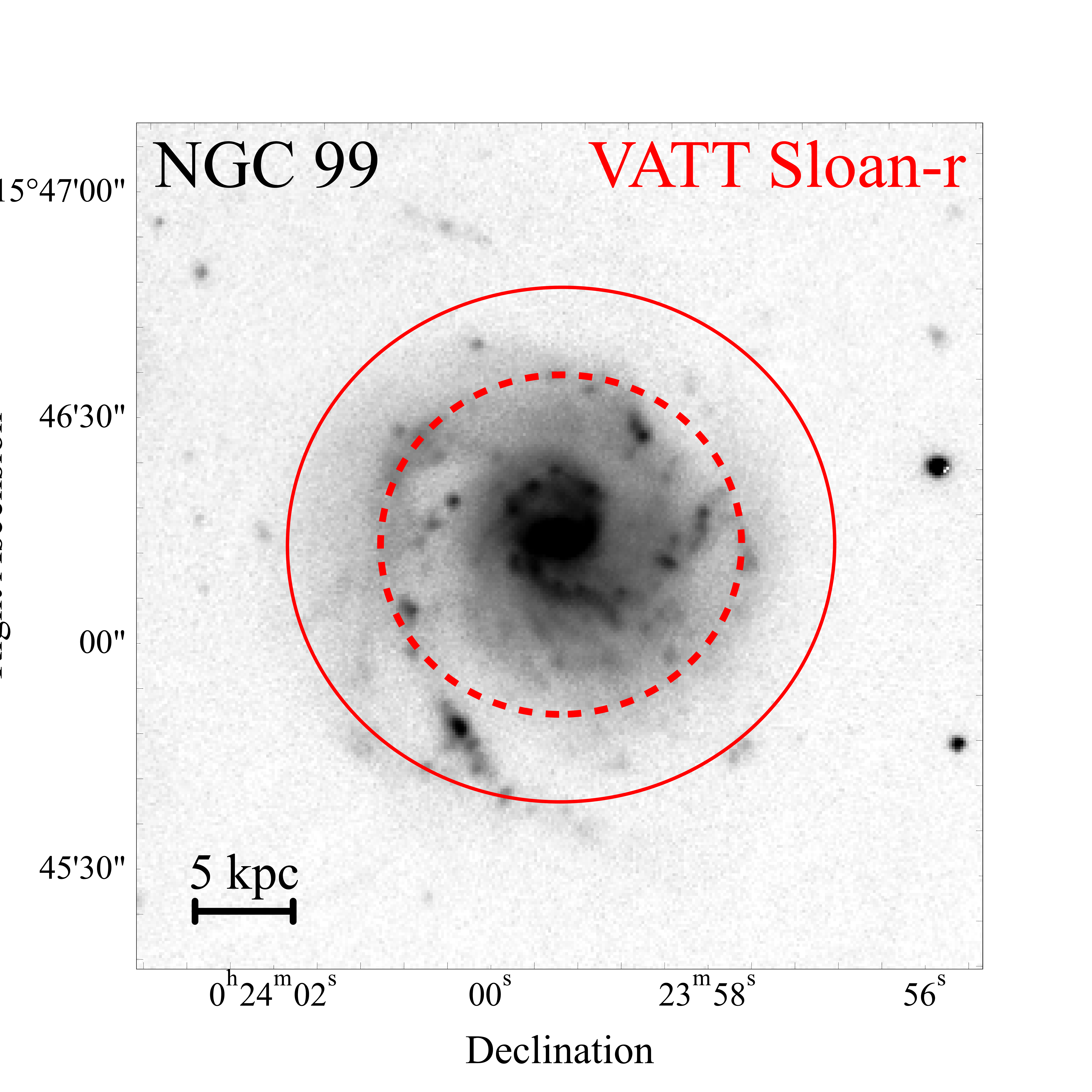}
\includegraphics[trim = 13cm 12.5cm 0cm 0cm, clip,scale=0.04]{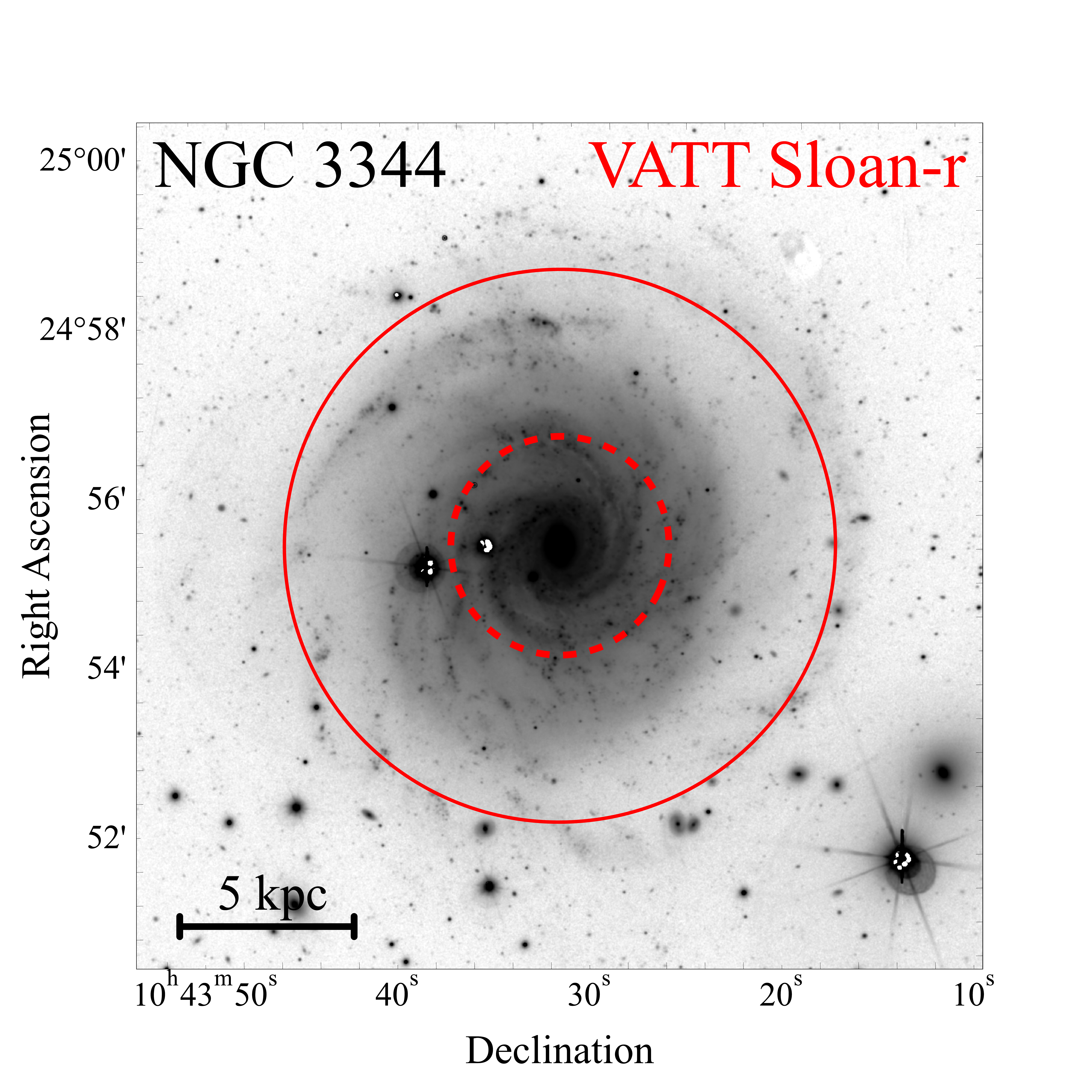}
\includegraphics[trim = 13cm 12.5cm 0cm 0cm, clip,scale=0.04]{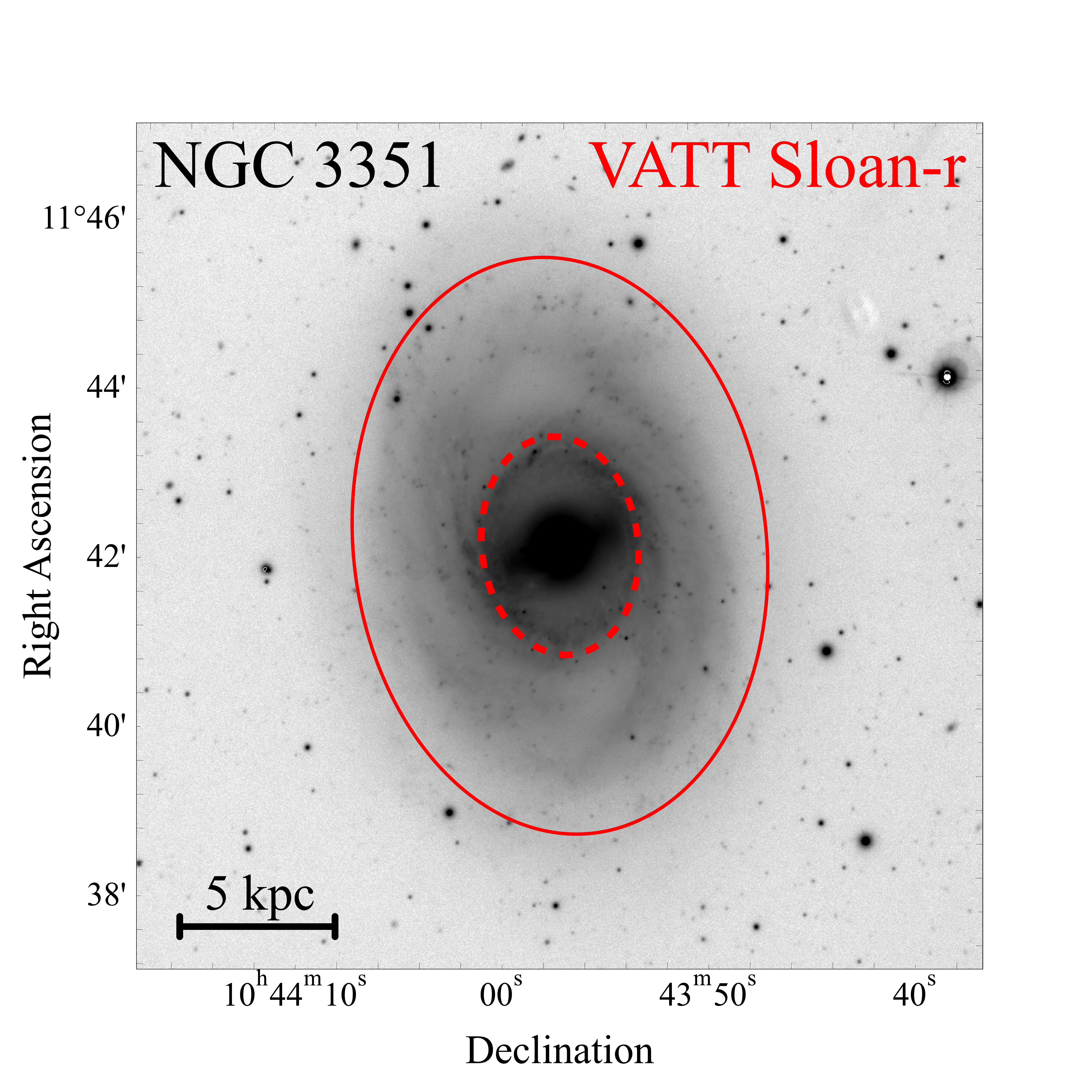}
\includegraphics[trim = 13cm 12.5cm 0cm 0cm, clip,scale=0.04]{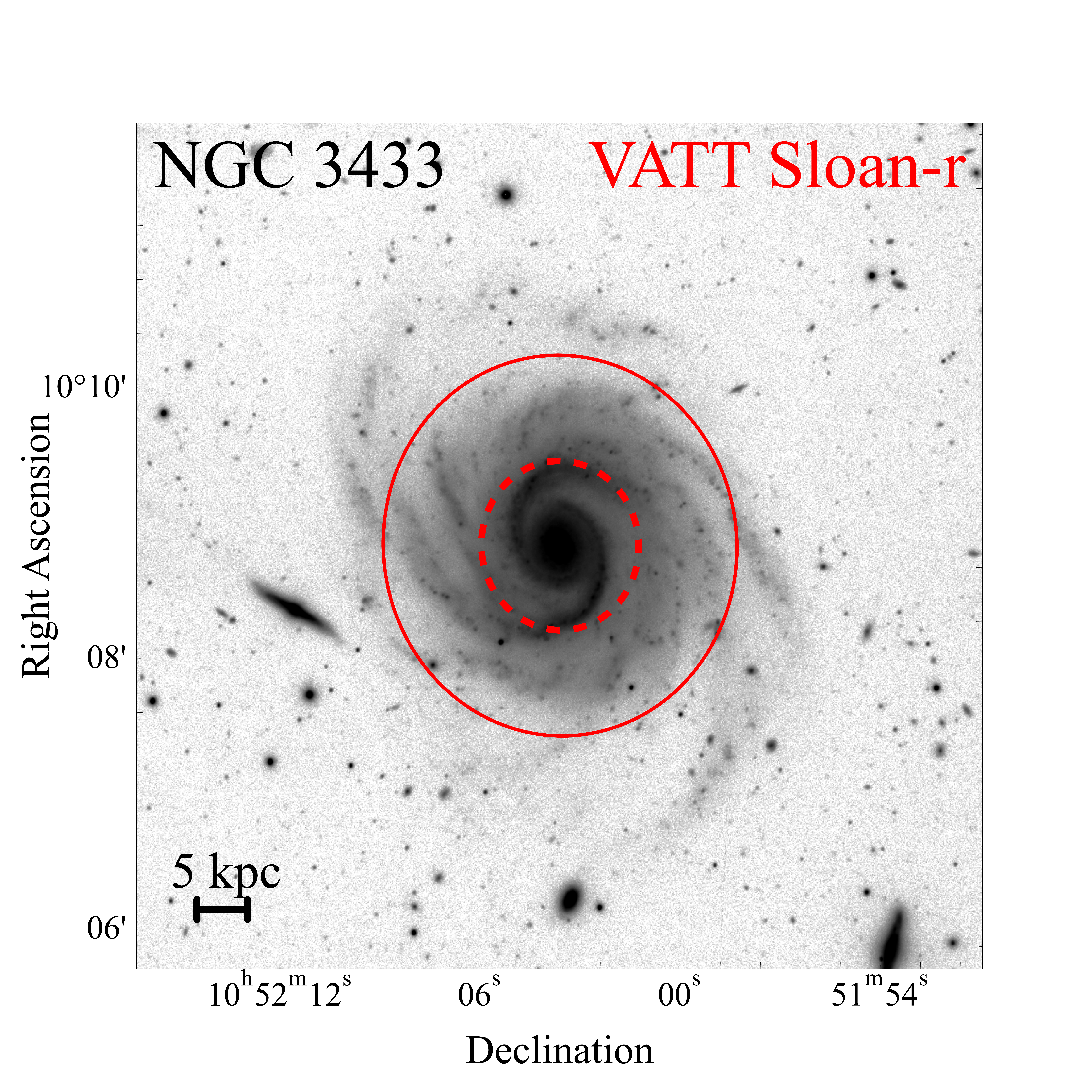}
\includegraphics[trim = 13cm 12.5cm 0cm 0cm, clip,scale=0.04]{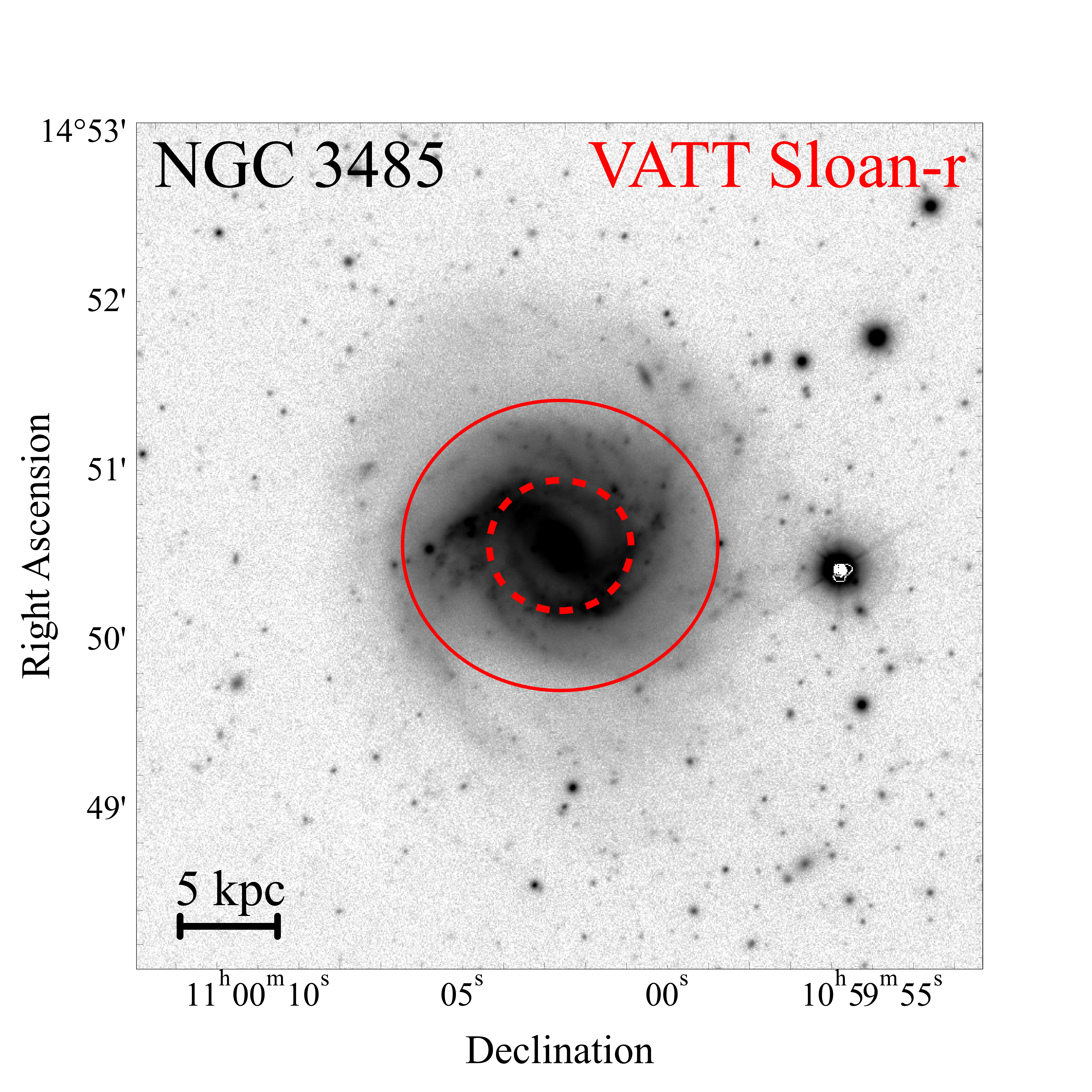}
\includegraphics[trim = 13cm 12.5cm 0cm 0cm, clip,scale=0.04]{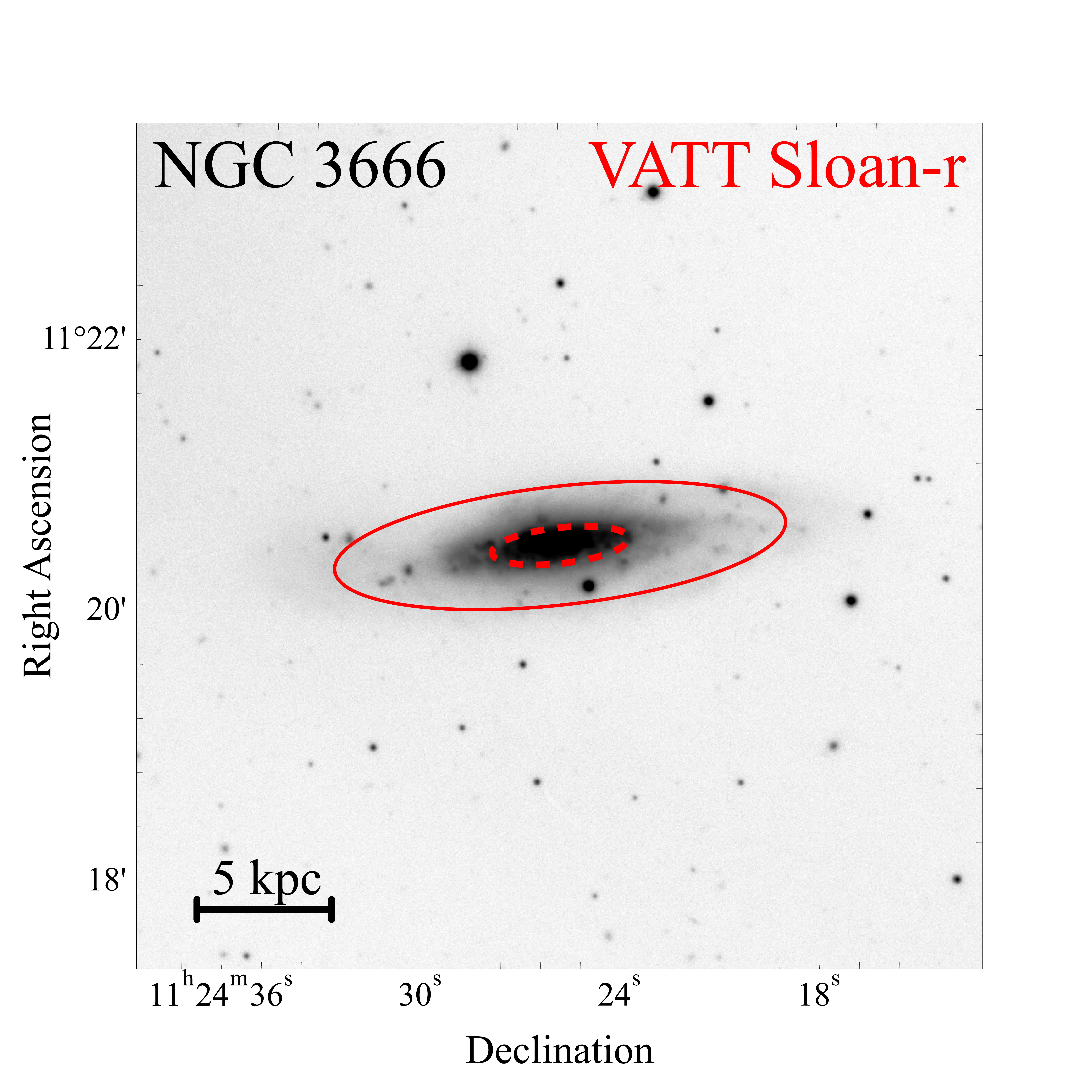}
\includegraphics[trim = 13cm 12.5cm 0cm 0cm, clip,scale=0.04]{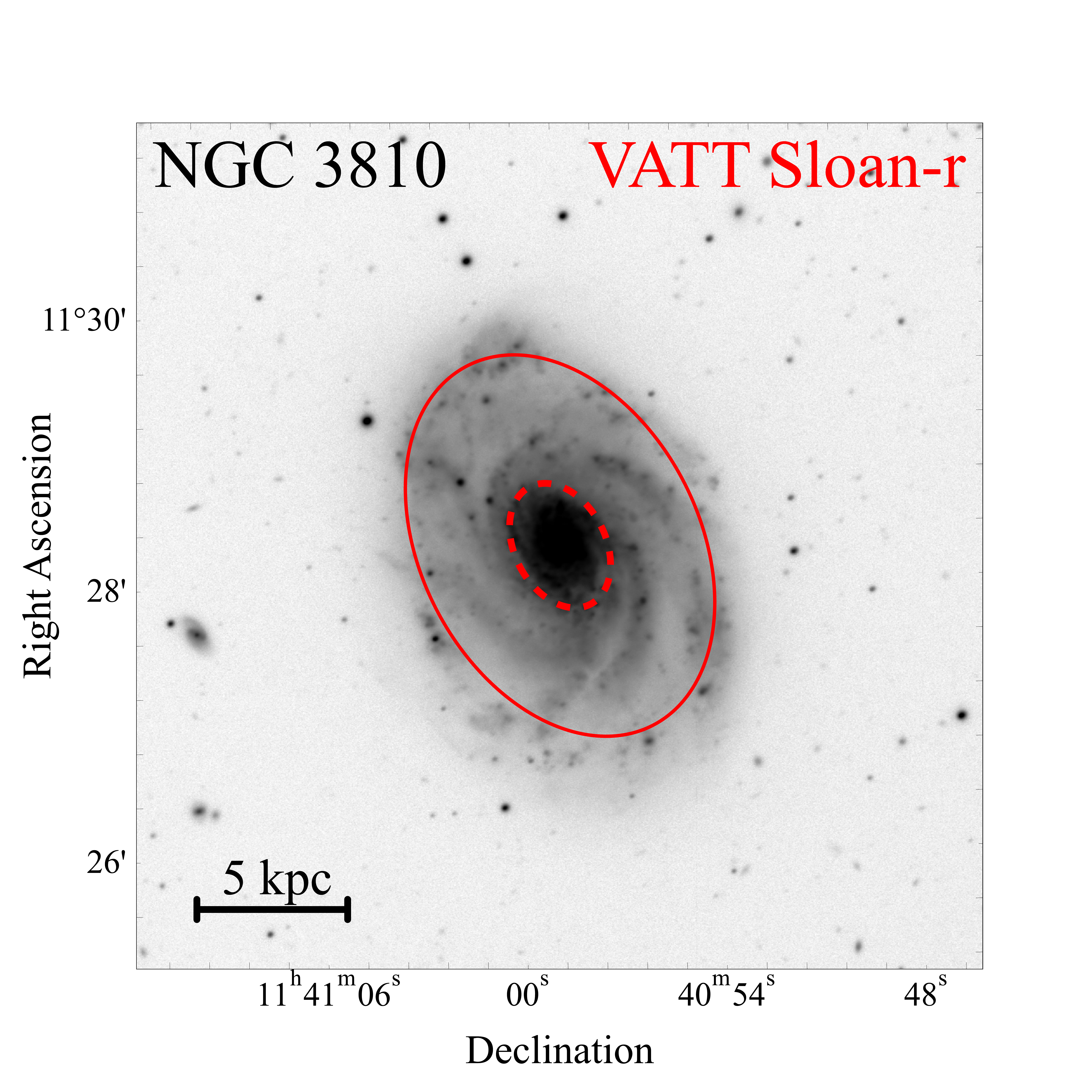}
\includegraphics[trim = 13cm 12.5cm 0cm 0cm, clip,scale=0.04]{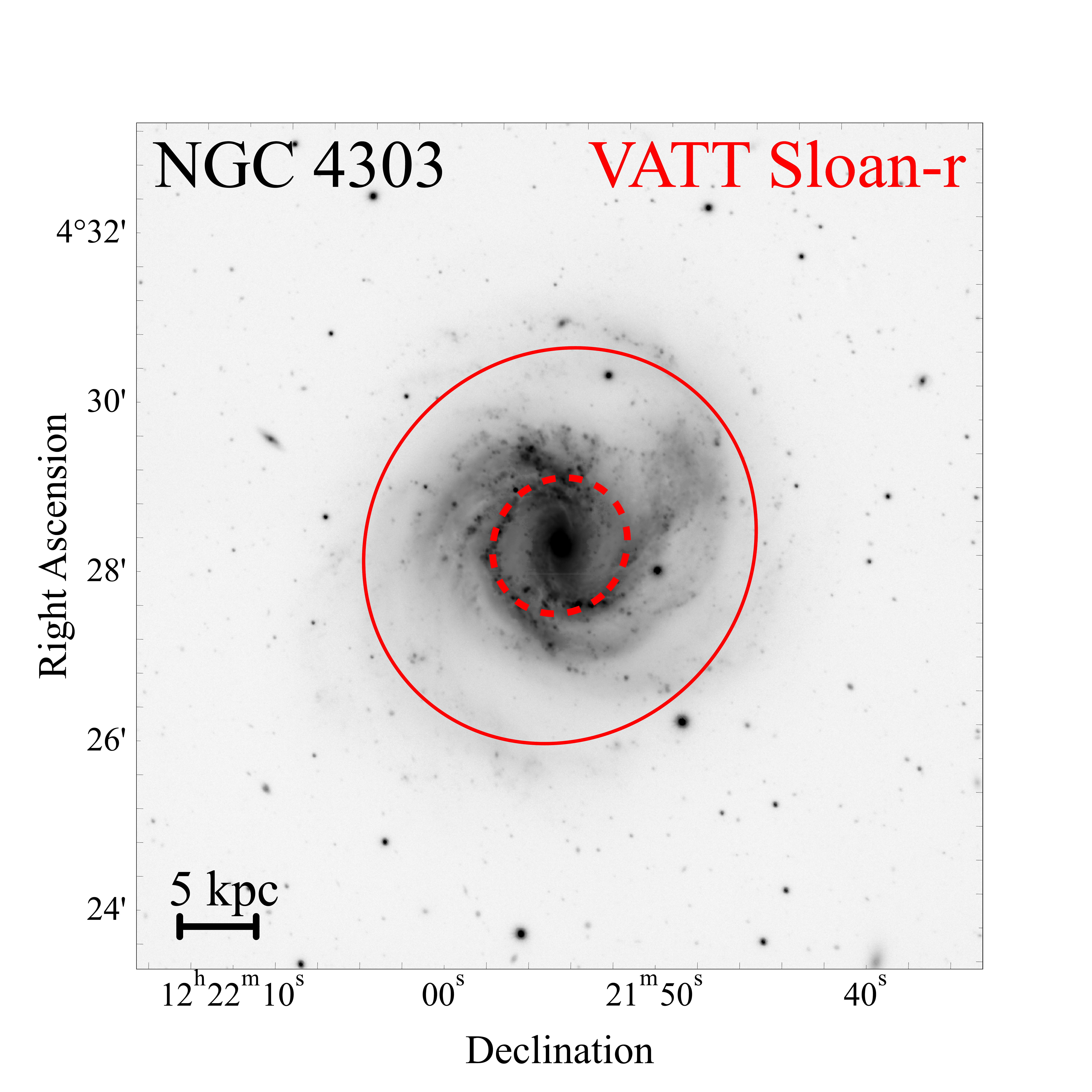}
\includegraphics[trim = 13cm 12.5cm 0cm 0cm, clip,scale=0.04]{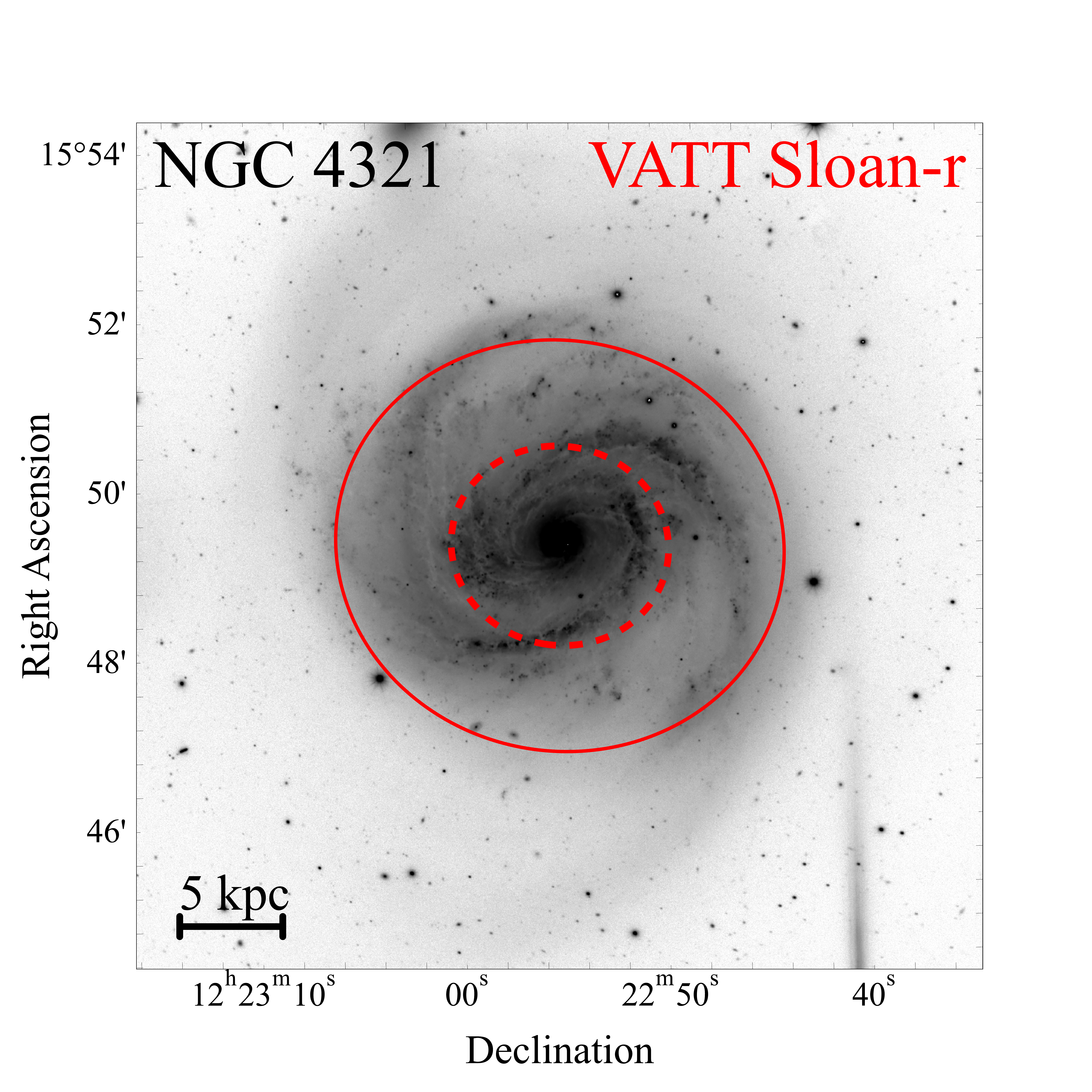}
\includegraphics[trim = 13cm 12.5cm 0cm 0cm, clip,scale=0.04]{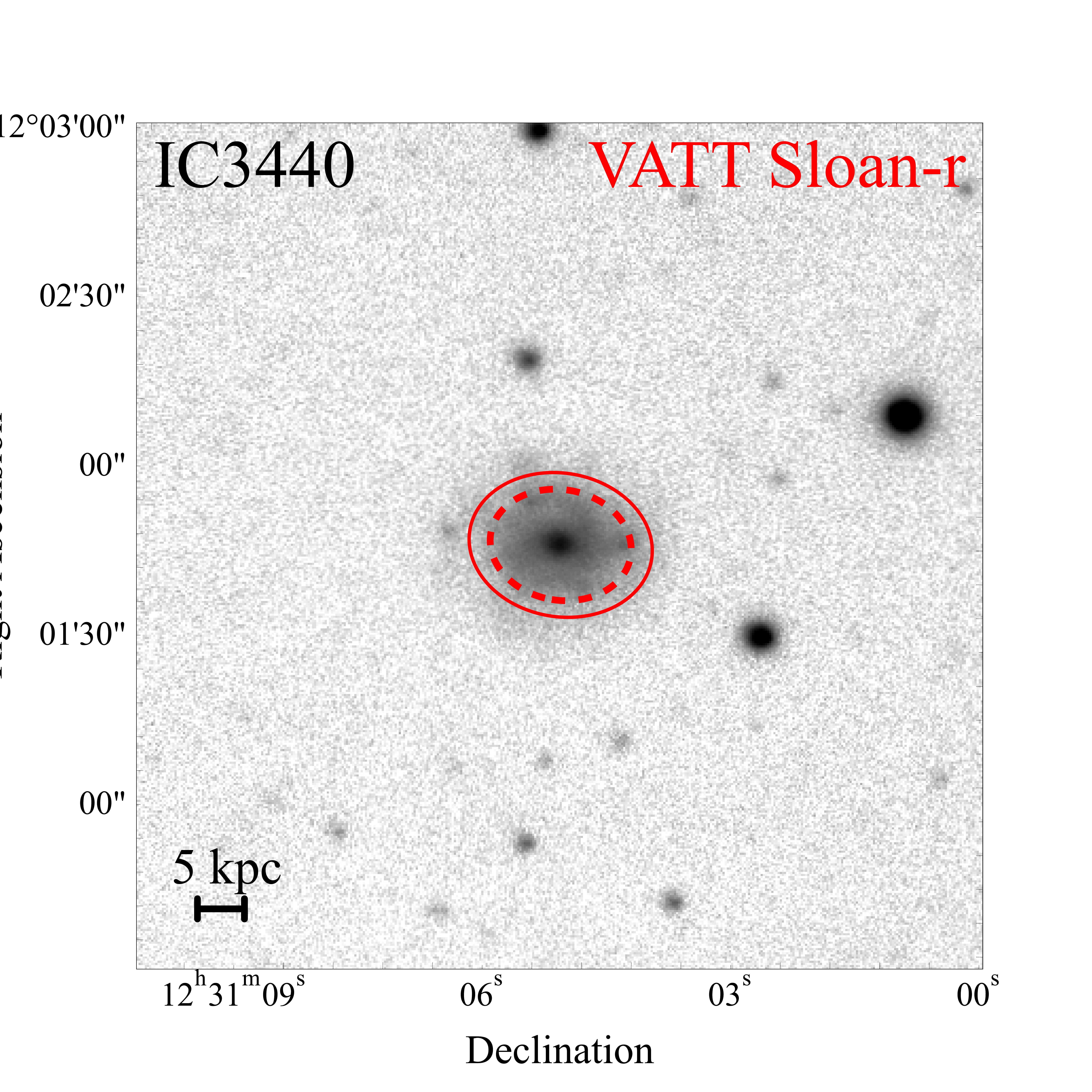}
\includegraphics[trim = 13cm 12.5cm 0cm 0cm, clip,scale=0.04]{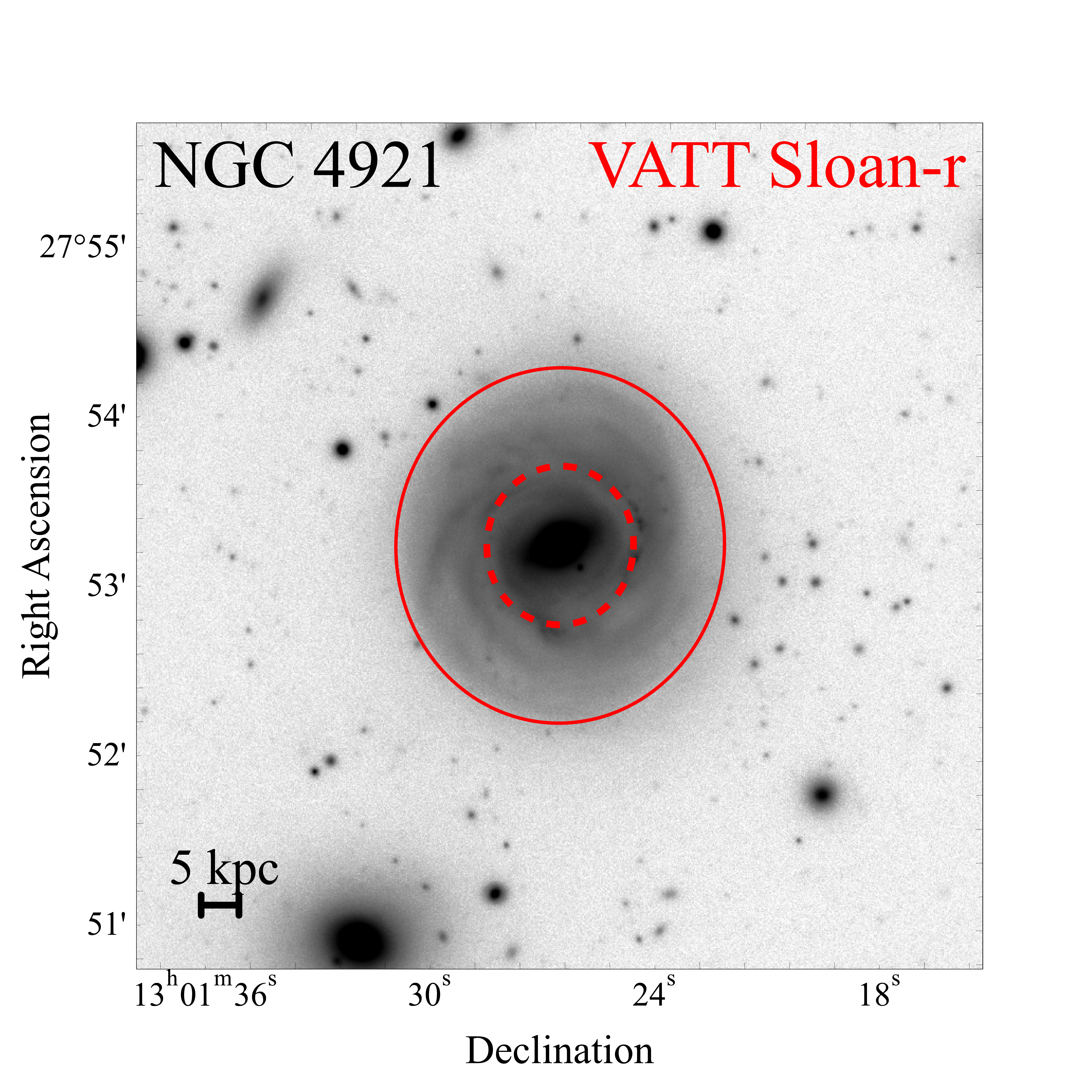}
\includegraphics[trim = 13cm 12.5cm 0cm 0cm, clip,scale=0.04]{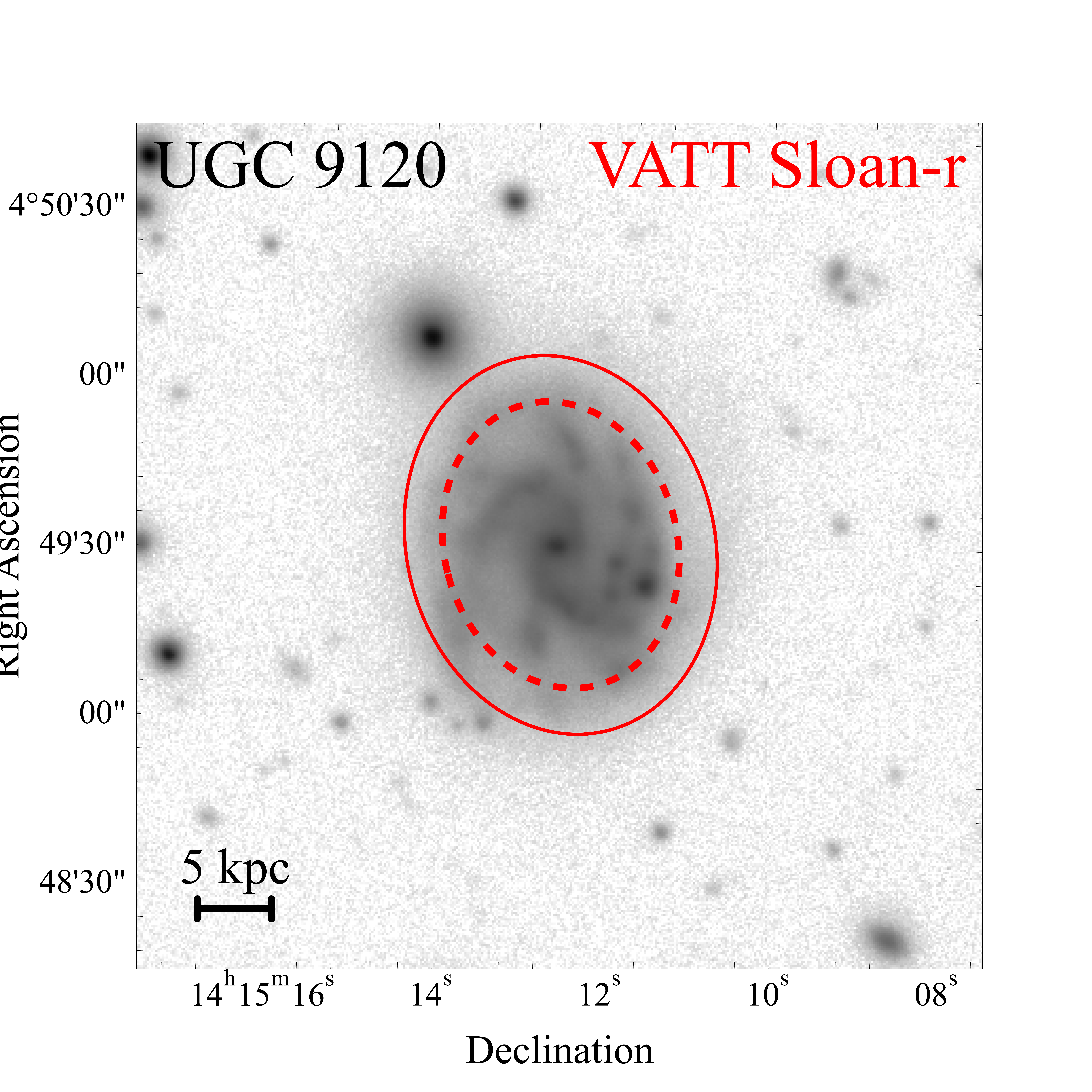}
\includegraphics[trim = 13cm 12.5cm 0cm 0cm, clip,scale=0.04]{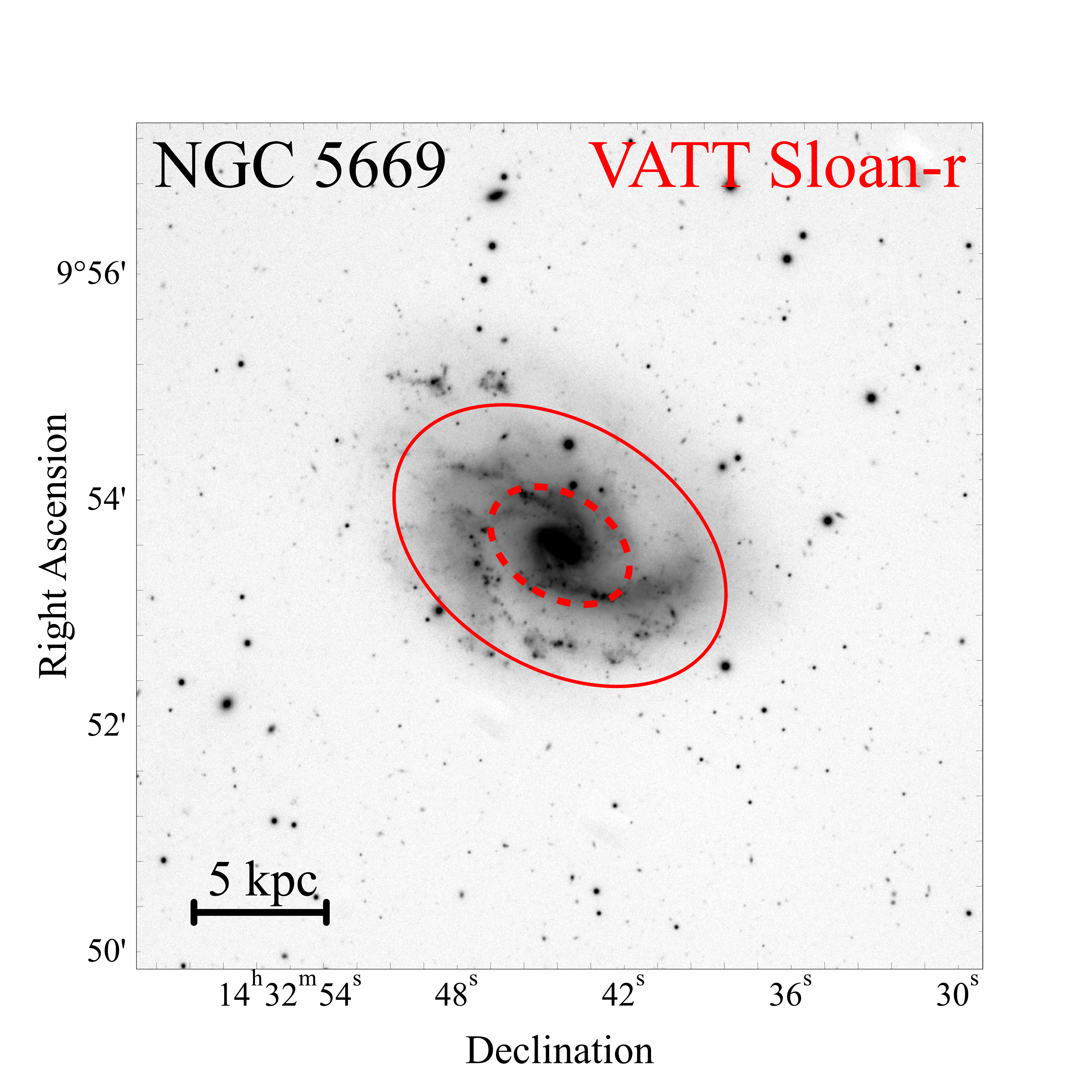}
\includegraphics[trim = 13cm 12.5cm 0cm 0cm, clip,scale=0.04]{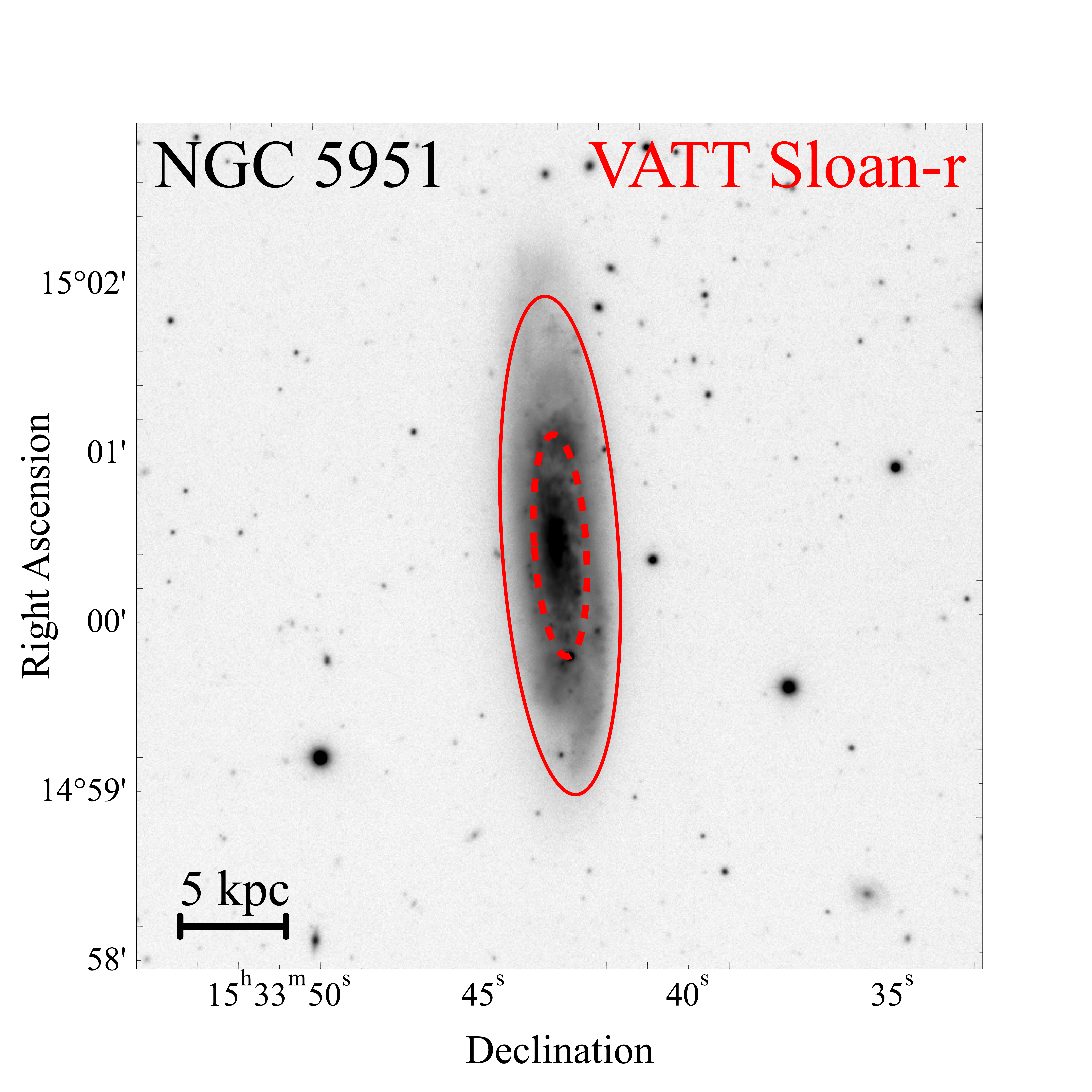}
\caption{Optical (Sloan-\sdssr) imagery of our galaxies. The red ellipses mark the apertures containing 50\% (dashed) and 90\% (solid) of the total \mstar\ (see Section \ref{sec:it} and \ref{sec:radii} for details). A physical scale of 5.0~kpc is shown at the bottom left corner of each map.}  \label{fig:opt}
\end{figure*}

Each galaxy was observed with a total integration time of 600 s in \sdssg\ and 1200 s in \sdssr. Bias subtracted and flat fielded (using both dome and twilight sky flats) science images were created using standard image processing routines in the Image Reduction and Analysis Facility (IRAF). The sky background level was estimated from the average intensity in the source-free regions of each science image and subtracted. The cosmic ray-induced signal was removed using the \texttt{L.A.COSMIC} 
routine \citep{vand01}. To create the final images in \sdssg\ and \sdssr\ for each galaxy, the science exposures were aligned using field stars, smoothed to match the resolution of the worst-seeing image, and exposure-time-weighted before combining. 
Flux calibration was performed using the Aperture Photometry Tool \citep{apt} and field stars. Instrumental magnitudes of the field stars were determined through aperture photometry from the \sdssg\ and \sdssr\ images and matched to the SDSS DR7 catalog to derive the photometric zero point of the images.
The photometric zero point of the \sdssg\ and \sdssr\ image was then calculated by comparing the instrumental magnitudes to their corresponding SDSS magnitudes for the same stars. 
The resulting \sdssr\ maps of our galaxies are shown in Figure \ref{fig:opt}. The 1$\sigma$ noise equivalent mean surface brightness of the final \sdssg\ and \sdssr\ maps are 25.57~mag\,arcsec$^{-2}$ and 25.44~mag\,arcsec$^{-2}$, respectively.

\subsection{Infrared Imaging data}
Archival \mips\ images taken with the Multiband Imaging Photometer for Spitzer \citep[MIPS;][]{mips} are available for six galaxies in our sample (NGC~3344, NGC~3351, NGC~3666, NGC~4303, NGC~4321, and NGC~5669). These maps have a resolution of 6\arcsec. We also utilize archival \wise\ imaging from the Wide-field Infrared Survey Explorer (WISE) All-Sky Data Release \citep{wright10} for the remaining galaxies. These maps have a resolution of 11\farcs8.  
All galaxies, except IC~3440, show emission in the \wise\ maps.

\newpage
\subsection{VLA H~I data}
The \hi\ observations were conducted using the VLA under the VLA-DIISC program (P.I. S. Borthakur; Gim et al. in prep.) and project codes 17A-090, 18A-006, 19B-183, and 20A-125. Most galaxies were observed in the VLA D-configuration (maximum baseline, $b_{\rm max}= 1$~km) with follow-up for some in B- and C-configurations ($b_{\rm max}= 11.1$km, and $b_{\rm max}= 3.4$km, respectively). No VLA observations are available for IC 3440. 
Data were reduced using the Common Astronomy Software Application \citep[CASA, ][]{CASA_2022}. A semi-automatic pipeline was used to solve for the delay, bandpass, amplitude, and phase gain calibrations. 
Hanning smoothing was implemented to mitigate the presence of Gibbs ringing artifacts, resulting in a reduction of the spectral (velocity) resolution to twice the original channel width. Specifically, the velocity resolution for the image cubes of NGC 99 and NGC 9120 was set at 2.2~km~s$^{-1}$, while the remaining image cubes maintained a velocity resolution of 3.3~km~s$^{-1}$, with the exception of NGC 4921. In the case of NGC 4921, the image cube was generated using a velocity resolution of 20~km~s$^{-1}$, strategically applied to enhance the signal-to-noise ratio through supplementary spectral axis smoothing. Data affected by Radio frequency interference were also edited out. The observation details, including project code, total integration time, flux density/bandpass and phase calibrators, synthesized beam sizes, velocity resolutions, and sensitivities of the resulting image cubes are summarized in Table~\ref{table:VLA_DIISC}.

\begin{deluxetable*}{lccccccc}[!ht]
\tablecaption{VLA observations and properties of the resulting H\,{\small I} image cubes \label{table:VLA_DIISC}}
\tablehead{
\colhead{Galaxy}  & \colhead{Project} & \colhead{Total}  & \colhead{Flux \&}  & \colhead{Phase}  &   \colhead{Synthesized} &  \colhead{Velocity} & \colhead{Sensitivity}\vspace{-0.2cm} \\
 & \colhead{Code} & \colhead{Integration}  & \colhead{Bandpass}  & \colhead{Calibrator} &   \colhead{beamsize} &  \colhead{Resolution} & \colhead{} \vspace{-0.2cm}\\
   \colhead{}  & \colhead{} & \colhead{}  & \colhead{Calibrator}  & \colhead{} &   \colhead{} &  \colhead{} & \colhead{}\vspace{-0.2cm}\\
 &  & \colhead{(hours)} & & & \colhead{(\arcsec$\times$\arcsec)} & \colhead{(km s$^{-1}$)} & \colhead{(mJy beam$^{-1}$}\vspace{-0.2cm} \\
  & & & & & & & \colhead{km~s$^{-1}$)}}
\startdata
NGC 99  &  19B-183 & 6.5 & 3C48  & J2340+1333 & 50.5$\times$47.5 & 2.2 & 0.98  \\ 
\multirow{3}{*}{NGC 3344}  & \multirow{2}{*}{AB365$^{1}$}  & \multirow{2}{*}{11.87} & \multirow{3}{*}{3C286} & 1108+201 & \multirow{3}{*}{6.7$\times$5.4} & \multirow{3}{*}{10.36} & \multirow{3}{*}{0.19} \\
  &  & & & 1040+123 & & &\\ 
  & 20A-125 & 48 & & J1021+2159 & & &\\ 
NGC 3351  & 17A-090 & 6.5  & 3C147 & J1120+1420 & 53.0$\times$51.0  & 3.3 & 0.70  \\
  NGC 3433   & 17A-090 & 6.5  & 3C147 & J1120+1420 & 65.5$\times$51.0  & 3.3  & 0.65 \\ 
\multirow{2}{*}{NGC 3485}  & 18A-006 & \multirow{2}{*}{8.5} & \multirow{2}{*}{3C286}  & \multirow{2}{*}{J1120+1420} & \multirow{2}{*}{54.0$\times$54.0} & \multirow{2}{*}{3.3} & \multirow{2}{*}{0.74} \\
  & 19B-183 & & & & & & \\ 
 \multirow{2}{*}{NGC 3666} & 18A-006 & \multirow{2}{*}{8.7} & \multirow{2}{*}{3C286} & \multirow{2}{*}{J1120+1420} & \multirow{2}{*}{60.5$\times$56.5} & \multirow{2}{*}{3.3} & \multirow{2}{*}{0.60} \\ 
  & 19B-183 & & & & & &\\
  \multirow{2}{*}{NGC 3810} & 17A-090 & \multirow{2}{*}{6.5}  & \multirow{2}{*}{3C286} & \multirow{2}{*}{J1120+1420} & \multirow{2}{*}{39.5$\times$30.0} & \multirow{2}{*}{3.3} & \multirow{2}{*}{0.63} \\ 
  & 18A-006 & & & & & & \\ 
 NGC 4303  &  19B-183 & 6 & 3C286 & J1150-0023 & 56.0$\times$48.0  & 3.3 & 0.70 \\
 NGC 4321   & 18A-006 & 4 & 3C286 & J1254+1141 & 44.0$\times$30.0 & 3.3 & 1.30 \\
 IC 3440   & ...  & ... &  ... & ... & ... & ...  & ... \\ 
 NGC 4921  & 17A-090 & 18.7 & 3C286 & 3C286 & 31.5$\times$28.0  & 20.0 & 0.19 \\ 
 UGC 9120  & 17A-090 & 6.5 & 3C286 & J1419+0628 & 69.0$\times$57.5 & 2.2 & 1.20 \\ 
NGC 5669  & 17A-090 & 14 & 3C286 & J143222+095551 & 62.5$\times$53.0 & 3.3 & 0.85 \\ 
NGC 5951 & 17A-090 & 6.5  & 3C286 & J1520+2016 & 73.0$\times$50.5  & 3.3 & 0.85 
\enddata
\flushleft
$^{1}$ VLA archival data (PI: F.H. Briggs)
\end{deluxetable*}

The spectral line visibilities were obtained using the CASA task {\it uvcontsub} by subtracting the continuum and used to produce \hi\ image cubes with the CASA task {\it tclean}. We applied the Clark point spread function and BARY spectral frame in general. We made the HI image cubes using the Briggs weighting function with a robust value of 0.5 for optimal sensitivity and synthesized beam size, except for NGC 4321. For NGC 4321, a uniform weighting function was used instead. More details about NGC 4321 and NGC 3344 can be found in \citet{gim21} and \citet{padave21}, respectively. 

\subsection{COS Absorption line Spectra}\label{sec:cos}
To probe the disk-CGM interface of the sample, UV-absorption line spectroscopic observations of background UV-bright quasars were carried out with G130M grating of COS \citep{green12} aboard HST. These observations were conducted as part of the COS-DIISC Survey \citep[HST Program ID: 14071;][in prep]{borthinprep}. The spectral resolution of the data was $\approx$ 20~\kms ~(R$\sim$15000), covering a wavelength range from 1140--1430~\AA\ in the observed frame. 
Data reduction and calibration were performed following the standard COS pipeline and the procedures described in the COS Data Handbook (Rafelski et al. 2018). 
The data covered multiple line transitions including \ion{H}{1} $\lambda$1215~\AA\ (\Lya), \ion{Si}{2} $\lambda$$\lambda$ 1190~\AA,1193~\AA, 1260~\AA, \ion{Si}{3} $\lambda$1206~\AA,  \ion{Si}{4} $\lambda\lambda$1393,1402~\AA, and \ion{C}{2} $\lambda$1334~\AA. Procedures detailing the absorption line fitting are discussed in \cite{borthinprep}, in prep. 

In this work, we mainly focus on \Lya\  absorption-line tracing the neutral hydrogen content of the CGM. In general, absorption-line strengths are found to strongly correlate with the distance of the QSO sightline from the galaxy's optical center, i.e., the impact parameter \citep{chen98}. The impact parameters, $\rho$, for our sightlines, range from 26--163~kpc. 
Therefore, before making any comparison between the CGM properties and the galaxies, we normalize the observed equivalent width (W$_{Ly\alpha}$) to that predicted for the same $\rho$. The predicted equivalent width is estimated by the best-fit line relating the equivalent width and normalized impact parameter, $\rho/R_{vir}$, derived by \citet{bort16} for the combined sample of COS-Halos and COS-GASS surveys spanning a stellar mass range of $\rm \sim 10^{10-11} M_{\odot}$. We adopt the $\rho$-corrected equivalent width defined as the ratio of the observed equivalent width and the predicted equivalent width ($\overline{\rm W}_{Ly\alpha}$), also termed as the W$_{Ly\alpha}$ excess. 
The impact parameters along with the W$_{Ly\alpha}$ excess values are tabulated in Table \ref{tbl:sample}.
\section{Derived Galaxy Properties} \label{sec:it}
We derive the SFR surface density, \sfr, from the FUV and \mips\ maps using the following prescriptions \citep{sali07, calz07, lero08}:
\begin{equation}\label{eq:1}
     \Sigma_{\rm SFR}\rm{(FUV)} = 0.17~{\rm I}_{\rm{FUV}}
\end{equation}
\begin{equation}\label{eq:2}
\Sigma_{\rm SFR} \rm{(FUV} + 24 \mu \rm{m)} = 0.081~I_{\rm{FUV}} + 0.0032~I_{24}
\end{equation}
where \sfr\ has units of M$_\odot$~yr$^{-1}$~kpc$^{-2}$ and I$_{\rm FUV}$ and I$_{24}$ are in the units of MJy~sr$^{-1}$. Equation \ref{eq:1} and \ref{eq:2} assume a \citep{salp55} and \citep{kroupa01} initial mass function (IMF), respectively. A factor of 1.59 is divided from equation \ref{eq:1} to correct for the IMF when comparing the values from the above prescriptions.   
The FUV flux, stemming from the photospheres of massive O and B stars, traces recent star formation \citep{kenn98} while the \mips\ flux density accounts for the dust-processed UV photons. 
For galaxies without \mips\ data, the dust-corrected \sfr\ is estimated using \wise\ flux instead of \mips. This may cause the inferred SFRs to be slightly lower, but comparable to SFRs estimated using FUV+\mips\ \citep[see][and references therein]{cass17}. Since the \wise\ flux accounts for UV photons reprocessed by dust, FUV+\wise\ is a better estimator of the total SFR than FUV alone. We follow the same procedure of combining the FUV and infrared (\wise\ or \mips) maps as \cite{padave21}. 

The global SFR estimates presented in Table \ref{tbl:sample} were made by summing over all the pixels belonging to the galaxy in the FUV and the infrared maps and using the dust-corrected \sfr\ prescription. We note that for IC~3440, the SFR is estimated using only the FUV map as the galaxy shows no emission in the \wise\ map. 
Identification of pixels belonging to a galaxy was done by creating a segmentation map. Each segmentation map was visually inspected to ascertain that any extended or low-surface brightness features were included. 
In certain cases, especially for galaxies spanning more than 8\arcmin\ in the sky, an elliptical region (defined by the galaxy's position angle and ellipticity) larger than the detected region was manually defined to include pixels containing diffuse emission from the galaxy. Foreground objects along the line of sight to each galaxy were separately masked.  

The stellar mass (\mstar) maps for our galaxies were created by using the relationship derived by \cite{yang07} for SDSS galaxies: 
\begin{align*}
\log\bigg[\frac{{\rm M}_{\star}}{h^{-2} {\rm M}_\odot}\bigg] = -0.406 \,+\,1.097\,(g-r)\,-\\
    \hspace{1.5cm} 0.4\,({\rm M}_r\,-\,5\log h\, -\,4.64)
\end{align*}
Here, M$_r$ is the absolute magnitude in \sdssr\ and $g-r$ is the color derived from the \sdssg\ and \sdssr\ maps at each pixel.  
While creating the \mstar\ maps, pixels were masked if $g-r$~$>3$ or $g-r$~$<-3$ mag, as might result from the subtraction of low S/N pixels or pixels containing sources unassociated with the galaxy. We adopt $h=H_0/(100~{\rm km~s}^{-1}{\rm Mpc}^{-1})=0.7$. Global estimates of \mstar\ were made by summing over the pixels belonging to the target galaxy. These estimates were found to be comparable to those from Two-micron All Sky Survey's $K$-band \citep[Equation A2 from][]{muno07}, but our images reach lower stellar density outer galaxy regions.  

The surface density of atomic gas, \mhi, for our galaxies, were estimated using the velocity-integrated flux density ($I_{tot}$) VLA \hi-21~cm maps and the relation from \cite{mullan13}:
\begin{equation}\label{eq:6}
    \Sigma_{\rm {H\,\small{I}}} = 1.0 \times 10^{4}\; \frac{I_{tot}}{\rm{A}_{\rm beam}}
\end{equation}
where $I_{tot}$ has units of Jy~beam$^{-1}$ km~s$^{-1}$ with the beam area, $A_{beam}$, expressed in arcsecond$^{2}$, and \mhi\ in M$_\odot$ pc$^{-2}$. The total \higas\ was found by integrating the total flux. However, some galaxies may have diffuse \hi\ in the outskirts which may not be probed by the VLA maps, causing the \higas\ estimates to be lower limits. Hence, we adopt the single dish Arecibo Legacy Fast ALFA \citep[ALFALFA;][]{alfalfa} survey measurements for estimating global \higas. Since IC 3440 has no observations with the VLA, we only have the global \higas\ measurement and no \mhi\ distribution.

\begin{figure}[!htb] 
 \centering
\includegraphics[trim = 0cm 0cm 0cm 0cm, clip,scale=0.25]{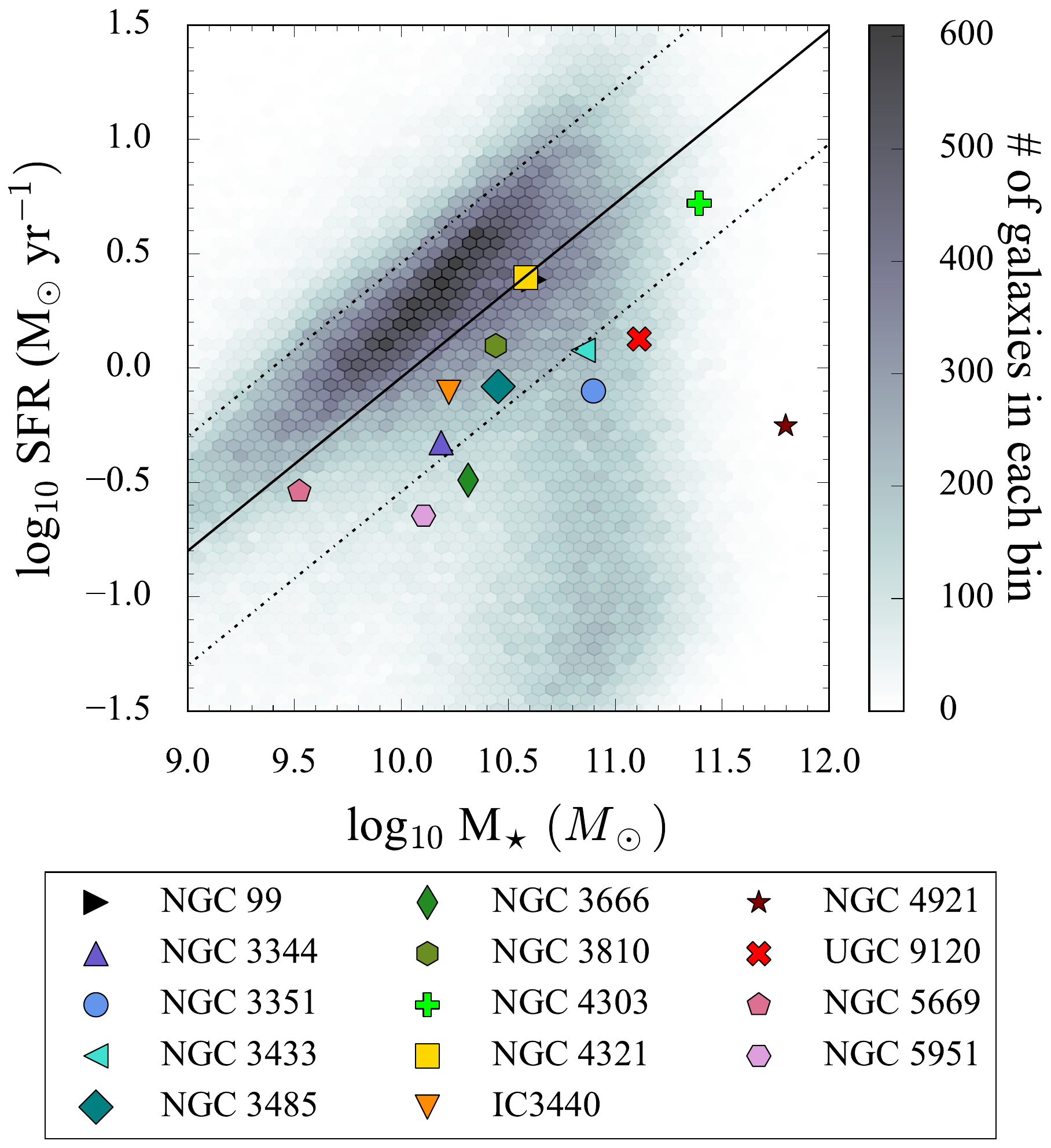}

\caption{The star formation rates and stellar masses of the sample plotted over the galaxy main sequence obtained from the MPA-JHU value-added catalog. The solid and dash-dotted lines represent the galaxy main sequence ridge line from \cite{renz15} and $\pm0.5$ dex from the main sequence, respectively. Each galaxy in the sample has a unique marker that is consistent throughout the paper. Most of our sample galaxies have sSFR$\gtrsim10^{-11}$~yr$^{-1}$ with the exception of NGC 4921 with sSFR$<10^{-11}$~yr$^{-1}$. }
\label{fig:sample}
\end{figure}

The measured SFRs, \mstar, and \higas\ for the sample are presented in Table \ref{tbl:sample}. In Figure \ref{fig:sample}, we show the distribution of our sample with respect to the star formation main sequence of galaxies. The background bins show the \mstar-SFR distribution of local galaxies obtained from the MPA-JHU value-added catalog \citep{kauff03, brinch04, sali07}. Most of our galaxies lie within 0.5 dex from the galaxy main sequence ridge line from \cite{renz15} in the star-forming sequence and the green valley region. 
Except NGC 4921, all galaxies have sSFR~$\gtrsim10^{-11}$~yr$^{-1}$ and are considered star-forming. NGC 4921  has sSFR~$=10^{-12.0}$~yr$^{-1}$ and is a member of the Coma cluster \citep{cramer21}. We note that there could be a discrepancy between our SFR and \mstar\ estimates and those from MPA-JHU. Therefore, the \mstar-SFR distribution of local galaxies in Figure \ref{fig:sample} are intended only for reference and illustration.  

\section{Results}\label{sec:prof}

\subsection{Radial Surface Brightness Profiles}

\begin{figure*}[!htb] 
\centering
\includegraphics[trim = 0cm 0.cm 0cm 0cm, clip,scale=0.13]{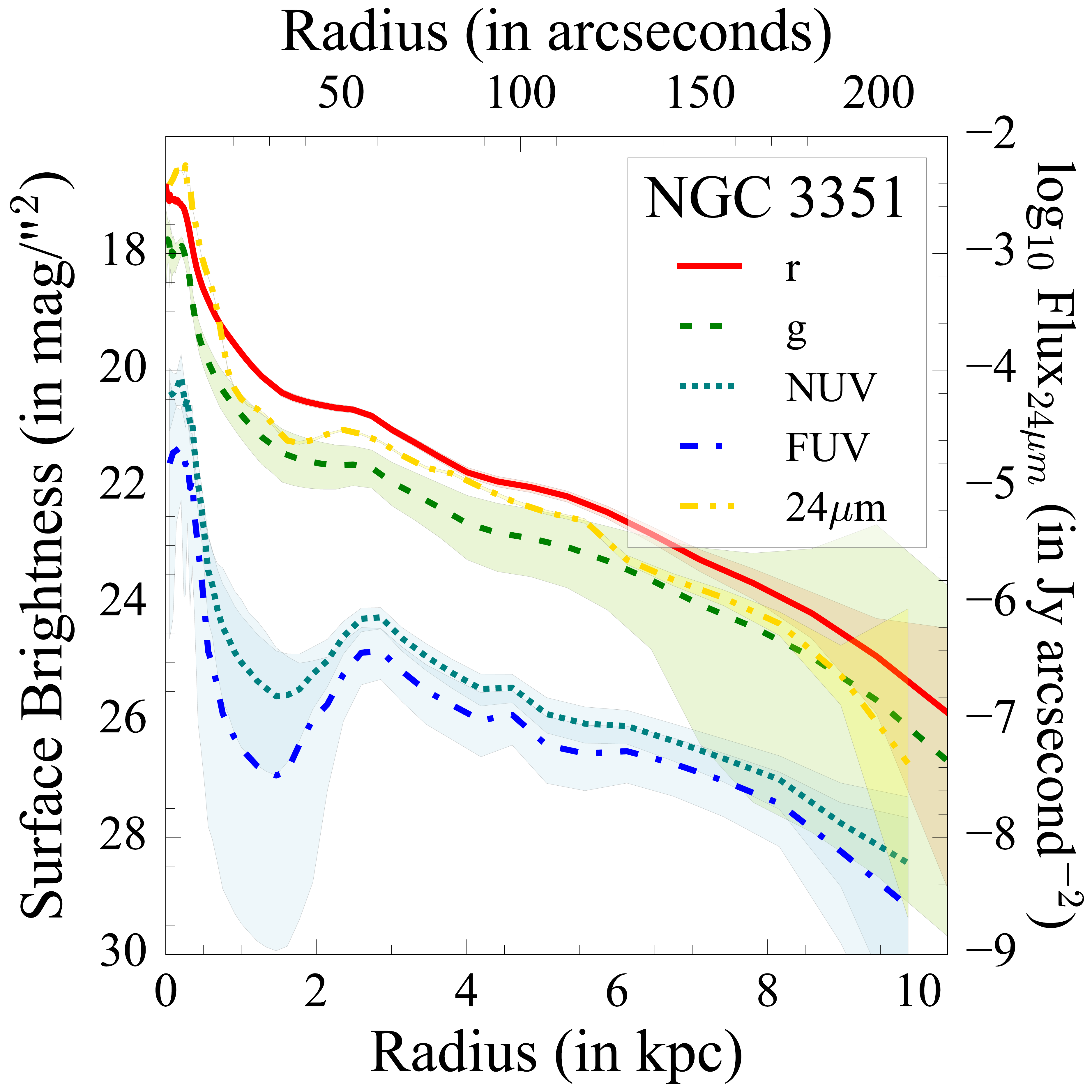}
\includegraphics[trim = 0cm 0.cm 0cm 0cm, clip,scale=0.13]{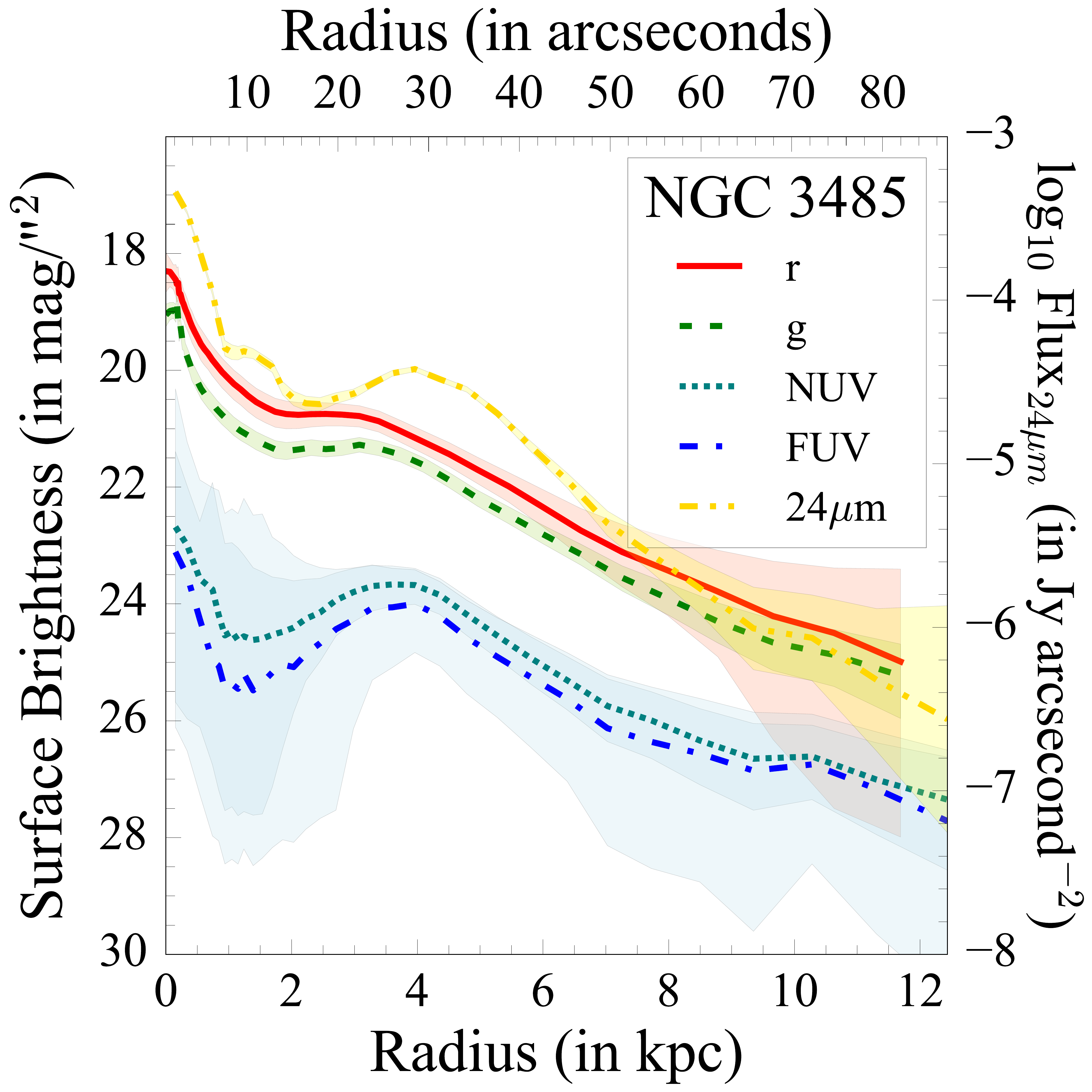}
\caption{Multi-wavelength profiles of surface brightness for the galaxies in our sample. The profiles for \sdssr, \sdssg, NUV, and FUV are presented in mag~arcsec$^{-2}$, while for \mips\ profiles are shown in Jy~arcsec$^{-2}$. The shaded region around each profile shows the 1$\sigma$ uncertainty in the surface brightness. The full figure set is shown in Appendix \ref{app:fig4}.}\label{fig:profiles}
\end{figure*}

Surface photometry of the panchromatic maps of our galaxies was performed by fitting elliptical isophotes with a fixed center using the IDL routine, 
{\texttt {galprof}}\footnote{\footnotesize{https://www.public.asu.edu/$\sim$rjansen/idl/galprof1.0/galprof.pro}}, where the routine estimates the position angle and ellipticity of the ellipse for each map. Figure \ref{fig:profiles} shows the surface brightness profiles in FUV, NUV, \sdssg, \sdssr, and \wise\ or \mips. 
The surface brightness in FUV, NUV, \sdssg, \sdssr, and \wise\ are estimated in the units of mag~arcsec$^{-2}$ and the 24\um\ flux is estimated in Jy~arcsec$^{-2}$. 
The profiles are extracted till a surface brightness limit of 3$\sigma$ is reached. Uncertainties in the surface brightness include the Poisson noise in the signal in a given annulus and error in the background subtraction.  
Disregarding the fluctuations in the inner part (likely caused by the presence of a bulge), we find that all profiles generally drop smoothly with radius. Some also show variations caused by asymmetries in the flux distribution and the presence of spiral arms. As evident from the surface brightness profiles, most of the \wise\ data is dominated by noise. 
In NGC 4921, a prominent break is observed in the FUV and NUV profiles at $\sim$10~kpc while the \sdssg\ and \sdssr\ profiles drop smoothly. This upturn in the UV profiles results from the bright star-forming feature in the spiral arm of the galaxy. Following this, the FUV emission shows a steep drop. 


\begin{figure*}[!t] 
\centering
\includegraphics[trim = 05mm 0.5cm 0cm 0cm, clip,scale=0.14]{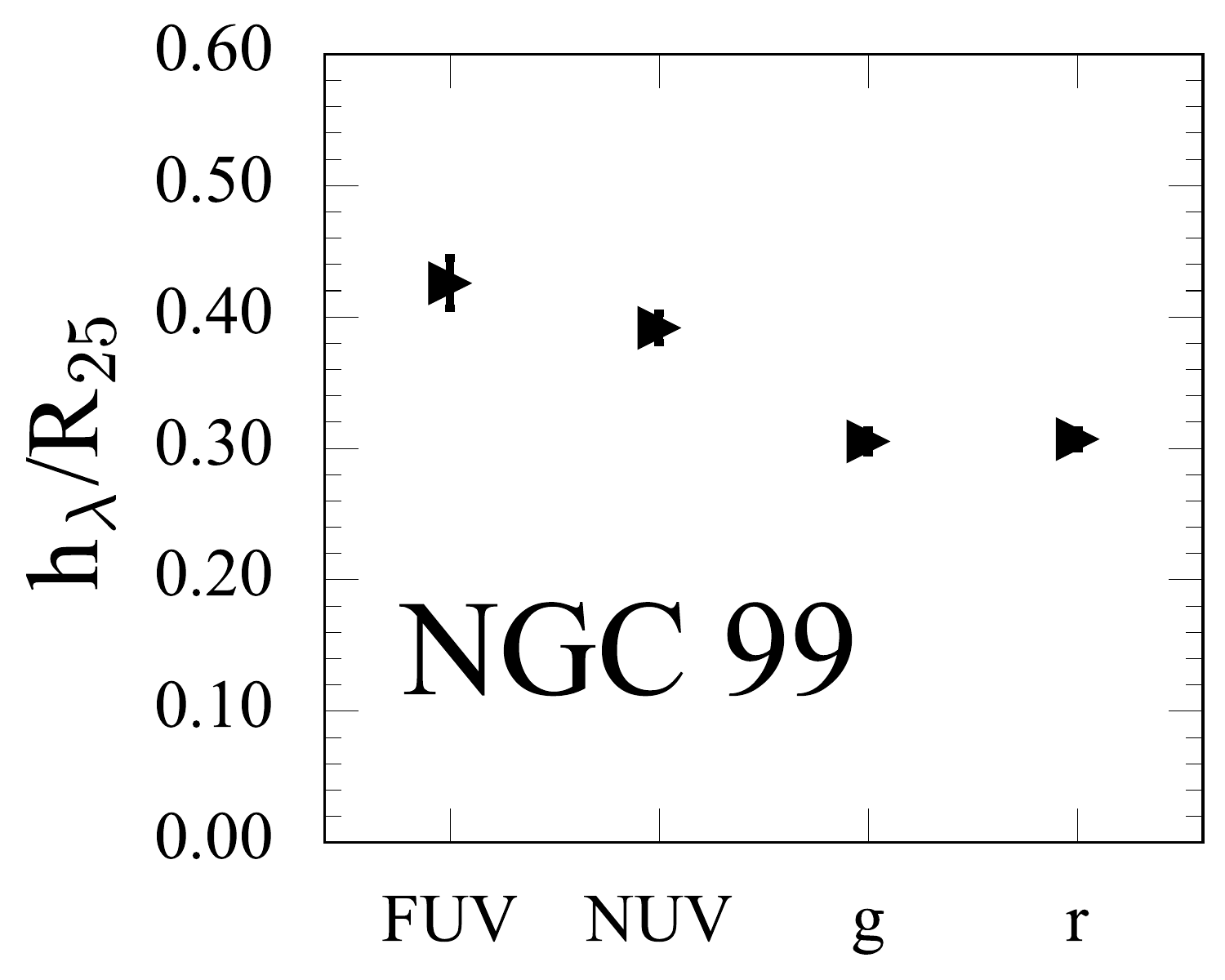}
\includegraphics[trim = 05mm 0.5cm 0cm 0cm, clip,scale=0.14]{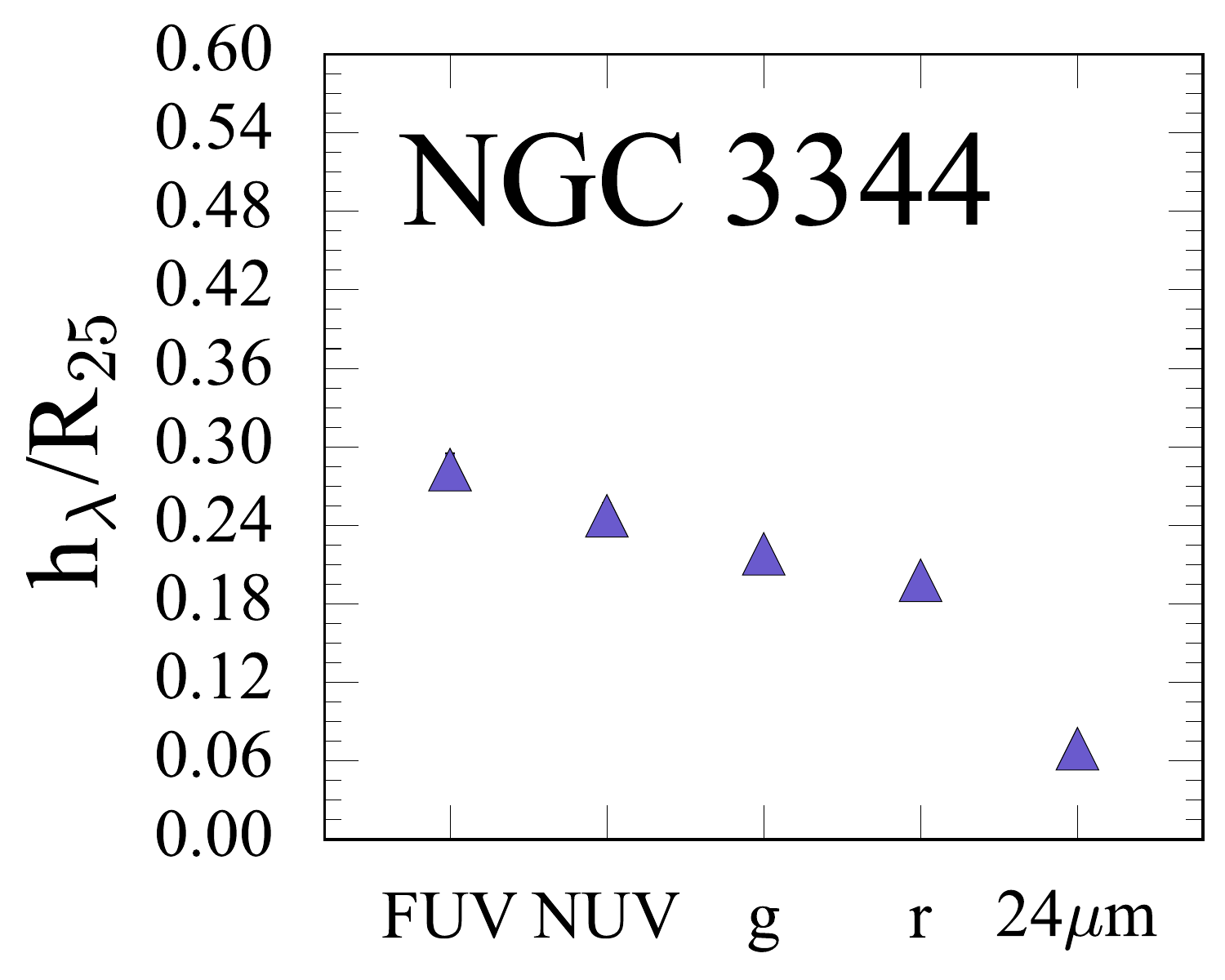}
\includegraphics[trim = 05mm 0.5cm 0cm 0cm, clip,scale=0.14]{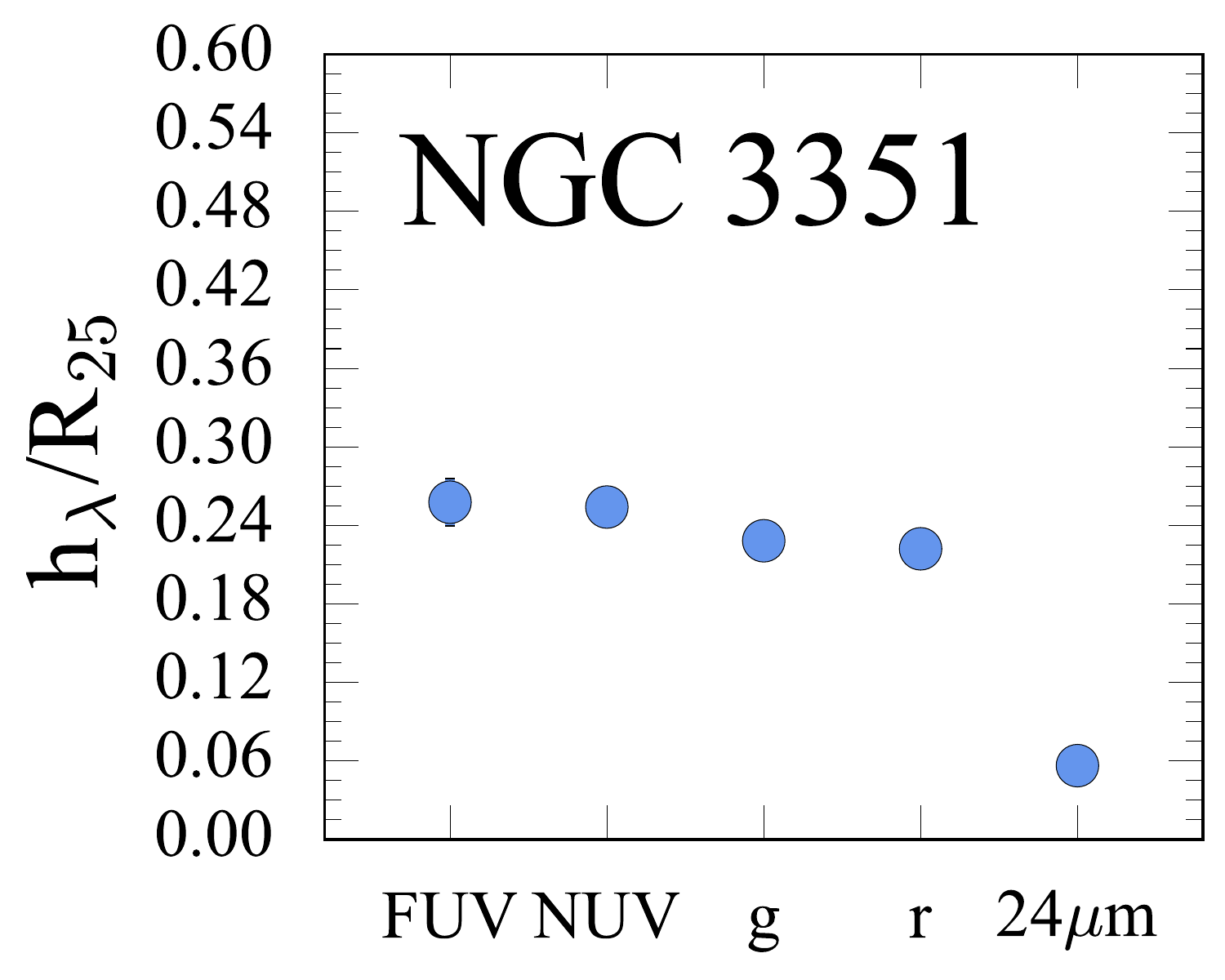}
\includegraphics[trim = 05mm 0.5cm 0cm 0cm, clip,scale=0.14]{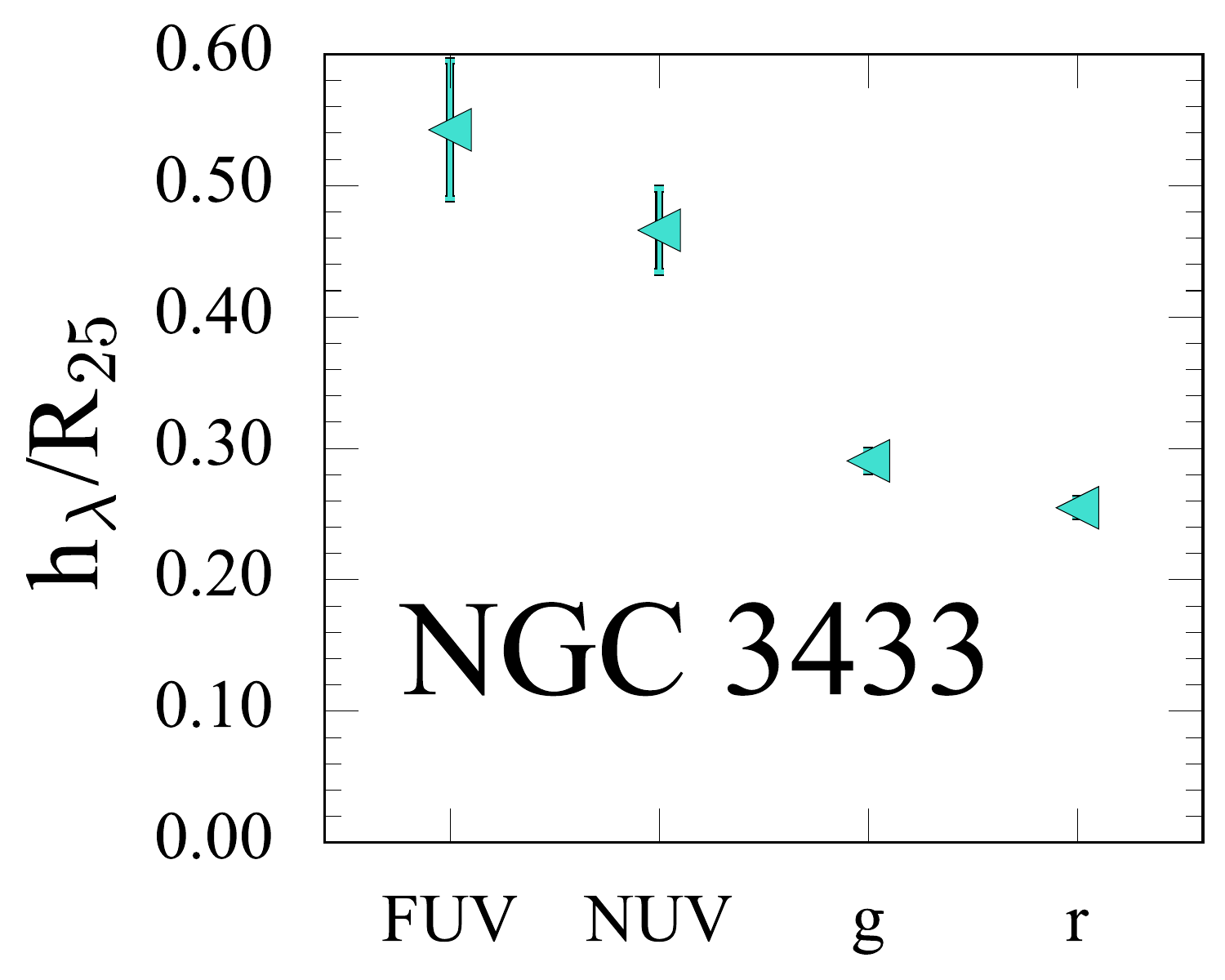}
\includegraphics[trim = 05mm 0.5cm 0cm 0cm, clip,scale=0.14]{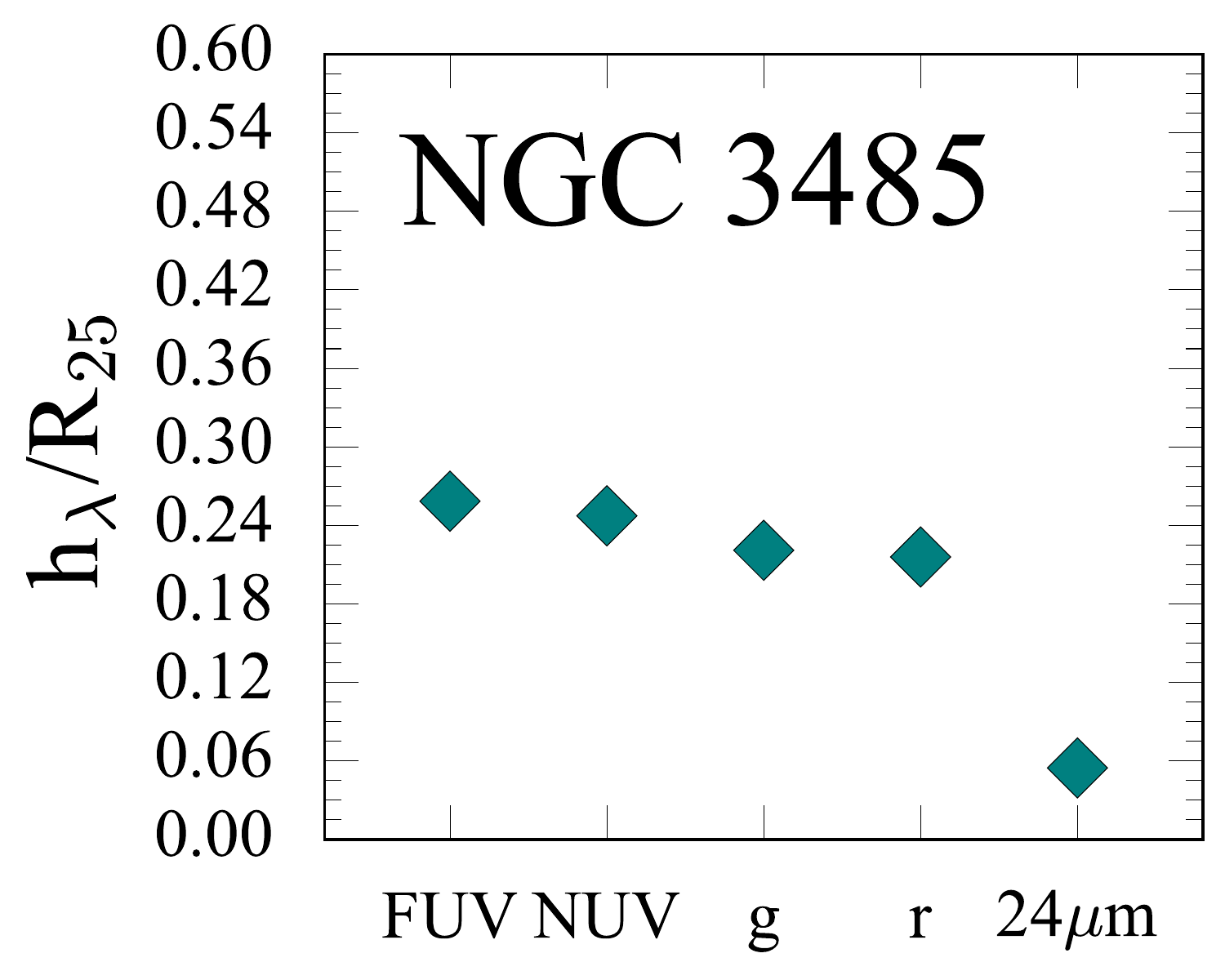}
\includegraphics[trim = 05mm 0.5cm 0cm 0cm, clip,scale=0.14]{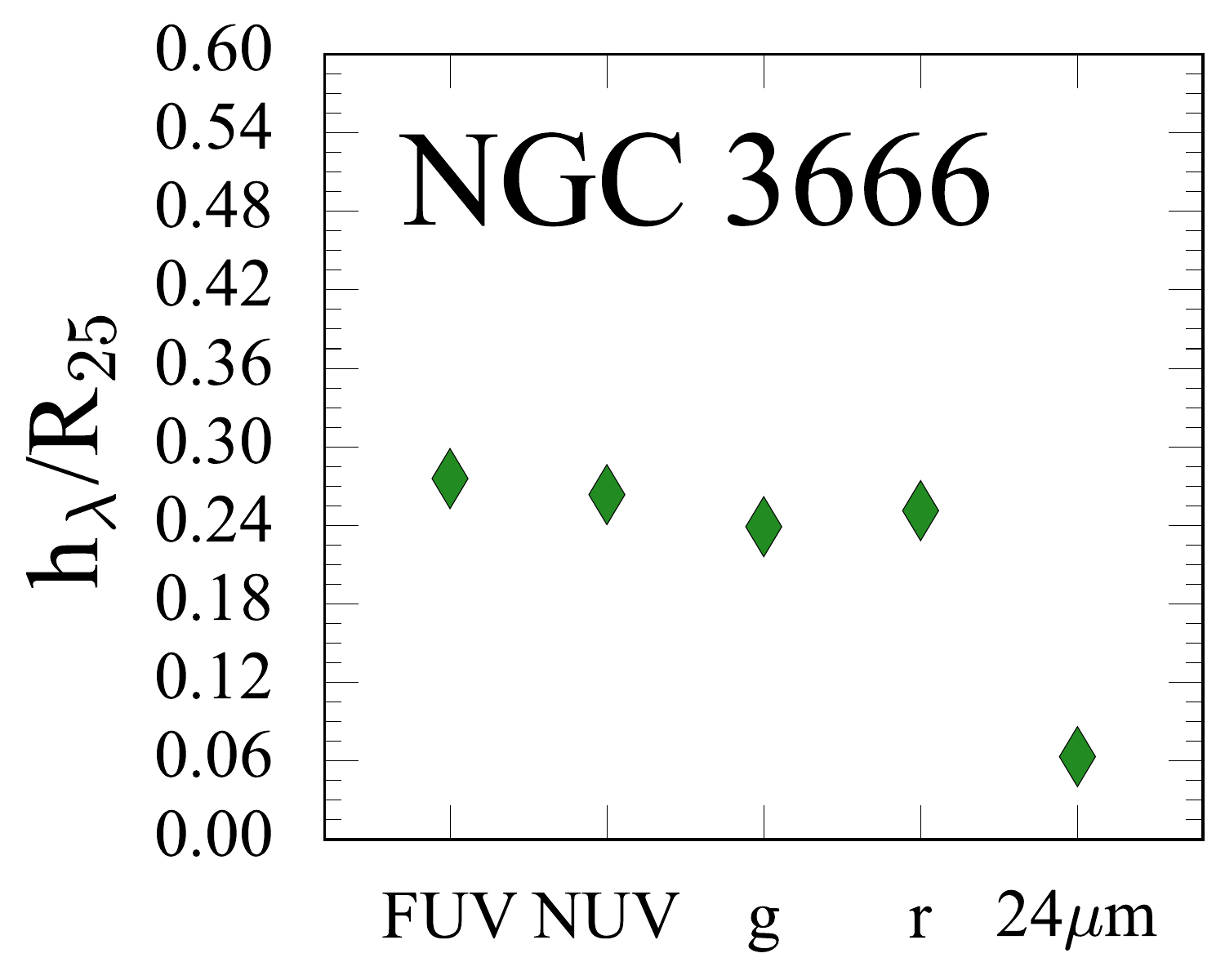}
\includegraphics[trim = 05mm 0.5cm 0cm 0cm, clip,scale=0.14]{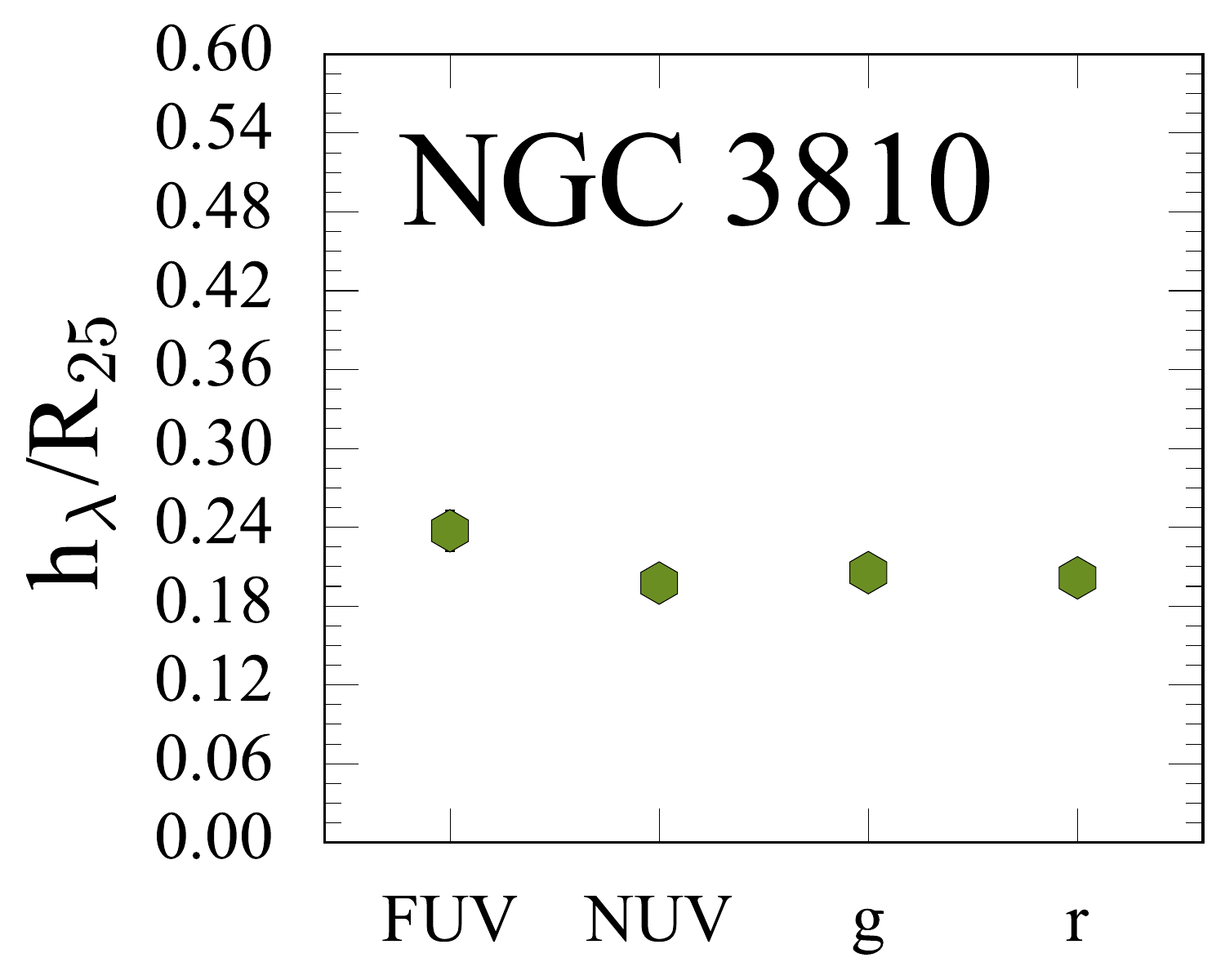}
\includegraphics[trim = 05mm 0.5cm 0cm 0cm, clip,scale=0.14]{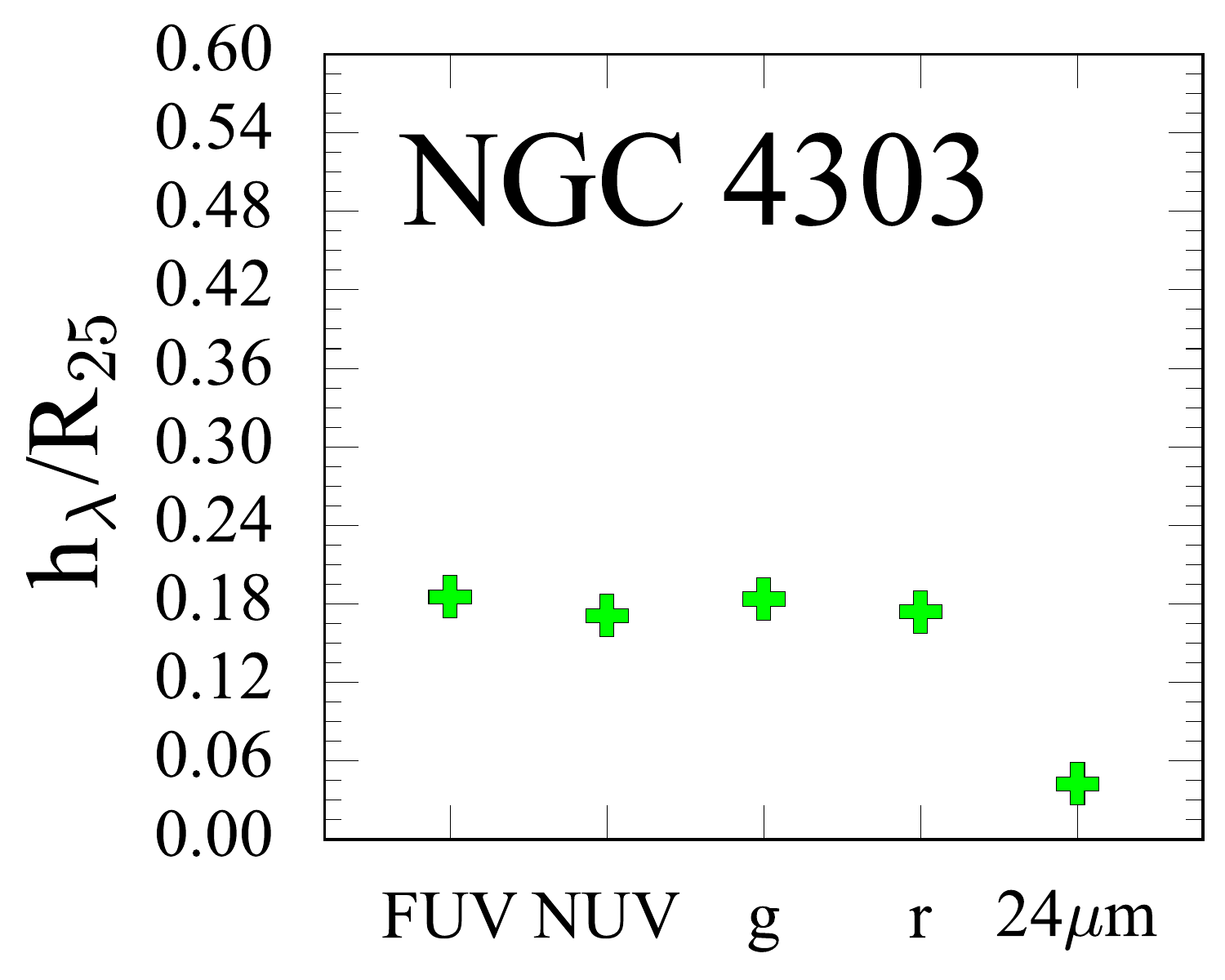}
\includegraphics[trim = 05mm 0.5cm 0cm 0cm, clip,scale=0.14]{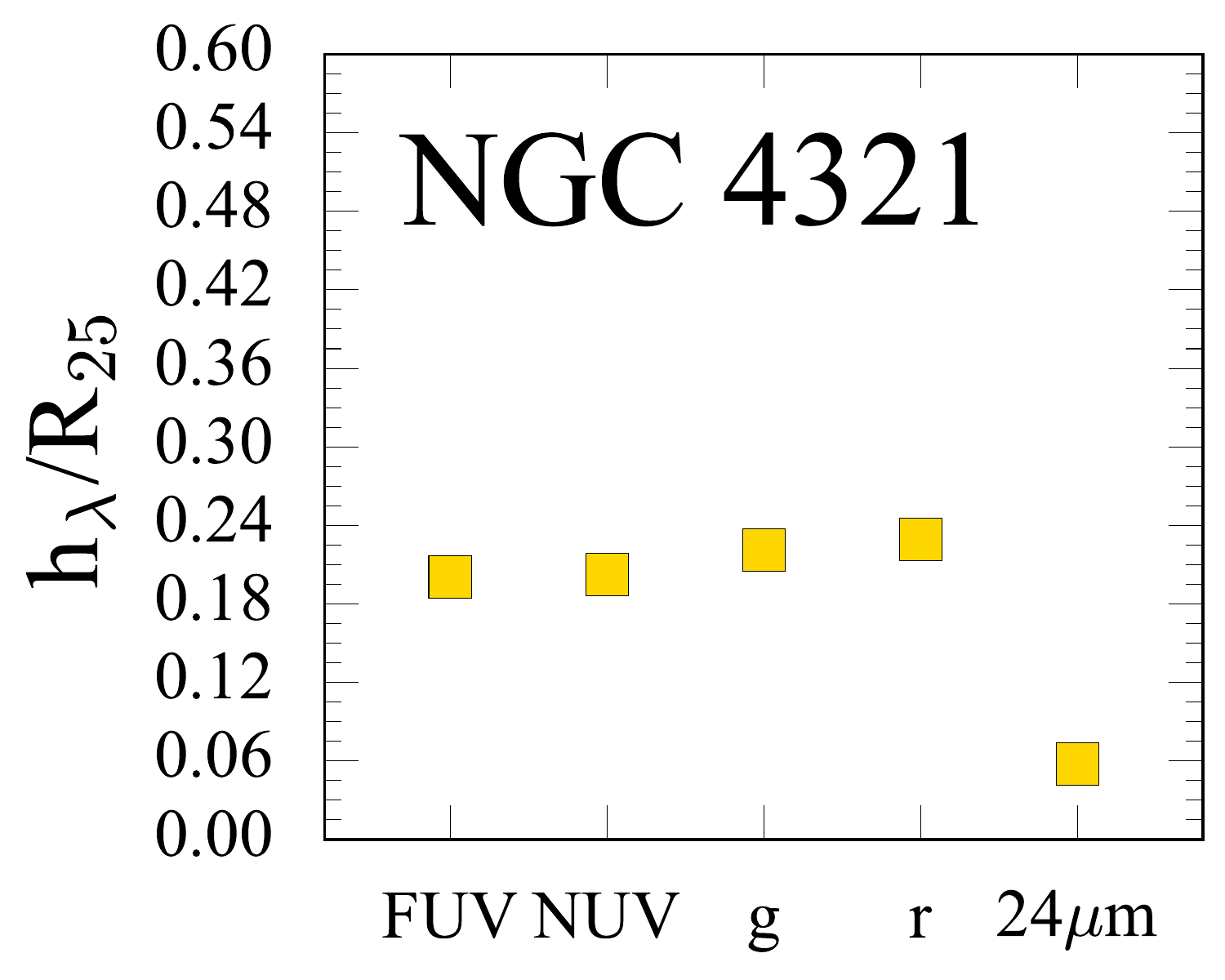}
\includegraphics[trim = 05mm 0.5cm 0cm 0cm, clip,scale=0.14]{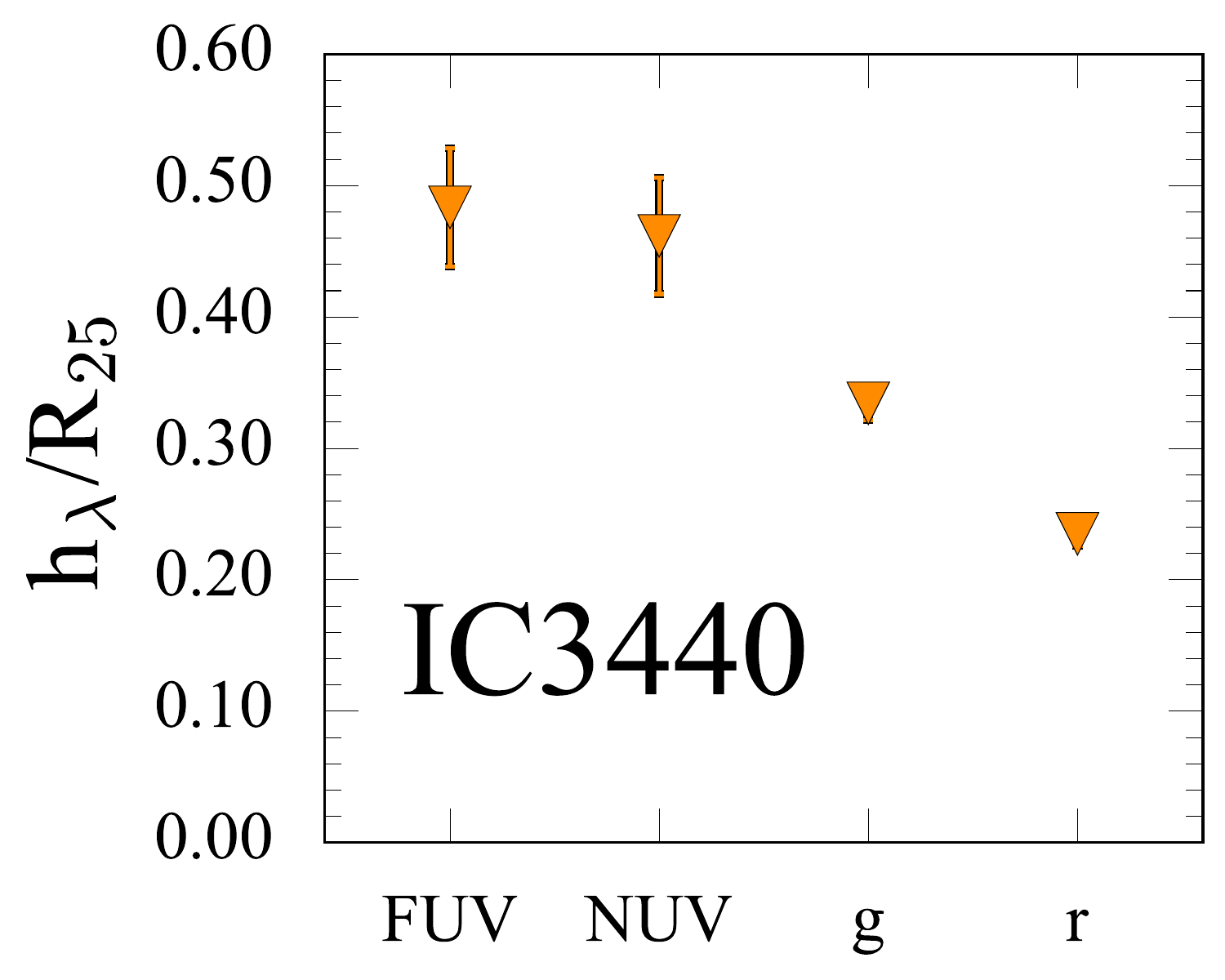}
\includegraphics[trim = 05mm 0.5cm 0cm 0cm, clip,scale=0.14]{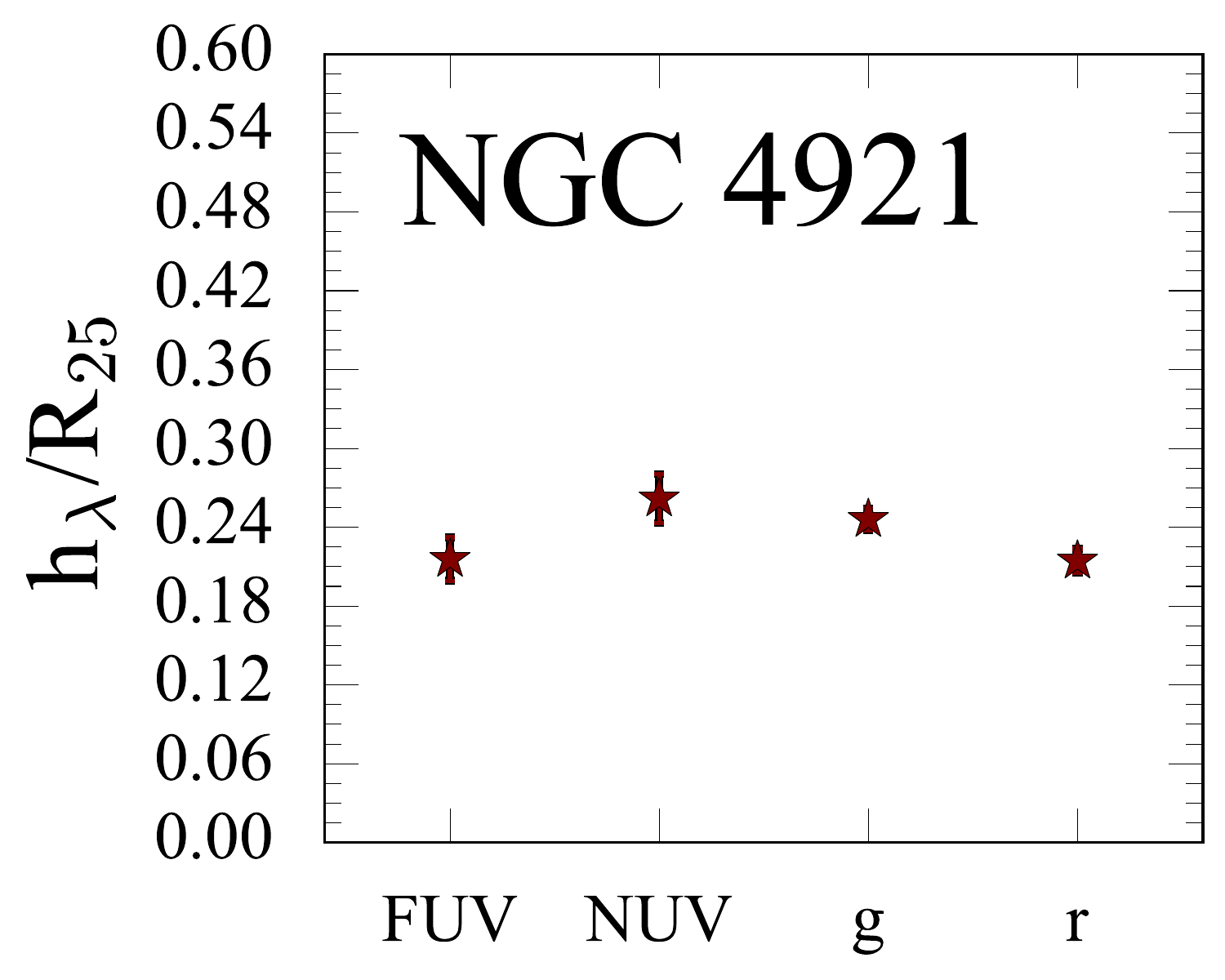}
\includegraphics[trim = 05mm 0.5cm 0cm 0cm, clip,scale=0.14]{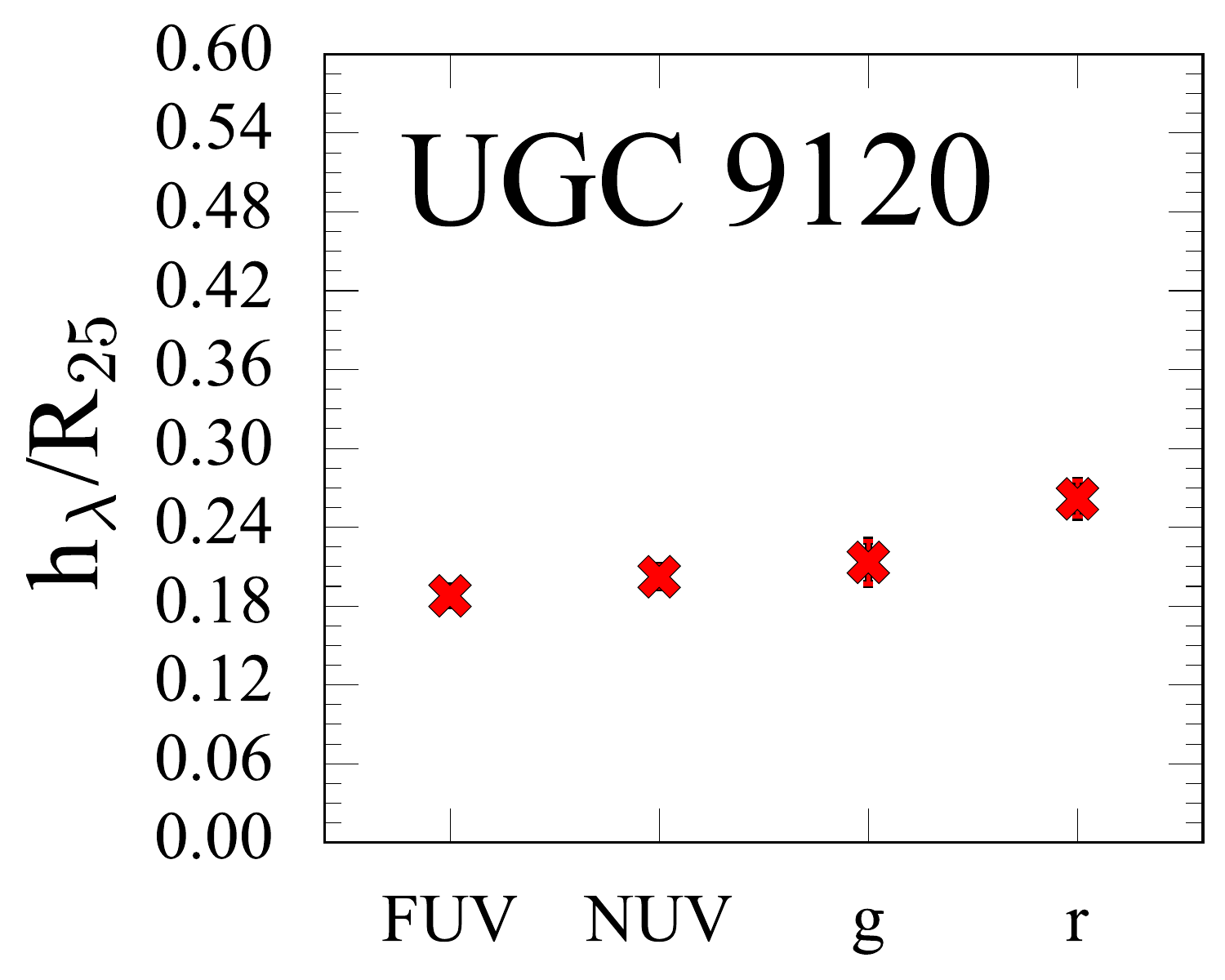}
\includegraphics[trim = 05mm 0.5cm 0cm 0cm, clip,scale=0.14]{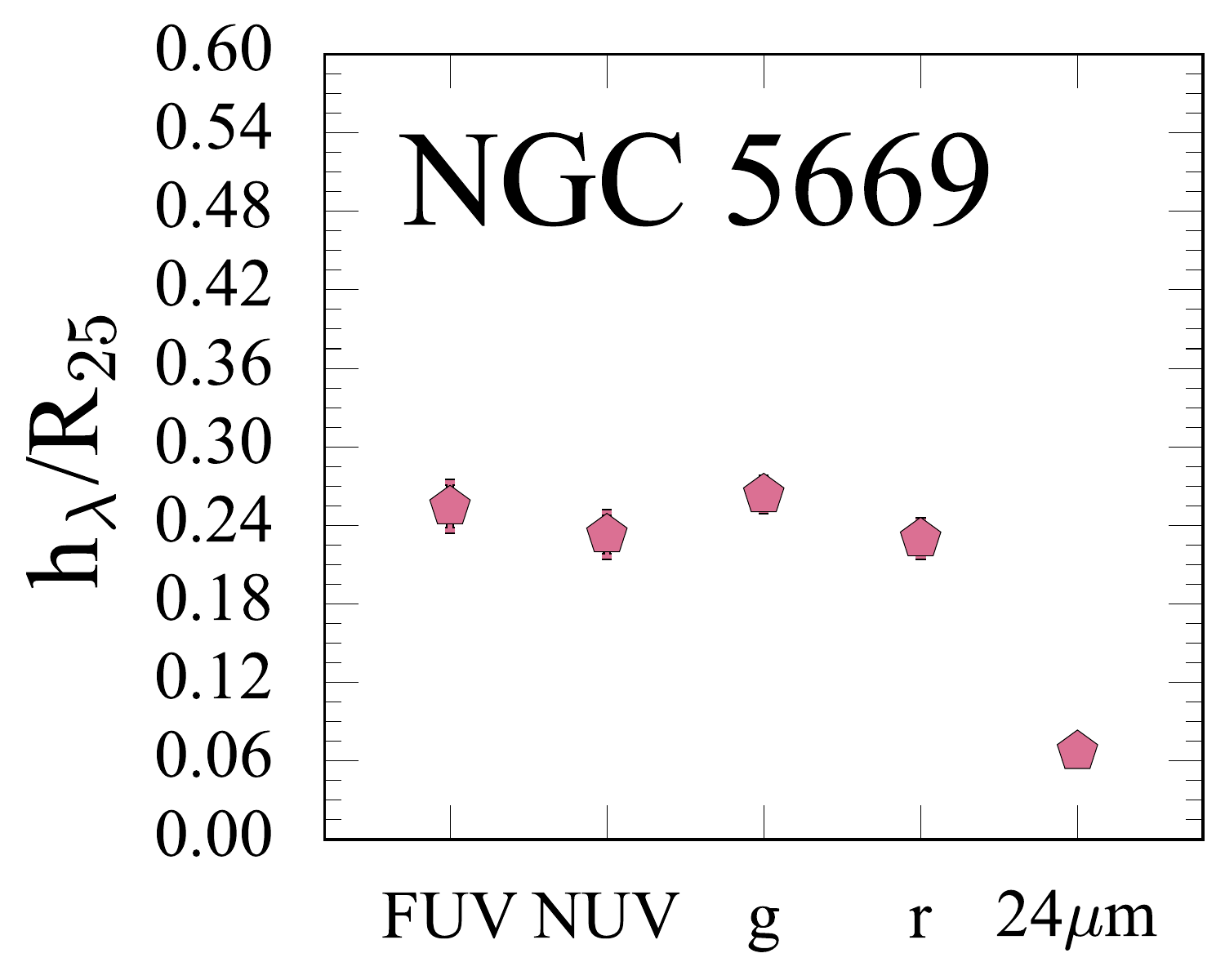}
\includegraphics[trim = 5mm 0.5cm 0cm 0cm, clip,scale=0.14]{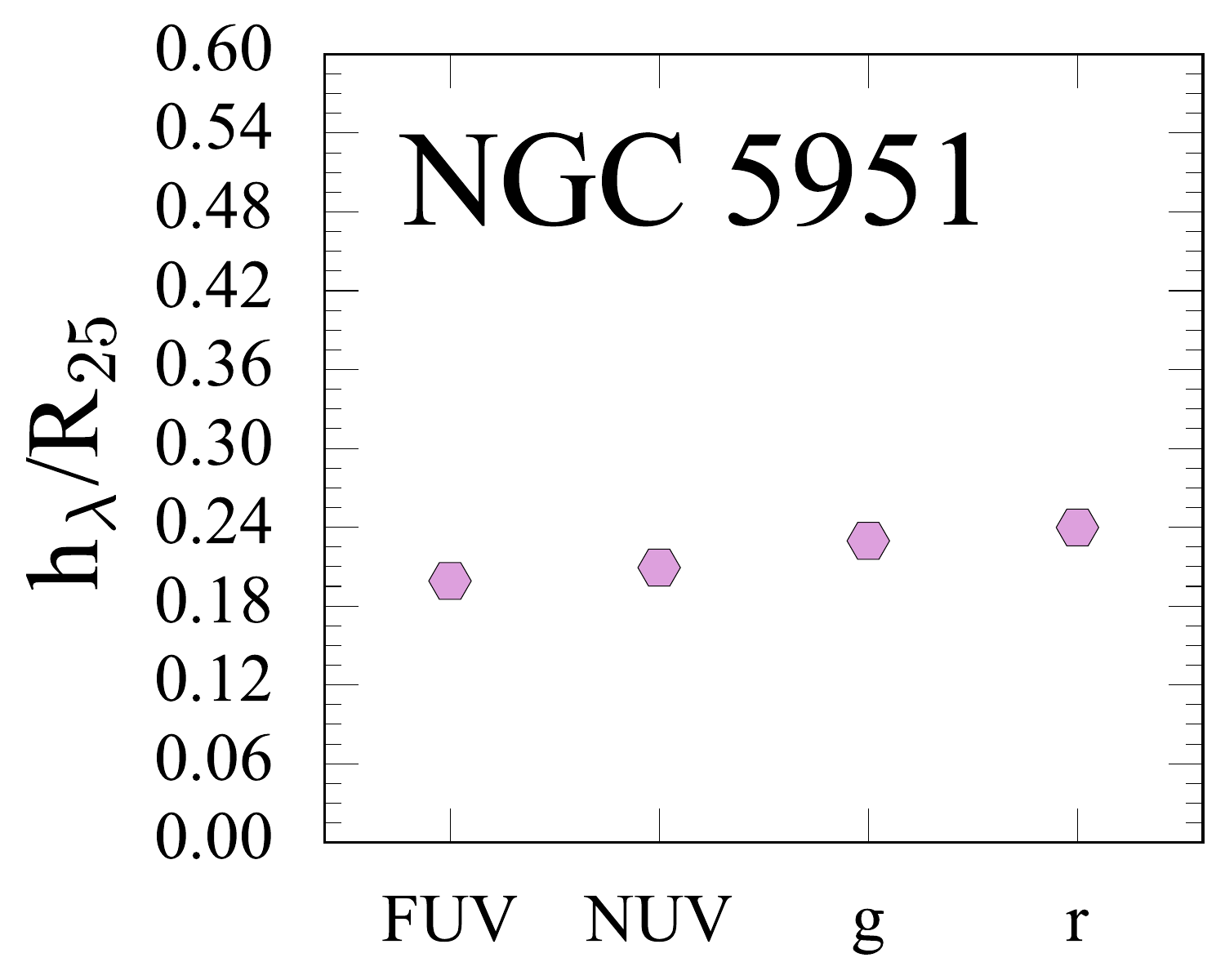}
\caption{Scale lengths in the FUV, NUV, \sdssg, \sdssr, and \mips\ normalized to R$_{25}$. The scale lengths are obtained by fitting an exponential function to the surface brightness profiles, excluding their central regions.} \label{fig:scale}
\end{figure*}

We fit an exponential curve to the surface brightness ($\mu$) profiles, $\mu = \mu_0 + 1.086(R/h_\lambda)$, where $\mu_0$ is the surface brightness at radius $R=0$ and $h_\lambda$ is the scale length \citep{devauc48}. The inner part of the radial profiles, which is dominated by emission from the bulge, was excluded from the fit. This bulge-dominated region was identified by guesstimating the bulge radius from the \sdssr-band map, followed by looking for a break in the \sdssr-band surface brightness profile near this region. Once we estimate the bulge radius, this value is then adopted for all profiles of a galaxy.
Figure \ref{fig:scale} shows the $h_\lambda/$R$_{25}$ measurements in FUV, NUV, \sdssg, \sdssr, and \mips\ of our galaxies. The R$_{25}$ values are obtained from a simulated B-band profile created using the \sdssg\ and \sdssr\ profiles and the transformation, $B=g+0.33(g-r)+0.20$ from \citep{jest05}. We do not make any scale length measurements for the \wise\ maps due to their poor resolution and shallow depths.
The estimated R$_{25}$ of the galaxies and $h_\lambda$ values are noted in Table \ref{tbl:scale}. Larger scale lengths imply an extended distribution and smaller values indicate a stronger concentration towards the center. In general, we find decreasing $h_\lambda$ toward longer wavelengths. For most galaxies, $h_{\rm FUV}>h_r$. 
Particularly, in NGC 3433, $h_{\rm FUV}/h_r\gtrsim2$ indicates that younger stars are twice as extended as older stellar populations. 
For NGC 4921, $h_{\rm FUV}\approx h_r$, and for NGC 4321, UGC 9120, and NGC 5951, $h_{\rm FUV}<h_r$. 

\begin{deluxetable*}{lccccccc}[!t]
\tablecaption{Exponential Scale-lengths, $h_\lambda$, expressed in kpc in different bands obtained from the Surface Brightness profiles. The Spearman rank correlation coefficient of each fit is noted in parentheses. \label{tbl:scale}}
\footnotesize
\tablehead{
\colhead{Galaxy} & \colhead{R$_{25}$}    & \colhead{$h_{FUV}$}  & \colhead{$h_{NUV}$}  & \colhead{$h_g$} & \colhead{$h_r$}  &  \colhead{$h_{24\mu m}$} \\
\colhead{}  & \colhead{(kpc)} & \colhead{(kpc)}  & \colhead{(kpc)} &  \colhead{(kpc)}  & \colhead{(kpc)} &  \colhead{(kpc)}} 
\startdata
 NGC 99  &  11.46  &  $4.88_{\pm0.22}$ (0.99)  &  $4.49_{\pm0.13}$ (1.0)  &  $3.5_{\pm0.1}$ (1.0)  &  $3.52_{\pm0.08}$ (1.0)  &  ...  \\
 NGC 3344  &  7.93  &  $2.24_{\pm0.08}$ (0.98)  &  $1.96_{\pm0.06}$ (0.98)  &  $1.73_{\pm0.03}$ (1.0)  &  $1.57_{\pm0.03}$ (1.0)  &  $0.55_{\pm0.02}$ (1.0)  \\
 NGC 3351  &  8.15  &  $2.1_{\pm0.13}$ (0.99)  &  $2.07_{\pm0.09}$ (0.99)  &  $1.86_{\pm0.08}$ (1.0)  &  $1.81_{\pm0.07}$ (1.0)  &  $0.46_{\pm0.02}$ (1.0)  \\
 NGC 3433  &  17.66  &  $9.58_{\pm0.93}$ (0.89)  &  $8.23_{\pm0.56}$ (0.99)  &  $5.13_{\pm0.14}$ (1.0)  &  $4.5_{\pm0.12}$ (1.0)  &  ...  \\
 NGC 3485  &  9.87  &  $2.55_{\pm0.16}$ (0.98)  &  $2.44_{\pm0.14}$ (0.99)  &  $2.18_{\pm0.07}$ (1.0)  &  $2.13_{\pm0.05}$ (1.0)  &  $0.54_{\pm0.02}$ (0.97)  \\
NGC 3666  &  4.9  &  $1.35_{\pm0.06}$ (0.98)  &  $1.29_{\pm0.04}$ (1.0)  &  $1.17_{\pm0.02}$ (1.0)  &  $1.23_{\pm0.03}$ (1.0)  &  $0.31_{\pm0.01}$ (1.0)  \\
NGC 3810  &  7.54  &  $1.79_{\pm0.1}$ (0.99)  &  $1.49_{\pm0.07}$ (0.99)  &  $1.55_{\pm0.07}$ (1.0)  &  $1.52_{\pm0.06}$ (1.0)  &  ... \\
NGC 4303  &  17.37  &  $3.22_{\pm0.19}$ (1.0)  &  $2.97_{\pm0.15}$ (1.0)  &  $3.19_{\pm0.09}$ (1.0)  &  $3.02_{\pm0.08}$ (1.0)  &  $0.74_{\pm0.02}$ (1.0)  \\
NGC 4321  &  12.85  &  $2.58_{\pm0.14}$ (0.98)  &  $2.6_{\pm0.12}$ (0.99)  &  $2.84_{\pm0.12}$ (1.0)  &  $2.95_{\pm0.11}$ (1.0)  &  $0.74_{\pm0.03}$ (1.0)  \\
 IC 3440  &  11.07  &  $5.35_{\pm0.5}$ (0.97)  &  $5.11_{\pm0.49}$ (0.99)  &  $3.7_{\pm0.14}$ (1.0)  &  $2.6_{\pm0.1}$ (1.0)  &  ...  \\
 NGC 4921  &  23.04  &  $4.97_{\pm0.38}$ (1.0)  &  $6.03_{\pm0.43}$ (1.0)  &  $5.67_{\pm0.19}$ (1.0)  &  $4.94_{\pm0.21}$ (1.0)  &  ...  \\
 UGC 9120  &  10.97  &  $2.06_{\pm0.08}$ (1.0)  &  $2.22_{\pm0.09}$ (0.99)  &  $2.34_{\pm0.18}$ (0.98)  &  $2.87_{\pm0.15}$ (1.0)  &  ...  \\
 NGC 5669  &  6.52  &  $1.66_{\pm0.12}$ (0.98)  &  $1.52_{\pm0.11}$ (0.99)  &  $1.72_{\pm0.08}$ (1.0)  &  $1.5_{\pm0.09}$ (1.0)  &  $0.44_{\pm0.03}$ (0.98)  \\
NGC 5951  &  6.88  &  $1.37_{\pm0.05}$ (0.98)  &  $1.44_{\pm0.05}$ (0.99)  &  $1.58_{\pm0.02}$ (1.0)  &  $1.65_{\pm0.03}$ (1.0)  &  ...  
\enddata
\end{deluxetable*}

\subsection{SFR, Stellar Mass, and HI Mass Radial Profiles}

\begin{figure*}[!t] 
\centering
\includegraphics[trim = 0.7cm 0.cm 0.5cm 0cm, clip,scale=0.165]{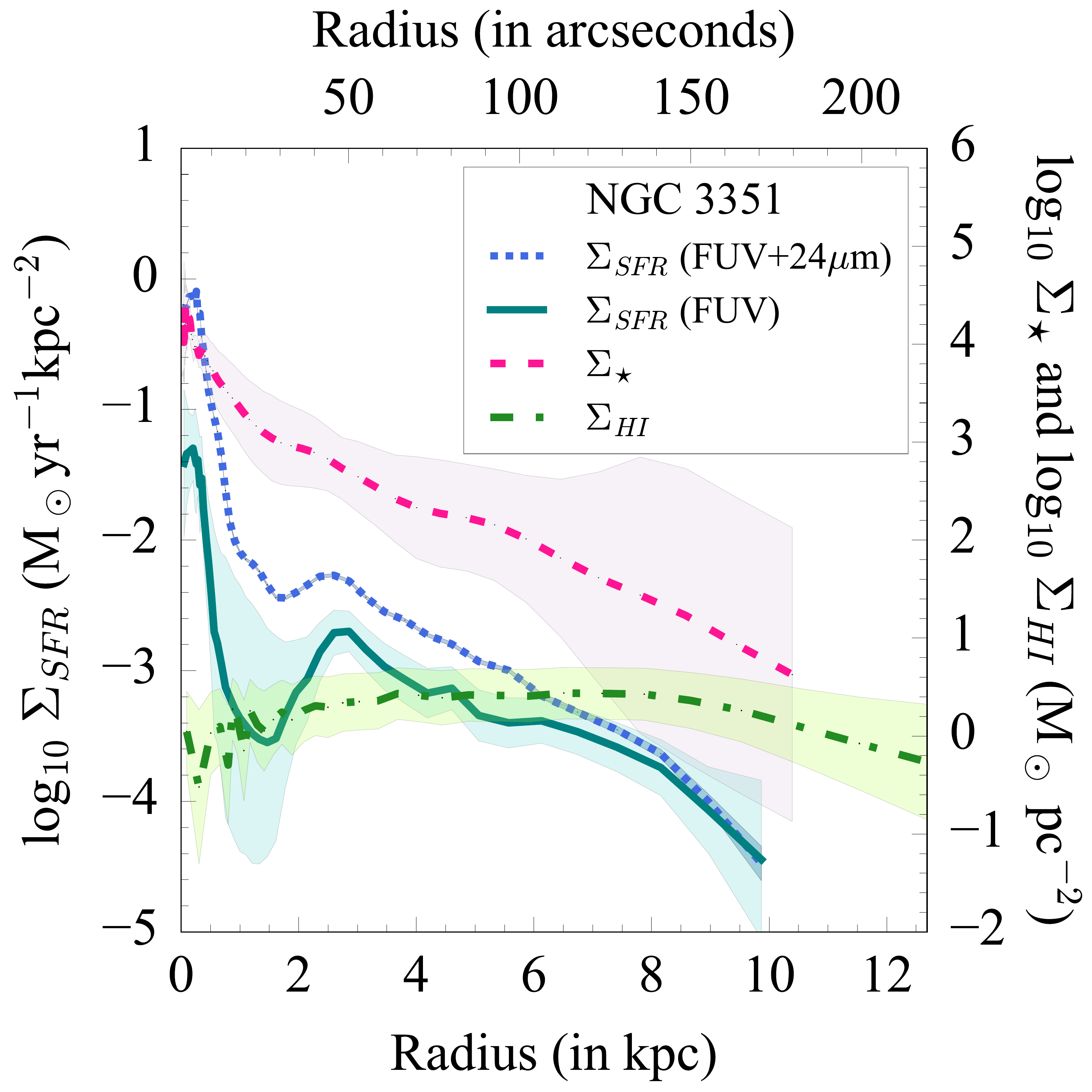}
\hspace{0.5cm}
\includegraphics[trim = 0.7cm 0.cm 0.5cm 0cm, clip,scale=0.165]{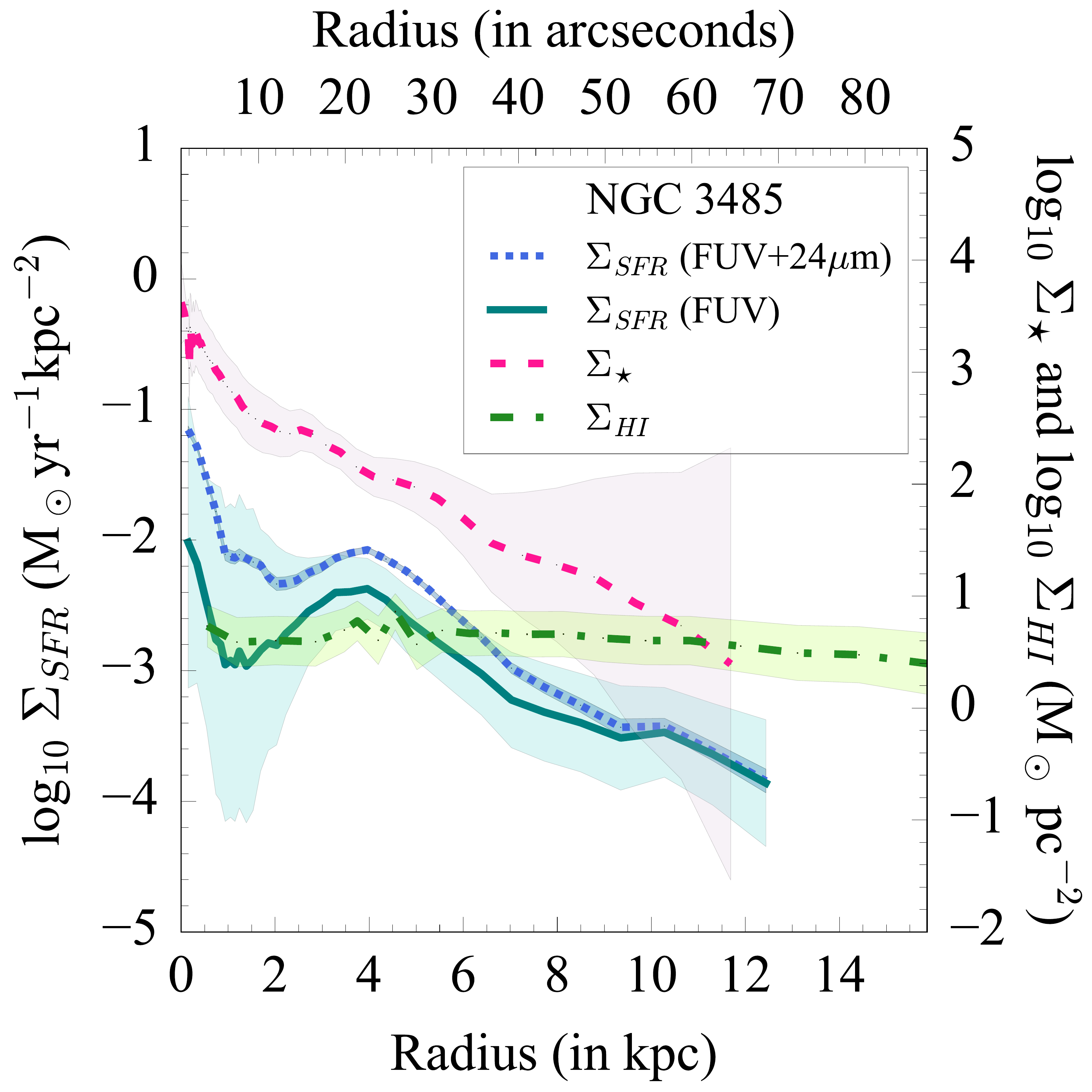}
\caption{Profiles of \sfr\ obtained from FUV and FUV+\mips\ (or FUV+\wise), \sm, and \mhi\ for the galaxies in our sample. The shaded region around each profile shows the 1$\sigma$ uncertainty in the respective quantities. The full figure set is shown in Appendix \ref{app:fig6}.}\label{fig:sigma_prof}
\end{figure*}

We also derive the radial profiles of \sfr, \sm, and \mhi\ using the surface brightness profiles and the prescriptions provided in Section \ref{sec:it}. Figure \ref{fig:sigma_prof} shows the radial variation in \sfr, \sm, and \mhi. 
The dust-corrected values of \sfr\ are systematically higher than the unobscured \sfr\ derived from FUV validating the inclusion of the mid-infrared data. The \sm\ profiles show a smooth decline with radius in all galaxies. 
The minor variations that come across as breaks in the profile could be due to asymmetries in the galactic structure in the outer parts of galaxies. The \mhi\ profile is seen to stay relatively flat over the stellar disk, reflecting, to a significant extent, the selection of the sample. 

\begin{figure*}[!ht] 
\centering
\includegraphics[trim =  0.cm 0.cm 0.cm 0cm, clip,scale=0.165]{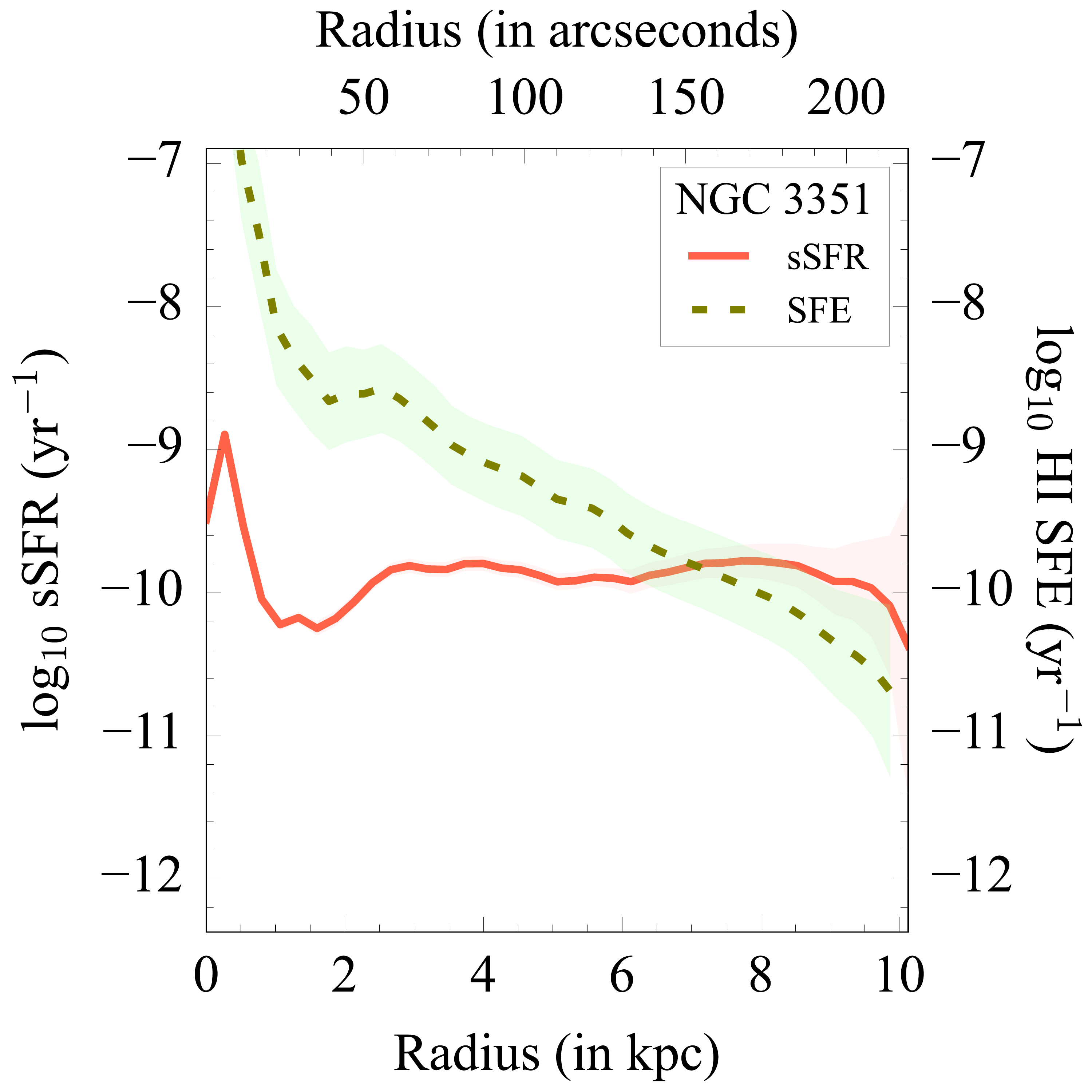}
\hspace{0.5cm}
\includegraphics[trim =  0.cm 0.cm 0.cm 0cm, clip,scale=0.165]{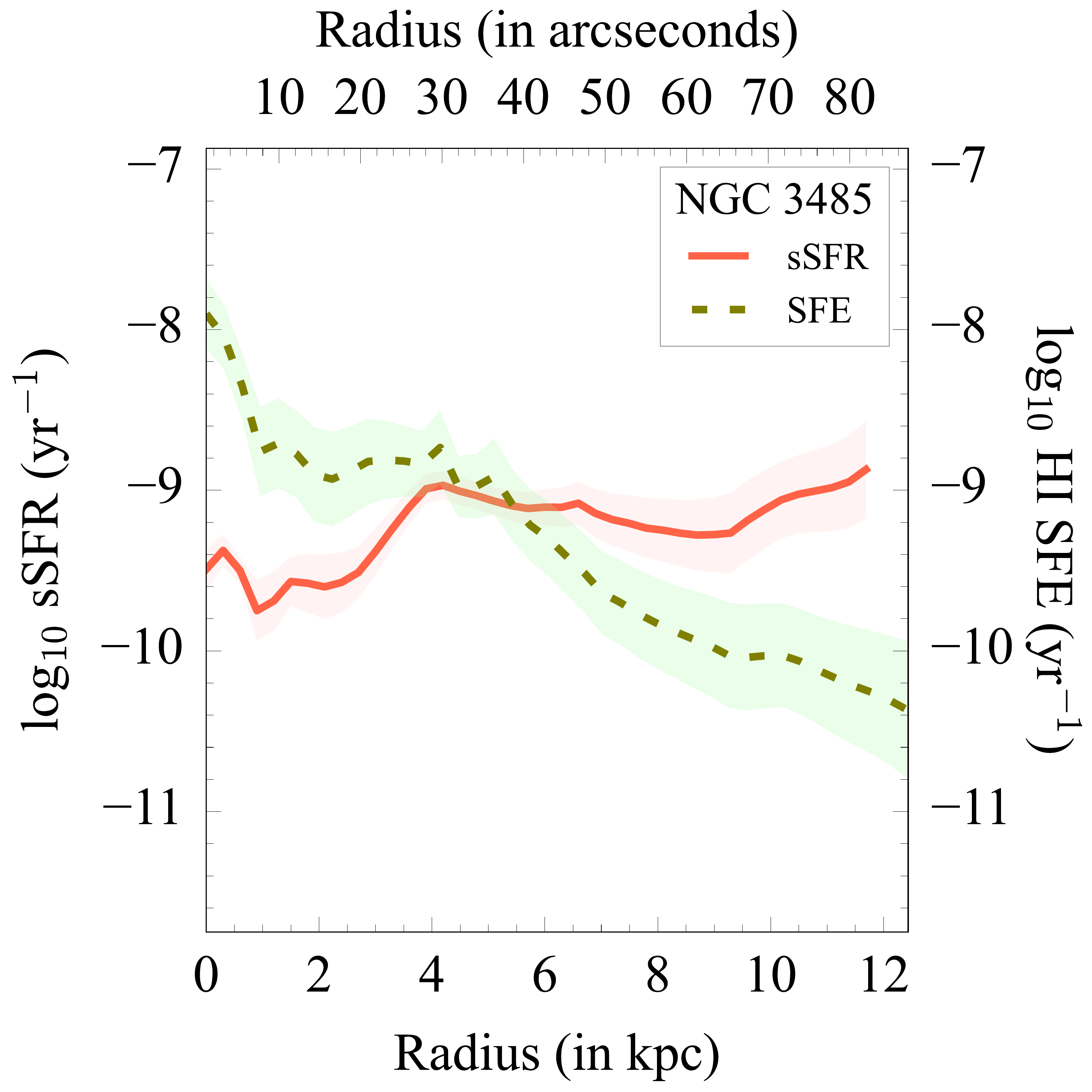}

\caption{Radial variation in specific SFR and \hi\ star formation efficiency for the galaxies in our sample. The shaded region around each profile shows the 1$\sigma$ uncertainty in the respective quantities. The full figure set is shown in Appendix \ref{app:fig7}.}\label{fig:ssfr_prof}
\end{figure*}

Using these profiles, we are able to derive the radial variations in sSFR (star formation rate per unit stellar mass) and \hi-based star formation efficiency (SFE; star formation rate per unit \hi\ mass). Due to the immense variation in the resolution of our \hi\ data, we do not convolve the \sfr, \sm, and \mhi\ profiles to a fixed resolution. Instead, we perform a 1D interpolation on the \sfr, \sm, and \mhi\ profiles and then measure the radial gradients in sSFR and SFE.  
We do not expect this to affect the overall radial gradients due to the overall averaging of fluxes. 
The sSFR and SFE profiles for our galaxies are estimated using the dust-corrected \sfr\ measurements, shown in Figure \ref{fig:ssfr_prof}. The sSFR profiles extend as far as the combined range of the \sm\ and \sfr\ profiles. In some cases, the discontinuation of \sm\ profiles occurs prior to that of \sfr\ profiles, particularly evident in NGC 3666 and IC 3440. 
Hence, we caution the reader that we are not always probing the full extent of the sSFR distribution in these galaxies.  

In general, the sSFR profile provides an excellent opportunity to investigate the contributions of young and old stars and it has been well established that the gradient of the sSFR profile is an indicator of both the growth of the stellar disk and its direction  \citep{muno07}. 
Across our sample, we find that the sSFR profiles exhibit various trends. While NGC~3344, NGC~3433, and IC 3440 show a steady increase in the sSFRs with radius, NGC~4321 shows a steady decline. A positive gradient in the radial profile provides evidence for inside-out disk growth while a negative gradient could be a result of outside-in formation, affected by the galaxy's environment. The sSFR profile for the remaining galaxies either stays relatively flat or shows pronounced breaks. 
Along with NGC~4321, a decline in the sSFR profiles is seen in NGC~3666 and UGC 9120, although both galaxies show breaks in their profiles.  

The \hi-based SFE profiles, however, show a steady drop with galactocentric radii for our galaxies, generally extending as far as the \sfr, except for NGC~4921. The decline suggests a radial increase in depletion time (\tdep~$\propto$~SFE$^{-1}$), which is the time taken by the current rate of star formation to consume the available supply of gas. We note that the peculiarity of the SFE profile of NGC~4921 is a result of its environment. While radial profiles inherently assume a symmetric distribution, the \hi\ disk of NGC~4921 is found to be truncated and asymmetric \citep{kenn15}, as a consequence of ram pressure stripping of its ISM due to the dense cluster environment. 

\begin{figure}[!t] 
\centering
    \includegraphics[trim = 0.0cm 0cm 0cm 0cm, clip,scale=0.3]{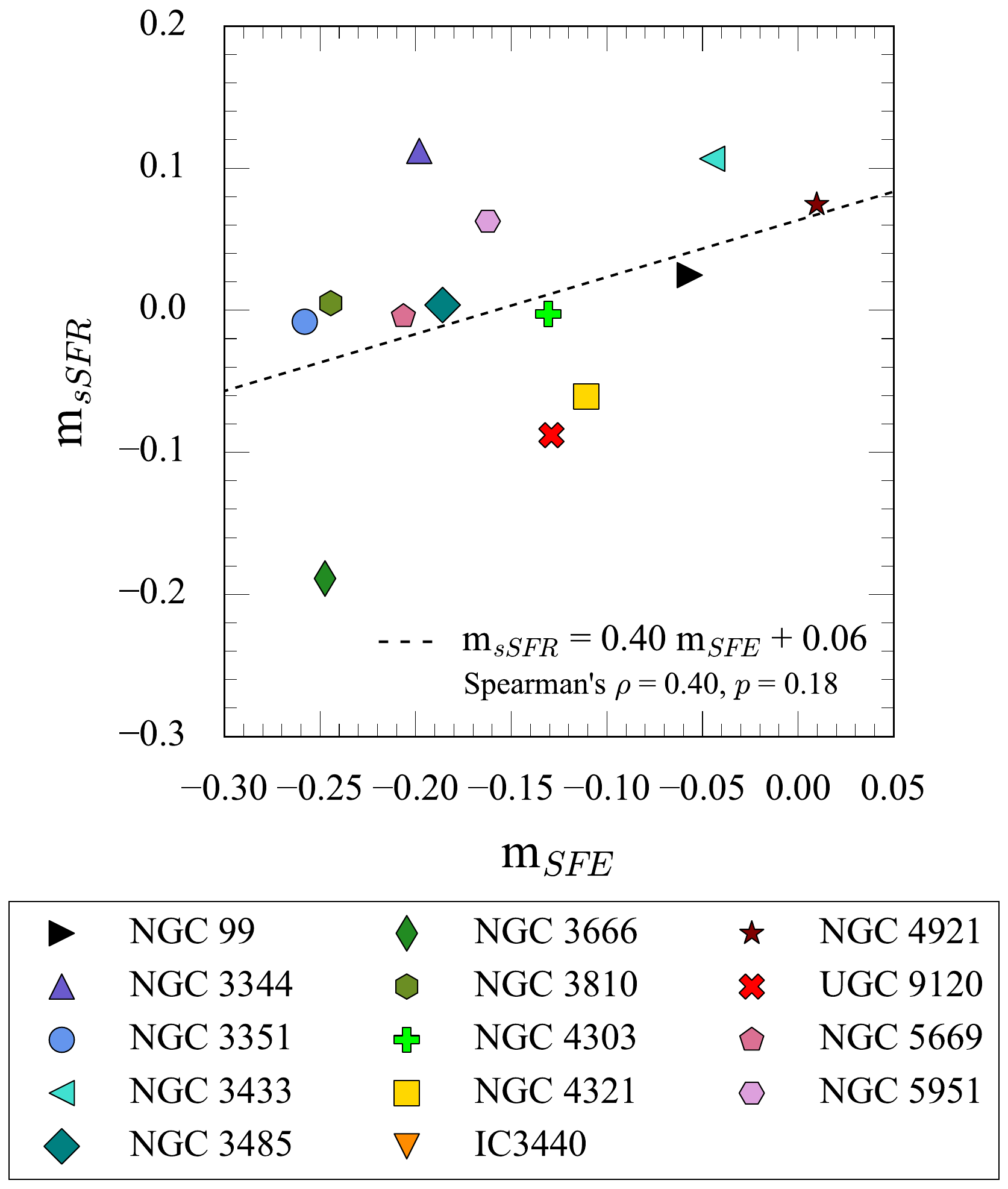}
 \caption{Slopes of the sSFR (m$_{sSFR}$) and SFE (m$_{SFE}$) profiles for the galaxies. The slopes are estimated by fitting a straight line to the corresponding radial profiles, excluding the bulge. The dashed line represents the line of best fit to the points. The equation defining the best-fit line is noted in the panel. The observed positive correlation links the direction of disk growth to the cold gas in the ISM and how efficiently it is being used to form stars.}
    \label{fig:slope}
\end{figure}

We measure the slopes of the sSFR and SFE profiles, m$_{sSFR}$ and m$_{SFE}$, respectively, by fitting a straight line function and excluding the bulge-dominated inner part in the fit. The m$_{sSFR}$, m$_{SFE}$ and intercepts, b$_{sSFR}$ and b$_{SFE}$, of the profiles along with the Spearman's rank correlation coefficient ($\rho$) are noted in Table \ref{tbl:slopes}. For each fit, we find $p<0.001$. In Figure \ref{fig:slope}, we plot m$_{sSFR}$ and m$_{SFE}$ of our galaxies. 
While the sSFR profiles show both positive and negative gradients, m$_{SFE}$ generally have negative values, except for NGC~4921. We find a positive relation between the slopes (Spearman's $\rho=0.4$), showing a flattening of the SFE profiles with increasing sSFR gradients. This flattening would suggest an almost constant \tdep\ in galaxy disks exhibiting a higher young-to-old star ratio in their outer parts. However, the slopes show a large scatter, and the trend is observed to be statistically insignificant ($p>0.1$). 

\begin{deluxetable*}{lcccccccc}[!ht]
\tablewidth{0pt}
\tablecaption{Slopes, $m$, and intercepts, $b$, of the sSFR and SFE profiles. The Spearman rank correlation coefficient ($\rho$) for each fit is noted in parentheses. \label{tbl:slopes}}
\tablehead{
 \colhead{Galaxy} & \colhead{$m_{sSFR}$}    & \colhead{$b_{sSFR}$} &   \colhead{Spearman's $\rho$} & \colhead{$m_{SFE}$}  & \colhead{$b_{SFE}$} & \colhead{Spearman's $\rho$}}
\startdata
NGC 99  &  $0.025_{\pm0.0003}$  &  $-9.082_{\pm0.003}$ & 0.54 & $-0.057_{\pm0.0003}$  &  $-8.443_{\pm0.003}$ & -0.77\\
NGC 3344  &  $0.112_{\pm0.0004}$  &  $-9.684_{\pm0.003}$ & 0.94 & $-0.198_{\pm0.0004}$  &  $-8.63_{\pm0.003}$ &-0.98\\
NGC 3351  &  $-0.008_{\pm0.0003}$  &  $-9.806_{\pm0.002}$ &  -0.16 & $-0.258_{\pm0.0002}$  &  $-7.999_{\pm0.002}$ & -0.99\\
NGC 3433  &  $0.107_{\pm0.0001}$  &  $-10.095_{\pm0.002}$ &  0.99 & $-0.045_{\pm0.0001}$  &  $-8.939_{\pm0.001}$ & -0.99\\
NGC 3485  &  $0.004_{\pm0.0005}$  &  $-9.147_{\pm0.004}$ &  0.06 & $-0.186_{\pm0.0004}$  &  $-8.181_{\pm0.003}$ & -0.99\\
NGC 3666  &  $-0.189_{\pm0.0013}$  &  $-9.066_{\pm0.007}$ & -0.76 &  $-0.247_{\pm0.0008}$  &  $-8.968_{\pm0.005}$ & -0.99\\
NGC 3810  &  $0.005_{\pm0.0006}$  &  $-8.96_{\pm0.003}$ &  0.21 & $-0.244_{\pm0.0005}$  &  $-8.137_{\pm0.004}$ & -0.99\\
NGC 4303  &  $-0.003_{\pm0.0003}$  &  $-9.235_{\pm0.003}$ & -0.12 & $-0.131_{\pm0.0002}$  &  $-7.972_{\pm0.003}$ & -0.99\\
NGC 4321  &  $-0.061_{\pm0.0001}$  &  $-8.525_{\pm0.001}$ & -0.97 & $-0.111_{\pm0.0003}$  &  $-8.277_{\pm0.003}$ & -0.93\\
IC 3440  &  $0.397_{\pm0.001}$  &  $-9.95_{\pm0.005}$ & 0.99 & $-0.089_{\pm0.0002}$  &  $-9.054_{\pm0.001}$ & -0.99\\
NGC 4921  &  $0.075_{\pm0.0003}$  &  $-11.726_{\pm0.007}$ & 0.78 & $0.01_{\pm0.0002}$  &  $-8.922_{\pm0.002}$ & 0.48\\
UGC 9120  &  $-0.088_{\pm0.0005}$  &  $-9.887_{\pm0.004}$ & -0.89& $-0.129_{\pm0.0002}$  &  $-8.024_{\pm0.002}$ & -0.99\\
NGC 5669  &  $-0.004_{\pm0.0012}$  &  $-8.862_{\pm0.007}$ &  -0.13&$-0.206_{\pm0.0007}$  &  $-8.865_{\pm0.004}$ & -0.97\\
NGC 5951  &  $0.063_{\pm0.0002}$  &  $-9.45_{\pm0.001}$ & 0.99& $-0.162_{\pm0.0004}$  &  $-9.045_{\pm0.002}$ & -0.99
\enddata
\end{deluxetable*}
 Assuming the trend in Figure \ref{fig:slope} to be plausible, we could speculate that if galaxies stabilize their \hi\ consumption and maintain a constant \tdep, their star formation then is likely proceeding in the outward direction, i.e., the galaxies are evolving inside-out. In nearby spiral galaxies, molecular gas shows a constant $\tau_{dep}\sim2$~Gyr \citep{lero08, bigi11} highlighting the importance of H$_2$ for star formation. The \hi\ SFEs, however, tend to be significantly lower in the outer parts than in the inner parts. This inefficiency is likely being driven by the low molecular gas content in the \hi-dominated phase \citep{lero08, krum09, rafel16}.
In this regard, \cite{lero08} observed a positive relation between SFE and \sm\ in \hi-dominated regions of nearby galaxies. This could imply that stellar potential well/feedback can allow for the conversion of \hi\ to H$_2$ -- corroborating our observation in Figure \ref{fig:slope} or simply hinting towards the current star formation following past star formation.

\subsection{Structural Properties of SFR and Stellar Mass}\label{sec:radii}
\begin{figure*}[!t] 
 \centering
\includegraphics[trim = 0mm 0cm 9cm 0cm, clip,scale=0.3]{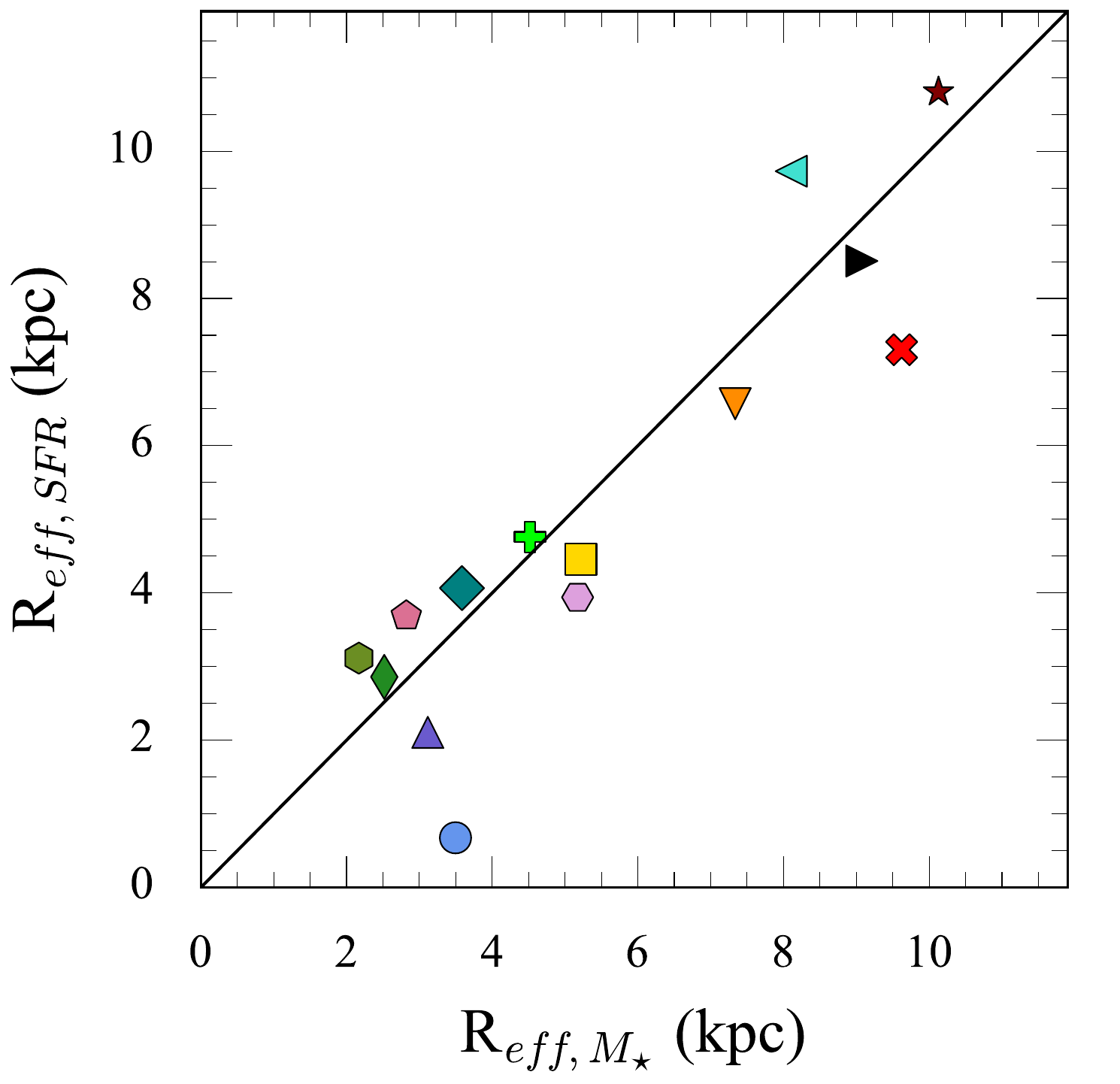}
\includegraphics[trim = 0mm 0cm 0cm 0cm, clip,scale=0.3]{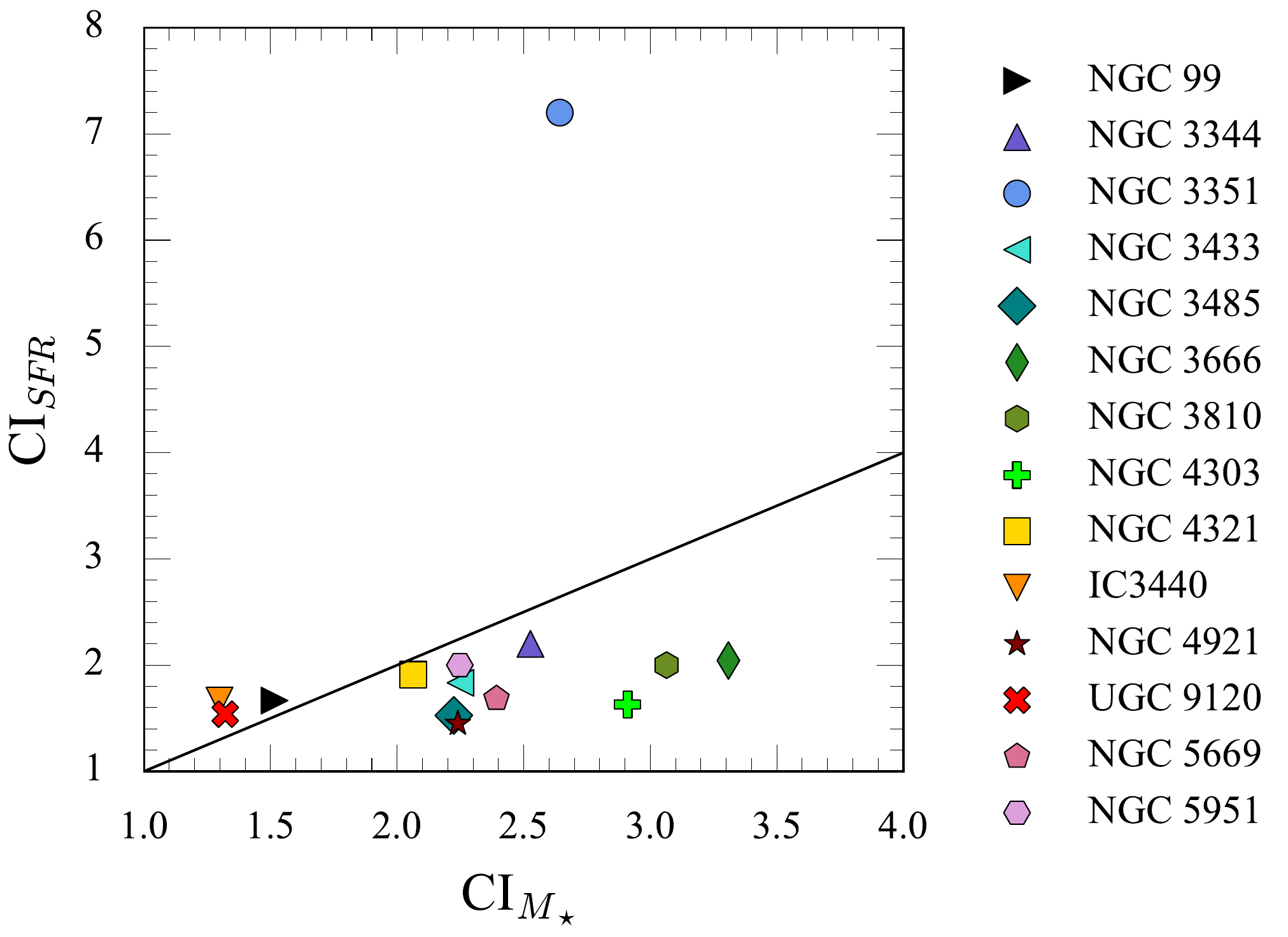}
 \caption{{\it (Left)} Effective radii, \re\ (or R$_{50}$), and {\it (Right)} the concentration index (\ci) of SFR and \mstar\ distribution for our galaxies. The black solid line denotes a one-to-one correspondence between the quantities. The majority of the galaxies show $\text{CI}_{SFR}<\text{CI}_{M_\star}$ indicative of star formation activity that is more spread out than the stellar mass already in place.}
 \label{fig:morph}
\end{figure*}

In the left panel of Figure \ref{fig:morph}, we show the effective radii of the SFR and \mstar\ maps for the galaxies. The effective radius is defined as the radius which contains 50\% of the galaxy's total emission for a particular wavelength or band. 
In our case, we define \re $_{,SFR}$ and \re $_{,M_\star}$ as the effective radii of the SFR and \mstar\ maps, respectively. The \re $_{,SFR}$ and \re $_{,M_\star}$ are also shown in Figure \ref{fig:galuv} and Figure \ref{fig:opt}, respectively, as the dashed elliptical apertures marking the area containing 50\% of the galaxy's total SFR and \mstar. The values are noted in Table \ref{tbl:param}. 

In general, galaxies with \re $_{,SFR}>$ \re $_{,M_\star}$ should have an extended star formation distribution indicating that growth is propagating outwards. However, effective radii are susceptible to intrinsic factors affecting star formation and its measurement. This can be seen as the larger deviations of galaxies from the one-to-one line toward the \re $_{,M_\star}$ side in Figure \ref{fig:morph} while the ones with an extended SFR distribution show \re $_{,SFR}\lesssim1.5\times$\re $_{,M_\star}$. 
This is most likely a result of estimating \re $_{,SFR}$ from the dust-corrected SFR maps.
Dust, which is concentrated in the center, adds a significant contribution to the SFR in the inner parts of the galaxy, reducing \re $_{,SFR}$. A concentrated SFR distribution could also result from a starburst activity in the center. The lower \re $_{,SFR}$ in NGC 3351 is the consequence of a bar-driven stellar activity in the circumnuclear region \citep{coli97, guil07, lea19}. In such cases, information on what is happening at large radii can be lost in the \re $_{,SFR}$-\re $_{,M_\star}$ measurements. Galaxies can still be actively forming stars in the outer regions, albeit at low levels, and not make a dominant contribution to the global SFR. 
For example, in the left panel of Figure \ref{fig:morph}, NGC 3344 is seen to have \re $_{,SFR}<$\re $_{,M_\star}$. However, the galaxy hosts an XUV disk which accounts for $\lesssim$15\% of the total SFR \citep{padave21}. 

To get a clear picture of the structure in SFR and \mstar\ at large radii, we examine the concentration indices of the respective maps in the right panel of Figure \ref{fig:morph}. The concentration index (CI) is measured as the ratio of radii containing 90\% and 50\% of the galaxy's total SFR or \mstar. Elliptical aperture containing 90\% of galaxy's total SFR and \mstar\ are marked in Figure \ref{fig:galuv} and Figure \ref{fig:opt}, respectively. The concentration indices are also noted in Table \ref{tbl:param}.
Interestingly, the values of CI$_{SFR}$ are quite similar for our galaxies and show less variation than CI$_{M_\star}$ demonstrating that star formation is generally more extended than the stellar mass distribution for the majority of the sample. We find that most galaxies show CI$_{M_\star}>$~CI$_{SFR}$ except NGC~99, NGC~3351, IC 3440, and UGC 9120. The significantly high CI$_{SFR}$ seen in NGC~3351 can be attributed to the central starburst, as discussed earlier. 

The extended distribution of young stellar populations traced by the concentration indices, scale lengths, and the radial profiles of \sfr\ and \mstar, together provide clues to the direction in which stellar growth may be occurring. 
While most galaxies in our sample show stellar activity proceeding toward the outer regions, some show signs of outside-in evolution. To investigate further, we quantify stellar disk growth in the next section. 

\newpage
\section{Discussion}\label{sec:discussion}

We have analyzed the SFR, \mstar, and \hi\ distribution of 14 low-z galaxies and see stellar growth occurring in the outskirts for most. Following \cite{pezz21}, we quantify the stellar disk growth as \dsfr/sSFR, where the numerator $\Delta_{sSFR}=\text{sSFR}_{out}-\text{sSFR}_{in}$ for which sSFR$_{out}$ and sSFR$_{in}$ are sSFR estimates outside and inside \re $_{,M_\star}$, respectively, and the denominator is the galaxy-integrated sSFR. 
The inner radius of \re $,_{M_\star}$ essentially divides the galaxy's old stellar population into two equal zones, allowing us to objectively probe star formation activity. Consequently, $\Delta_{sSFR}>0$ would suggest stellar growth propagating in the outward direction while $\Delta_{sSFR}<0$ indicates enhanced SFR in the inner disk and/or suppressed star formation in the outskirts. 
In our sample, we find that most galaxies show sSFR$_{out}\gtrsim$~sSFR$_{in}$. UGC 9120 shows the lowest value of  \dsfr/sSFR$\approx-1.04$. Jointly with the negative sSFR gradient, it is likely that this galaxy may be evolving outside-in. We note that the \dsfr\ measurements can be affected by the active galactic nuclei as it would bias central concentration measurements, and the presence of strong bars in galaxies could alter the sSFR signatures. These were not investigated and lie beyond the scope of this study. The sSFR$_{in}$ and sSFR$_{out}$ values are tabulated in Table \ref{tbl:param}.

In this section, we will scrutinize how \dsfr/sSFR is associated with cold gas in the ISM and CGM and how stellar growth is affected by a galaxy's environment.  

\subsection{Connection of Disk Growth with Cold Gas in the ISM \& GCM}\label{sec:hi}

\begin{figure*}[!ht]
\includegraphics[trim = 0cm 0cm 9cm 0cm, clip,scale=0.29]{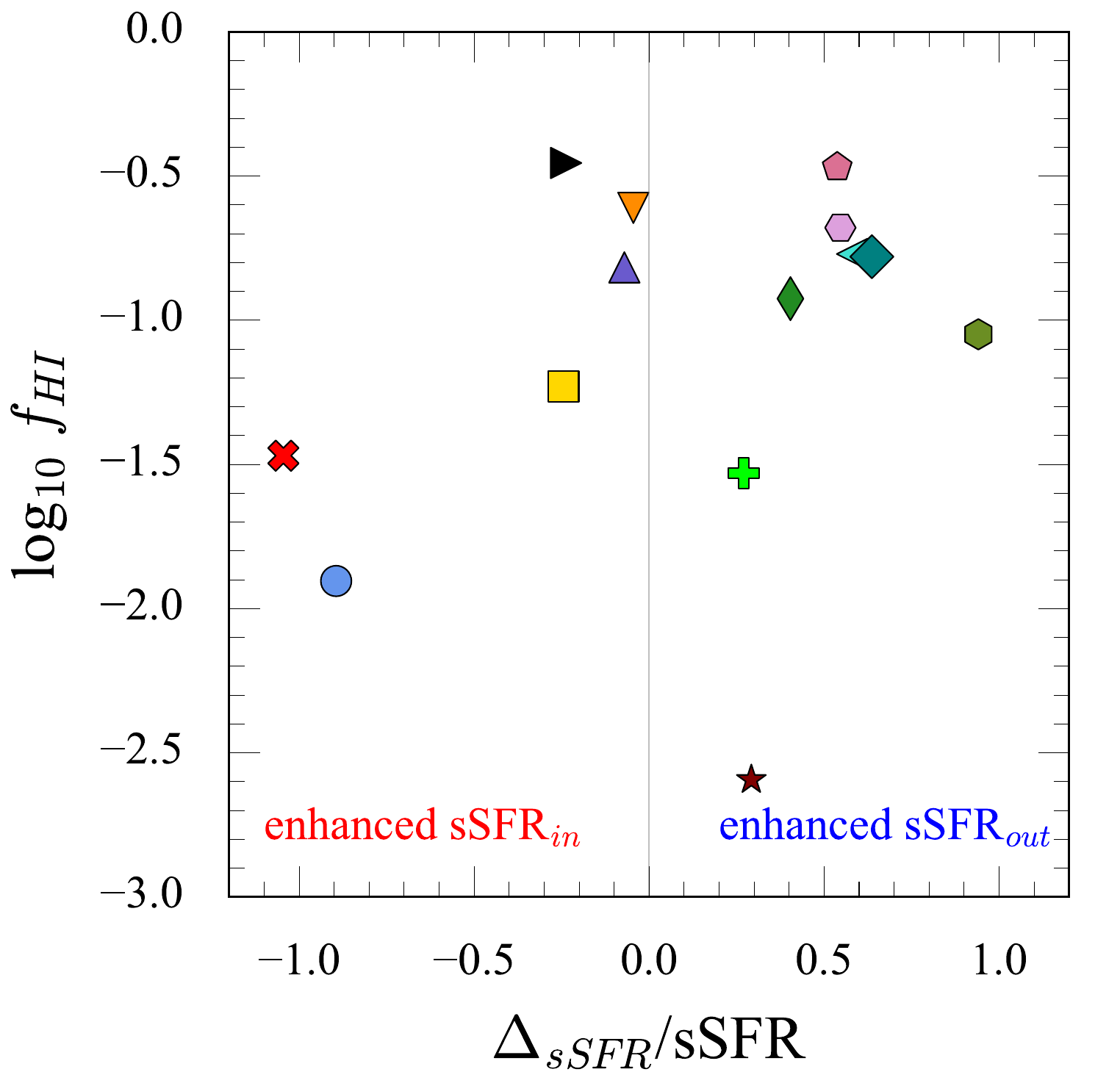}
\includegraphics[trim = 0cm 0cm 0cm 0cm, clip,scale=0.29]{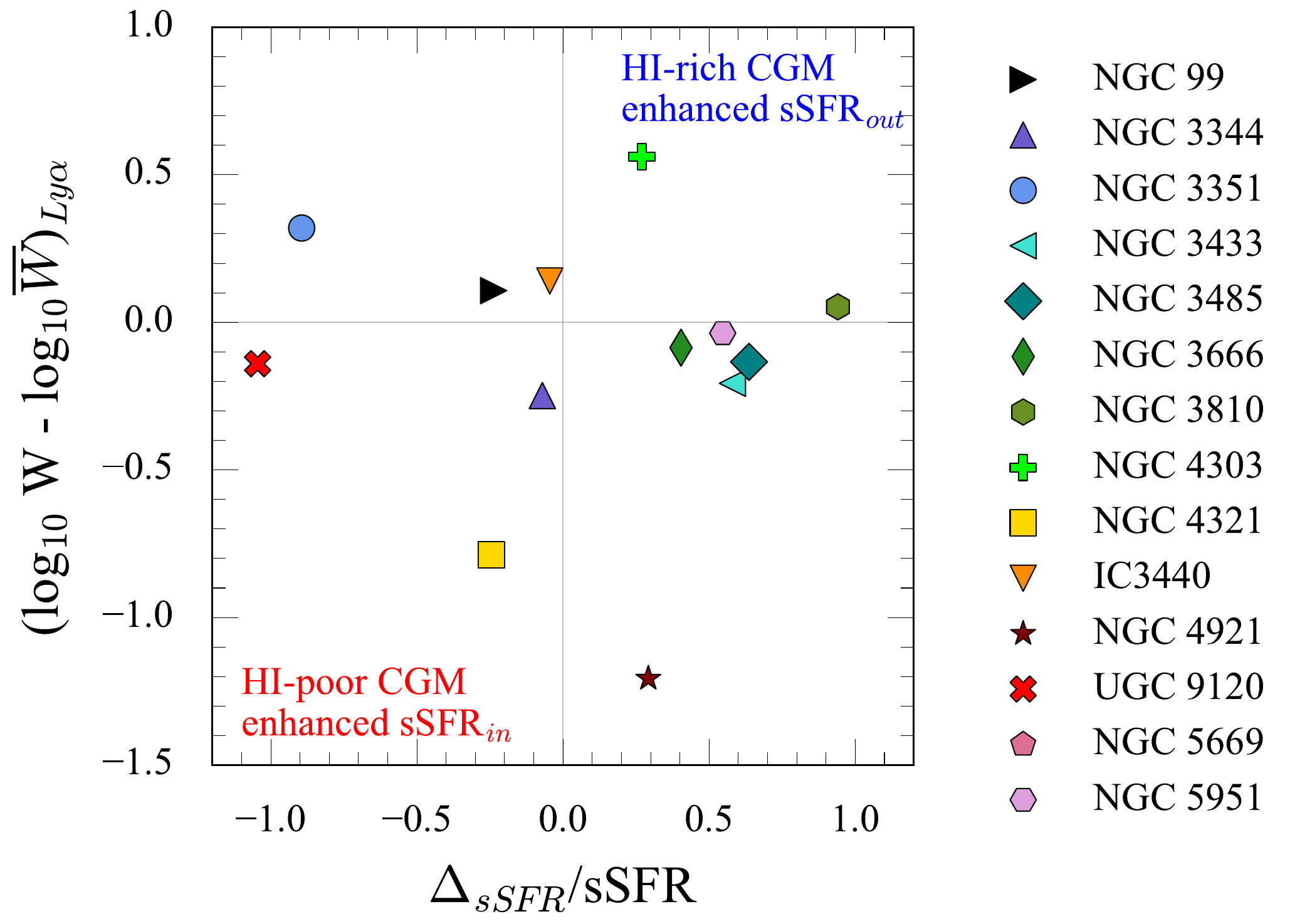}
\caption{The \hi\ gas fractions {\em{(left)}} and W$_{Ly\alpha}$ excess {\em{(right)}} for the galaxies plotted as a function of \dsfr/sSFR. The solid lines denote \dsfr~$=0$ and $\log_{10}~\text{W}=\log_{10}~\overline{W}$.}\label{fig:gf_ssfr}
\end{figure*}

In the left panel of Figure \ref{fig:gf_ssfr}, we show the relation between \dsfr/sSFR and the \hi\ gas fraction, $f_{HI}=$~\higas/(\higas$+$\mstar). 
All galaxies have moderate $f_{HI}\gtrsim0.01$ except the lowest value of $f_{HI}=2.5\times10^{-3}$ for NGC~4921 by virtue of its location in the Coma cluster and stripped ISM. We observe a weak positive correlation (Spearman's $\rho=0.30$)  between \dsfr/sSFR and $f_{HI}$ for our galaxies with a 1$\sigma$ scatter of 0.56 dex. 
\cite{wang11} had observed a positive trend between sSFR$_{out}$/sSFR$_{in}$ and $f_{HI}$ for \hi-rich galaxies at $0.025<z<0.05$ with \mstar~$>10^{10}$~M$_\odot$. However, this relation was studied for different stellar mass bins and the positive trend differed for each bin. Meanwhile, our sample spans a larger stellar mass range and no binning is performed due to the small size of the sample. The observed scatter is likely a sample selection effect.

While the increase in sSFR with higher $f_{HI}$ is expected \citep{coldism22}, a positive correlation between $f_{HI}$ and \dsfr/sSFR accentuates the direction in which star formation is likely to propagate. Galaxies with higher $f_{HI}$ tend to have higher young-to-old star ratios in their outskirts -- an indication of a rapidly growing outer disk. 
Besides, a positive correlation can also be thought of as a semblance of the correlation between \hi\ and SFR in the outer parts of a galaxy. In the \hi-dominated low-star forming outer parts, \mhi\ and \sfr\ are known to be correlated \citep{bigi10a,padave21}, while the correlation is lacking in the H$_2$-dominated inner parts \citep{bigi08}. As a result, $f_{HI}$ would correlate with the outer SFR, and subsequently, with \dsfr/sSFR. 
Studying a larger sample size may expound on these observed relations, indicating that the increase in the radial sSFR gradient, and thereby, the direction of disk growth can be linked to the cold gas in the ISM and how efficiently it is being used for star formation. 

The right panel of Figure \ref{fig:gf_ssfr} shows the  W$_{Ly\alpha}$ excess, described in Section \ref{sec:cos}, of the detected absorbers in the CGM of our galaxies and disk growth probed by \dsfr/sSFR. We again bring to the reader's attention that the W$_{Ly\alpha}$ excess takes into account the inhomogeneous radial distribution of the impact parameter in the sample. 
Our observations show no dependence between W$_{Ly\alpha}$ excess and \dsfr/sSFR. Additionally, we also do not see W$_{Ly\alpha}$ excess to show any relation with sSFR and $f_{HI}$ for our galaxies. In general, the W$_{Ly\alpha}$ excess is known to strongly correlate with sSFR and $f_{HI}$ \citep{bort15, bort16}. These were seen for the combined COS-GASS+COS-Halos and COS-GASS samples, respectively. 
It should be noted that the selection criteria for both the COS-GASS and COS-Halos samples are statistically motivated, probing stellar masses between $10^{10-11}$~M$_\odot$ with redshifts between $0.025\gtrsim z\gtrsim 0.05$ and $0.1\gtrsim z\gtrsim 0.2$, respectively, while the DIISC sample consists of galaxies with known \hi-disk size and QSO sightlines within $\sim3.5\times {\rm R}_{HI}$, i.e., closer to the disk-halo interface. Hence, the inconsistency between our observation and those from \cite{bort15, bort16} may not really be a contradiction and can be attributed to a small sample size along with the differences in sample selection. 
As a result, we infer that the enhancement and suppression of neutral \hi\ gas in the CGM has no effect on the direction in which star formation proceeds in a galactic disk or vice-versa.

\subsection{Connection of Disk Growth with Environment}\label{sec:env}

The local environment also affects a galaxy's ability to sustain cold gas, as well as star formation \citep{peng10, muzz12, peng15, behn15, behn16, stein16, coen19}. 
In this section, we investigate how various properties of our galaxies depend on the environment. 
We make use of the NYU Value-Added Galaxy Catalog \citep[NYU-VAGC]{blant05} derived from SDSS Data Release 7 \citep{sdssdr7} to define a magnitude-limited sample of galaxies to study the environment. We consider sources with M$_r\leq-15.97$ which were chosen based on the faintest source in the catalog at the farthest redshift of ${\rm z}=0.026$ in our sample. Subsequently, we search for the nearest neighbors to our galaxies that have relative velocities, $|v_{neighbor} - v_{galaxy}| \leq 300$~\kms. We then calculate the galaxy's projected distance to its first and fifth nearest L$_\star$ neighbor, d$_{NN_1}$ and d$_{NN_5}$, respectively. We note the d$_{NN_1}$ and d$_{NN_5}$ in Table \ref{tbl:param}. 

\begin{figure*}[!ht] 
\centering
\hspace{-1cm}
\includegraphics[trim = 0cm 0cm 9.5cm 0cm, clip,scale=0.29]{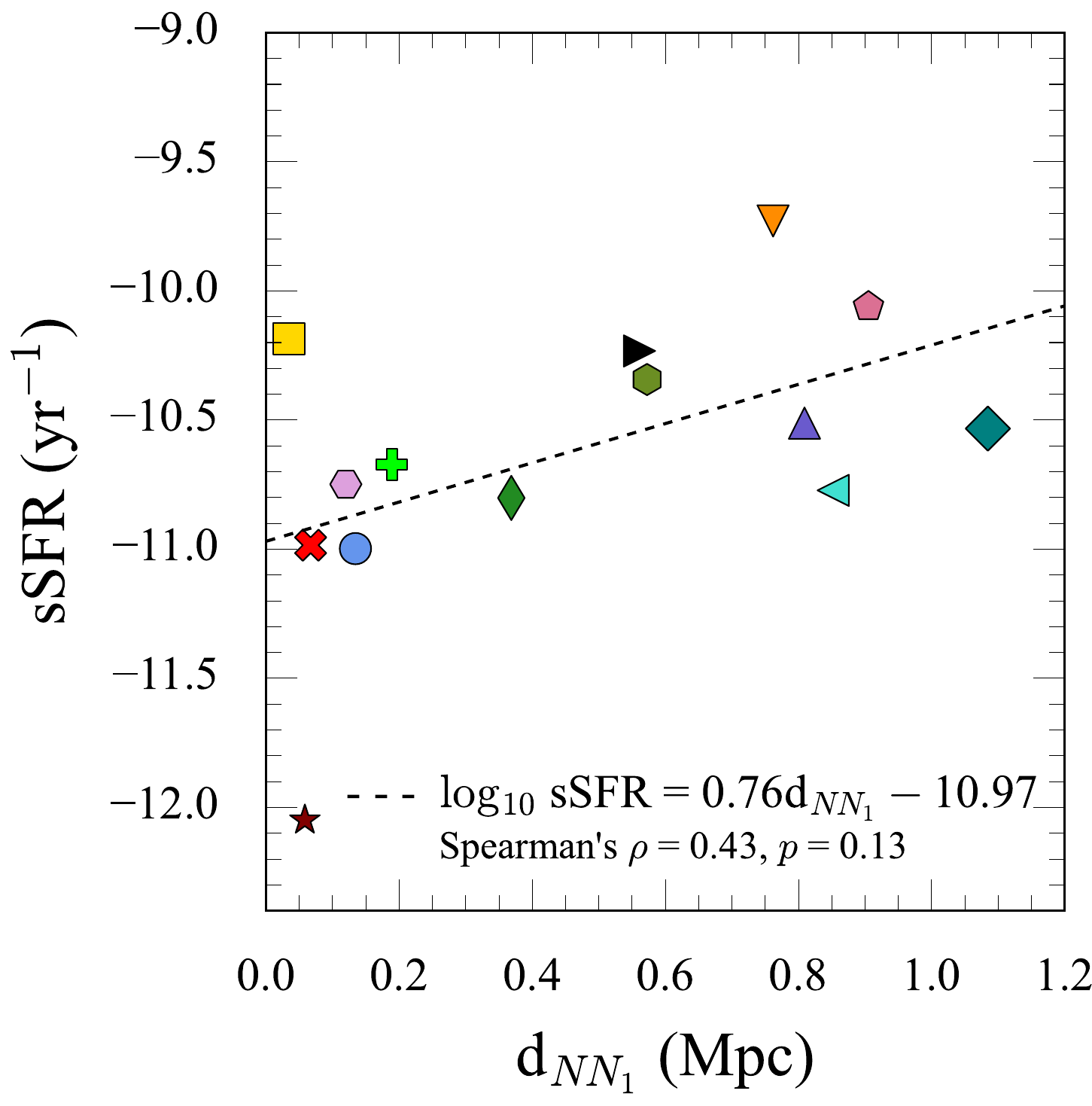}
\hspace{1cm}
\includegraphics[trim = 0cm 0cm 9.5cm 0cm, clip,scale=0.29]{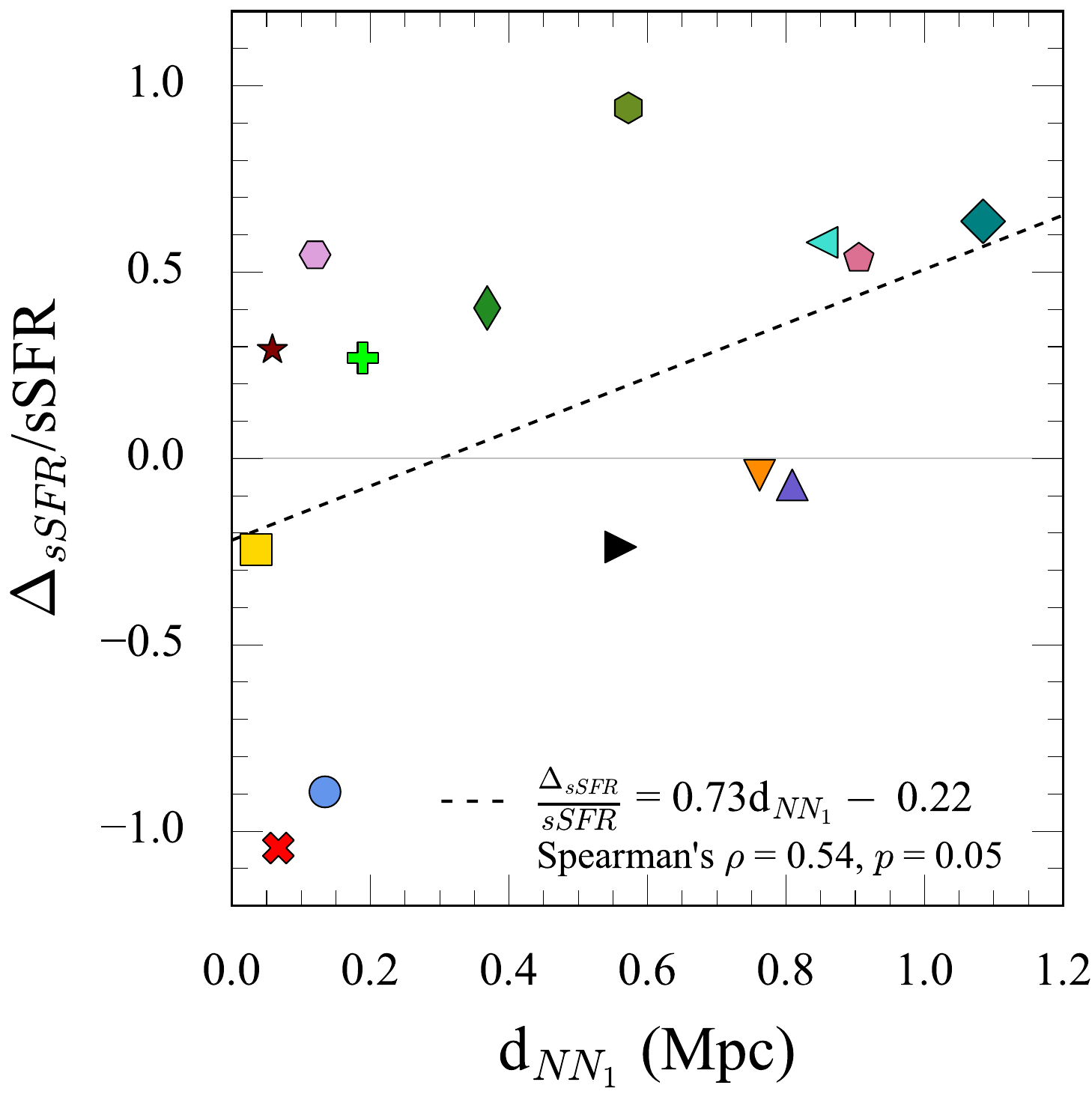}
\centering
\includegraphics[trim = 0cm 0cm 0cm 0cm, clip,scale=0.29]{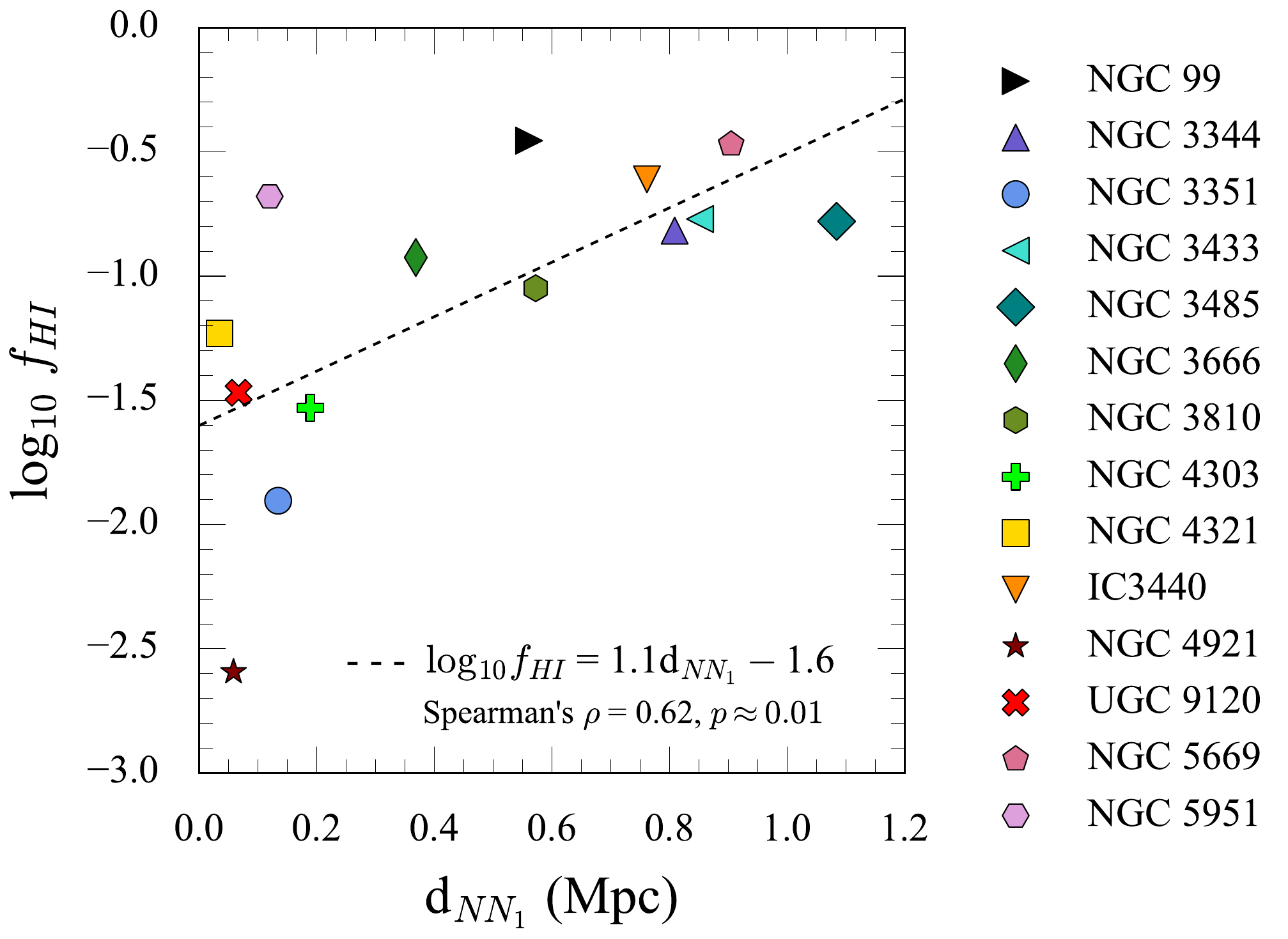}

\caption{The sSFR {\em{(top left)}}, \dsfr/sSFR {\em{(top right)}}, and \hi\ gas fractions {\em{(bottom)}} as a function of the galaxy's distance from the first nearest neighbor, d$_{NN_1}$. The solid gray line denotes \dsfr~$=0$. The lines of best fit for the plots are depicted in the black dashed line. Parameters describing the best-fit lines are noted in the plots.}
    \label{fig:nn1}
\end{figure*}

\begin{figure*}[!ht] 
\centering

\includegraphics[trim = 0cm 0cm 9.5cm 0cm, clip,scale=0.29]{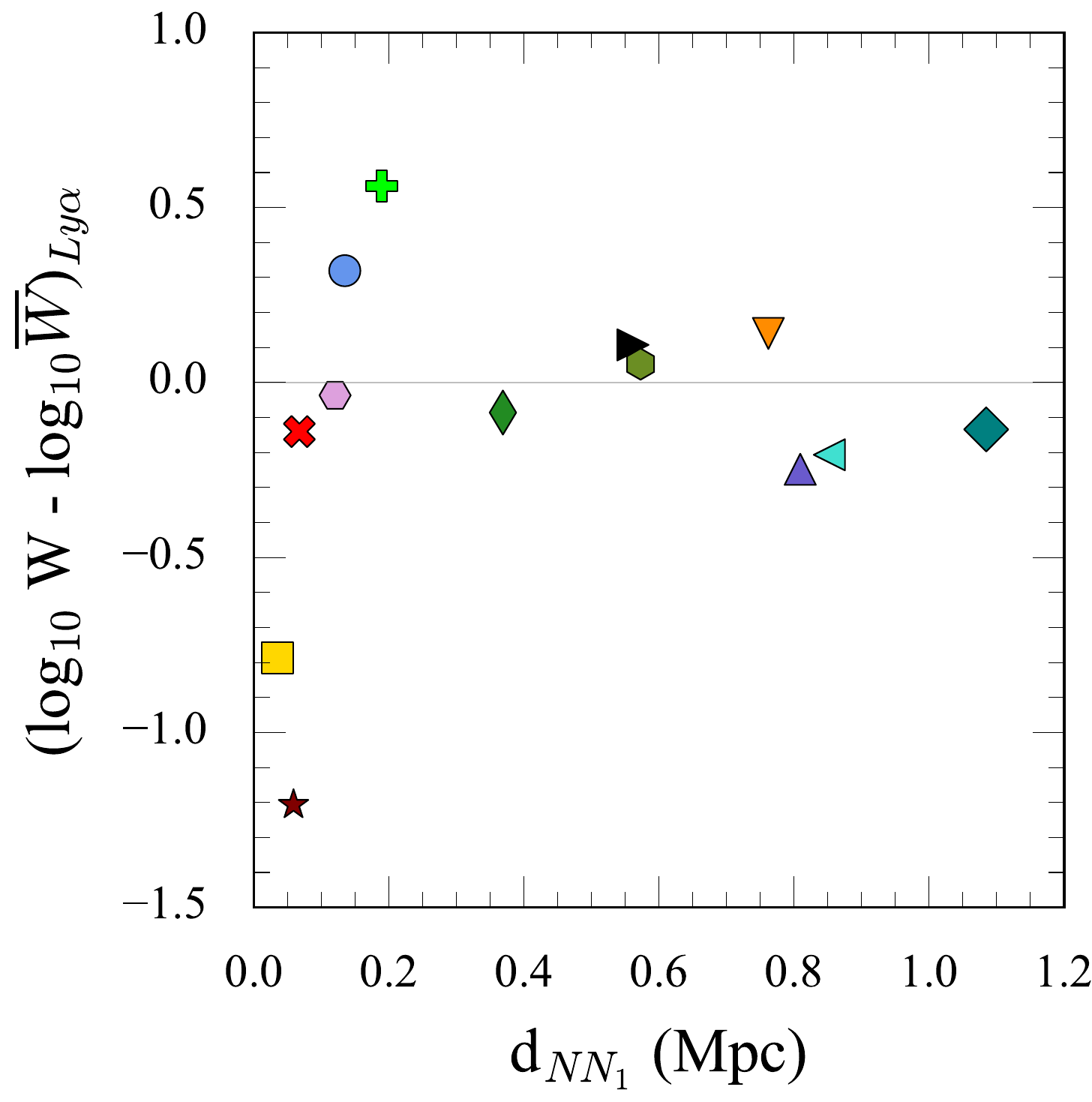}
\includegraphics[trim = 0cm 0cm 0cm 0cm, clip,scale=0.29]{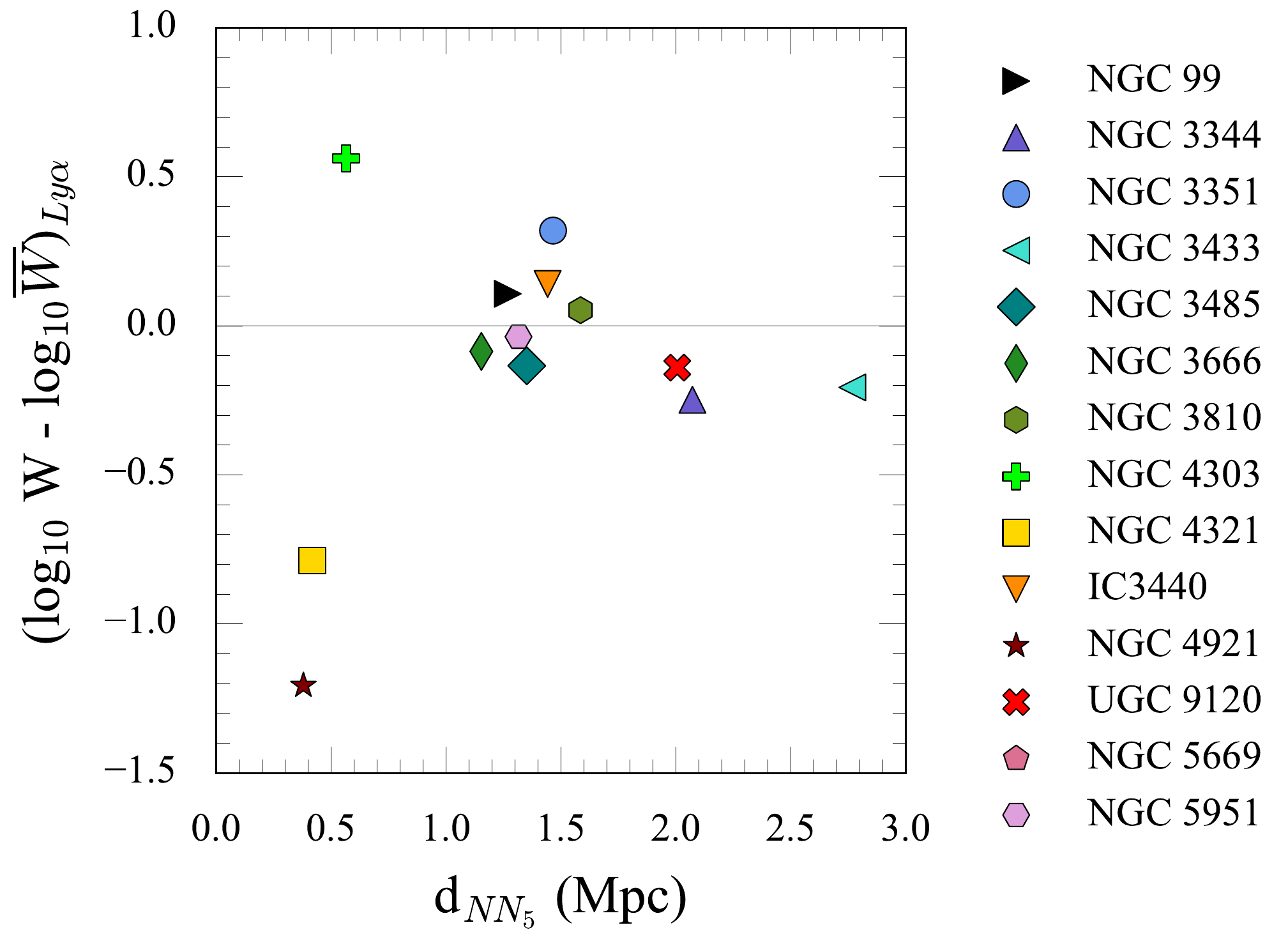}

\caption{ The W$_{Ly\alpha}$ excess as a function of the galaxy's distance from the first and fifth nearest neighbor, d$_{NN_1}$ {\em{(left)}} and d$_{NN_5}$ {\em{(right)}}, respectively. The gray solid line denotes $\log_{10}~\text{W}=\log_{10}~\overline{W}$.}
    \label{fig:nn5}
\end{figure*}

\begin{deluxetable*}{lcccccccc}[!t]
\tablecaption{Measured parameters for the sample galaxies \label{tbl:param}}
\tablewidth{0pt}
\tablehead{
 \colhead{Galaxy} & \colhead{\re $_{,SFR}$} &  \colhead{\re $_{,M_\star}$} & \colhead{CI$_{SFR}$}   & \colhead{CI$_{M_\star}$}  & \colhead{log$_{10}$ sSFR$_{in}$}  & \colhead{log$_{10}$ sSFR$_{out}$} & \colhead{d$_{NN_1}$}& \colhead{d$_{NN_5}$} \\
 \colhead{} & \colhead{(kpc)} &   \colhead{(kpc)}  & \colhead{} & \colhead{}  & \colhead{}  & \colhead{} &   \colhead{(Mpc)} &   \colhead{(Mpc)} \\
 \colhead{(1)} & \colhead{(2)} &   \colhead{(3)}  & \colhead{(4)} & \colhead{(5)}  & \colhead{(6)}  & \colhead{(7)} &   \colhead{(8)} &   \colhead{(9)}} 
\startdata
NGC 99  & 8.51 & 9.08 & 1.67 & 1.52 & -10.18 & -10.29 & 0.56 & 1.27 \\ NGC 3344  & 2.11 & 3.12 & 2.20 & 2.53 & -10.5 & -10.53 & 0.81 & 2.07 \\ 
 NGC 3351  & 0.68 & 3.50 & 7.20  & 2.64 & -10.84 & -11.26 & 0.13 & 1.46 \\ 
NGC 3433  & 9.73 & 8.11 & 1.83 & 2.25 & -10.92 & -10.66 & 0.85 & 2.77 \\ 
 NGC 3485  & 4.07 & 3.58 & 1.52 & 2.22 & -10.7 & -10.41 & 1.08 & 1.35\\ 
  NGC 3666 & 2.86 & 2.52 & 2.04 & 3.31 & -10.9 & -10.72 & 0.37 & 1.15 \\ 
 NGC 3810 & 3.12 & 2.17 & 2.00 & 3.06 & -10.62 & -10.18 & 0.57 & 1.58 \\ 
  NGC 4303  & 4.76 & 4.52 & 1.63 & 2.91 & -10.74 & -10.62 & 0.19 & 0.56 \\ 
 NGC 4321  & 4.46 & 5.22 & 1.91 & 2.06 & -10.13 & -10.24 & 0.03 & 0.42\\ 
 IC 3440  & 6.57 & 7.34 & 1.67 & 1.30 & -10.47 & -10.60  & 0.76  & 1.44\\ 
 NGC 4921  & 10.80 & 10.13 & 1.45 & 2.24 & -12.12 & -11.99 & 0.06 & 0.38\\
 UGC 9120  & 7.3 & 9.62 & 1.54 & 1.32 & -10.8 & -11.31 & 0.07 & 2.01\\ 
 NGC 5669  & 3.69 & 2.82 & 1.69 & 2.40 & -10.2 & -9.96 & 0.91 & 1.85\\ 
  NGC 5951 & 3.94 & 5.17 & 2.00 & 2.25 & -10.89 & -10.64 & 0.12 & 1.31
\enddata
\flushleft 
\tablecomments
{Column (1): Galaxies in our sample. Columns (2) and (3): Effective radii of the SFR and \mstar\ maps expressed in kpc described in Section \ref{sec:radii}. Columns (4) and (5): Concentration indices of the SFR and \mstar\ maps described in Section \ref{sec:radii}. Columns (6) and (7): sSFR  estimated inside and outside \re $_{,M_\star}$ used to estimate \dsfr\ presented in Section \ref{sec:discussion}. Columns (8) and (9): Distance to the 1$^{\rm st}$ and 5$^{\rm th}$ nearest neighbor of the target galaxy expressed in Mpc and described in Section \ref{sec:env}.}
\end{deluxetable*}

In the left panel of Figure \ref{fig:nn1}, we find the sSFR to increase with increasing d$_{NN_1}$ for our galaxies (Spearman's $\rho=0.43$). 
We also see a positive correlation between \dsfr/sSFR and d$_{NN_1}$ (top right panel of Figure \ref{fig:nn1}), although with more scatter. These observations suggest that an isolated galaxy is more likely to build its stellar content in an inside-out fashion. Interestingly, this is corroborated by the even tighter correlation between $f_{HI}$ and d$_{NN_1}$ highlighting that these galaxies are able to sustain their cold gas reservoir in sparse environments, which in turn allows them to form stars. 
The correlations of sSFR, \dsfr/sSFR, $f_{HI}$ with d$_{NN_1}$ are tighter compared to those with d$_{NN_5}$.

In Figure \ref{fig:nn5}, we show the W$_{Ly\alpha}$ excess as a function of d$_{NN_1}$ and d$_{NN_5}$. We find that excluding NGC~4321 and NGC~4921, we see a steady decline in W$_{Ly\alpha}$ excess with increasing d$_{NN_1}$ and d$_{NN_5}$, the anti-correlation quantified with a Spearman's $\rho$ of -0.36 and -0.62, respectively. This anti-correlation indicates the suppression of the CGM \hi\ content in isolated environments. While d$_{NN_1}$ simply depicts how far the closest L$_\star$ neighbor from our galaxy of interest is, d$_{NN_5}$ can trace the local environment density. A shorter d$_{NN_5}$ indicates a denser environment. So, the stronger anti-correlation of W$_{Ly\alpha}$ excess with d$_{NN_5}$ likely indicates that the suppression of W$_{Ly\alpha}$ in the CGM is mainly affected by the local environment density than by the presence of close neighbors. 
However, the suppression of W$_{Ly\alpha}$ could also result from galaxies undergoing ram-pressure stripping \citep{abadi99, stein16, coenda19} causing them to lose their cold gas or strangulate \citep{peng15}. NGC~4321 and NGC~4921 show a strong suppression of \Lya\ in their CGM, likely due to their location in Virgo and Coma clusters, respectively. 
In contrast, NGC 4303 and IC 3440 also reside in the Virgo cluster but NGC 4303 possesses abundant \hi\ in its CGM. \cite{yoon13}'s investigation of the background quasars in and around the Virgo cluster showed that the CGM covering fractions of the \Lya\ absorbers are lower for the Virgo galaxies indicating that galaxies in the denser environment have suppressed CGM, compared to circumcluster and field galaxies with an abundant CGM. Additionally, NGC 4321's interaction with its companion, NGC 4322, could have also resulted in the suppression of \Lya. The highly suppressed CGM in M100 is suggestive of the galaxy making its way to the red sequence \citep{gim21}. 
 
We note that DIISC galaxies being HI-rich are likely to be found in a gas-rich large-scale environment \citep{bort22}.  And,  galaxy density in the vicinity might play a critical role in dictating the neutral gas content of the CGM and the prevalence of star formation.  Further investigation using a larger sample is recommended.

\section{Conclusions}\label{sec:conc}
In this work, we investigate the radial profiles and distribution of star formation, stellar mass, \hi\ mass in 14 nearby galaxies and look for clues to disk growth and stellar build-up. We quantify growth as \dsfr/sSFR, where \dsfr~$={\rm sSFR}_{out}-{\rm sSFR}_{in}$, where sSFR$_{out}$ and sSFR$_{in}$ are sSFR outside and inside the effective radius of the stellar mass map and allows us to comment on the direction of growth and probe the star formation activity. We investigate the connections of \dsfr/sSFR with the atomic gas in the ISM and CGM and the local environment.  We summarize our findings below:

\begin{itemize}
    \item[1.] The FUV surface brightness profiles are more extended than those of \sdssr-band for most galaxies as indicated by their exponential scale lengths. 
    About 70\% of our galaxies also show $\text{CI}_{SFR}<\text{CI}_{M_\star}$ indicative of an actively star-forming outer disk. 
    \item[2.] We find the sSFR profiles exhibit both increasing and decreasing behavior as a function of galactocentric radius, the increase pointing towards an inside-out growing disk. The sSFR gradients show a positive relation (Spearman $\rho=0.4$) with the \hi\ SFE gradients. A bolder interpretation of this observation could be that galaxies are likely to evolve inside-out if they can stabilize their \hi\ consumption rate to maintain constant \tdep\ over their disk.   
    \item[3.] A positive correlation between \dsfr/sSFR and $f_{HI}$ (Spearman $\rho=0.30$) is observed with a 1$\sigma$ scatter of 0.56 dex. The result highlights the connection of direction of propagation of star formation with \hi, indicating an inside-out growth scenario for galaxies with high $f_{HI}$. The atomic gas content in the CGM, probed by W$_{Ly\alpha}$ excess, however, does not show any relation with the direction of stellar growth. 
    \item[4.] Both \dsfr/sSFR and $f_{HI}$ show a strong positive correlation with the distance to the nearest L$_\star$ neighbor showing that galaxies are able to hold onto their ISM cold gas reservoir in the absence of a close neighbor and subsequently, proceed to evolve inside-out. 
    \item[5.] The cold gas content of the CGM probed by W$_{Ly\alpha}$ excess is seen to be suppressed in dense cluster environments (as expected) but also in isolated environments. The latter is seen as a tight anti-correlation between W$_{Ly\alpha}$ excess and the distance to the 5$^{\rm th}$ nearest L$_\star$ neighbor with Spearman $\rho$ of -0.62. 
    Interestingly, the suppression is affected more by the local environment density than the presence of a close neighbor. 
\end{itemize}
Our study sheds light on how environment, and \hi\ in the CGM and ISM affect the stellar growth and evolution of a galaxy. Our results imply that the direction in which stellar growth occurs is likely governed by the distribution of \hi\ and \mstar\ of the galaxy while the environment controls the cold gas content. 
In isolated environments, galaxies are likely to evolve inside-out as they are able to sustain their cold gas content in the ISM, which in turn is fed by the CGM via accretion. 

As galaxies form new stars at larger radii, subsequent stellar age gradients also arise in that direction. In the future, we will be investigating the age and metallicity gradients and star formation histories for these galaxies with deep, multiband optical imaging. We will investigate the properties of the nearest neighbors to further explore the role the environment plays in controlling the gas content in the ISM and CGM.


\section*{ACKNOWLEDGEMENTS}
MP, SB, RJ, and DT are supported by NASA ADAP grant 80NSSC21K0643, SB and HG are also supported by NSF Award Number 2009409, and SB, HG, and TH are supported by
\emph{HST} grant HST-GO-14071 administrated by STScI which is operated by AURA
 under contract NAS\,5-26555 from NASA. 
MP, SB, RJ, and JM acknowledge the land and the native people that Arizona State University's campuses are located in the Salt River Valley. The ancestral territories of Indigenous peoples, including the Akimel O’odham (Pima) and Pee Posh (Maricopa) Indian Communities, whose care and keeping of these lands allow us to be here today.

We thank the referee for their constructive feedback.
We thank the staff at the Space Telescope Science Institute, the National Radio Astronomy Observatory (NRAO) Array Operations Center at Socorro,  the Steward Observatory, and the Vatican Advanced Technology Telescope for their help and support on this project. All GALEX and HST data presented in this paper were obtained from the Mikulski Archive for Space Telescopes (MAST) at the Space Telescope Science Institute. The specific observations analyzed can be accessed via\dataset[https://doi.org/10.17909/rf6m-3y73]{https://doi.org/10.17909/rf6m-3y73}.

GALEX is a NASA Small Explorer, launched in 2003 April.
We gratefully acknowledge NASA’s support for the construction,
operation, and science analysis of the GALEX mission, developed in cooperation with the
Centre National d’´Etudes Spatiales (CNES) of France and the
Korean Ministry of Science and Technology.

Based on observations made with the NASA/ESA Hubble Space Telescope, obtained from the data archive at the Space Telescope Science Institute. STScI is operated by the Association of Universities for Research in Astronomy, Inc. under NASA contract NAS 5-26555.

This work is also partly based on observations with the VATT: the Alice P. Lennon Telescope and the Thomas J. Bannan Astrophysics Facility.

The National Radio Astronomy Observatory is a facility of the National Science Foundation operated under cooperative agreement by Associated Universities, Inc.

This work is based [in part] on observations made with the Spitzer Space Telescope, which was operated by the Jet Propulsion Laboratory, California Institute of Technology under a contract with NASA

This publication makes use of data products from the Wide-field Infrared Survey Explorer, which is a joint project of the University of California, Los Angeles, and the Jet Propulsion Laboratory/California Institute of Technology, funded by the National Aeronautics and Space Administration.

%

\facilities{HST, GALEX, Sloan, Spitzer, VATT, VLA, WISE}





\bibliography{DIISC_growth}{}
\bibliographystyle{aasjournal}
\appendix
\section{Surface Brightness Profiles}\label{app:fig4}
\begin{figure*}[!ht] 
 \setcounter{figure}{3}
\centering
\includegraphics[trim = .cm 0.cm 0cm 0cm, clip,scale=0.12]{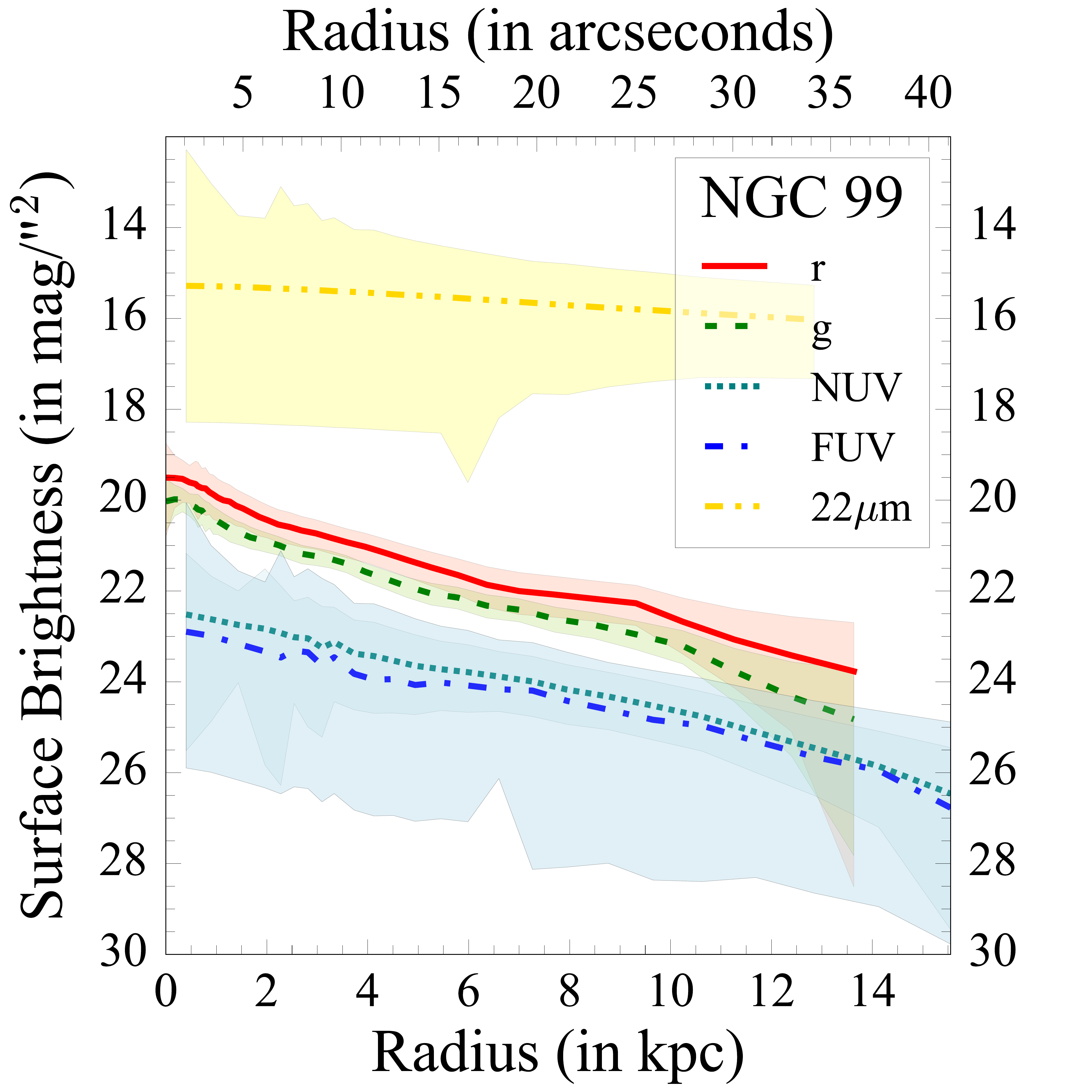}\hspace{0.5cm}
\includegraphics[trim = .cm 0.cm 0cm 0cm, clip,scale=0.12]{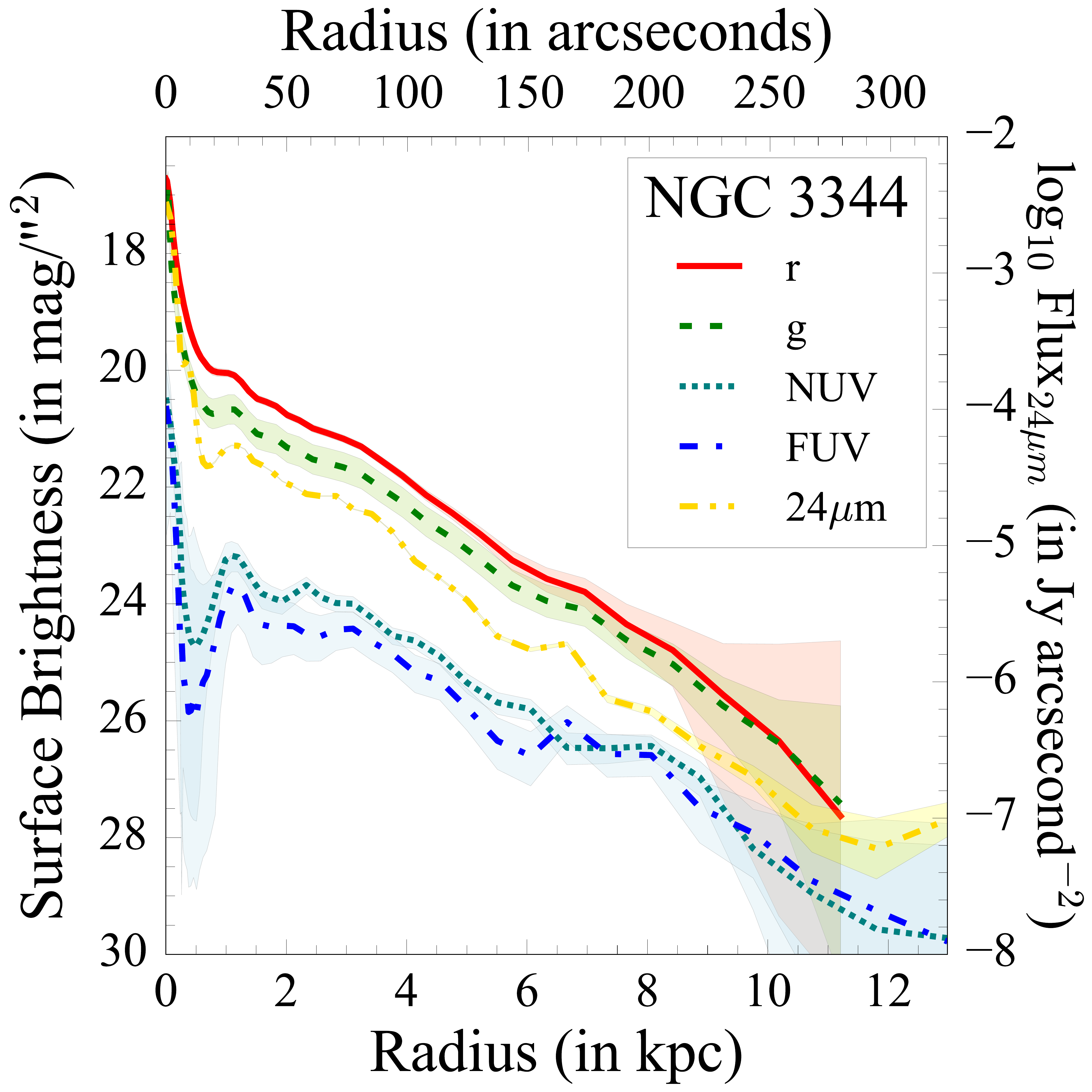}
\includegraphics[trim = .cm 0.cm 0cm 0cm, clip,scale=0.12]{figures/f4_3.pdf}\hspace{0.5cm}
\includegraphics[trim = .cm 0.cm 0cm 0cm, clip,scale=0.12]{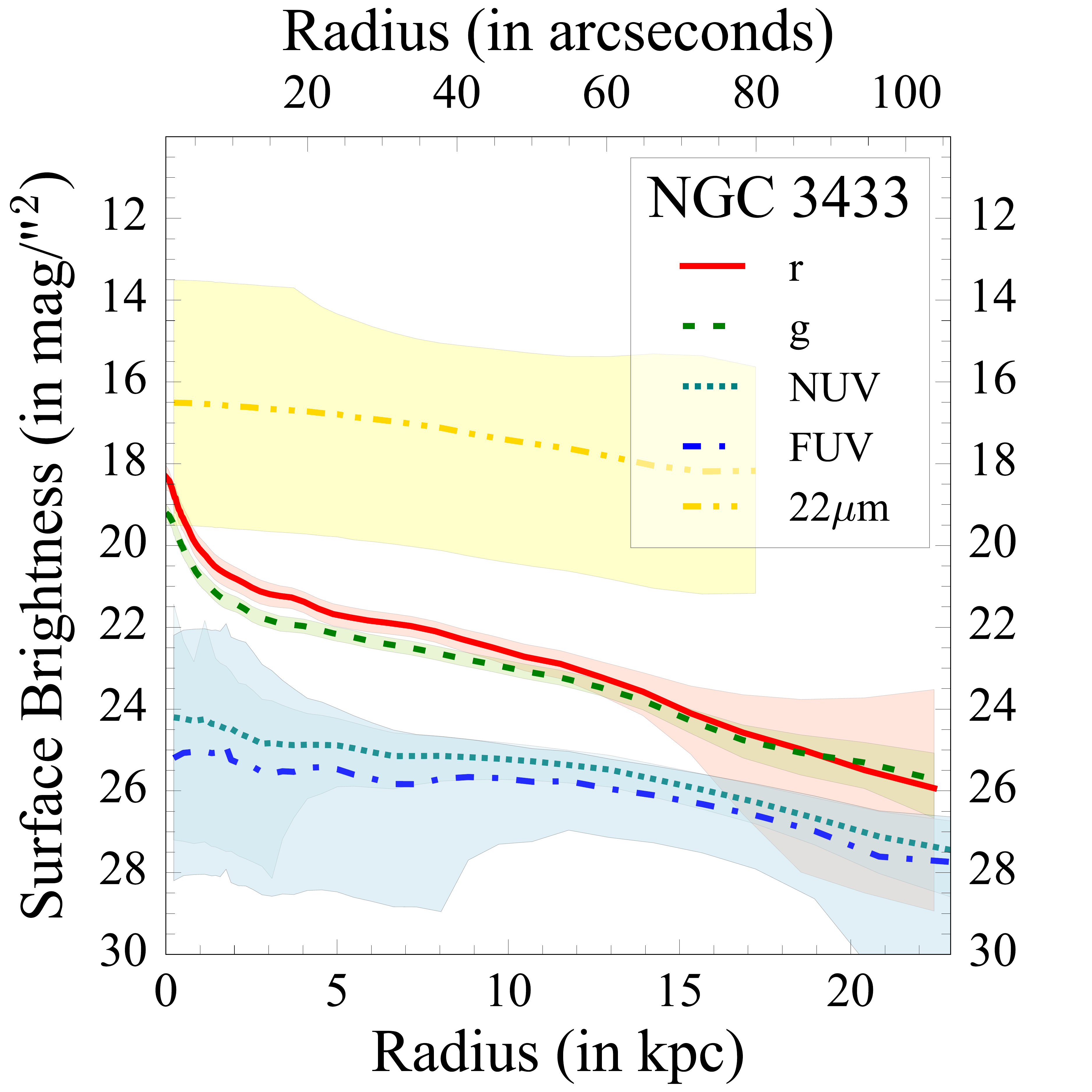}
\caption{Multi-wavelength profiles of surface brightness for the galaxies in our sample. The profile for SDSS-$r$, SDSS-$g$, NUV, FUV, and 22$mu$m (where available) are presented in mag~arcsec$^{-2}$, while for 24$\mu$m (where available) profile is shown in Jy~arcsec$^{-2}$. The shaded region around each profile shows the 1$\sigma$ uncertainty in the surface brightness. }
\end{figure*}

\begin{figure*}[!ht] 
\setcounter{figure}{3}

\centering
\includegraphics[trim = 0 0.cm 0cm 0cm, clip,scale=0.12]{f4_5.pdf}\hspace{0.5cm}
\includegraphics[trim = 0 0.cm 0cm 0cm, clip,scale=0.12]{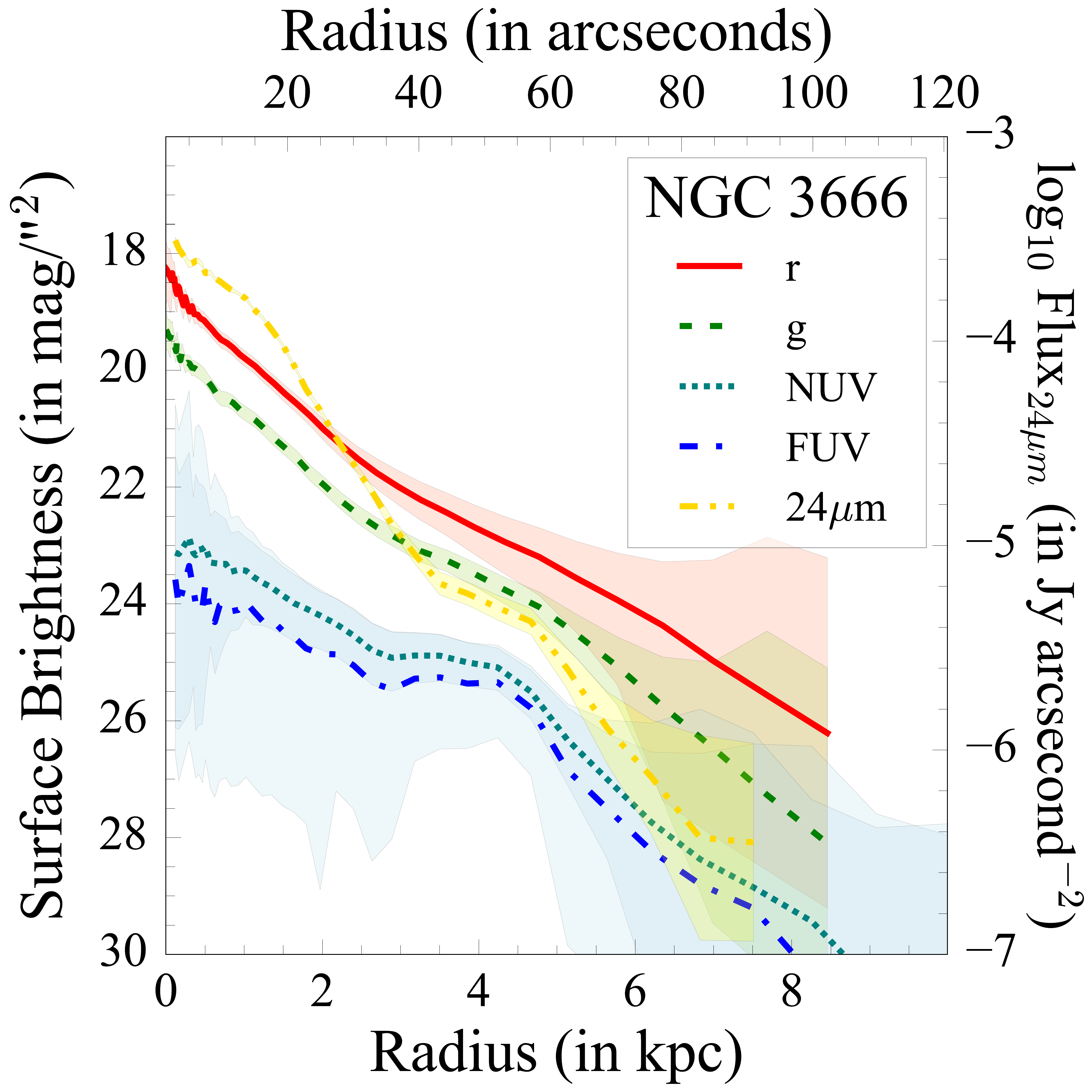}
\includegraphics[trim = 0 0.cm 0cm 0cm, clip,scale=0.12]{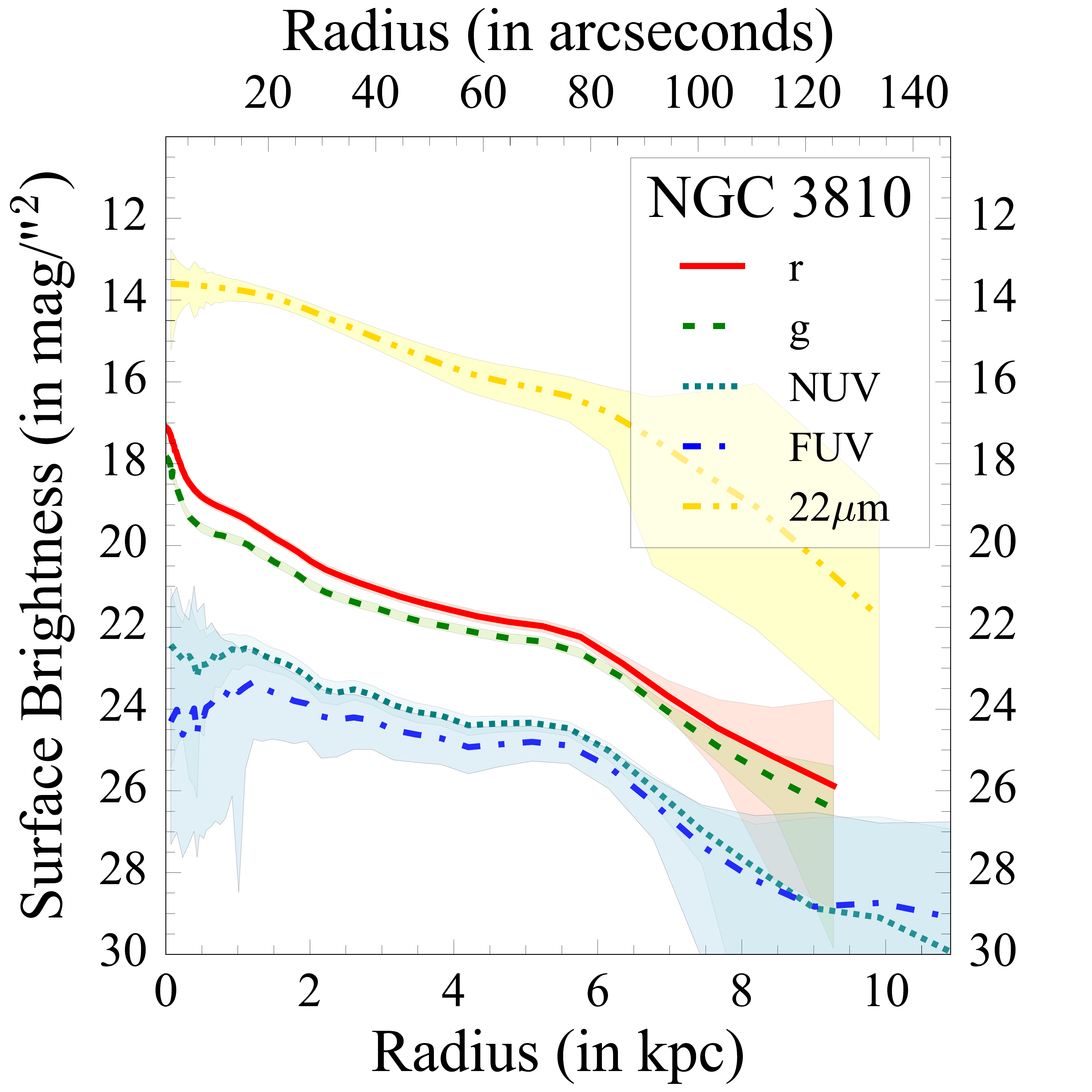}\hspace{0.5cm}
\includegraphics[trim = 0 0.cm 0cm 0cm, clip,scale=0.12]{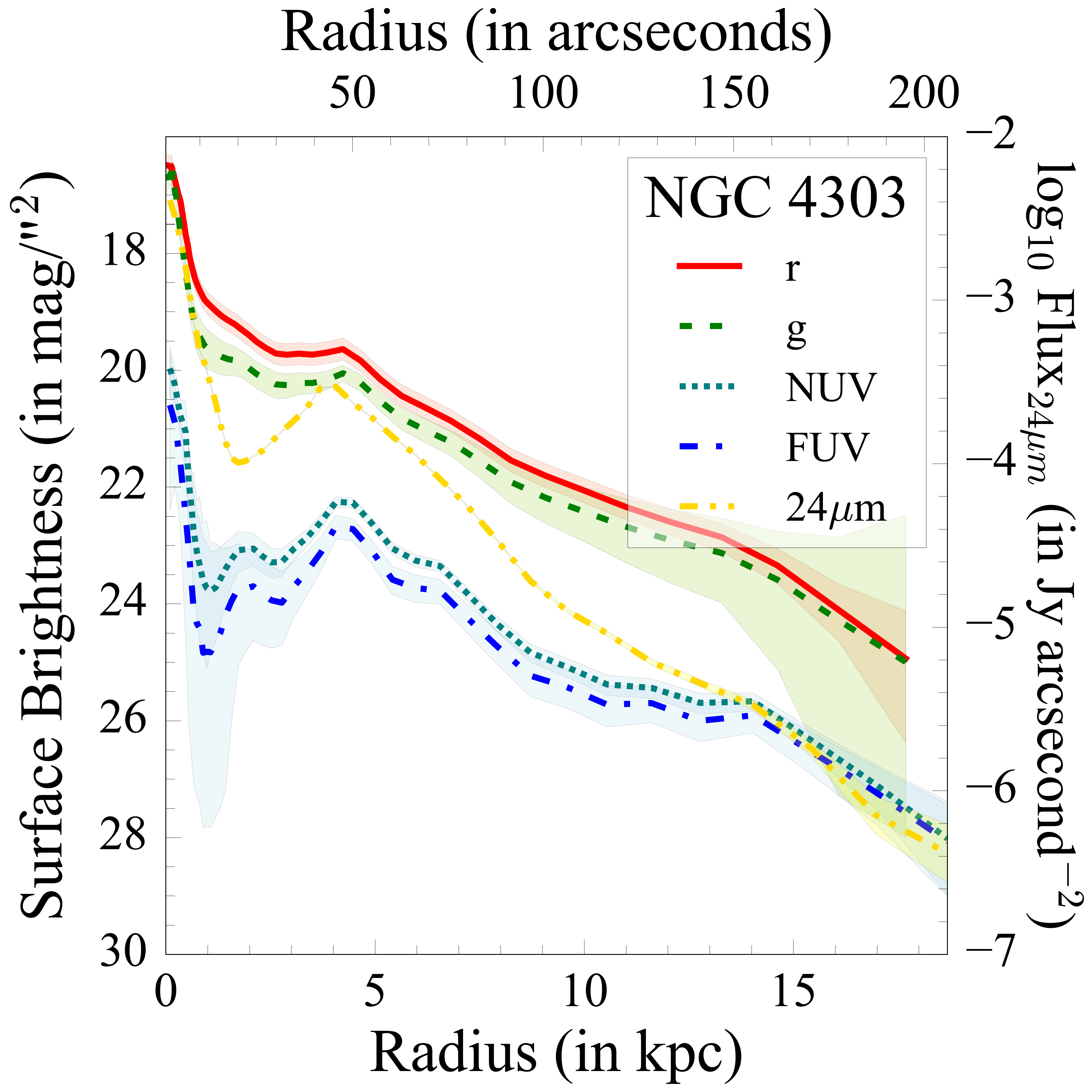}
\caption{Multi-wavelength profiles of surface brightness for the galaxies in our sample. The profile for SDSS-$r$, SDSS-$g$, NUV, FUV, and 22$mu$m (where available) are presented in mag~arcsec$^{-2}$, while for 24$\mu$m (where available) profile is shown in Jy~arcsec$^{-2}$. The shaded region around each profile shows the 1$\sigma$ uncertainty in the surface brightness. }
\end{figure*}

\begin{figure*}[!ht] 
\setcounter{figure}{3}
\centering
\includegraphics[trim = 0cm 0.cm 0cm 0cm, clip,scale=0.12]{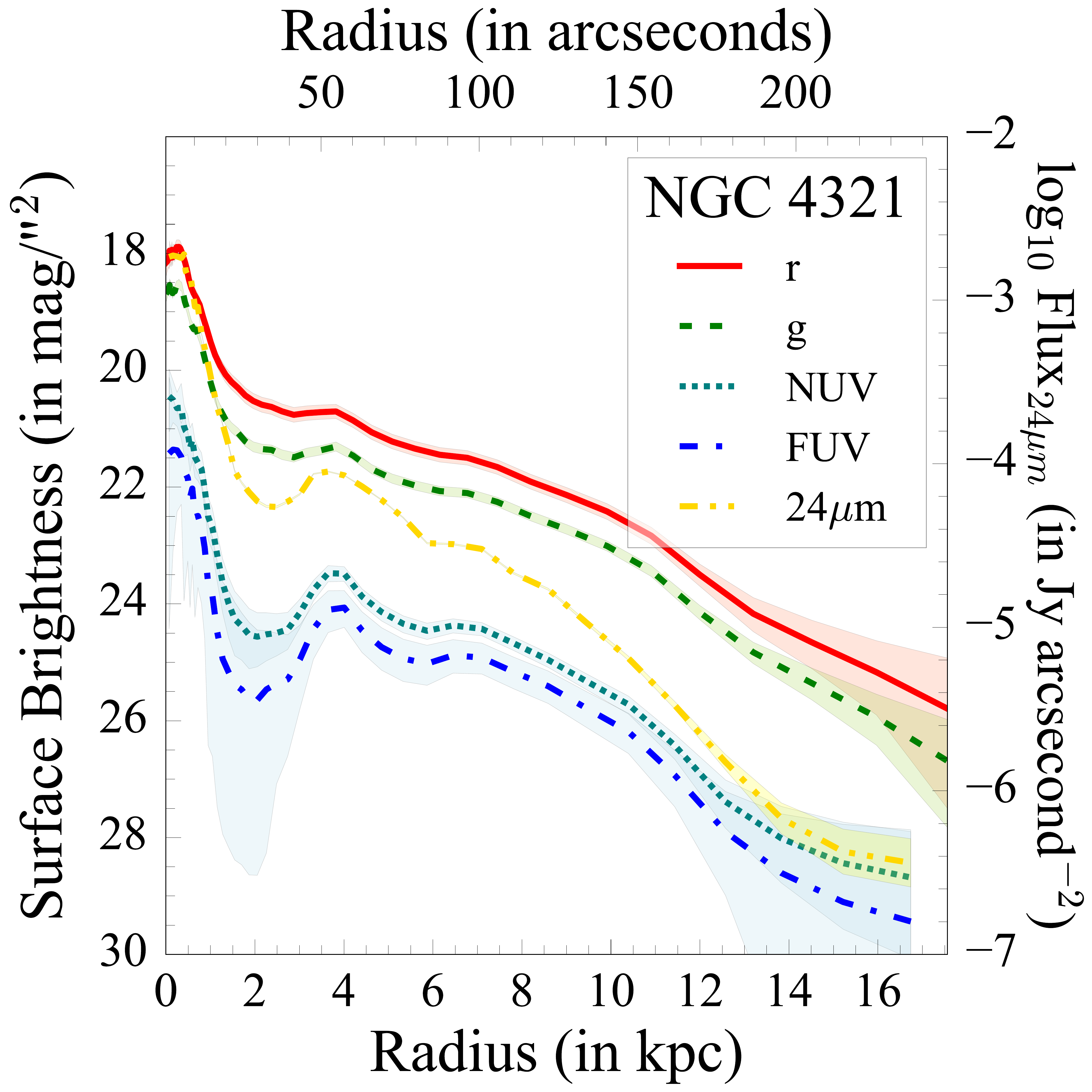}\hspace{0.5cm}
\includegraphics[trim = 0 0.cm 0cm 0cm, clip,scale=0.12]{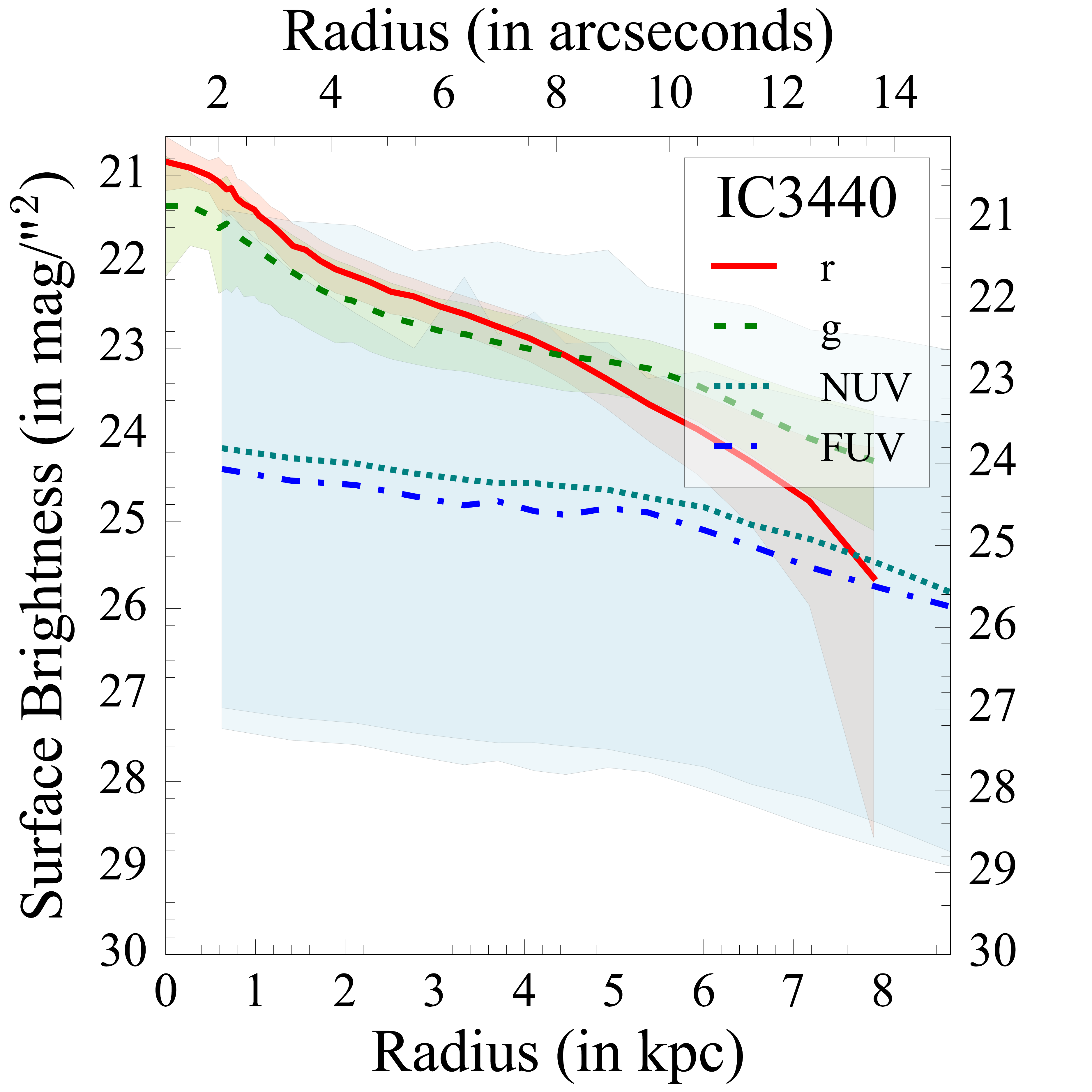}
\includegraphics[trim = 0 0.cm 0cm 0cm, clip,scale=0.12]{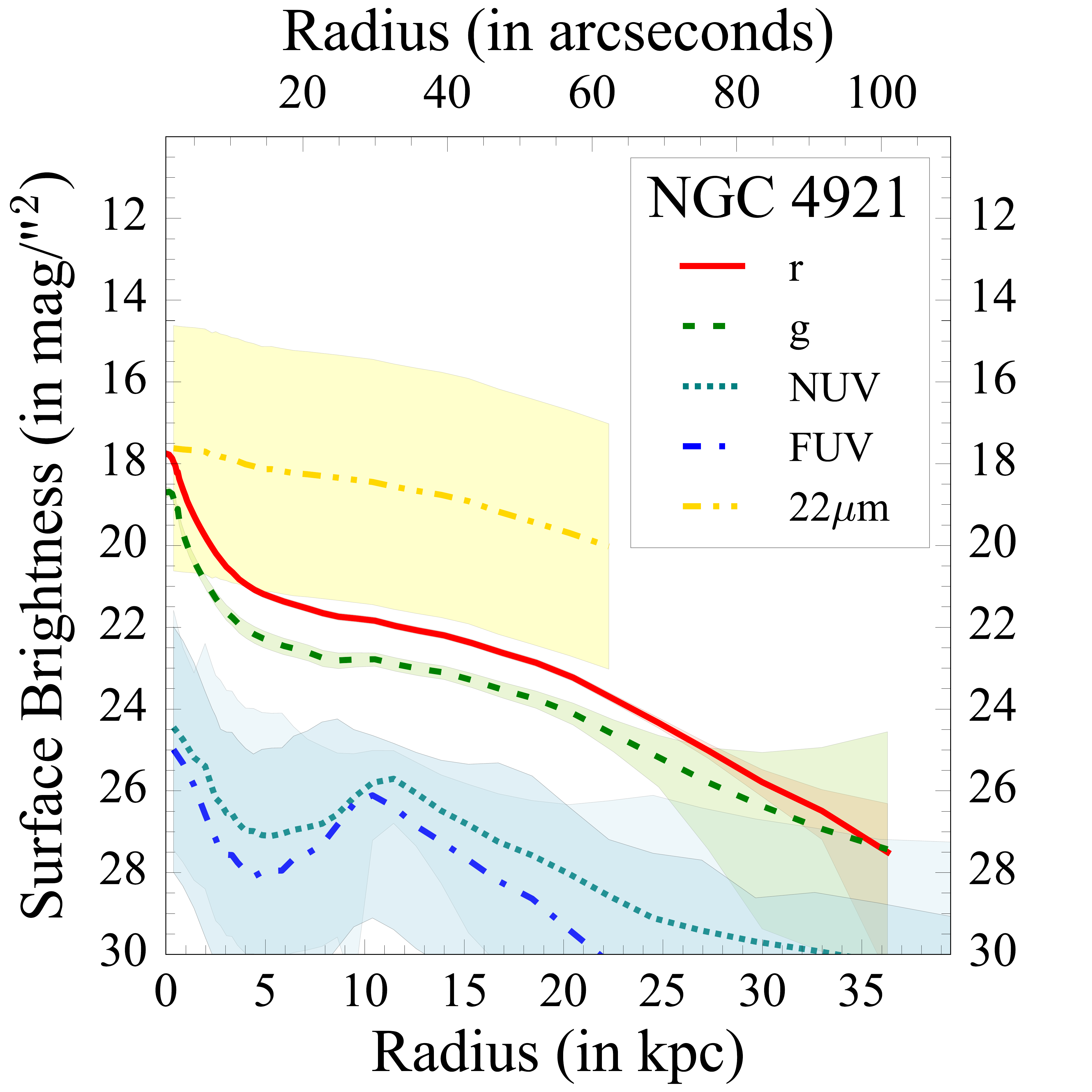}\hspace{0.5cm}
\includegraphics[trim = 0 0.cm 0cm 0cm, clip,scale=0.12]{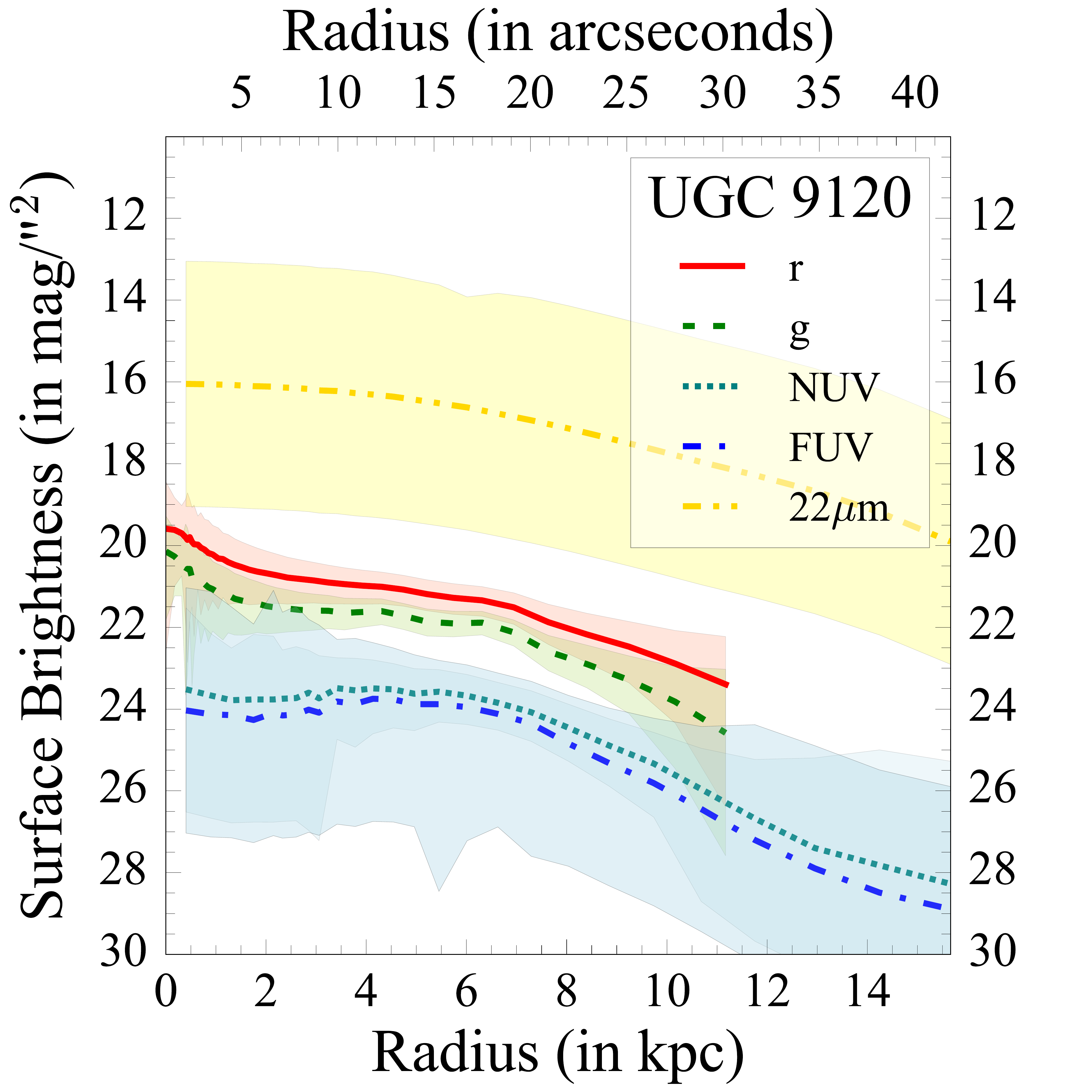}
\caption{Multi-wavelength profiles of surface brightness for the galaxies in our sample. The profile for SDSS-$r$, SDSS-$g$, NUV, FUV, and 22$mu$m (where available) are presented in mag~arcsec$^{-2}$, while for 24$\mu$m (where available) profile is shown in Jy~arcsec$^{-2}$. The shaded region around each profile shows the 1$\sigma$ uncertainty in the surface brightness. }
\end{figure*}
\clearpage
\begin{figure*}[!ht] 
\setcounter{figure}{3}
\centering
\includegraphics[trim = 0 0.cm 0cm 0cm, clip,scale=0.12]{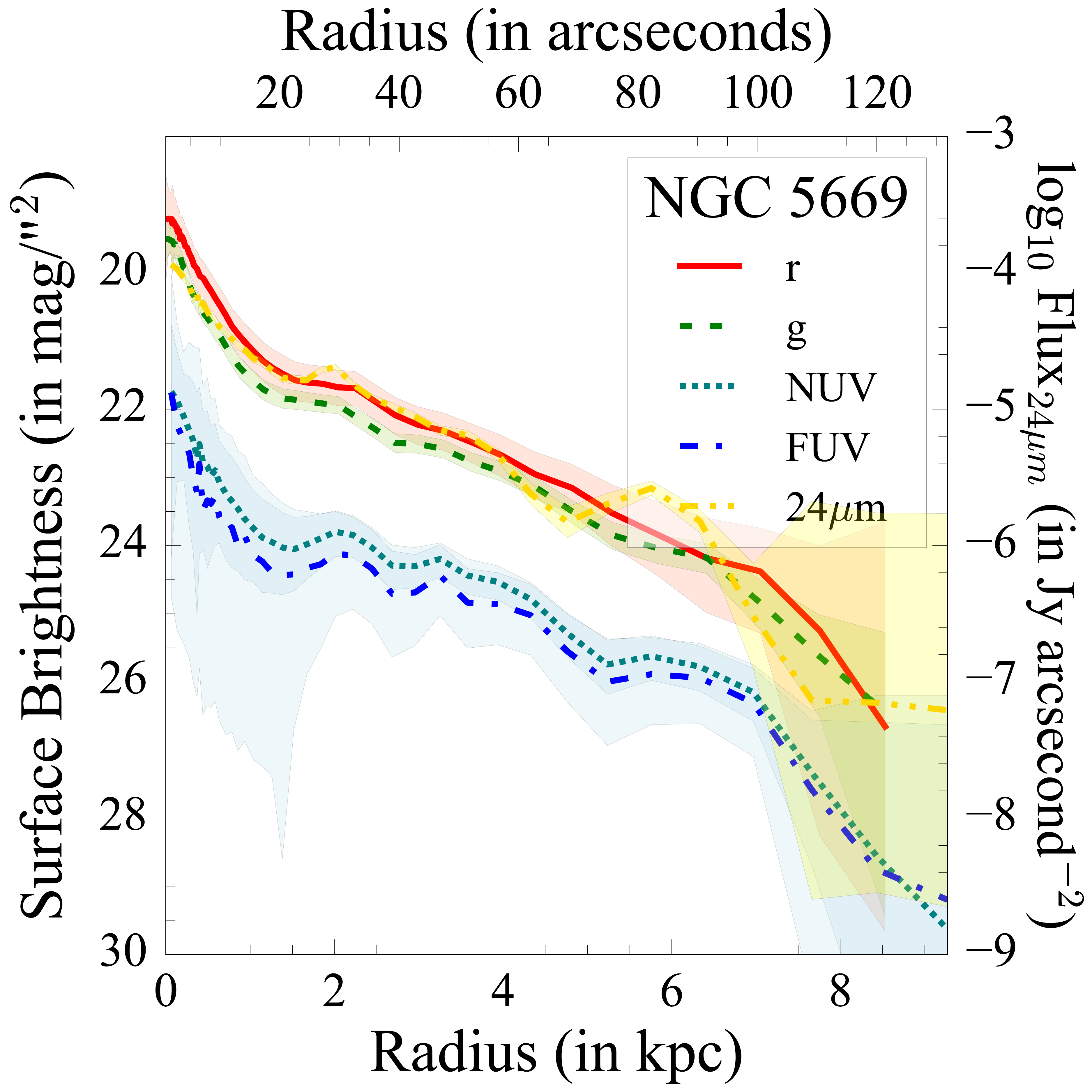}\hspace{0.5cm}
\includegraphics[trim = 0 0.cm 0cm 0cm, clip,scale=0.12]{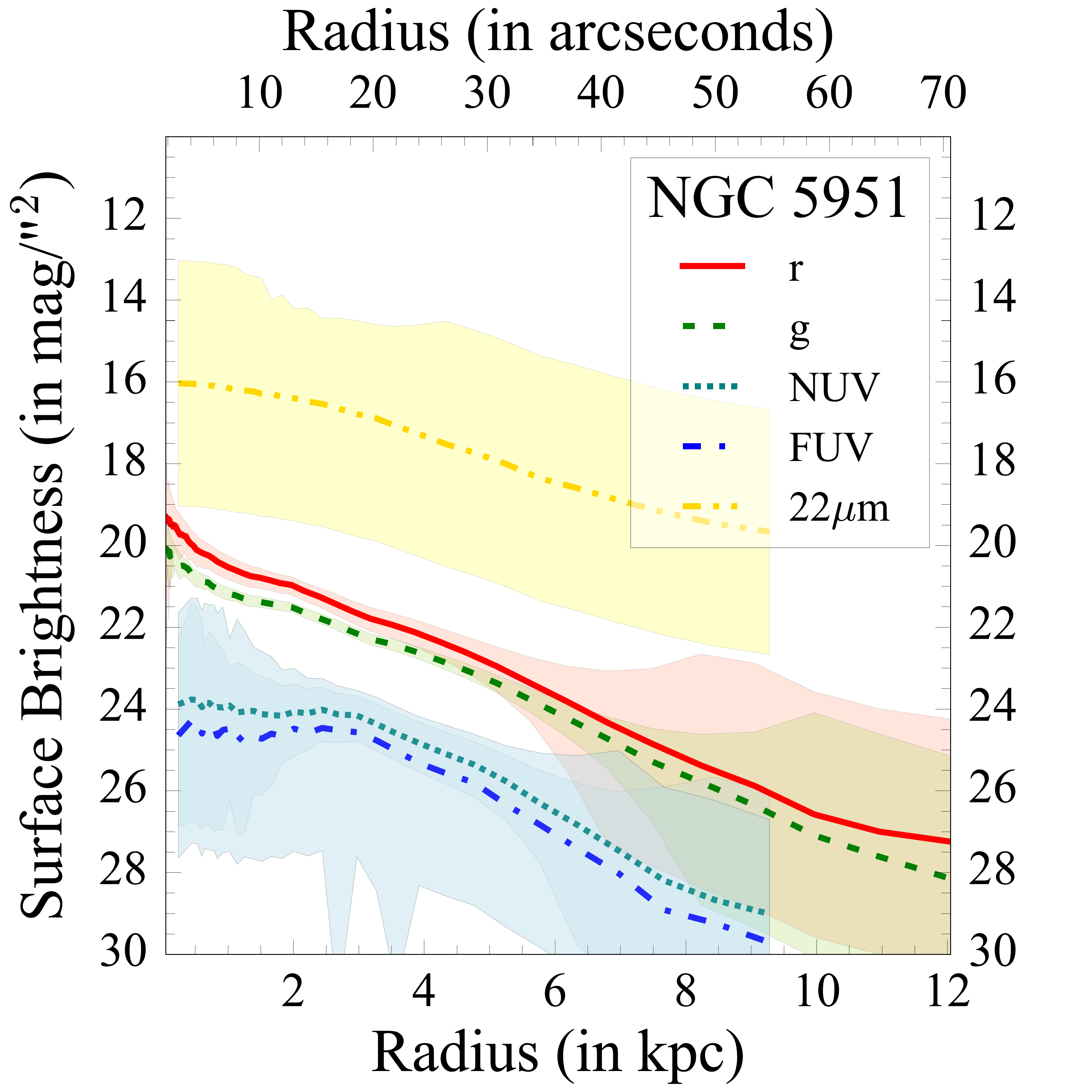}
\caption{Multi-wavelength profiles of surface brightness for the galaxies in our sample. The profile for SDSS-$r$, SDSS-$g$, NUV, FUV, and 22$mu$m (where available) are presented in mag~arcsec$^{-2}$, while for 24$\mu$m (where available) profile is shown in Jy~arcsec$^{-2}$. The shaded region around each profile shows the 1$\sigma$ uncertainty in the surface brightness. }
\end{figure*}

\section{Profiles of SFR, Stellar Mass, \& HI Surface Densities}\label{app:fig6}

\begin{figure*}[!ht] 
\setcounter{figure}{5}
\centering
\includegraphics[trim = 0.cm 0.cm 0cm 0cm, clip,scale=0.15]{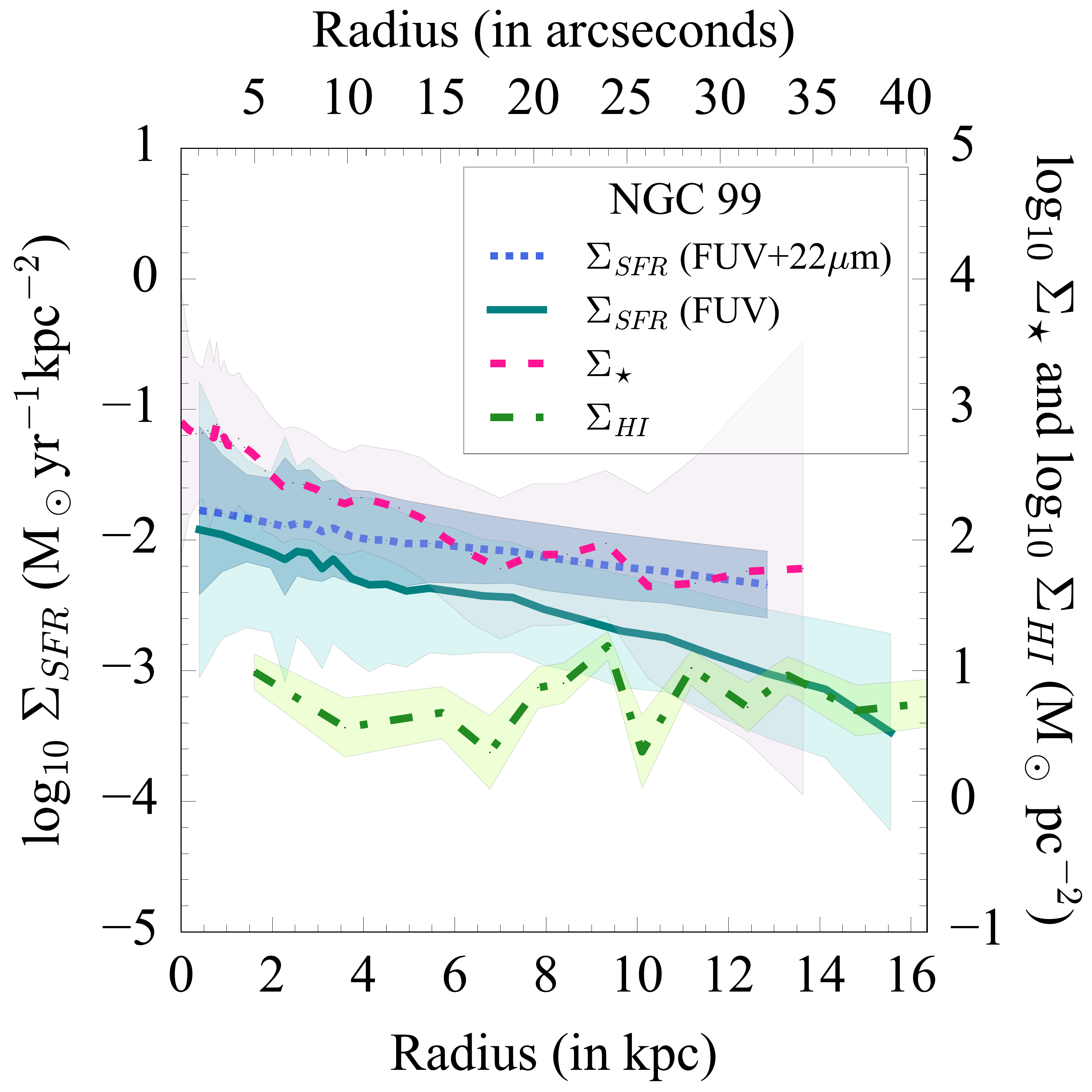}\hspace{1cm}
\includegraphics[trim = 0.cm 0.cm 0cm 0cm, clip,scale=0.15]{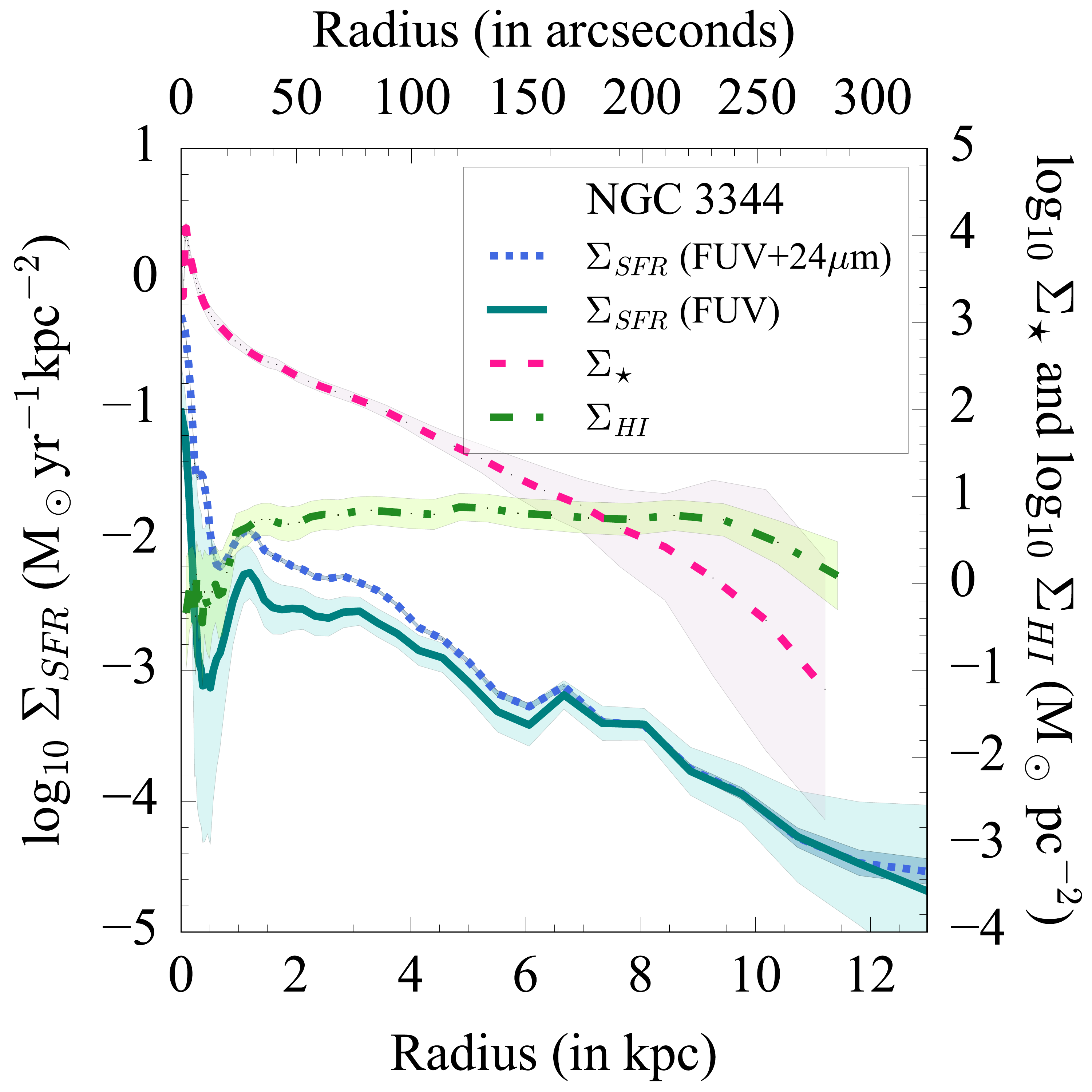}

\caption{Profiles of \sfr\ obtained from FUV and FUV+\mips\ (or FUV+\wise), \sm, and \mhi\ for the galaxies in our sample. The shaded region around each profile shows the 1$\sigma$ uncertainty in the respective quantities.}
\end{figure*}

\begin{figure*}[!t] 
\setcounter{figure}{5}
\centering

\includegraphics[trim = 0.cm 0.cm 0cm 0cm, clip,scale=0.15]{f6_3.pdf}\hspace{1cm}
\includegraphics[trim = 0.cm 0.cm 0cm 0cm, clip,scale=0.15]{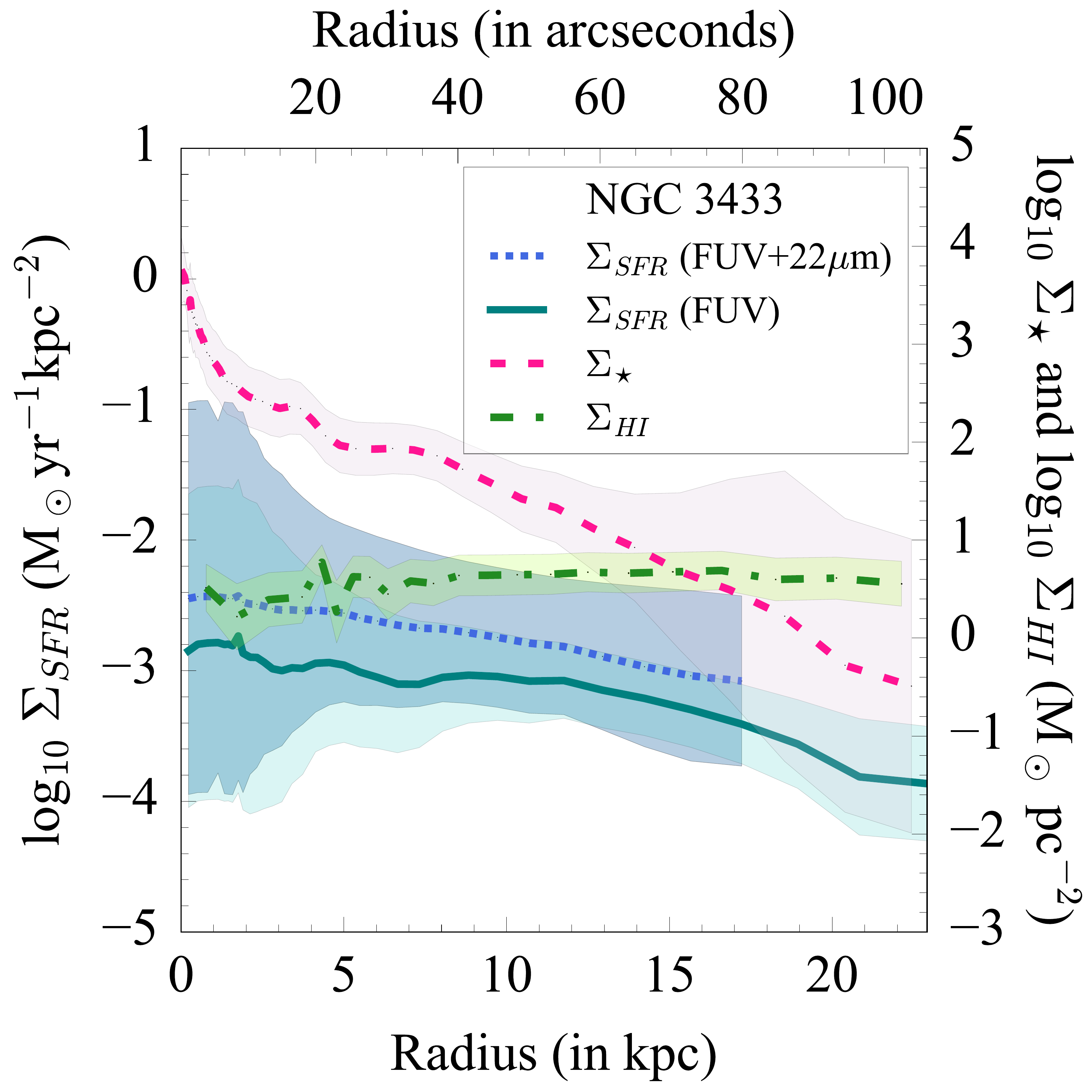}
\includegraphics[trim = 0.cm 0.cm 0cm 0cm, clip,scale=0.15]{f6_5.pdf}\hspace{1cm}
\includegraphics[trim = 0.cm 0.cm 0cm 0cm, clip,scale=0.15]{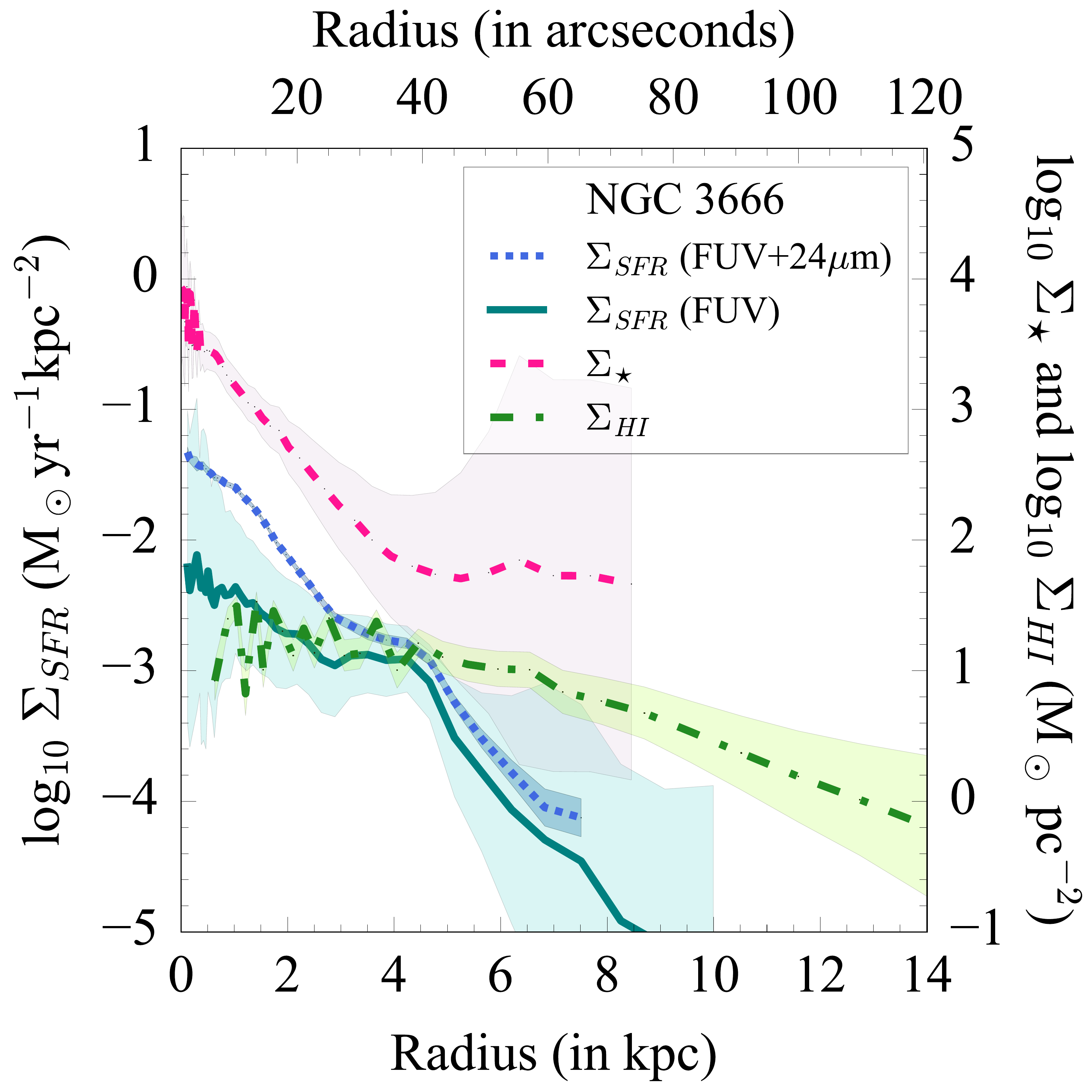}
\caption{Profiles of \sfr\ obtained from FUV and FUV+\mips\ (or FUV+\wise), \sm, and \mhi\ for the galaxies in our sample. The shaded region around each profile shows the 1$\sigma$ uncertainty in the respective quantities.}
\end{figure*}

\begin{figure*}[!ht] 
\setcounter{figure}{5}
\centering
\includegraphics[trim = 0.cm 0.cm 0cm 0cm, clip,scale=0.15]{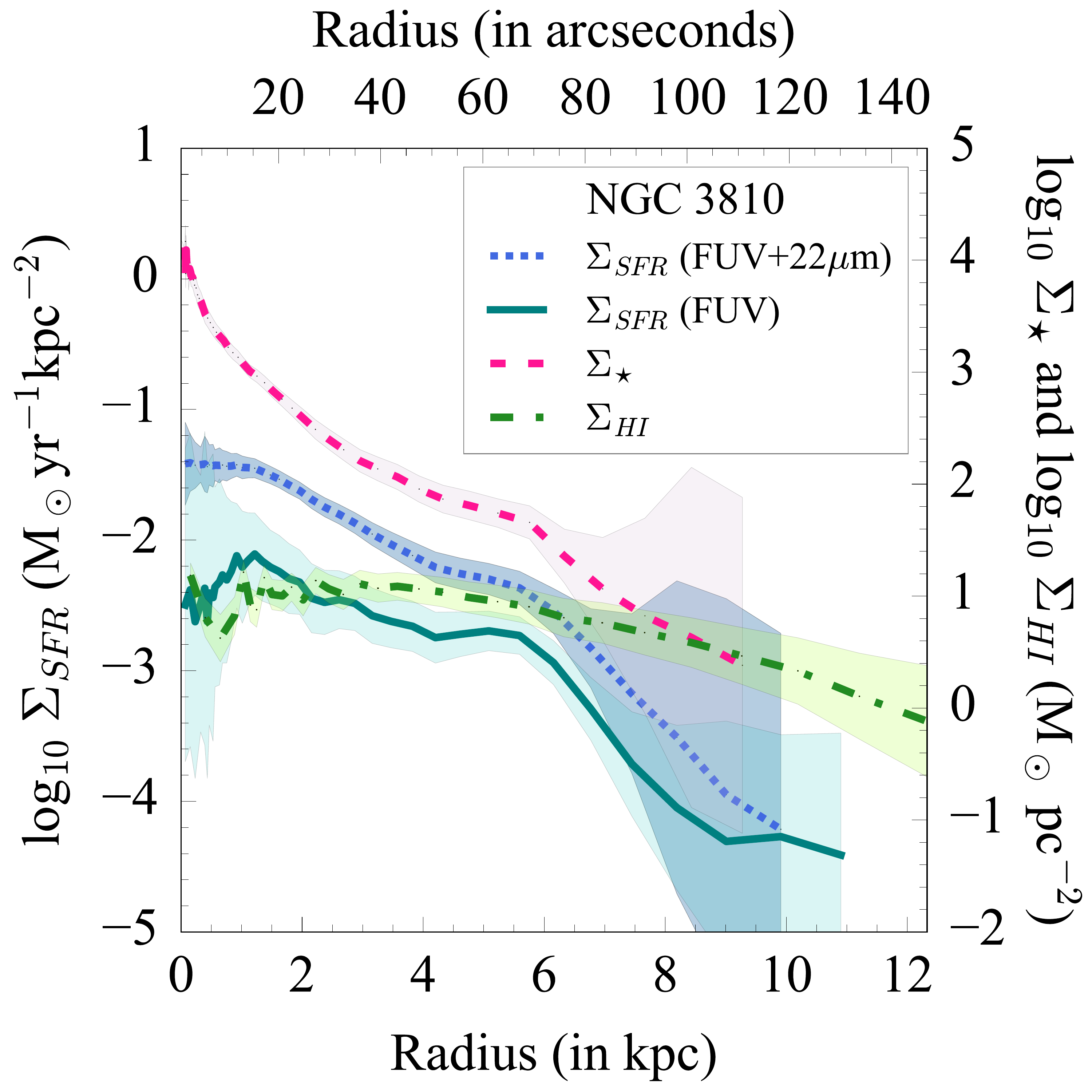}\hspace{1cm}
\includegraphics[trim = 0.cm 0.cm 0cm 0cm, clip,scale=0.15]{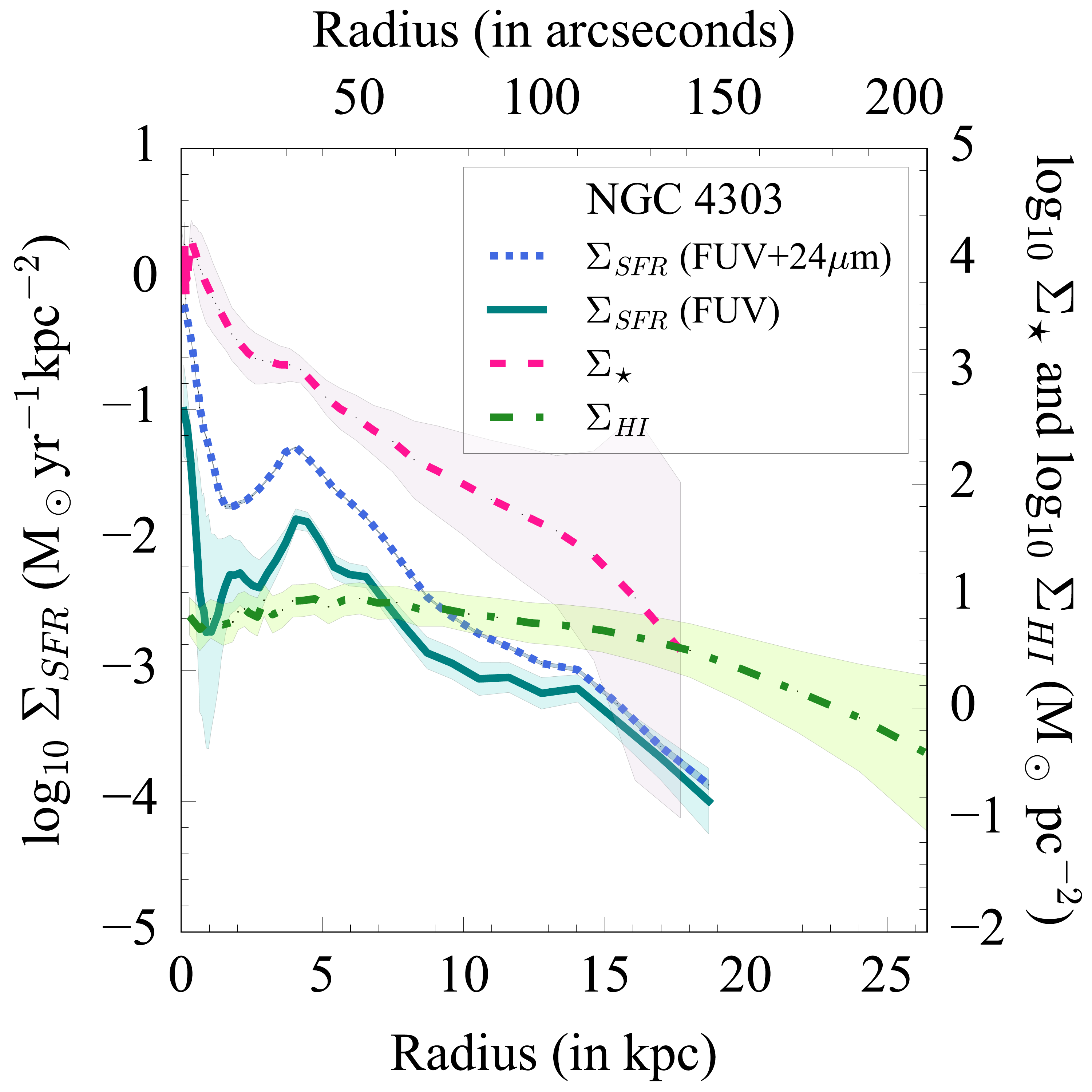}
\includegraphics[trim = 0.cm 0.cm 0cm 0cm, clip,scale=0.15]{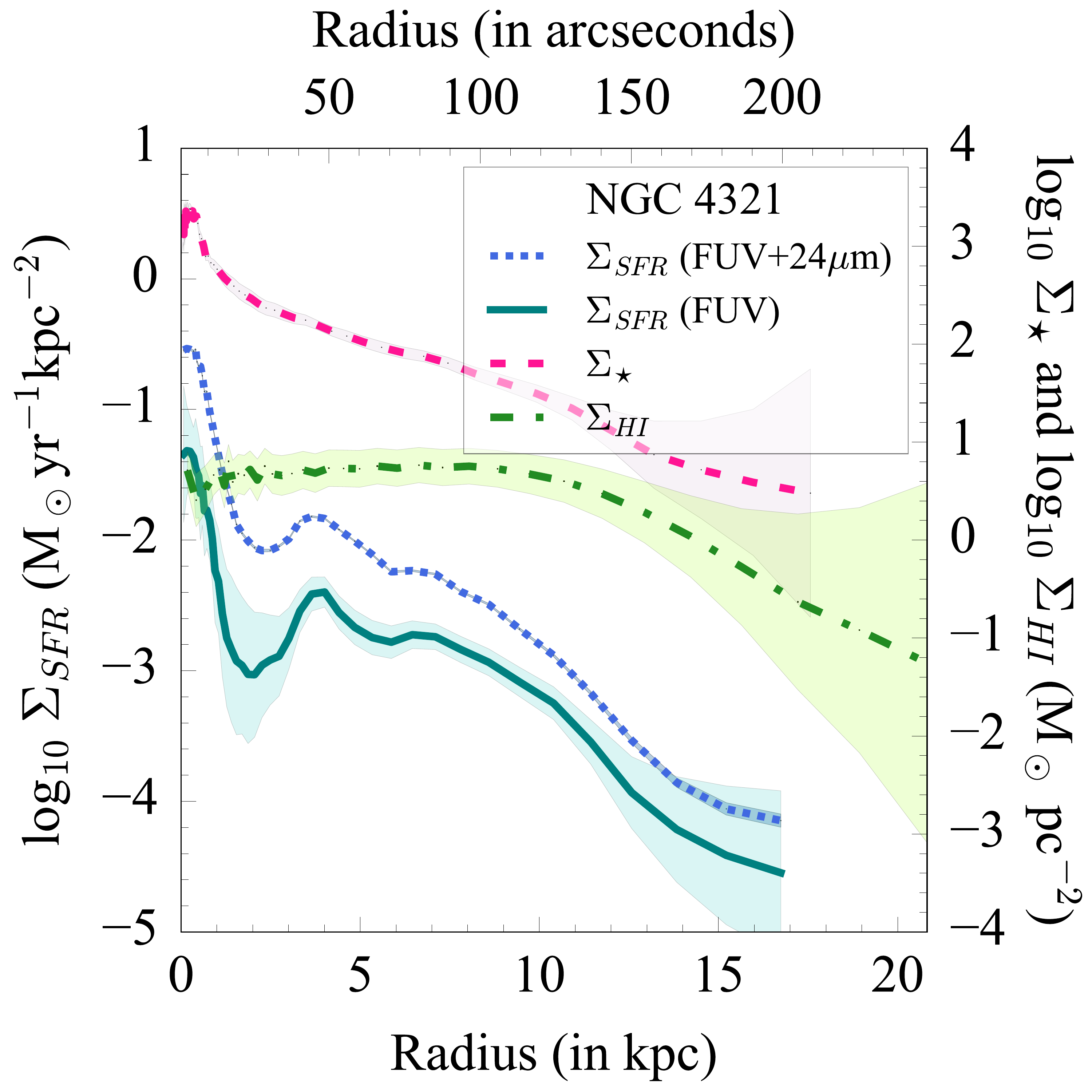}\hspace{1cm}
\includegraphics[trim = 0.cm 0.cm 0cm 0cm, clip,scale=0.15]{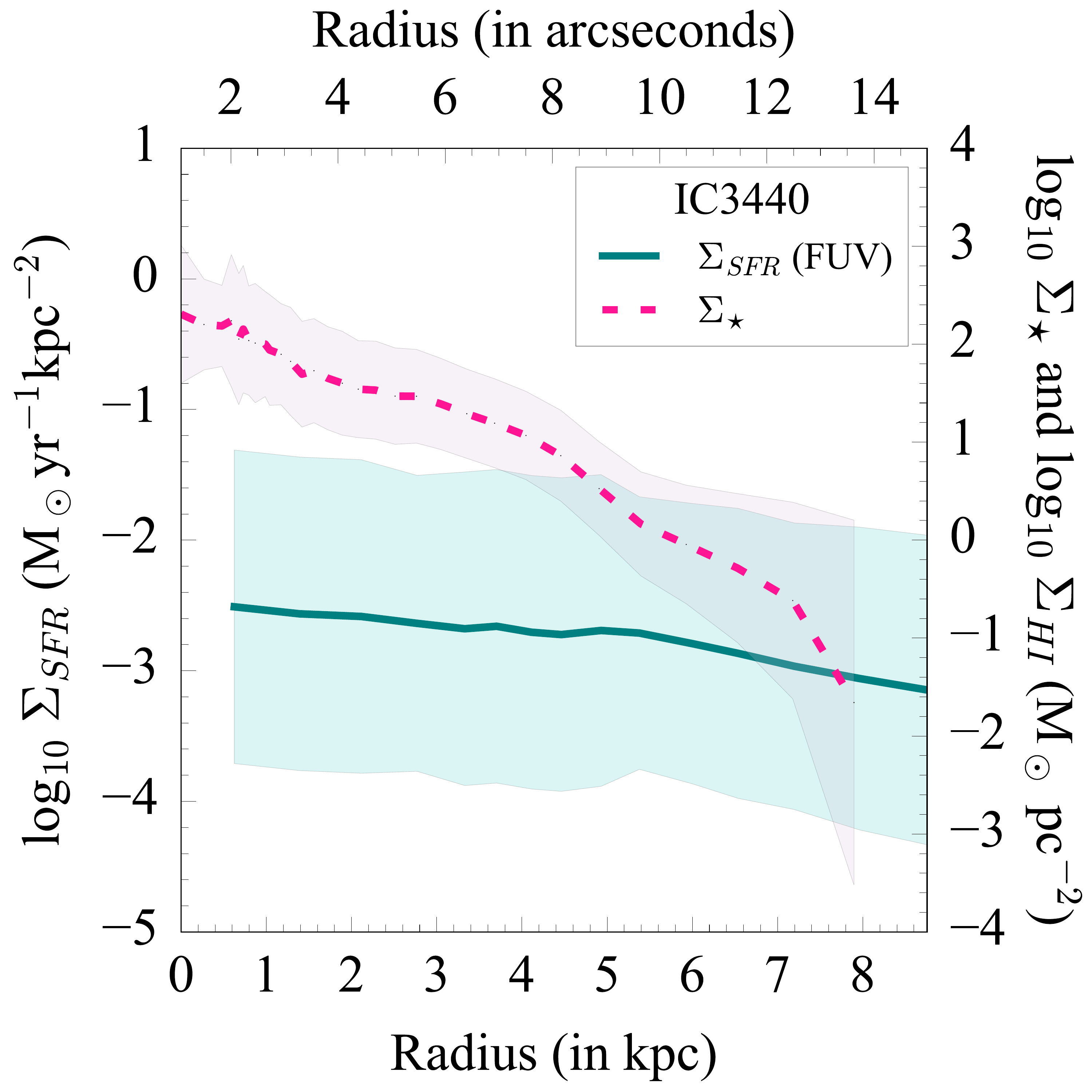}
\caption{Profiles of \sfr\ obtained from FUV and FUV+\mips\ (or FUV+\wise), \sm, and \mhi\ for the galaxies in our sample. The shaded region around each profile shows the 1$\sigma$ uncertainty in the respective quantities.}
\end{figure*}

\begin{figure*}[!ht] 
\setcounter{figure}{5}
\centering
\includegraphics[trim = 0.cm 0.cm 0cm 0cm, clip,scale=0.15]{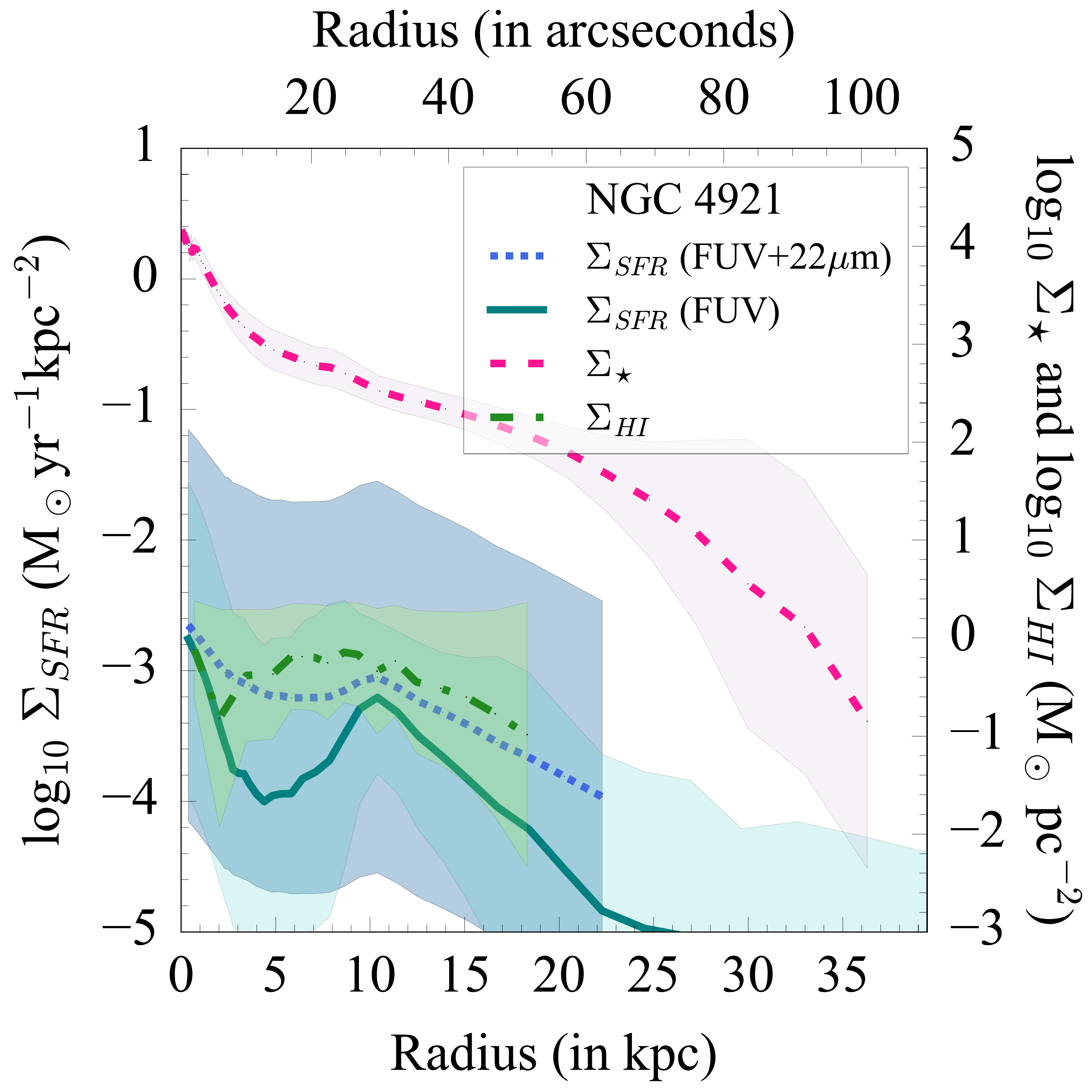}\hspace{1cm}
\includegraphics[trim = 0.cm 0.cm 0cm 0cm, clip,scale=0.15]{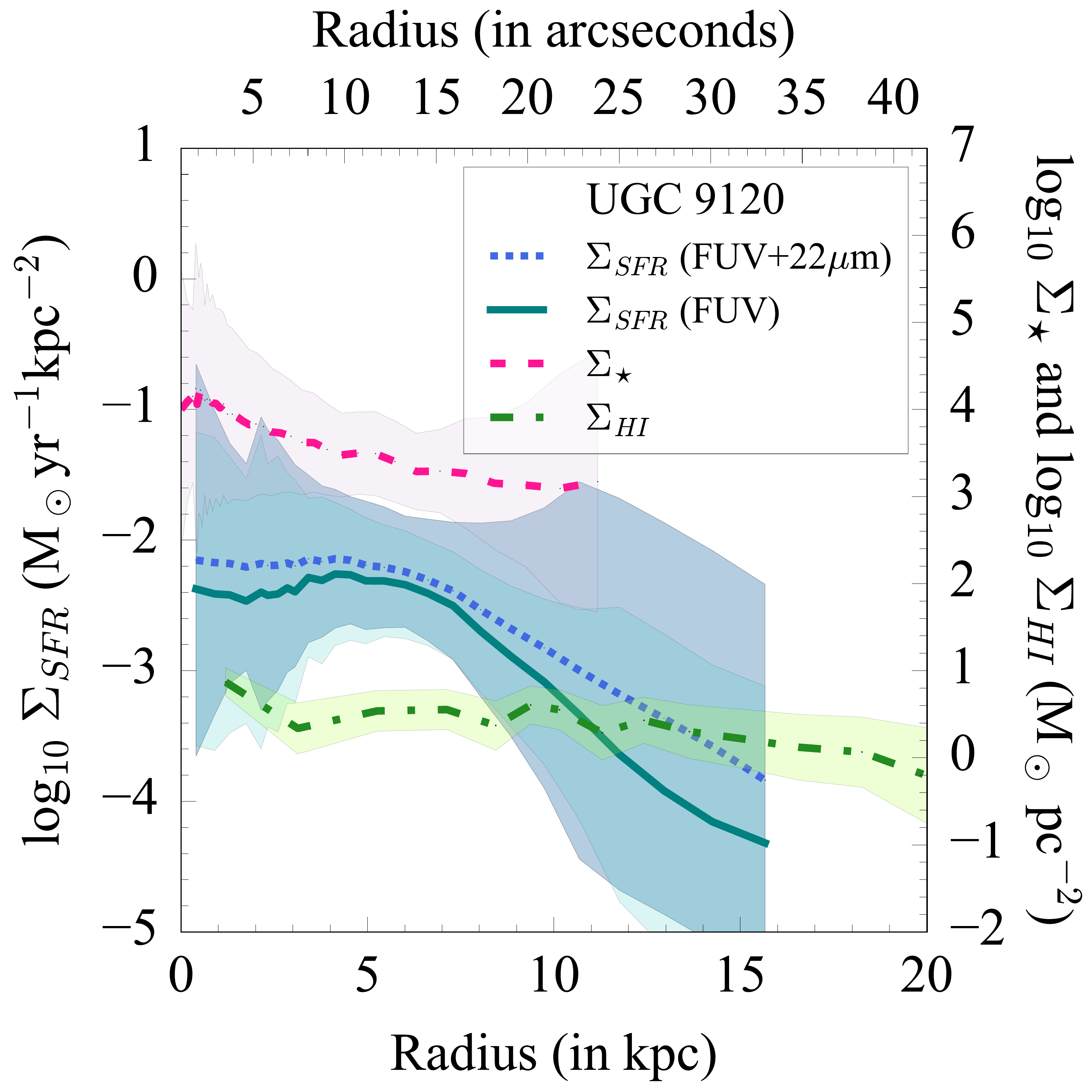}

\includegraphics[trim = 0.cm 0.cm 0cm 0cm, clip,scale=0.15]
{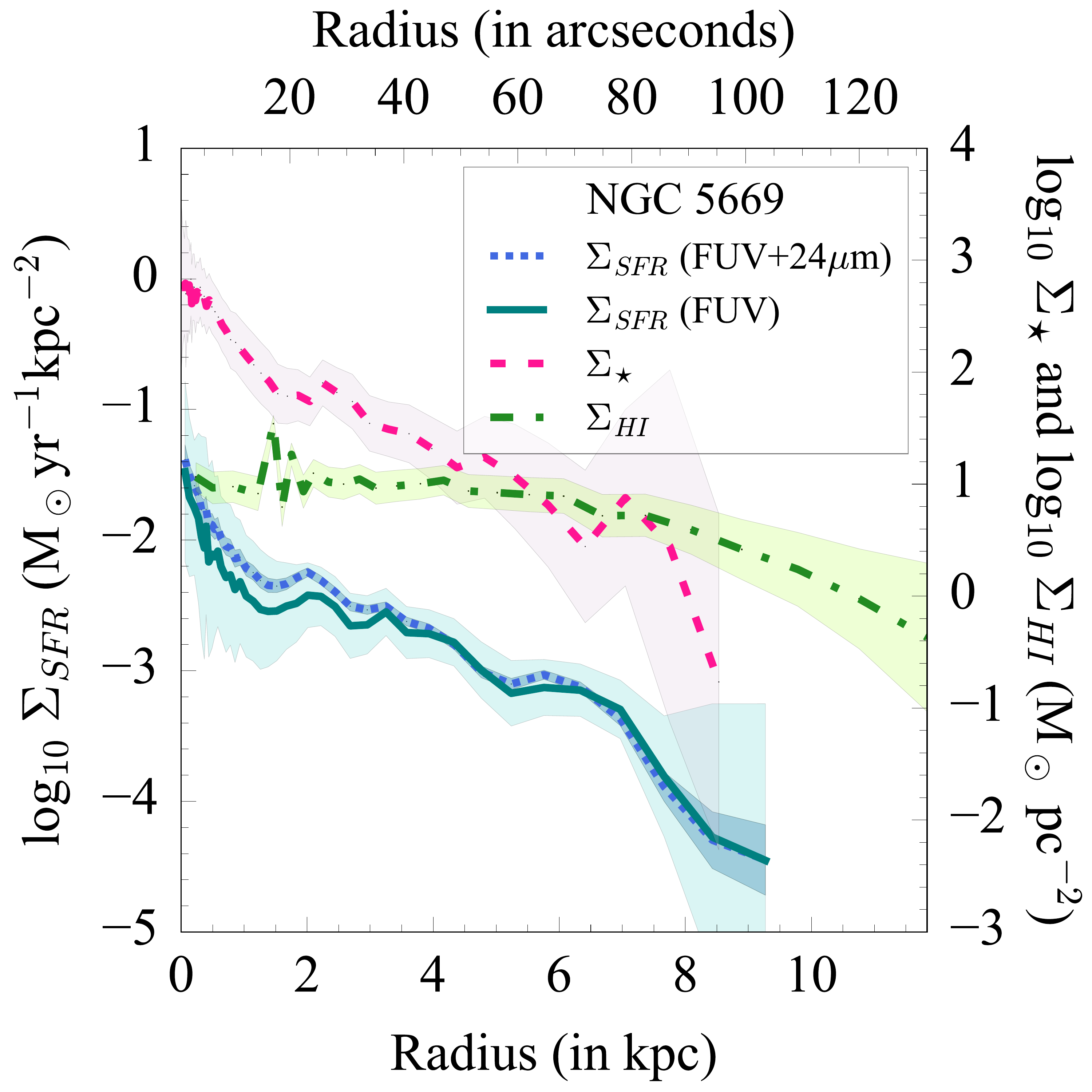}\hspace{1cm}
\includegraphics[trim = 0.cm 0.cm 0cm 0cm, clip,scale=0.15]{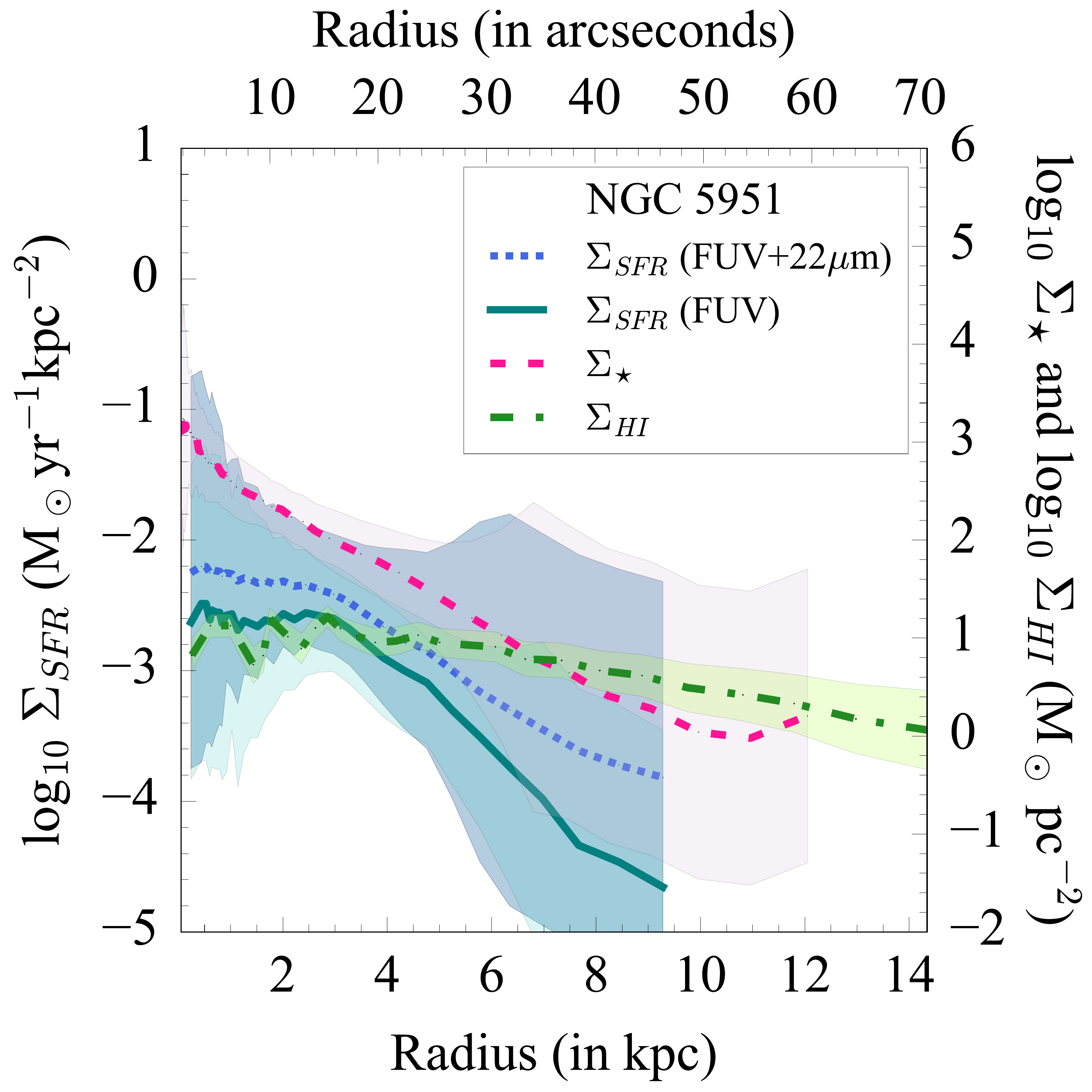}
\caption{Profiles of \sfr\ obtained from FUV and FUV+\mips\ (or FUV+\wise), \sm, and \mhi\ for the galaxies in our sample. The shaded region around each profile shows the 1$\sigma$ uncertainty in the respective quantities.}
\end{figure*}

\clearpage
\newpage
\section{Profiles of specific Star Formation rates and HI Star Formation Efficiency}\label{app:fig7}

\begin{figure*}[!ht] 
\setcounter{figure}{6}
\centering
\includegraphics[trim = 0.cm 0.cm 0cm 0cm, clip,scale=0.15]{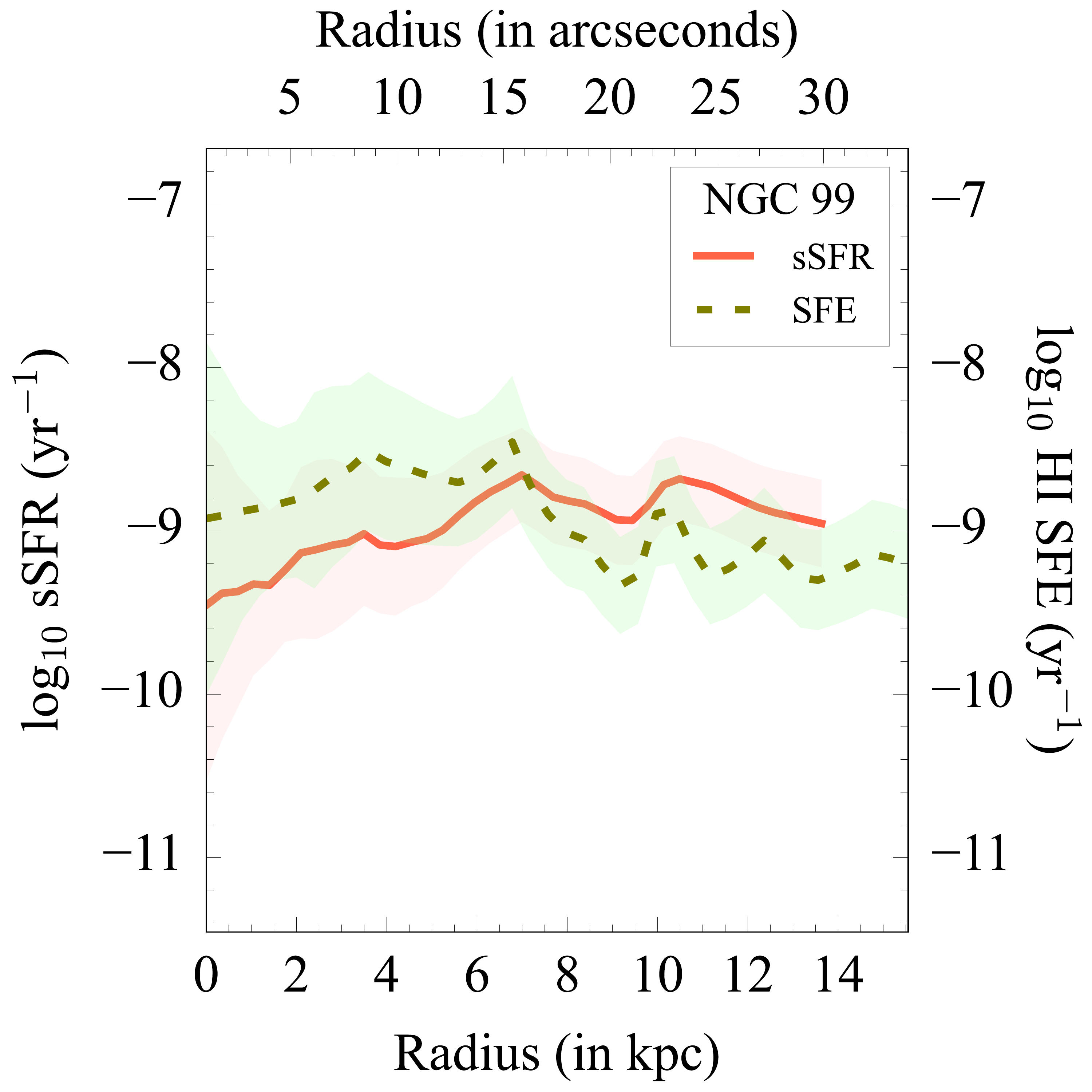}\hspace{1cm}
\includegraphics[trim = 0.cm 0.cm 0cm 0cm, clip,scale=0.15]{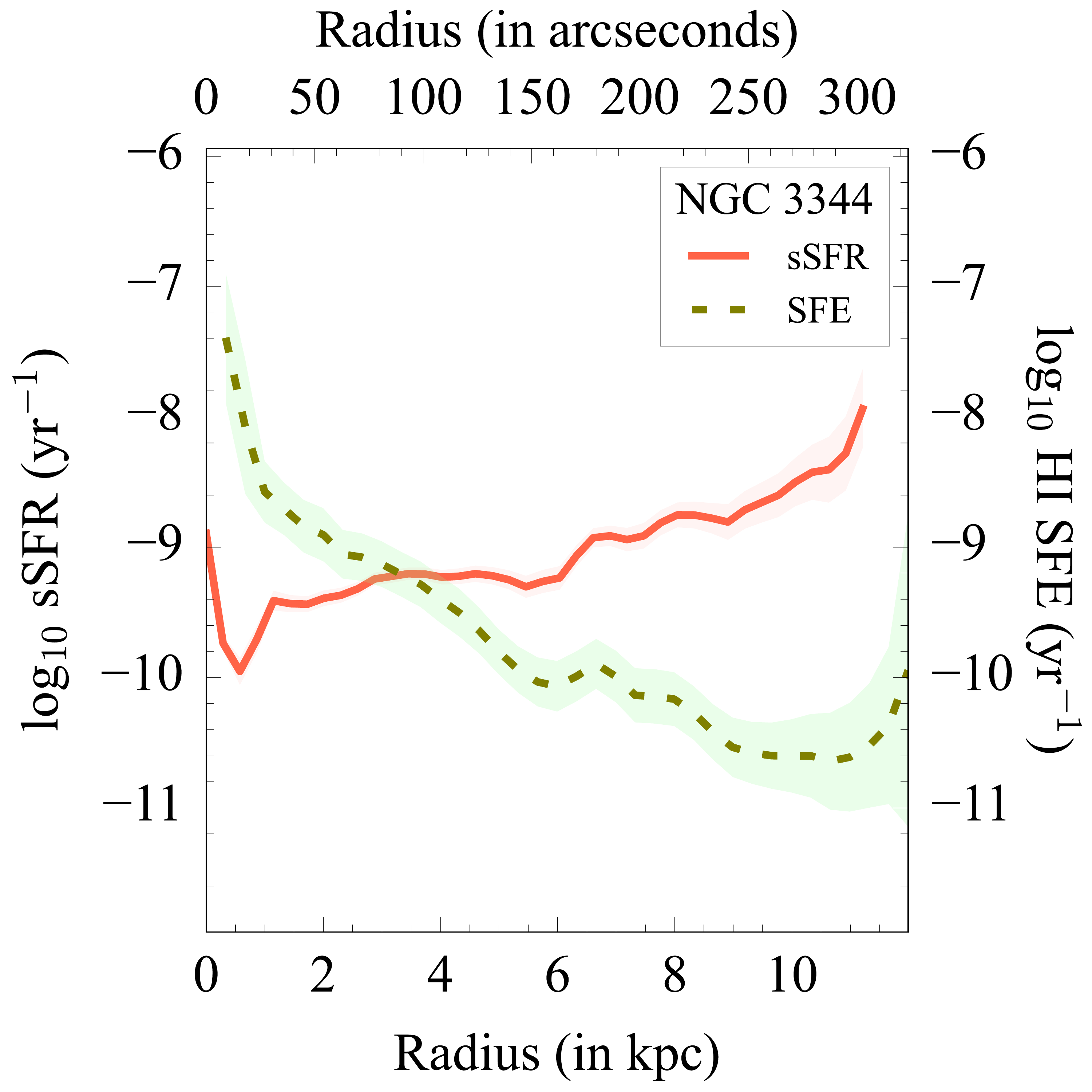}
\includegraphics[trim = 0.cm 0.cm 0cm 0cm, clip,scale=0.15]{f7_3.pdf}\hspace{1cm}
\includegraphics[trim = 0.cm 0.cm 0cm 0cm, clip,scale=0.15]{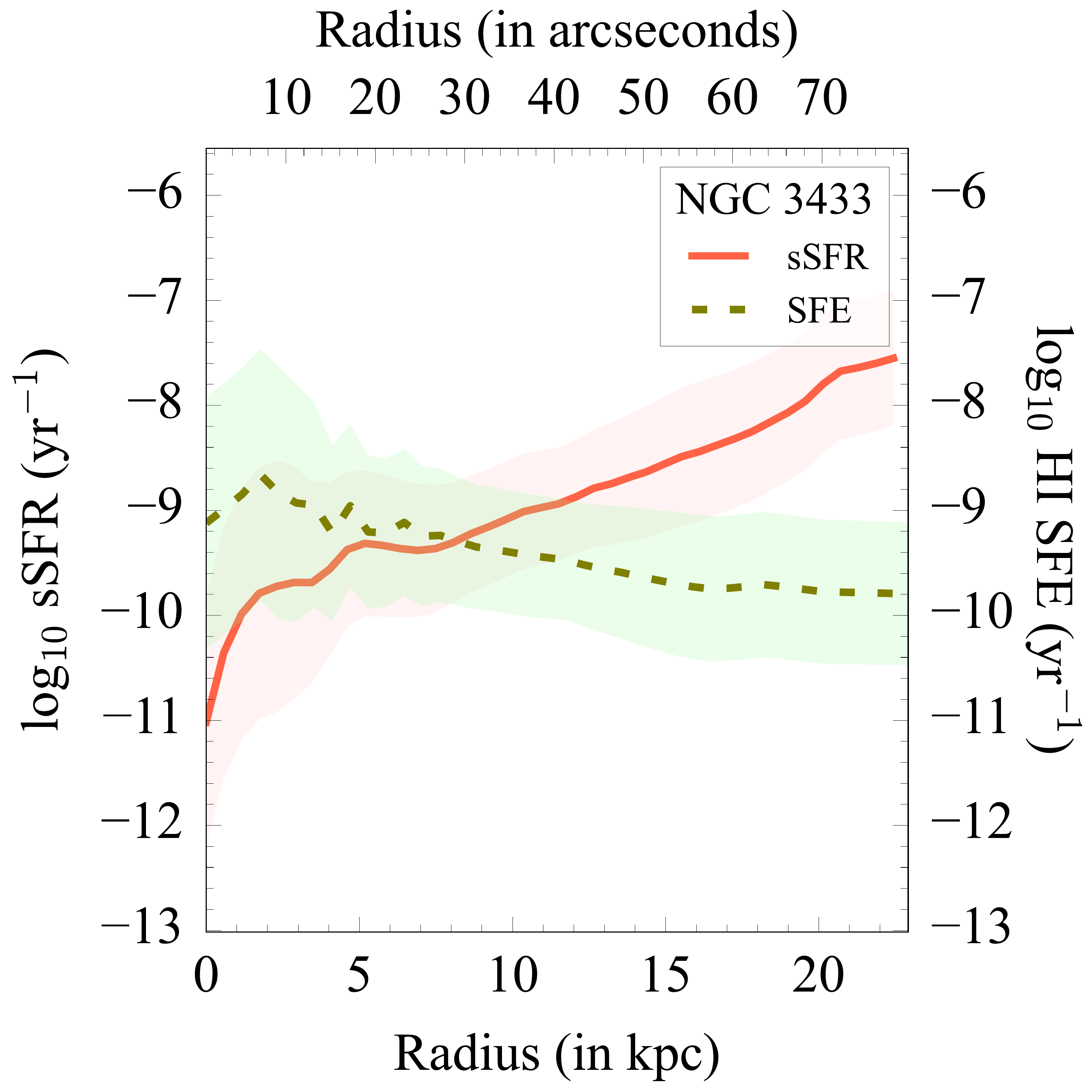}

\caption{Radial variation in specific SFR and \hi\ star formation efficiency for the galaxies in our sample. The shaded region around each profile shows the 1$\sigma$ uncertainty in the respective quantities.}
\end{figure*}

\begin{figure*}[!t] 
\setcounter{figure}{6}
\centering
\includegraphics[trim = 0.cm 0.cm 0cm 0cm, clip,scale=0.15]{f7_5.pdf}\hspace{1cm}
\includegraphics[trim = 0.cm 0.cm 0cm 0cm, clip,scale=0.15]{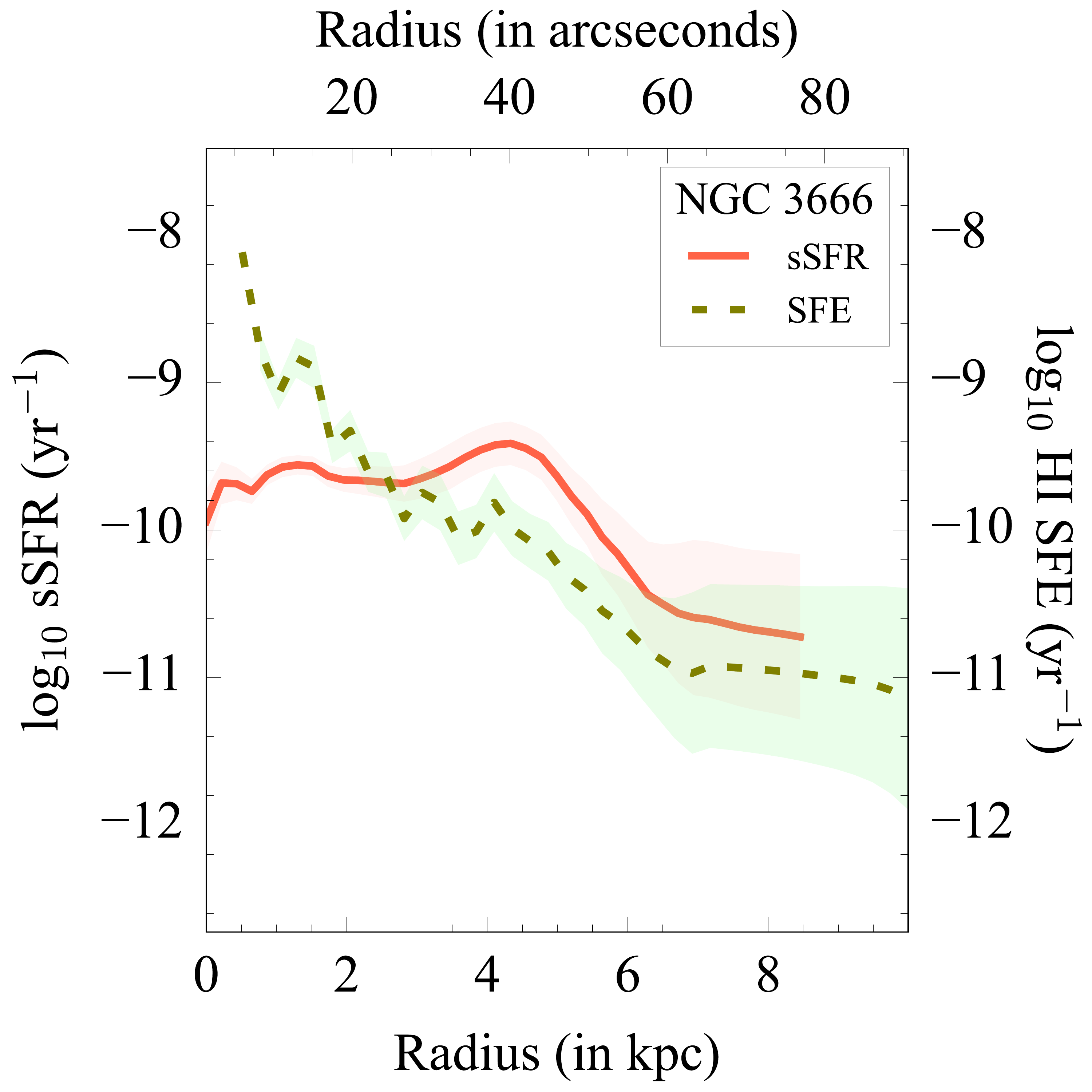}
\includegraphics[trim = 0.cm 0.cm 0cm 0cm, clip,scale=0.15]{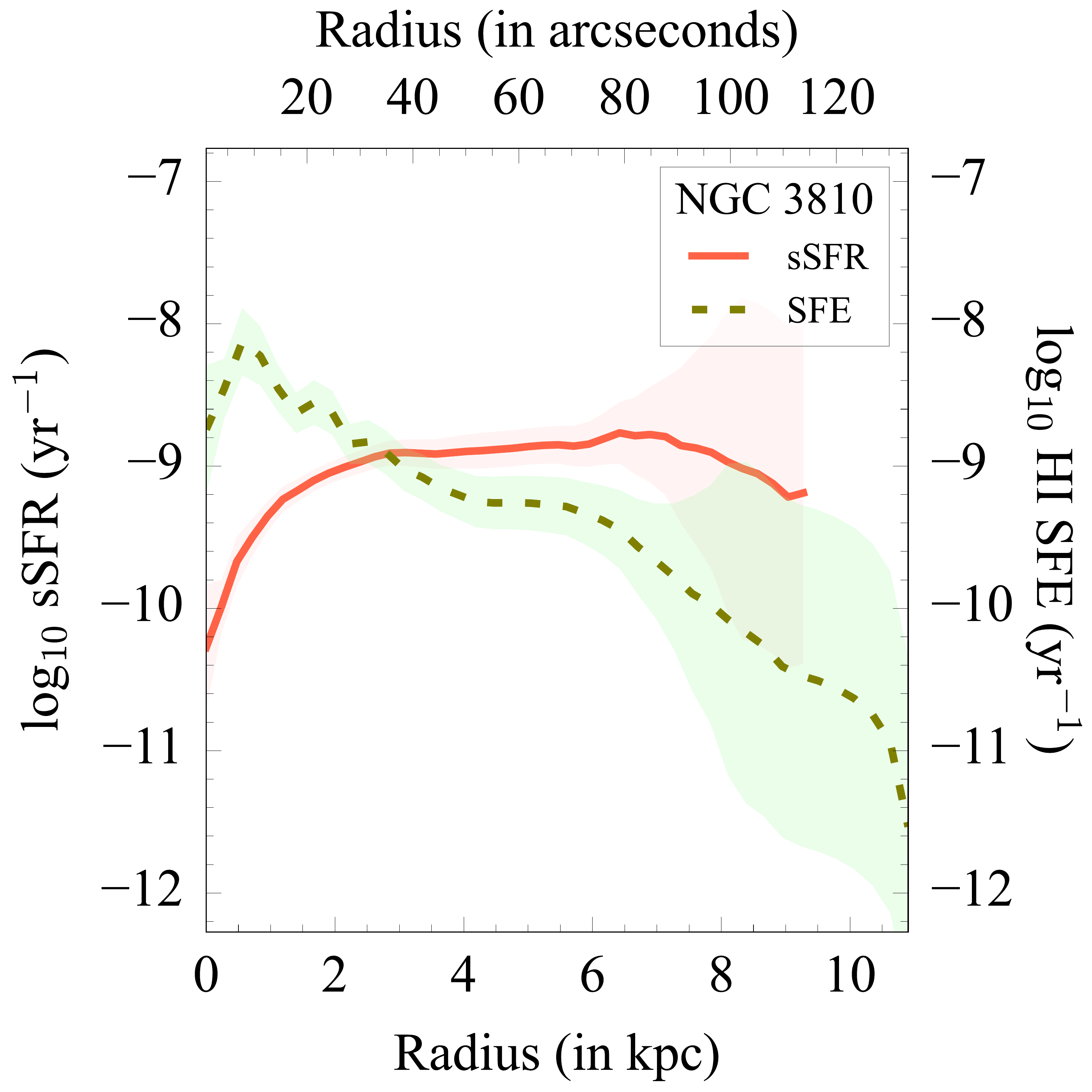}\hspace{1cm}
\includegraphics[trim = 0.cm 0.cm 0cm 0cm, clip,scale=0.15]{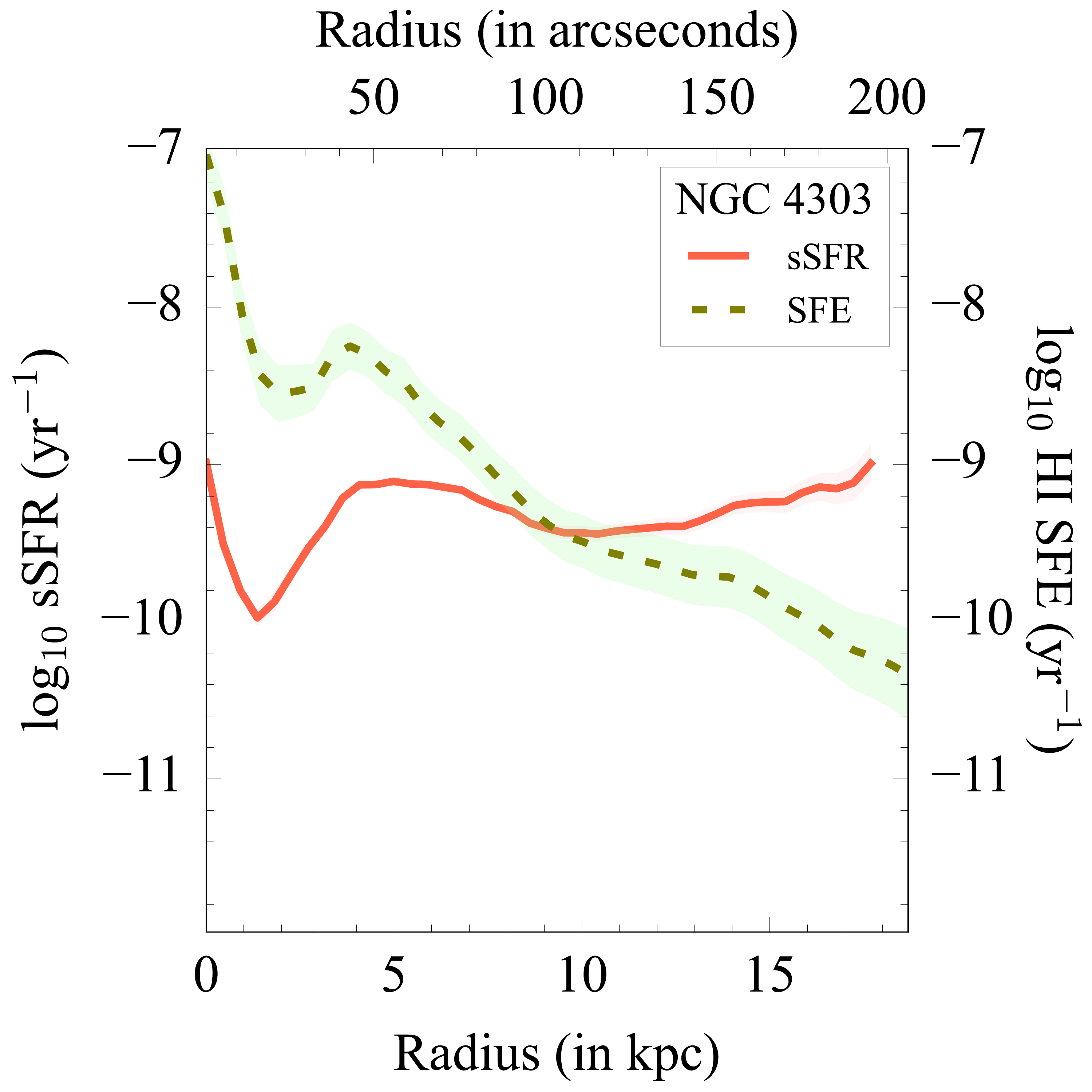}

\caption{Radial variation in specific SFR and \hi\ star formation efficiency for the galaxies in our sample. The shaded region around each profile shows the 1$\sigma$ uncertainty in the respective quantities.}
\end{figure*}

\begin{figure*}[!ht] 
\setcounter{figure}{6}
\centering
\includegraphics[trim = 0.cm 0.cm 0cm 0cm, clip,scale=0.15]{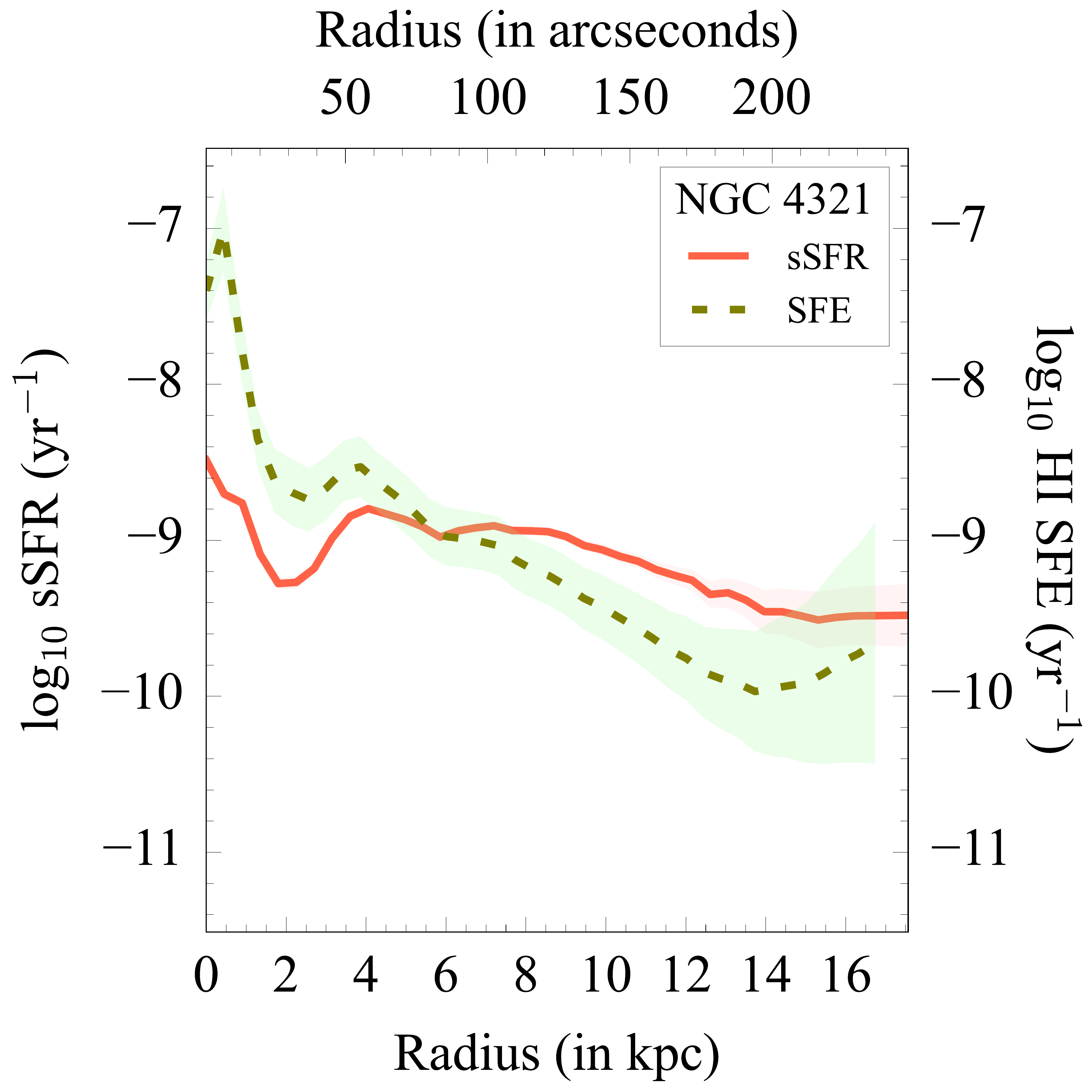}\hspace{1cm}
\includegraphics[trim = 0.cm 0.cm 0cm 0cm, clip,scale=0.15]{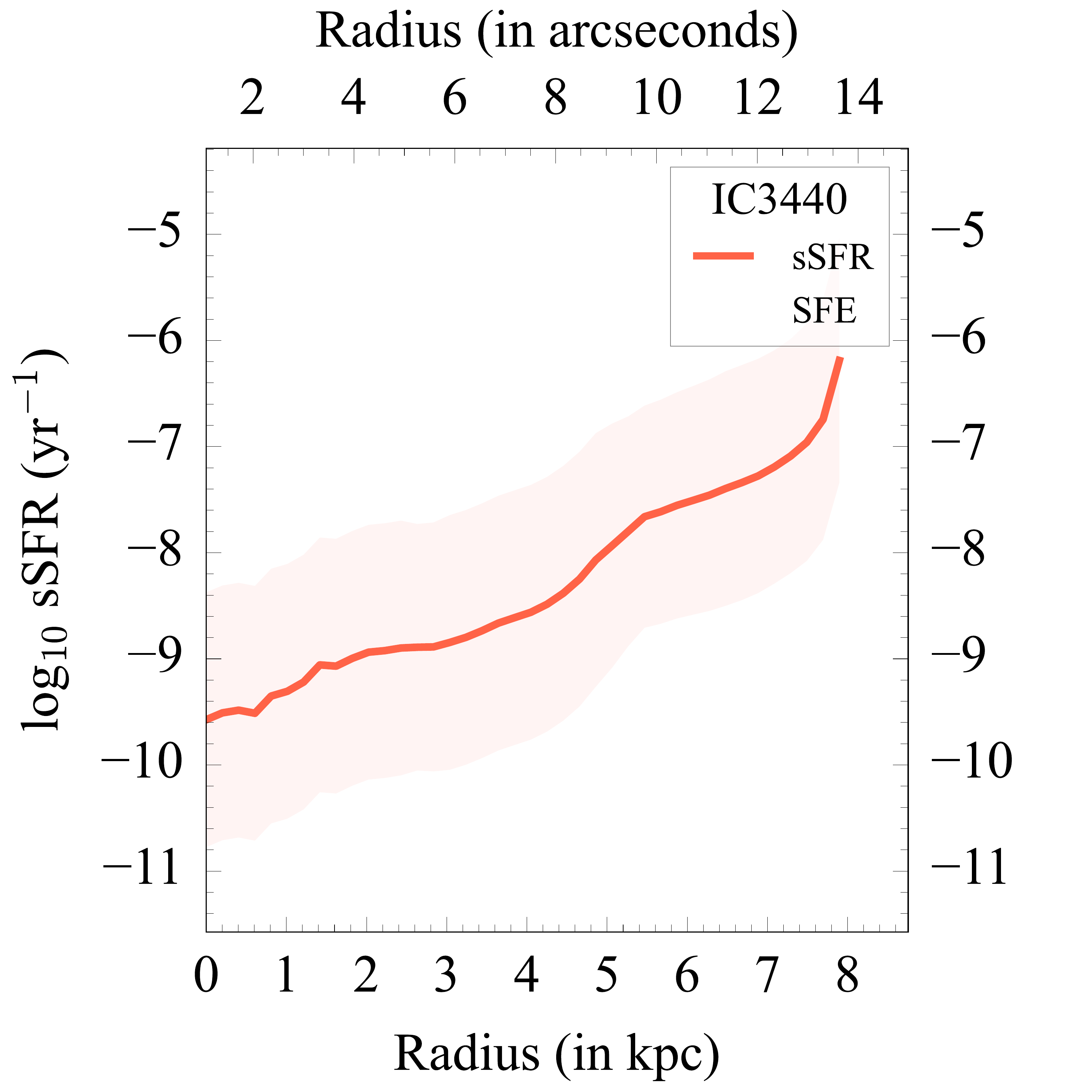}
\includegraphics[trim = 0.cm 0.cm 0cm 0cm, clip,scale=0.15]{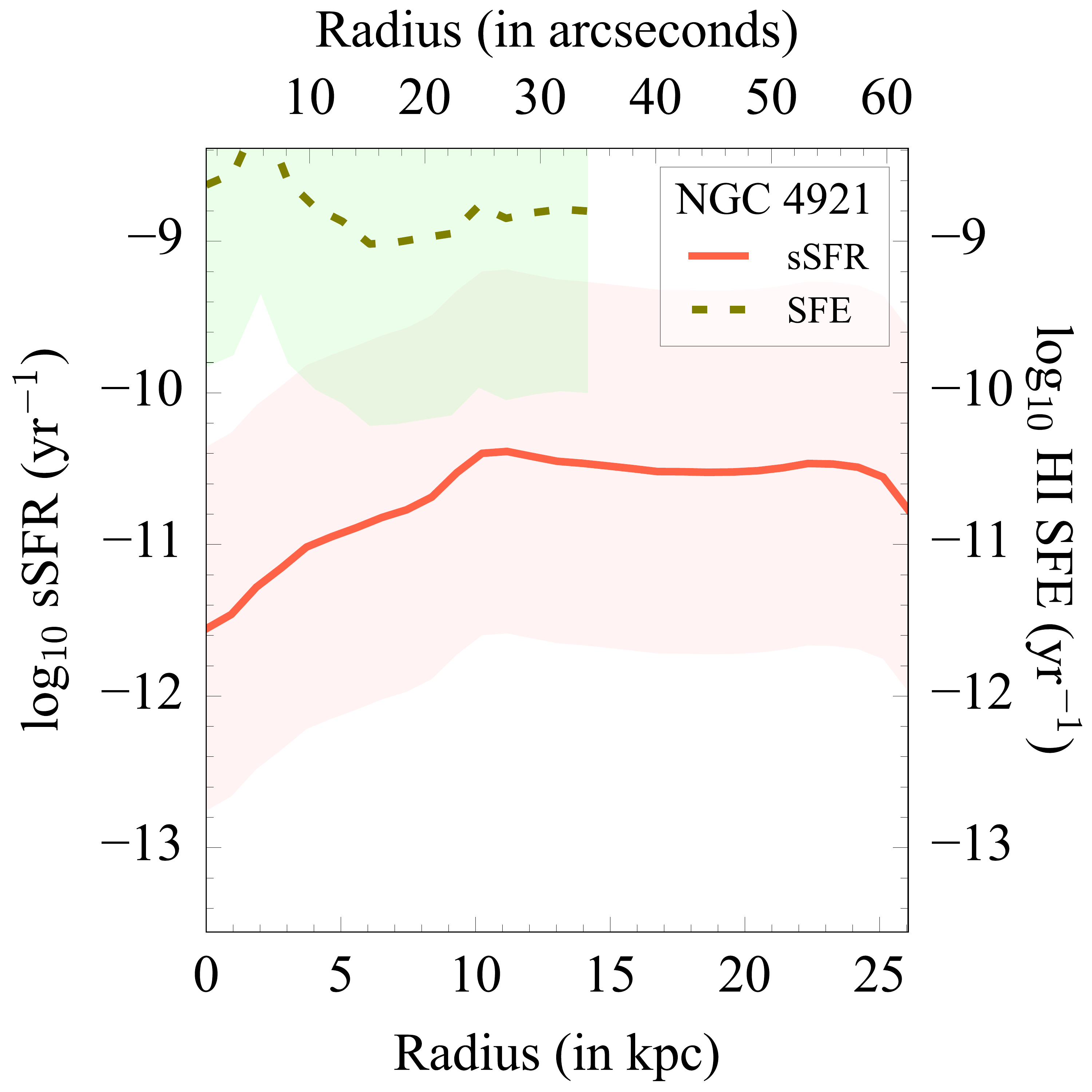}\hspace{1cm}
\includegraphics[trim = 0.cm 0.cm 0cm 0cm, clip,scale=0.15]{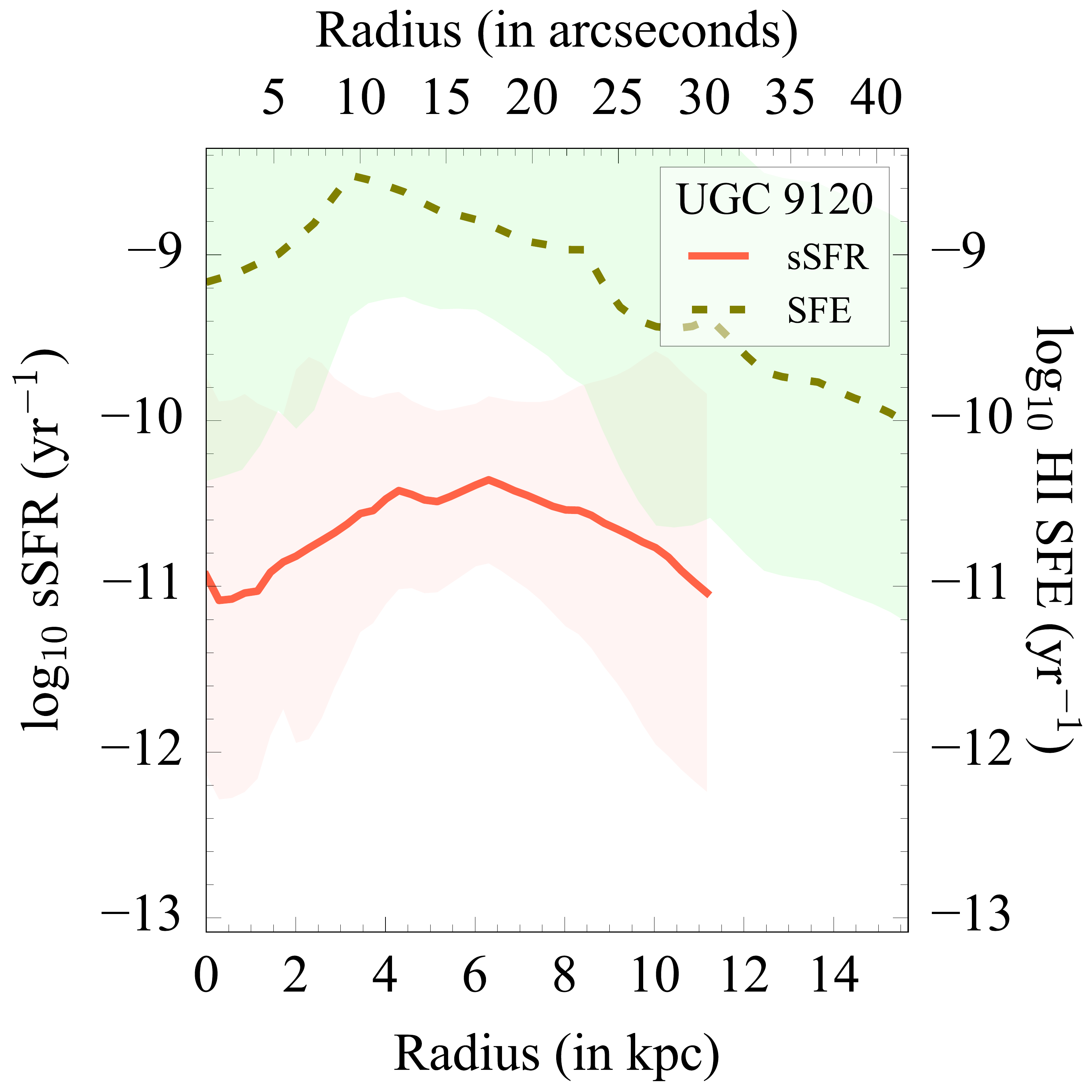}
\caption{Radial variation in specific SFR and \hi\ star formation efficiency for the galaxies in our sample. The shaded region around each profile shows the 1$\sigma$ uncertainty in the respective quantities.}
\end{figure*}

\begin{figure*}[!ht] 
\setcounter{figure}{6}
\centering
\includegraphics[trim = 0.cm 0.cm 0cm 0cm, clip,scale=0.15]{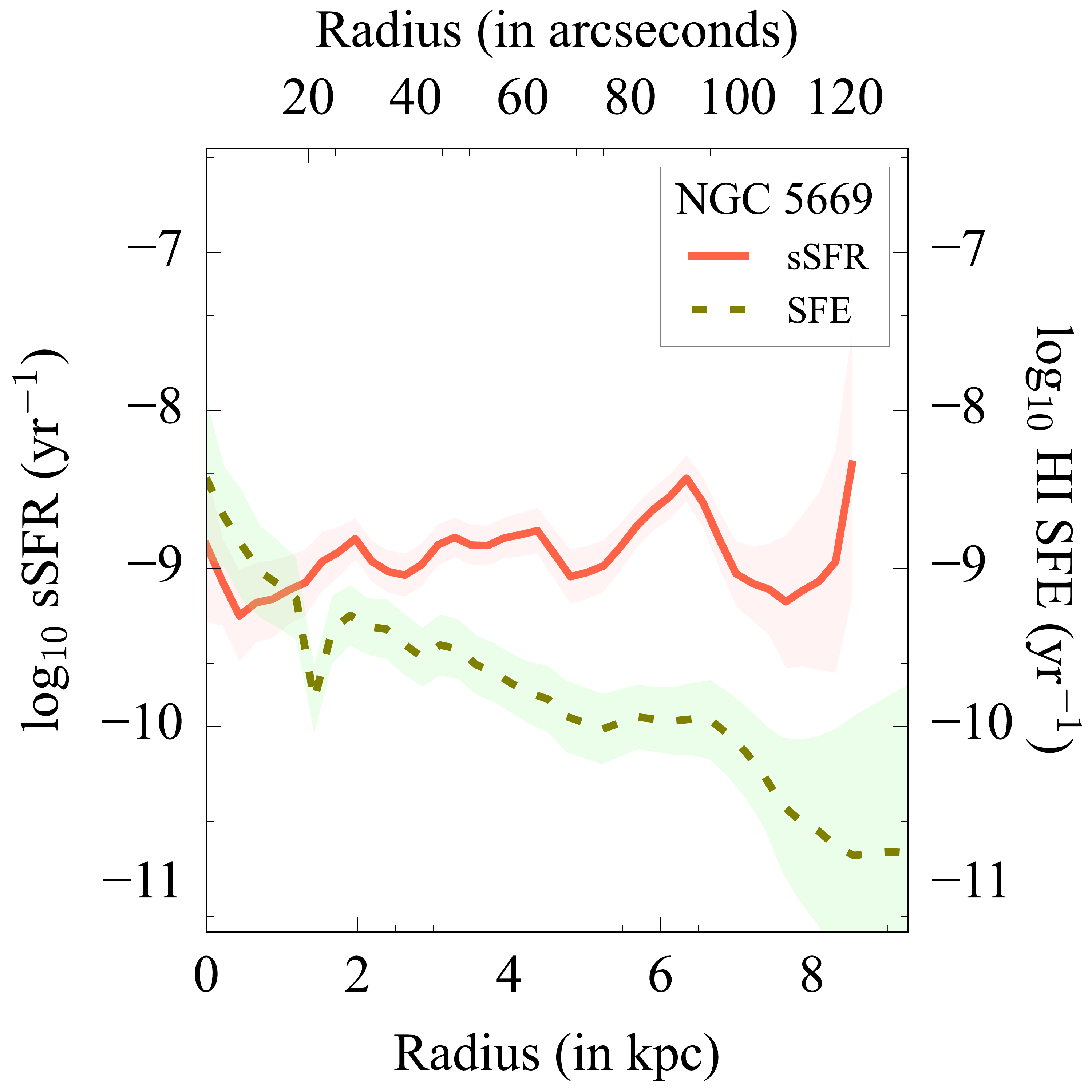}\hspace{1cm}
\includegraphics[trim = 0.cm 0.cm 0cm 0cm, clip,scale=0.15]{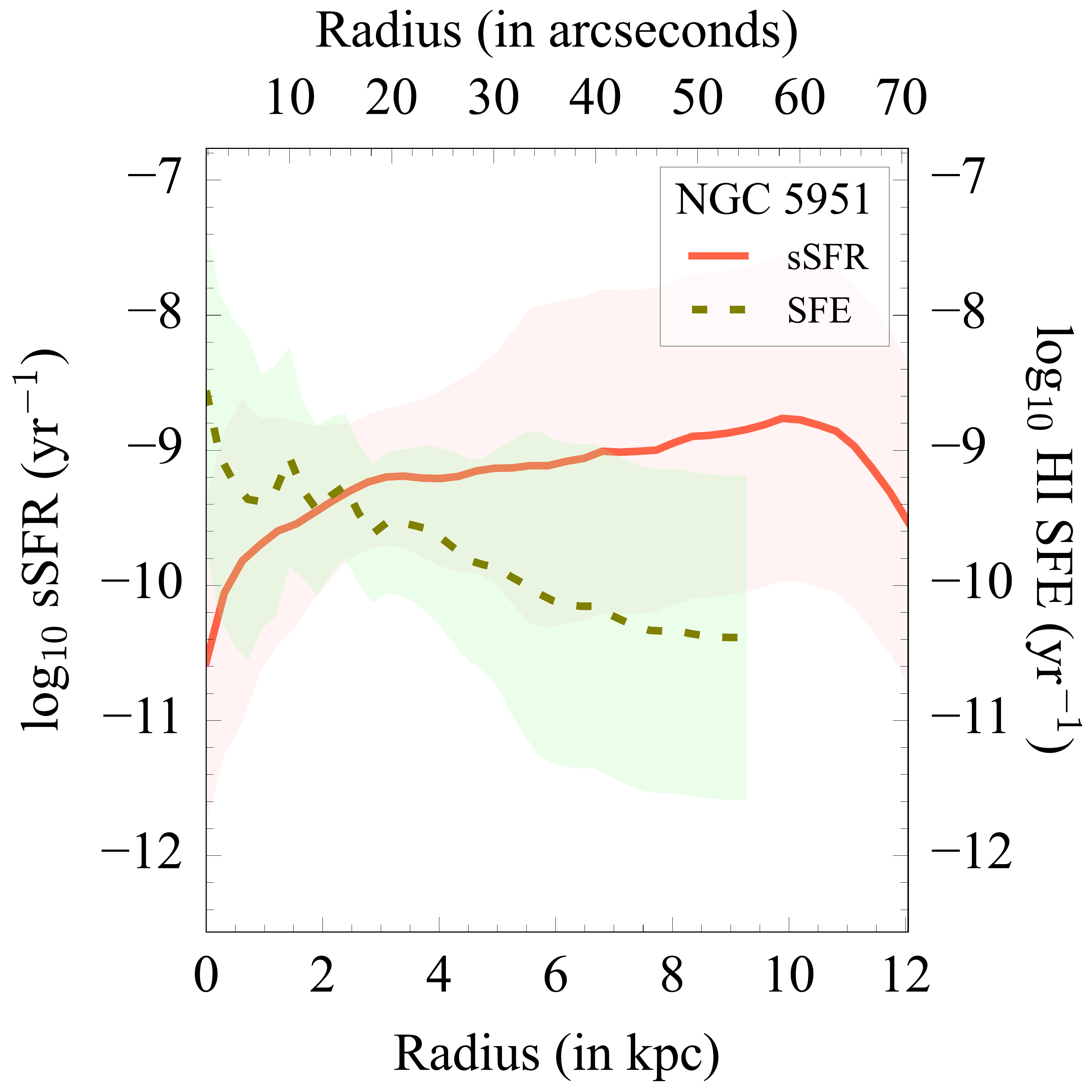}
\caption{Radial variation in specific SFR and \hi\ star formation efficiency for the galaxies in our sample. The shaded region around each profile shows the 1$\sigma$ uncertainty in the respective quantities.}
\end{figure*}




\end{document}